\documentclass[
12pt, 
english, 
singlespacing, 
headsepline, 
]{MastersDoctoralThesis} 

\usepackage[utf8]{inputenc} 
\usepackage[T1]{fontenc} 
\usepackage{tikz}
\usepackage{array}
\usepackage{xfrac}
\usepackage{xcolor}
\usepackage{amsthm}
\usepackage{sidecap}
\usepackage{amsmath}
\usepackage{tikz-cd}
\usepackage{wrapfig}
\usepackage{amssymb}
\usepackage{lmodern}
\usepackage{epigraph}
\usepackage{makecell}
\usepackage{mathrsfs}
\usepackage{amsfonts}
\usepackage{footmisc}
\usepackage{booktabs}
\usepackage{enumitem}
\usepackage{multirow}
\usepackage{multicol}
\usepackage{graphicx}
\usepackage{mathtools}
\usepackage{algorithm}
\usepackage{subcaption}
\usepackage{algorithmicx}
\usepackage[normalem]{ulem}
\usepackage[small]{titlesec}
\usepackage[noend]{algpseudocode}
\usepackage{pstricks,pst-node,pst-text,pst-3d}

\DeclareUnicodeCharacter{1EF3}{\`y}
\DeclareUnicodeCharacter{0301}{\' }
\DeclareUnicodeCharacter{02BC}{'}
\usetikzlibrary{patterns, patterns.meta}

\makeatletter
\newenvironment{breakablealgorithm}
{ 
  \begin{center}
     \refstepcounter{algorithm}
     \rule{\linewidth}{1pt} 
     \renewcommand{\caption}[2][\relax]{
       {\raggedright\textbf{\ALG@name~\thealgorithm} ##2\par}%
       \ifx\relax##1\relax 
         \addcontentsline{loa}{algorithm}{\protect\numberline{\thealgorithm}##2}%
       \else 
         \addcontentsline{loa}{algorithm}{\protect\numberline{\thealgorithm}##1}%
       \fi
       \vspace{-9pt}
      \rule{\linewidth}{0.5pt} 
      \vspace{-15pt}
     }
  }{
     \vspace{-9pt}
     \rule{\linewidth}{0.5pt} 
   \end{center}
 }
\makeatother

\renewcommand{\Return}{\State \textbf{return} }

\newcommand{\ssq}{
	\vcenter{\hbox{\scalebox{0.65}{$\;\mathbin{ \square }\;$}}}
}

\usepackage[bibstyle=ieee, citestyle=numeric, backend=biber, sortcites=false, sorting=nyt]{biblatex} 
\DeclareNameAlias{default}{family-given}

\usepackage[autostyle=true]{csquotes} 

\newcommand{\bbA}{\mathbb{A}}
\newcommand{\bbB}{\mathbb{B}}
\newcommand{\bbC}{\mathbb{C}}
\newcommand{\bbH}{\mathbb{H}}

\newcommand{\bbN}{\mathbb{N}}
\newcommand{\bbR}{\mathbb{R}}
\newcommand{\bbU}{\mathbb{U}}
\newcommand{\bbX}{\mathbb{X}}

\newcommand{\bfone}{\mathbf{1}}

\newcommand{\calA}{\mathcal{A}}
\newcommand{\calB}{\mathcal{B}}
\newcommand{\calC}{\mathcal{C}}
\newcommand{\calD}{\mathcal{D}}
\newcommand{\calE}{\mathcal{E}}
\newcommand{\calF}{\mathcal{F}}
\newcommand{\calH}{\mathcal{H}}
\newcommand{\calI}{\mathcal{I}}
\newcommand{\calK}{\mathcal{K}}
\newcommand{\calL}{\mathcal{L}}
\newcommand{\calM}{\mathcal{M}}
\newcommand{\calN}{\mathcal{N}}
\newcommand{\calP}{\mathcal{P}}
\newcommand{\calR}{\mathcal{R}}
\newcommand{\calS}{\mathcal{Q}}
\newcommand{\calT}{\mathcal{T}}
\newcommand{\calU}{\mathcal{U}}
\newcommand{\calV}{\mathcal{V}}
\newcommand{\calVE}{\mathcal{VE}}

\newcommand{\frakS}{\mathfrak{S}}

\newcommand{\conv}{\text{conv}}
\newcommand{\domS}{\ \text{dom}_S \ }
\newcommand{\dom}{\ \text{dom} \ }
\newcommand{\rk}{\text{rk}}
\newcommand{\slab}{\text{sl}}

\newtheorem{proposition}{Proposition}[section]
\newtheorem*{proposition*}{Proposition}
\newtheorem{theorem}[proposition]{Theorem}
\newtheorem*{minimax-theorem}{The Minimax Theorem}
\newtheorem*{minimax-theoreme}{Le théorème du Minimax}
\newtheorem*{BS-theorem}{The Bondareva-Shapley Theorem}
\newtheorem*{BS-theoreme}{Le théorème de Bondareva-Shapley}

\newtheorem{lemma}[proposition]{Lemma}

\theoremstyle{definition}
\newtheorem{example}[proposition]{Example}
\newtheorem{definition}[proposition]{Definition}
\newtheorem*{definition*}{Definition}

\newtheorem{remark}[proposition]{Remark}
\newtheorem*{remark*}{Remark}

\AtBeginEnvironment{example}{%
  \pushQED{\qed}%
}
\AtEndEnvironment{example}{\popQED\endexample}

\setlist{topsep=8pt, itemsep=1pt}

\thesistitle{Geometry of set functions in game theory: combinatorial and computational aspects} 
\supervisor{Michel \textsc{Grabisch} \\ Peter \textsc{Sudh{\"o}lter}} 
\examiner{} 
\degree{\textsc{Docteur de l'Universit{\'e} Paris I Panth{\'e}on-Sorbonne}} 
\author{Dylan \textsc{Laplace Mermoud}} 
\addresses{} 

\subject{Math{\'e}matiques appliqu{\'e}es} 
\university{Université Paris 1 Panthéon-Sorbonne} 
\department{Université Paris 1 Panthéon-Sorbonne} 
\group{Centre d'{\'E}conomie de la Sorbonne} 
\faculty{Centre d'{\'E}conomie de la Sorbonne} 

\AtBeginDocument{
\hypersetup{pdftitle=\ttitle} 
\hypersetup{pdfauthor=\authorname} 
\hypersetup{pdfkeywords=\keywordnames} 
}

\begin{document}

\numberwithin{equation}{section}

\frontmatter 

\pagestyle{plain} 


\begin{titlepage}
\begin{center}
\
\vspace{-2cm}
\begin{center}
\includegraphics[scale=0.35]{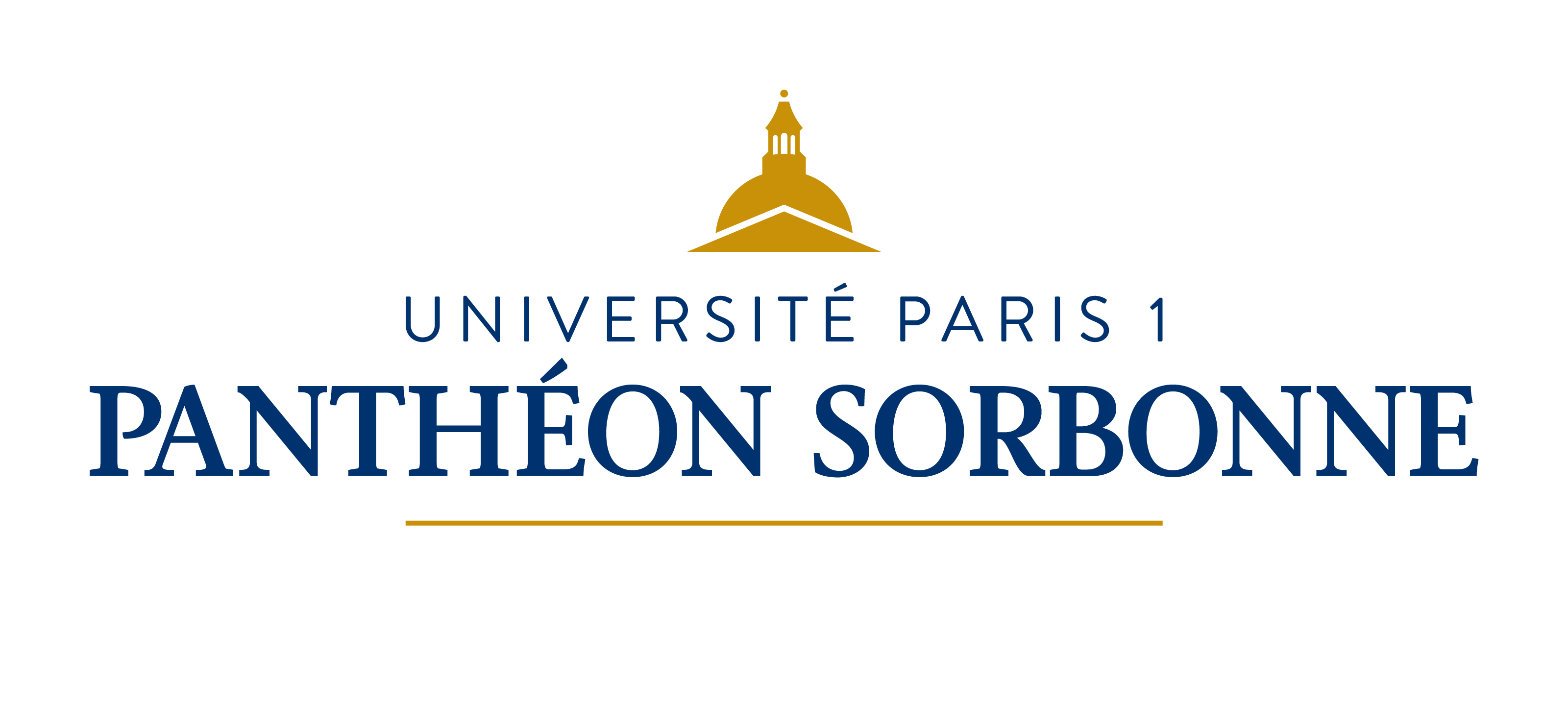}
\end{center}
\textsc{\large Th{\`e}se de doctorat}\\[1.3cm] 
\textit{présentée pour obtenir le grade de}\\[0.3cm]
{\large \degreename} \\[0.3cm]
{{\'E}cole doctorale ED-465} \\[0.3cm]
{Spécialité : \subjectname} \\[1.3cm]

\hrulefill \\[0.4cm] 
{\Large \textsc{\ttitle}\par}\vspace{0.4cm} 
\hrulefill \\[1.3cm] 

\textit{par} \\[0.3cm]
{\large \authorname} \\[2cm]
 
\textit{soutenue le 8 septembre 2023 devant le jury composé de} \\[0.3cm]

\begin{center}
\hrulefill \\[0.2cm]
\begin{tabular}{llr}
\textsc{Philippe Bich} & Universit{\'e} Paris I Panth{\'e}on-Sorbonne \quad & Président \\[0.1cm]
\textsc{Michel Grabisch} & Universit{\'e} Paris I Panth{\'e}on-Sorbonne & Directeur \\[0.1cm]
\textsc{Jean-Jacques Herings} \quad & Tilburg University & Examinateur \\[0.1cm]
\textsc{Marina Nu{\~n}ez} & Universitat de Barcelona & Rapportrice \\[0.1cm]
\textsc{Philippe Solal} & Universit{\'e} Jean Monnet & Examinateur \\[0.1cm]
\textsc{Tam{\`a}s Solymosi} & Corvinus University of Budapest & Rapporteur \\[0.1cm]
\textsc{Peter Sudh{\"o}lter} & University of Southern Denmark & Directeur \\[0.1cm]
\end{tabular} 
\leavevmode\\[0.1cm]
\hrulefill
\end{center}

 
\vfill

 
\vfill
\end{center}
\end{titlepage}



 
 



\begin{acknowledgements}
\addchaptertocentry{\acknowledgementname} 
I would like to express my deepest gratitude to my supervisors, Michel Grabisch and Peter Sudh{\"o}lter. The uncountable and invaluable meetings I had with you, both made me grow as a mathematician and as a person. I have never learned as much as I have in the last few years, and for that I am extremely grateful. I would like to extend my sincere thanks to Marina Nu{\~n}ez and Tam{\`a}s Solymosi, who accepted to review the work you are about to read, as well as Philippe Bich, Jean-Jacques Herings and Philippe Solal, to be part of my committee. 

\medskip

I am also thankful to Agnieszka Rusinowska for her comforting help and advice, along with Arianna Novaro and Alexandre Skoda. I would also have a word for Nadim, Pierre, Victor, Thibault and Thomas, with whom I had such a great time learning and discussing about mathematics among many other things, which has been the mother of all motivations to pursue a career in mathematics.

\medskip

Tot slot zou deze reis niet hetzelfde zijn geweest zonder de dagelijkse steun van Rosa, met wie ik graag zo veel samenwerk. Enfin, un dernier mot pour ma mère, qui a travaillé plus dur que moi pour que je puisse devenir docteur. Ces quelques pages de mathématiques font pâle figure à côté de ses immenses efforts durant 26 années.  
\end{acknowledgements}


\begin{abstract}
\addchaptertocentry{\abstractname} 
The main ambition of this thesis is to contribute to the development of cooperative game theory towards combinatorics, algorithmics and discrete geometry. Therefore, the first chapter of this manuscript is devoted to the highlighting of the geometric nature of the coalition functions of transferable utility games, and spotlights the existing connections with the theory of submodular set functions and polyhedral geometry. 

\medskip 

To deepen the links with polyhedral geometry, we define a new family of polyhedra, called the \emph{basic polyhedra}, on which we can apply a generalized version of the Bondareva-Shapley Theorem to check their nonemptiness. To allow a practical use of these computational tools, we present an algorithmic procedure generating the minimal balanced collections, based on Peleg's recursive method. Subsequently, we apply the generalization of the Bondareva-Shapley Theorem to design a collection of algorithmic procedures able to check properties or generate specific sets of coalitions. 

\medskip 

In the next chapter, the connections with combinatorics are investigated. First, we prove that the balanced collections form a combinatorial species, and we construct the one of $k$-uniform hypergraphs of size $p$, as an intermediary step to construct the species of balanced collections. Afterwards, a few results concerning resonance arrangements distorted by games are introduced, which gives new information about the space of preimputations and the facial configuration of the core. 

\medskip 

Finally, we address the question of core stability using the results from the previous chapters. Firstly, we present an algorithm based on Grabisch and Sudh{\"o}lter's nested balancedness characterization of games with a stable core, which extensively uses the generalization of the Bondareva-Shapley Theorem introduced in the second chapter. Secondly, a new necessary condition for core stability is described, based on the application of the aforementioned generalization to a specific cone. Finally, we study the domination relation between preimputations using projectors, and we provide explicit formulas and algorithms to project on polytopes and intersections of affine subspaces. 
\end{abstract}




\chapter*{Résumé en français} 

\label{summary} 

\addchaptertocentry{Résumé en français}




\setlength{\epigraphwidth}{0.5\textwidth}
\epigraph{\emph{La totalité est plus que la somme des parties.}}{Aristote, \emph{Métaphysique}}

Les mathématiques des sciences sociales ont ceci de particulier que l'objet d'étude de leurs équations sont ceux qui les écrivent. En particulier, le comportement décrit par ces équations peut changer uniquement parce qu'elles sont explicitement écrites, s'il y a un intérêt pour quelqu'un à s'en écarter. De plus, les problèmes d'optimisation rencontrés dans la théorie des jeux sont souvent subjectifs, et les paramètres de la fonction objectif d'une personne peuvent être les variables de la fonction objectif de quelqu'un d'autre. 

\medskip

À cet égard, Émile Borel a déclaré que la théorie des jeux « sera une nouvelle science, où la psychologie ne sera pas moins utile que les mathématiques ».\footnote{\fullcite{borel1924propos}} Il est l'un des rares, avec Nicolas de Condorcet et Antoine Augustin Cournot, à avoir développé une théorie mathématique des interactions stratégiques avant la seconde guerre mondiale et John von Neumann. L'objectif des travaux de Borel sur la théorie des jeux était de trouver, pour tout jeu à somme nulle entre deux joueurs (le gain de l'un est la perte de l'autre), la « meilleure » stratégie pour chaque joueur.\footnote{\fullcite{leonard2010neumann}\label{foot: leonard-fr}} 

\medskip 

Sans surprise, Borel s'est appuyé sur la théorie des probabilités dans sa recherche de la stratégie parfaite, affirmant que, quelle que soit la qualité d'une stratégie unique et déterministe, si l'adversaire la connaît, il peut jouer une stratégie qui peut être mauvaise en général, mais qui contre parfaitement la « meilleure » stratégie utilisée par le premier joueur. Ceci s'illustre parfaitement lors de parties d'échecs ou de poker, où l'un peut s'écarter de la « meilleure » stratégie pour profiter des faiblesses apparentes de l'adversaire. Borel a insisté sur les limites des mathématiques dans la théorie des jeux, compte tenu de la complexité psychologique des jeux réels.\footref{foot: leonard-fr}${}^{,}$\footnote{\fullcite{von1959role}}

\medskip 

L'importance de la psychologie dans la théorie des jeux distinguait von Neumann de Borel. Dans une note envoyée à Borel en 1928, et présentée par ce dernier à l'Académie des Sciences\footnote{\fullcite{von1928academie}}, von Neumann a expliqué qu'il travaillait indépendamment sur la même question, sans avoir connaissance des travaux de Borel. Dans cette note, von Neumann affirme également avoir trouvé la solution au problème de la « meilleure » stratégie, avec le théorème du Minimax, publié plus tard la même année.\footnote{\fullcite{von1959theorie}} 

\medskip

Le théorème du Minimax, parfois appelé théorème fondamental de la théorie des jeux à deux joueurs, est le premier d'une longue série de théorèmes du minimax en analyse fonctionnelle. Il a probablement inspiré le théorème de la dualité forte en programmation linéaire et les travaux de Nash sur les jeux à somme non nulle à $n$ personnes, qui ont abouti au concept bien connu d'équilibre de Nash. 

\begin{minimax-theoreme}
Soient $X \subseteq \bbR^p$ et $Y \subseteq \bbR^q$ deux ensembles compacts et convexes, et $f : X \times Y \to \bbR$ une fonction continue telle que
\begin{itemize}
\item pour tout $y \in \bbR^q$, la fonction $f( \cdot, y) : X \to \bbR$ est concave, 
\item pour tout $x \in \bbR^p$, la fonction $f(x, \cdot ) : Y \to \bbR$ est convexe.
\end{itemize}
Alors la fonction $f$ satisfait
\[
\max_{x \in X} \min_{y \in Y} f(x, y) = \min_{y \in Y} \max_{x \in X} f(x, y). 
\]
\end{minimax-theoreme}

Le théorème du Minimax est beaucoup plus général que nécessaire pour résoudre les jeux à somme nulle à deux joueurs, car $x$ et $y$ n'ont besoin que d'être des distributions de probabilités discrètes, et $f$ bilinéaire. Dans ce contexte, on peut simplement réécrire $f(x, y)$ comme $y^\top A x$, où $A$ représente $f$ dans les bases canoniques de $\bbR^p$ et $\bbR^q$. Si le joueur $1$ a $p$ stratégies possibles, et le joueur $2$ a $q$ stratégies, le coefficient $a_{ij}$ de $A$ représente le gain du joueur $1$ lorsqu'il choisit la stratégie $i \in \{1, \ldots, p\}$ et que le joueur $2$ choisit la stratégie $j \in \{1, \ldots, q\}$. Comme il s'agit d'un jeu à somme nulle, le gain du joueur $2$ est $-a_{ij}$. Les joueurs sont autorisés à jouer des combinaisons convexes de leurs stratégies, ce qui peut être interprété comme une randomisation du jeu, ou comme des fréquences si le jeu est répété, et les coefficients des combinaisons convexes sont les entrées des vecteurs $x$ et $y$, ce qui nous rappelle le point de vue de Borel sur ce que devrait être une bonne stratégie.

\medskip 

Une conséquence du théorème du Minimax est que, pour chaque joueur, il existe toujours une « meilleure » stratégie dans un jeu à somme nulle à deux joueurs. Expliquons enfin ce que nous entendons par « meilleure » stratégie. Comme il s'agit d'un jeu à somme nulle, le joueur $1$ essaiera toujours de maximiser l'image de $f$, tandis que le joueur $2$ essaiera toujours de la minimiser. Ainsi, un bon début pour le joueur $1$ est d'avoir une stratégie $x_0$ qui lui assure une certaine valeur, indépendamment des actions du joueur $2$, pour qui l'objectif est de minimiser $f(x_0, \cdot) : Y \to \bbR$. Désignons par $\gamma$ la valeur assurée par la stratégie $x_0$, donnée par $\gamma = \max_{x \in X} \min_{y \in Y} f(x,y)$. Le joueur $2$ suit exactement le même processus, et trouve une stratégie $y_0$ lui assurant $\delta = - \min_{y \in Y} \max_{x \in X} f(x, y)$. Par le théorème du Minimax, nous avons que 
\[
\gamma + \delta = \max_{x \in X} \min_{y \in Y} f(x, y) - \min_{y \in Y} \max_{x \in X} f(x, y) = 0, 
\]
ce qui confirme l'optimalité des stratégies $x_0$ et $y_0$. En effet, si le joueur $1$ n'applique plus $x_0$, sa stratégie ne maximisera plus la fonction $\min_{y \in Y} f(\cdot, y) : Y \to \bbR$, et son gain sera alors inférieur à $\gamma$. De plus, comme $\gamma + \delta$ vaut zéro, il n'y a pas de gain supplémentaire à partager entre les deux joueurs et, comme nous l'avons montré précédemment, tout écart par rapport aux stratégies $x_0$ ou $y_0$ peut être sanctionné par l'adversaire. Parce que le nombre $\gamma$ détermine complètement le gain du jeu lorsque les deux joueurs jouent les stratégies définies plus tôt, $\gamma$ est appelé la \emph{valeur} du jeu. Ce résultat est considéré par beaucoup comme le point de départ de la théorie des jeux et a été généralisé par John Nash\footnote{\fullcite{nash1951non}} aux jeux à somme non nulle à $n$ joueurs. Un équilibre de Nash est le choix d'une stratégie par chaque joueur de telle sorte qu'aucun joueur n'a intérêt à dévier unilatéralement de la stratégie choisie. 

\medskip 

Dans le même article, von Neumann a discuté de la possibilité d'étendre ce résultat à un ensemble plus grand de $n$ joueurs, noté $N$. Suivant la même idée que pour le cas à deux personnes, von Neumann voulait donner à chaque ensemble $S \subseteq N$ de joueurs une valeur, similaire à $\gamma$ dans l'exemple précédent, interpreté comme un gain qu'ils pourraient s'assurer si nous supposons que les joueurs formant l'ensemble $S$ coordonnent leurs actions. L'idée a été développée quelques années plus tard, avec son collègue de l'Institute of Advanced Study Oskar Morgenstern, dans le célèbre ouvrage intitulé \citetitle{von1944theory}.\footnote{\fullcite{von1944theory}}

\medskip 

À partir de maintenant, nous commençons à nous écarter des définitions de von Neumann et Morgenstern, qui ont fait de fortes hypothèses sur les valeurs des ensembles de joueurs. La seule hypothèse que nous conservons est que la valeur de l'ensemble vide de joueurs est de $0$, et nous abandonnons toutes les hypothèses de somme nulle. Nous pouvons interpréter la valeur d'un ensemble de joueurs $S$ comme étant le maximum qu'ils peuvent s'assurer lorsque tous les joueurs du complément de $S$ essaient de minimiser leur gain. Nous désignons par $\calN$ l'ensemble de tous les sous-ensembles non vides de $N$, que nous appelons \emph{coalitions} et par $v$ la fonction associant à chaque coalition $S \in \calN$ sa valeur $v(S)$ telle que $v(\emptyset) = 0$. La fonction $v$ est appelée \emph{fonction coalitionnelle} du jeu $(N, v)$. Nous supposons également que les valeurs des coalitions sont toutes exprimées dans la même unité et qu'elles peuvent être transférées d'un joueur à l'autre, d'une coalition à l'autre. 

\medskip 

La question est maintenant de savoir, en fonction de $v$, quelles coalitions se formeront. Une condition nécessaire à la formation de la grande coalition $N$ est que nous puissions diviser sa valeur $v(N)$ en $n$ réels $x_1, \ldots, x_n$ tels que, pour chaque coalition $S \in \calN$, nous ayons $\sum_{i \in S} x_i \geq v(S)$. En effet, si les joueurs de $S$ reçoivent au total moins que ce qu'ils peuvent obtenir seuls, ils n'ont aucun intérêt à rejoindre la grande coalition $N$. Par commodité, nous désignons par $\bbR^N$ le produit cartésien de $n$ copies de $\bbR$, indexées par les éléments de $N$. Nous avons alors $x \in \bbR^N$ et le réel $x_i$ est appelé le \emph{paiement} en $x$ du joueur $i$. Le paiement en $x$ d'une coalition $S \in \calN$, noté $x(S)$ est défini par $x(S) = \sum_{i \in S} x_i$. Le vecteur de paiement $x \in \bbR^N$ est appelé \emph{préimputation} s'il satisfait $x(N) = v(N)$, c'est-à-dire s'il divise exactement la valeur de la grande coalition $N$ en $n$ paiements, un pour chaque joueur. Enfin, nous appelons \emph{le cœur} du jeu $(N, v)$, noté $C(v)$, le sous-ensemble de $\bbR^N$ constitué de toutes les préimputations donnant à toute coalition un paiement au moins aussi important que sa valeur. 

\medskip 

La définition intuitive et pertinente du cœur en fait l'un des concepts de solution les plus étudiés en théorie des jeux et en économie. Même si sa définition moderne et sa reconnaissance en tant que concept de solution, pour la première fois par Shapley,\footnote{\fullcite{shubik1992game}}${}^{,}$\footnote{\fullcite{zhao2018three}} n'a que 70 ans,\footnote{\fullcite{shapley1955markets}} l'idée d'allocations mutuellement profitables issus d'un marché ou d'une coopération a déjà été étudiée depuis au moins la fin du XIX$^{\text{e}}$ siècle. En 1881, le philosophe et économiste politique anglo-irlandais Francis Edgeworth\footnote{\fullcite{edgeworth1881mathematical}} a étudié les résultats d'un marché d'échange de deux biens et a défini un concept de solution similaire, appelé \emph{courbe des contrats}. 

\medskip 

En 1963, Olga Bondareva a trouvé une condition nécessaire et suffisante\footnote{\fullcite{bondareva1963some}} pour que le cœur d'un jeu soit non vide, impliquant un ensemble d'objets appelés \emph{recouvrements}, découlant naturellement de l'application du théorème de dualité de la programmation linéaire à la définition du cœur. Pour chaque recouvrement, nous pouvons construire une forme linéaire sur l'espace vectoriel des jeux, et l'image du jeu pour chacune de ces formes linéaires doit être inférieure ou égale à la valeur de la grande coalition. Indépendamment, mais en suivant la même idée, Lloyd Shapley a publié en 1967 un article\footnote{\fullcite{shapley1967balanced}} présentant la même condition nécessaire et suffisante, les recouvrements étant appelées \emph{collections équilibrées}. Il a également prouvé que seul un petit sous-ensemble de ces collections était nécessaire, les collections équilibrées \emph{minimales}. 

\medskip 

Intuitivement, une collection équilibrée représente la façon dont les joueurs de $N$ peuvent être organisés en une ou plusieurs coalitions afin de maximiser leur valeur totale. Les joueurs peuvent former toutes les coalitions possibles, en répartissant leur temps entre plusieurs coalitions si nécessaire, à condition de respecter deux règles : 
\begin{enumerate}
\item[1.)] Chaque joueur est « actif » pendant exactement une unité de temps ;
\item[2.)] Si certains joueurs forment une coalition, ils y passent tous le même temps. 
\end{enumerate}
Dans une collection équilibrée, chaque coalition est associée à un poids correspondant au temps individuel passé par les joueurs dans la coalition. Les joueurs peuvent également faire partie de coalitions unipersonnelles, travaillant seuls pendant une certaine fraction du temps. Comme chaque joueur doit être actif pendant exactement une unité de temps, la somme des poids des coalitions auxquelles appartient un joueur donné doit être égale à un. Plus formellement, nous avons la définition suivante.

\begin{definition*}
Une collection de coalitions $\calB \subseteq \calN$ $\calB$ est une \emph{collection équilibrée} s'il existe un ensemble $\lambda = \{\lambda_S \mid S \in \calB\}$ de réels positifs tels que
\[
\sum_{S \in \calB} \lambda_S \bfone^S = \bfone^N, 
\]
avec $\bfone^S_i = 1$ si $i \in S$ et $\bfone^S_i = 0$ sinon. Une \emph{collection équilibrée minimale} est une collection équilibrée qui ne contient pas de sous-collections équilibrées. 
\end{definition*}

La définition d'une collection équilibrée est étroitement liée à celle de \emph{hypergraphes réguliers}. Les partitions de $N$ sont des exemples de collections minimales équilibrées, dans lesquelles tous les joueurs restent dans la même coalition pendant toute l'unité de temps. Lorsque les joueurs de $N$ sont organisés selon la collection équilibrée $\calB$, la valeur totale est égale à 
\[
\sum_{S \in \calB} \lambda_S v(S).
\]
Si, pour une collection équilibrée $\calB$, la valeur totale $\sum_{S \in \calB} \lambda_S v(S)$ est supérieure à $v(N)$, les joueurs préfèrent être organisés comme décrit par $\calB$ plutôt que comme décrit par la collection triviale $\{N\}$. Naturellement, les joueurs de $N$ recherchent la collection équilibrée qui maximise la somme pondérée des valeurs, et la grande coalition $N$ n'est formée que si, pour toutes les collections équilibrées $\calB \subseteq \calN$, nous avons 
\[
\sum_{S \in \calB} \lambda_S v(S) \leq v(N), 
\]
qui ressemble à la citation d'Aristote  « \emph{La totalité est plus que la somme des parties} ». Cette condition est également équivalente à la non-vacuité du cœur. 

\begin{BS-theoreme}
Soit $(N, v)$ un jeu. Le cœur $C(v)$ est non-vide si et seulement si, pour toute collection équilibrée minimale $\calB \subseteq \calN$, nous avons
\[
\sum_{S \in \calB} \lambda_S v(S) \leq v(N). 
\]
\end{BS-theoreme}

Ce théorème affirme qu'il existe des préimputations partageant la valeur de la grande coalition $N$ parmi les joueurs et donnant un paiement supérieur à leur valeur à toutes les coalitions si et seulement si la plus grande valeur réalisable par les joueurs dans $N$ est obtenue en formant $N$. De plus, ce théorème établit un lien entre les propriétés géométriques de polytopes particuliers et les propriétés des fonctions d'ensemble. 

\medskip 

Le lien entre les polytopes et les fonctions d'ensemble a été largement étudié dans le domaine de l'optimisation combinatoire, plus particulièrement dans la théorie des fonctions sous-modulaires, notamment par Jack Edmonds, Satoru Fujishige, L{\'a}szl{\'o} Lov{\'a}sz et Alexander Schrijver parmi d'autres. À chaque fonction d'ensemble $\xi$ sur $N$ satisfaisant $\xi(\emptyset) = 0$, on associe un polytope, appelé \emph{polytope de base}, noté ${\rm B}(\xi)$ et défini par ${\rm B}(\xi) = \{x \in \bbR^N \mid x(N) = \xi(N) \text{ and } x(S) \leq \xi(S), \forall S \in \calN\}$.
La fonction d'ensemble $\xi$ est souvent supposée être \emph{sous-modulaire}, c'est-à-dire que pour tout $S$ et $T$ dans $\calN$, nous avons $\xi(S \cup T) + \xi(S \cap T) \leq \xi(S) + \xi(T)$. 

\medskip 

Jack Edmonds a étudié  les fonctions de rang des matroïdes, une classe particulière de fonctions d'ensemble sous-modulaires.\footnote{\fullcite{edmonds1970submodular}} Il a défini une famille de polytopes héritant de propriétés similaires à celles des matroïdes, appelés \emph{polymatroïdes}, sur lesquelles il a développé de célèbres algorithmes d'optimisation. Il donne également une description détaillée de la structure faciale de ces polytopes, qui a inspiré certains résultats sur les cœurs des jeux convexes\footnote{\fullcite{shapley1971cores}} de Shapley. Tout polytope de base est un polymatroïde à une translation près, et les deux sont des cœurs de jeux convexes. 

\medskip

Des polytopes similaires ont attiré l'attention des combinatoriciens au cours des quinze dernières années. Les \emph{permutoèdres generalisés},\footnote{\fullcite{postnikov2009permutohedra}} définis par Postnikov, sont une généralisation du permutoèdre, qui a déjà été utilisé dans les mathématiques des sciences sociales par \citeauthor{guilbaud1963analyse},\footnote{\fullcite{guilbaud1963analyse}} qui l'ont nommé ainsi. Le permutoèdre, noté $\Pi_N$, est l'enveloppe convexe d'un ensemble de $n!$ points définis en permutant les coordonnées du vecteur $(1, \ldots, n) \in \bbR^N$. Cependant, certains auteurs définissent le permutoèdre comme l'enveloppe convexe des permutations de n'importe quel vecteur dont les coordonnées sont distinctes, ce qui donne un polytope ayant les mêmes propriétés combinatoires : ils ont le même treillis de faces et le même « normal fan ». 

\medskip 

À tout permutoèdre, on peut appliquer une déformation qui consiste en des translations parallèles de certaines de ses facettes sans croiser aucun sommet, c'est-à-dire sans changer la direction des arêtes. Ces déformations sont les permutoèdres généralisés. Il s'avère que les permutoèdres généralisés,\footnote{\fullcite{castillo2022deformation}} les polytopes de base des fonctions sous-modulaires, et les polymatroïdes (à une translation près), sont les mêmes polytopes. Ils coïncident tous avec des cœurs de jeux convexes, et grâce à cela la théorie des jeux a souvent tiré parti des résultats de la combinatoire ou de l'optimisation combinatoire. 

\medskip 

Le but de cette thèse est de contribuer au développement de la théorie des jeux (coopératifs) dans la direction décrite ci-dessus, vers la combinatoire et la géométrie. Cette thèse se compose de quatre chapitres distincts, couvrant différents aspects de mon travail effectué lors de mon doctorat. 

\medskip 

Le premier chapitre est consacré principalement aux notations et définitions utilisées dans cette thèse. En particulier, les principaux sous-espaces affines de $\bbR^N$ d'intérêt sont définis, et une nouvelle extension continue de jeux dans $\bbR^N$ est présentée, sous la forme d'un polynôme tropical, qui renforce encore les connections entre la théorie des jeux coopératifs et la géométrie discrète. Cette extension s'applique à tous les jeux dits « totalement équilibrés », c'est-à-dire les jeux pour lesquels le cœur est non-vide, et dont tous les sous-jeux ont également un cœur non-vide. Les sous-jeux d'un jeu $(N, v)$ sont définis en restreignant la fonction coalitionelle $v$ à un sous-ensemble de $N$. 

\medskip 

Le deuxième chapitre s'inspire en grande partie de l'article que j'ai rédigé avec mes deux directeurs de thèse,\footnote{\fullcite{laplace2023minimal}} dans lequel nous construisons un algorithme générant les collections équilibrées minimales récursivement sur $N$, à partir d'une méthode décrite par Bezalel Peleg.\footnote{\fullcite{peleg1965inductive}} Deuxièmement, nous décrivons une collection d'outils algorithmiques basés sur des implémentations du théorème de Bondareva-Shapley. Nous étendons le champ d'application de ces méthodes algorithmiques à une nouvelle classe de polyèdres, appelée \emph{polyèdres basiques}. Ces polyèdres, qui généralisent les polytopes évoqués plus tôt, apparaissent fréquemment en théorie des jeux, et leur non-vacuité implique souvent l'existence de solutions, ou de certaines propriétés satisfaites par un certain nombre d'objets, en particulier par les coalitions de $N$. Plus formellement, un polyèdre basique est un polyèdre $P \subseteq \bbR^N$ qui peut s'écrire comme 
\[
P \coloneqq \left\{ x \in \bbR^N \mid x(N) = b_N, \; A_1 x \leq b_1, \; A_2 x < b_2, \; A_3 \geq b_3, \text{ and } A_4x > b_4 \right\},
\]
avec les coefficients des matrices $\{A_i\}_{i = 1, \ldots, 4}$ appartenant à $\{0, 1\}$. L'un des résultats principaux de ce chapitre est l'adaptation du théorème de Bondareva-Shapley à ces polyèdres, qui permet ensuite de développer des outils algorithmiques vérifiant l'existence de certains types de solutions, ou la satisfaction de certaines propriétés. 

\medskip 

Dans le troisième chapitre, les liens entre la combinatoire et les fonctions des ensembles sont étudiés. Tout d'abord, certaines similitudes entre les collections équilibrées et les hypergraphes uniformes ou réguliers sont mises en évidence. Ensuite, il est démontré que les collections équilibrées (minimales) forment une \emph{espèce combinatoire}, concept introduit par André Joyal\footnote{\fullcite{joyal1981theorie}} afin de fournir une méthode abstraite et systématique pour identifier les fonctions génératrices des structures discrètes. Par la suite, je construis l'\emph{espèce des hypergraphes $k$-uniformes de taille $p$}, pour $k$ et $p$ deux entiers strictement positifs donnés, comme étape intermédiaire pour construire l'espèce des collections équilibrées. Dans la deuxième partie de ce chapitre, j'étudie les liens entre la théorie des arrangements d'hyperplans, dont l'arrangement de \emph{résonance}, et la théorie des jeux coopératifs. 

\medskip 

Le dernier chapitre est consacré aux applications des deux chapitres précédents. Tout d'abord, un algorithme basé sur une caractérisation de Grabisch et Sudh{\"o}lter\footnote{\fullcite{grabisch2021characterization}} pour la stabilité du cœur d'un jeu donné est présentée. Deuxièmement, une nouvelle condition nécessaire pour la stabilité du cœur est donnée, basée sur la vérification de la non-vacuité d'un cône particulier. Enfin, une collection d'outils est présentée pour étudier les projections entre les préimputations, sur des sous-espaces affines d'intérêt. Une formule explicite est donnée, ainsi que quelques procédures algorithmiques. Nous terminons ce chapitre par une application de ces projecteurs aux défaillances du marché, en utilisant le modèle des \emph{jeux de marché} introduits par Shapley et Shubik.\footnote{\fullcite{shapley1969market}}


 
\setcounter{tocdepth}{1}
\tableofcontents 













\begin{symbols}{lll} 

$N$ & p. \pageref{sym: N} & finite set of $n$ players \\
$\calN$ & p. \pageref{sym: calN} & set of coalitions, i.e., nonempty subsets of $N$ \\
$\bbR^N$ & p. \pageref{sym: RN} & Cartesian product of $\lvert N \rvert$ copies of $\bbR$, indexed by $N$ \\
$v$ & p. \pageref{sym: v} & coalition function of a game \\
$\bbX(v) \qquad$ & p. \pageref{def: preimputation-upper} & set of preimputations of the game $(N,v)$ \\
$C(v)$ & p. \pageref{sym: core} & core of the game $(N,v)$ \\
$\bbA_S(v)$ & p. \pageref{sym: AS} & affine subspace $\bbA_S = \{x \in \bbX(v) \mid x(S) = v(S)\}$ \\
$e(S, x)$ & p. \pageref{sym: excess} & excess of $S$ at $x$: $e(S, x) = v(S) - x(S)$ \\
$X_\calS(v)$ & p. \pageref{sym: region} & region associated with the feasible collection $\calS$ \\
$\eta^S$ & p. \pageref{prop: eta} & normal to $\bbA_S$ in $\bbX(v)$ \\
$\Sigma$ & p. \pageref{sym: sigma} & set of side payments $\Sigma = \{x \in \bbR^N \mid x(N) = 0\}$ \\
$\pi_K$ & p. \pageref{prop: first-proj} & projector $\pi_K: \bbR^N \to K$ \\
${\rm aff}(P)$ & p. \pageref{sym: aff} & affine hull of the polytope $P \subseteq \bbR^N$ \\
$\calL(P)$ & p. \pageref{sym: calL(P)} & face lattice of the polytope $P \subseteq \bbR^N$ \\
$\calK_P$ & p. \pageref{sym: calK_P} & normal fan of the polytope $P \subseteq \bbR^N$ \\
$\mathfrak{S}_N$ & p. \pageref{sym: symmetric-group} & group of permutations of $N$ \\
$\Pi_N(x)$ & p. \pageref{sym: permutohedron} & permutohedron of $x$: $\Pi_N(x) = {\rm conv} \{x^\sigma \mid \sigma \in \mathfrak{S}_N\}$ \\
$\calE(v)$ & p. \pageref{sym: calE(v)} & set of coalitions $S$ such that $C(v) \subseteq \bbA_S$ \\
$(\calF_P, v_P)$ & p. \pageref{sym: calF_P-v_P} & game associated with the polytope $P \subseteq \bbR^N$ \\
$\calVE(v)$ & p. \pageref{sym: sve} & set of strictly vital-exact coalitions \\
$\varsigma^{[p]}$ & p. \pageref{sym: varsigma} & combinatorial species of $p$-multisets \\
$\wp^{[k]}$ & p. \pageref{sym: wp} & combinatorial species of $k$-subsets \\
${\normalfont \textsc{Hyp}}_{k,p}$ & p. \pageref{prop: comb-equation} & combinatorial species of $k$-uniform hypergraphs of size $p$ \\
$\calA_R$ & p. \pageref{def: resonance} & resonance arrangement of hyperplanes \\
$\calL_C(v)$ & p. \pageref{sym: lcv} & face lattice of the core $C(v)$ \\
$\phi(x)$ & p. \pageref{sym: aggrieved} & set of coalitions $\phi(x) = \{S \in \calVE(x) \mid x(S) < v(S)\}$ \\
$\delta_S(x)$ & p. \pageref{sym: dom-cone} & domination cone of $x$ with respect to $S$ \\
${\rm Aug}(x)$ & p. \pageref{sym: aug} & augmentation cone of $x$ \\
$\langle \calS \rangle$ & p. \pageref{sym: span} & subspace of $\Sigma$ spanned by $\{\eta^S \mid S \in \calS\}$ \\
$G_\calS$ & p. \pageref{sym: gram} & Gram matrix of collection $\calS$ \\

\end{symbols}

\vfill

In general, we use `blackboard bold' characters to denote geometric spaces such as $\bbA_S$ or fields such as $\bbR$; the capital letters are used mainly for coalitions, sets or matrices; the calligraphic capital letters for sets of sets, sets of coalitions; and the lowercase letters for preimputations, vectors, etc. 




\mainmatter 

\pagestyle{thesis} 



\chapter*{Introduction} 
\markboth{Introduction}{}

\label{Intro} 

\addchaptertocentry{Introduction}




\setlength{\epigraphwidth}{0.5\textwidth}
\epigraph{\emph{The whole is more than the sum of its parts.}}{Aristotle, \emph{Metaphysics}}

The mathematics of social sciences has this particularity that the object of study of its equations are those who write them. It means that the behavior captured in the equations can change only because these equations are explicitly written down, if there is an interest for someone to deviate from it. Moreover, the optimization problems found in game theory are often subjective, and the parameters of someone's objective function can be the variables of someone else's objective function. 

\medskip

To this regard, Émile Borel said that game theory `will be a new science, where psychology will be no less useful than mathematics'.\footnote{\fullcite{borel1924propos}} He is one of the few, with Nicolas de Condorcet and Antoine Augustin Cournot, to have developed a mathematical theory of strategic interactions between people in the pre-von Neumann era. The goal of Borel's work with game theory was to find, for any two-person zero-sum game (the gain of one is the loss of the other), the `best' strategy for each player.\footnote{\fullcite{leonard2010neumann}\label{foot: leonard}} 

\medskip 

Unsurprisingly, Borel relied on probability theory in his pursuit of the perfect strategy, saying that, no matter how good a unique, deterministic strategy was, if the opponent knows it, she can play a strategy which can be bad in general, but perfectly counters the `best' one used by the first player. This is perfectly illustrated by the mind game between two players in chess or in poker, where one could deviate from the `best' strategy to take advantage of the opponent's apparent weaknesses. Borel insisted on the limitation of mathematics in game theory, given the psychological complexity of real games.\footref{foot: leonard}${}^{,}$\footnote{\fullcite{von1959role}}

\medskip 

The importance of psychology distinguished von Neumann from Borel. In a note sent to Borel in 1928, and presented by the latter to the French Acad{\'e}mie des Sciences,\footnote{\fullcite{von1928academie}} von Neumann explained that he was independently working on the same question, without being aware of the work of Borel. In this note, von Neumann also claimed that he found the solution to the `best' strategy problem, with the Minimax Theorem, published later in the same year.\footnote{\fullcite{von1959theorie}} 

\medskip

The Minimax Theorem, sometimes called the fundamental theorem of two-person game theory, is the first of a long sequence of `minimax' theorems in functional analysis, and has probably inspired the strong duality theorem in linear programming and the works of Nash on $n$-person general-sum games, leading to the well-known concept of Nash equilibrium. 

\begin{minimax-theorem}
Let $X \subseteq \bbR^p$ and $Y \subseteq \bbR^q$ be compact convex sets, and let $f : X \times Y \to \bbR$ be a continuous function such that 
\begin{itemize}
\item for all $y \in \bbR^q$, the map $f( \cdot, y) : X \to \bbR$ is concave, 
\item for all $x \in \bbR^p$, the map $f(x, \cdot ) : Y \to \bbR$ is convex.
\end{itemize}
Then we have that 
\[
\max_{x \in X} \min_{y \in Y} f(x, y) = \min_{y \in Y} \max_{x \in X} f(x, y). 
\]
\end{minimax-theorem}

The Minimax Theorem is a lot more general than required to solve two-person zero-sum games, because $x$ and $y$ need only to be discrete probability distributions, and $f$ bilinear. In this context, we can simply rewrite $f(x, y)$ as $y^\top A x$, where $A$ represents $f$ in the canonical bases of $\bbR^p$ and $\bbR^q$. If player $1$ has $p$ possible strategies, and player $2$ has $q$ strategies, the coefficient $a_{ij}$ of $A$ represents the payoff of player $1$ when she chooses strategy $i \in \{1, \ldots, p\}$ and player $2$ chooses strategy $j \in \{1, \ldots, q\}$. Because it is a zero-sum game, the payoff of player $2$ is $-a_{ij}$. Players are allowed to play convex combinations of their strategies, which can be interpreted as a randomization of the play, or as frequencies if the game is repeated, and the coefficients of the convex combinations are the entries of the vectors $x$ and $y$, which reminds us of Borel's view on what a good strategy should be.

\medskip 

A consequence of the Minimax Theorem is that, for each player, there always exists a `best' strategy in a two-person zero-sum game. Let us finally explain what we understand by `best' strategy. Because it is a zero-sum game, player $1$ will always try to maximize the image of $f$, while player $2$ will always try to minimize it. So, a good start for player $1$ is to have a strategy $x_0$ that ensures him a certain value, independently of the actions of player $2$, for whom the objective is to minimize $f(x_0, \cdot): Y \to \bbR$. Let us denote by $\gamma$ the value ensured by strategy $x_0$, given by $\gamma = \max_{x \in X} \min_{y \in Y} f(x,y)$. The player $2$ follows exactly the same process, and finds a strategy $y_0$ ensuring $\delta = - \min_{y \in Y} \max_{x \in X} f(x, y)$. By the Minimax Theorem, we have that 
\[
\gamma + \delta = \max_{x \in X} \min_{y \in Y} f(x, y) - \min_{y \in Y} \max_{x \in X} f(x, y) = 0, 
\]
which asserts the optimality of the strategies $x_0$ and $y_0$. Indeed, if player $1$ deviates from $x_0$, its strategy will no longer maximize the map $\min_{y \in Y} f(\cdot, y) : Y \to \bbR$, and then her payoff will be lower than $\gamma$. Moreover, because $\gamma$ and $\delta$ sum to zero, there is no extra payoff left to split between the two players, and, as we have previously showed, any deviation from $x_0$ or $y_0$ could be punished by the opponent. Because the number $\gamma$ completely determines the payoff of the game when both players play the strategies defined above, $\gamma$ is called the \emph{value} of the game. This result is considered by many as the starting point of game theory, and has been generalized by John Nash\footnote{\fullcite{nash1951non}} to $n$-person non-zero-sum games. A Nash equilibrium is the choice of a strategy by each player such that no player has an interest to unilaterally deviate from her chosen strategy. 

\medskip 

In the same paper, von Neumann discussed the possibility to extend this result to a larger set of $n$ players, denoted by $N$. Following the same idea as for the two-person case, von Neumann wanted to give to each set $S \subseteq N$ of players a \emph{worth}, similar to the value $\gamma$ in the previous example, that they could ensure for themselves, if we assume that the players forming the set $S$ are coordinating their actions. The idea was developed a few years later, with his colleague from the Institute of Advanced Study Oskar Morgenstern, in the well-known \citetitle{von1944theory}.\footnote{\fullcite{von1944theory}}

\medskip 

Starting from now, we deviate from the definitions of von Neumann and Morgenstern, who made strong assumptions on the values of the sets of players. The only assumption we keep is that the value of the empty set of players is $0$, and we drop all the zero-sum assumptions. We can interpret the worth of a set of players $S$ as being the maximum they can ensure for themselves when all the players in the complement of $S$ try to minimize their payoff. We denote by $\calN$ the set of all nonempty subsets of $N$, which we call \emph{coalitions} and by $v$ the map associating to each coalition $S \in \calN$ its worth $v(S)$ such that $v(\emptyset) = 0$. The map $v$ is called the \emph{characteristic function} of the game $(N, v)$. Also, we assume that the worths of the coalitions are all in the same unit, and can be transferred from one player to another, from one coalition to another. 

\medskip 

The question now is to find out, according to the coalition function $v$, which coalitions will form. A necessary condition for the grand coalition $N$ to form is that we can split its value $v(N)$ into $n$ numbers $x_1, \ldots, x_n$ such that, for every coalition $S \in \calN$, we have $\sum_{i \in S} x_i \geq v(S)$. Indeed, if the players in $S$ received in total less than what they achieve on their own, they have no interest to join the grand coalition $N$. For convenience, we denote by $\bbR^N$ the Cartesian product of $n$ copies of $\bbR$, indexed by the elements of $N$. Then, we have $x \in \bbR^N$ and the real number $x_i$ is called the \emph{payment} at $x$ of player $i$. The payment at $x$ of a coalition $S \in \calN$, denoted by $x(S)$ is defined by $x(S) = \sum_{i \in S} x_i$. The payment vector $x \in \bbR^N$ is called a \emph{preimputation} if it satisfies $x(N) = v(N)$, i.e., if it splits exactly the worth of the grand coalition $N$ into $n$ payments, one for each player. Finally, we call \emph{the core} of the game $(N, v)$, denoted by $C(v)$, the subset of $\bbR^N$ consisting of all preimputations giving to any coalition a payment at least as large as its worth. 

\medskip 

The intuitive and meaningful definition of the core makes it one of the longest-studied solution concepts. Even if its modern definition and its consideration as a solution concept, first by Shapley,\footnote{\fullcite{shubik1992game}}${}^{,}$\footnote{\fullcite{zhao2018three}} is only 70 years old,\footnote{\fullcite{shapley1955markets}} the idea of mutually profitable outcomes coming from a market or cooperation was already studied since at least the end of the XIX$^{\text{th}}$ century. In 1881, the Anglo-Irish philosopher and political economist Francis Edgeworth\footnote{\fullcite{edgeworth1881mathematical}} studied the outcomes of an exchange market of two commodities and defined a similar solution concept, called the \emph{contract curve}. 

\medskip 

In 1963, Olga Bondareva found a necessary and sufficient condition\footnote{\fullcite{bondareva1963some}} for the core of a game to be nonempty, involving a set of objects called \emph{coverings}, arising naturally from the application of the duality theorem of linear programming to the definition of the core. From each covering, we can construct a linear form on the vector space of games, and the image of the game for each of these linear forms should be lower than or equal to the worth of the grand coalition. Independently but following the same idea, Lloyd Shapley published in 1967 a paper\footnote{\fullcite{shapley1967balanced}} presenting the same necessary and sufficient condition, with the coverings being called \emph{balanced collections}. He also proved that only a small subset of these was required, the \emph{minimal} balanced collections. 

\medskip 

Roughly speaking, a balanced collection represents how the players in $N$ can be organized in one or several coalitions in order to maximize their total worth. The players can form any possible coalition, even split their time in several coalitions if needed, as long as they follow two natural rules: 
\begin{enumerate}
\item Each player is `active' for exactly one unit of time, 
\item If some players form a coalition, they all spend the same time in it. 
\end{enumerate}
In a balanced collection, each coalition is associated to a weight corresponding to the individual time spent by the players in that coalition. Players can also be in one-person coalitions, working on its own for a certain fraction of time. Because each player must be active for exactly one unit of time, the sum of the weights of the coalitions to which a particular player belongs must be one. More formally, we have the following definition. 

\begin{definition*}
Let $\calB \subseteq \calN$ be a collection of coalitions. We say that $\calB$ is a \emph{balanced collection} if there exists a set $\lambda = \{\lambda_S \mid S \in \calB\}$ of positive real numbers such that
\[
\sum_{S \in \calB} \lambda_S \bfone^S = \bfone^N, 
\]
where $\bfone^S_i = 1$ if $i \in S$ and $\bfone^S_i = 0$ otherwise. A \emph{minimal balanced collection} is a balanced collection for which no subcollection is balanced. 
\end{definition*}

The definition of a balanced collection is closely related to the one of \emph{regular hypergraphs}. Notice that the partitions of $N$ are minimal balanced collections, in which all players stay in the same coalition during the whole unit of time. When the players in $N$ are organized according to the balanced collection $\calB$, the total worth is equal to 
\[
\sum_{S \in \calB} \lambda_S v(S).
\]
If, for a balanced collection $\calB$, the total worth $\sum_{S \in \calB} \lambda_S v(S)$ is greater than $v(N)$, the players prefer to be organized as described by $\calB$ rather than as described by the trivial collection $\{N\}$. Naturally, the players in $N$ are looking for the balanced collection which maximizes the weighted sum of worths, and the grand coalition $N$ is formed only if, for all balanced collections $\calB \subseteq \calN$, we have 
\[
\sum_{S \in \calB} \lambda_S v(S) \leq v(N), 
\]
which resembles Aristotle's ``\emph{The whole is more than the sum of its parts}''. This condition is also equivalent to the nonemptiness of the core. 

\begin{BS-theorem}
Let $(N, v)$ be a game. Then the core $C(v)$ is nonempty if and only if, for all minimal balanced collections $\calB \subseteq \calN$, we have 
\[
\sum_{S \in \calB} \lambda_S v(S) \leq v(N). 
\]
\end{BS-theorem}

This theorem states that there exist preimputations sharing the worth of the grand coalition $N$ and giving a payment greater than their worth to all coalitions if and only if the greatest worth achievable by the players in $N$ is obtained by forming $N$. Moreover, this theorem establishes a connection between geometrical properties of specific polytopes, and properties of set functions. 

\medskip 

The connection between polytopes and set functions was extensively studied in the field of combinatorial optimization, more specifically in submodular functions theory, notably by Jack Edmonds, Satoru Fujishige, L{\'a}szl{\'o} Lov{\'a}sz and Alexander Schrijver among others. For each grounded set function $\xi$ on $N$, i.e., $\xi: 2^N \to \bbR$ satisfying $\xi(\emptyset) = 0$, we associate a polytope, called the \emph{base polytope}, denoted by ${\rm B}(\xi)$ and defined by 
\[
{\rm B}(\xi) = \{x \in \bbR^N \mid x(N) = \xi(N) \text{ and } x(S) \leq \xi(S), \forall S \in \calN\}. 
\]
The set function $\xi$ is often assumed to be \emph{submodular}, i.e., for all $S$ and $T$ in $\calN$, we have
\[
\xi(S \cup T) + \xi(S \cap T) \leq \xi(S) + \xi(T). 
\]
Jack Edmonds studied a particular class of submodular set functions, the matroid rank functions.\footnote{\fullcite{edmonds1970submodular}} He defined a family of polytopes inheriting properties similar to those of matroids, called the \emph{polymatroids}, on which he developed well-known optimization algorithms. He also gives a detailed description of the facial structure of these polytopes, which inspired some results about cores of convex games\footnote{\fullcite{shapley1971cores}} from Shapley. Any base polyhedron is a polymatroid up to a translation, and both are cores of convex games. 

\medskip

Similar polytopes have attracted the attention of the combinatorialists for the last fifteen years. The \emph{generalized permutohedra},\footnote{\fullcite{postnikov2009permutohedra}} defined by Postnikov, are a generalization of the permutohedron, which has already been used in the mathematics of social sciences by \citeauthor{guilbaud1963analyse},\footnote{\fullcite{guilbaud1963analyse}} who gave it its name. The (standard) permutohedron, denoted by $\Pi_N$, is the convex hull of a set of $n!$ points defined by permuting the coordinates of the vector $(1, \ldots, n) \in \bbR^N$. However, some authors define the permutohedron to be the convex hull of the permutations of any vector with distinct coordinates, which yields a polytope with the same combinatorial properties: they have the same face lattice and the same normal fan. 

\medskip 

To any permutohedron, a deformation can be applied, which consists of parallel translations of some of its facets without crossing any vertices, i.e., without changing any edge direction. These deformations are the generalized permutohedra. It turns out that generalized permutohedra,\footnote{\fullcite{castillo2022deformation}} base polytopes of submodular functions, and polymatroids (up to translation), are the same polytopes. They all coincide with cores of convex games, and game theory has often taken advantage of the results in combinatorial optimization or combinatorics. 

\medskip 

The aim of this thesis is to contribute to the development of (cooperative) game theory in the direction described above, towards combinatorics and geometry. This thesis consists of four distinct chapters, covering different aspects of my doctoral work. 

\medskip 

The first chapter consists of the preliminaries, where the vocabulary and the notation are presented. The second chapter is based on the paper written with my supervisors,\footnote{\fullcite{laplace2023minimal}} in which we construct an algorithm generating the minimal balanced collections, based on a method described by Bezalel Peleg.\footnote{\fullcite{peleg1965inductive}} Secondly, we describe a collection of computational tools based on algorithmic implementations of the Bondareva-Shapley Theorem. We extend the scope of applications of these computational tools to a new class of polyhedron, called the \emph{basic polyhedra}. 

\medskip 

In the third chapter, the links between combinatorics and set functions are investigated. First, some similarities with the balanced collections and other objects studied in combinatorics, in particular the uniform and regular hypergraphs, are highlighted. Then, I prove that the (minimal) balanced collections form a \emph{combinatorial species}, concept introduced by André Joyal\footnote{\fullcite{joyal1981theorie}} to provide an abstract, systematic method for deriving the generating functions of discrete structures. Subsequently, I build the \emph{species of $k$-uniform hypergraphs of size $p$}, for $k$ and $p$ two given positive integers, as an intermediary step to construct the species of balanced collections. In the second part of this chapter, I present and study the links between the theory of hyperplane arrangements, in particular the \emph{resonance arrangement}, and cooperative game theory. 

\medskip 

The last chapter is devoted to the applications of the two previous chapters. First, an algorithm based on a characterization of \citeauthor{grabisch2021characterization}\footnote{\fullcite{grabisch2021characterization}} for stablity of the core of a given game is presented. Secondly, a new necessary condition for core stability is given, based on the verification of the nonemptiness of a specific cone. Finally, a collection of tools is introduced to study projections between preimputations, onto affine subspaces of interest. An explicit formula is given, as well as a few algorithmic procedures. We finish the chapter with an application of these projection operators to market failures, using the model of \emph{market games} introduced by \citeauthor{shapley1969market}.\footnote{\fullcite{shapley1969market}}



\chapter{The geometry of a game} 

\label{ChapterA} 




This chapter mainly consists of the preliminaries of the thesis. First, we recall some common notation from game theory, linear algebra, polyhedra theory and Euclidean geometry. We also introduce some new notation of geometric nature, in the aim of spatially representing specific features of a game. 

\medskip 

In the second part, we introduce the core and the balanced collections, which are the main subjects of study in this thesis. We highlight deep connections between the core of convex games and polytopes defined in neighboring areas of cooperative game theory, such as combinatorial optimization or combinatorics. Following this idea, we introduce and give some examples of totally balanced games, and investigate the connections between these and discrete mathematics, as has been done for convex games.

\section{The Euclidean space of payment vectors} \label{sec: euclidean space}

\label{sym: N} \label{sym: power} \label{sym: set-function} \label{sym: v} \label{sym: empty}
Let $N$ be a finite nonempty set of cardinality $n$, and denote by $2^N$ its power set, i.e., the set of all subsets of $N$. A (real) \emph{set function} on $N$ is a mapping $\xi: 2^N \to \bbR$, assigning a real number $\xi(S)$ to any subset $S$ of $N$. According to \textcite{von1959theorie}, a (cooperative transferable utility) \emph{game} is an ordered pair $(N, v)$, with $N$ a finite set whose elements are called the \emph{players}, and $v$ is a set function satisfying $v(\emptyset) = 0$ called the \emph{coalition function}, or the \emph{characteristic function} of the game. 

\medskip

\label{sym: calN}
We denote by $\calN$ the set of nonempty subsets of $N$, called \emph{coalitions}. The number $v(S)$ is called the \emph{worth} of $S \in \calN$. Throughout this thesis, we denote by $\xi$ a typical set function and by $v$ a typical coalition function of a game. In this manuscript we follow the interpretation of \textcite{peleg2007introduction} of a coalition function: if a coalition $S$ forms, then its members get the amount $v(S)$ of money.

\medskip 

A large part of cooperative game theory is interested in the rules of allocation of $v(S)$ among the members of $S$. Another question in cooperative game theory, which interests us in this thesis, is to know at which scale the cooperation may occur. More specifically, we want to know at which level the commonly acquired money can benefit any subcoalition, and subsequently whether the players in specific subcoalitions can threaten to stop cooperating to ensure a quantity that satisfies them. These two properties, called \emph{balancedness} and \emph{stability}, are defined in the sequel.

\medskip 

\label{sym: RN}
In this manuscript, we denote by $\bbR^N$ the Euclidean vector space built as the Cartesian product of $n$ copies of $\bbR$, one for each player in $N$. We use the same notation $\bbR^S$ which only considers the players belonging to the coalition $S$. An element $x \in \bbR^N$ is called a \emph{payment vector} and associates a real number $x_i$ to any player $i \in N$. We denote by $x(S)$ the sum of the coordinates associated with the players of $S$: $x(S) = \sum_{i \in S} x_i$, which is called the \emph{payment} of $S$ at $x$. 

\begin{definition} \label{def: preimputation-upper}
A payment vector $x \in \bbR^N$ is called 
\begin{itemize}
\item a \emph{preimputation} if $N$ is \emph{effective} for $x$, i.e., if $x(N) = v(N)$. We denote the set of preimputations by $\bbX(v)$.
\item an \emph{upper vector} if, for all $S \in \calN$, we have $x(S) \geq v(S)$. We denote the set of upper vectors by $U(v)$. 
\end{itemize}
\end{definition}

\label{sym: AS}
The set $\bbX(v)$ forms a hyperplane of $\bbR^N$, and the set $U(v)$ is a polyhedral set unbounded from above \parencite{van1999prosperity}. We denote by $\bbX^\leq(v)$ the set of \emph{affordable} payment vectors for $N$, i.e., $\bbX^\leq(v) \coloneqq \{x \in \bbR^N \mid x(N) \leq v(N)\}$. An element of $\bbX^\leq(v)$ represents an allocation which can be achieved with the worth of $N$. A preimputation represents an allocation of the worth of the \emph{grand coalition} $N$ among its players, while an upper vector is an allocation that benefits any coalition. Let $S \in \calN \setminus \{N\}$ be a coalition. We denote by $\bbA_S(v)$ the set of preimputations for which $S$ is effective, i.e., 
\[
\bbA_S(v) \coloneqq \{x \in \bbX(v) \mid x(S) = v(S)\}. 
\]
If there is no risk of confusion about the game, we denote these sets by $\bbA_S$. Notice that these sets are affine subspaces of dimension $n-2$, and have the same normals for any game $v$, i.e., they are always orthogonal to the same set of vectors. The game is only responsible for the relative positions of these affine subspaces, as increasing the worth $v(S)$ of a coalition will solely shift the subspace along their normals. 

\medskip

\label{sym: core}
Each affine subspace $\bbA_S$ cuts the hyperplane $\bbX(v)$ in two halves, denoted by \[
\bbA^\geq_S \coloneqq \{x \in \bbX(v) \mid x(S) \geq v(S)\} \qquad \text{and} \qquad \bbA^\leq_S \coloneqq \{x \in \bbX(v) \mid x(S) \leq v(S)\}. 
\]
$\bbA^\geq_S$ is the set of preimputations benefiting $S$, and $\bbA^\leq_S$ is the set of preimputations affordable for $S$. The preimputations that benefit every coalition, i.e., preimputations being also upper vectors, are said to be \emph{coalitionally rational}, and their set, denoted by
\[
C(v) \coloneqq \bigcap_{S \in \calN} \bbA^\geq_S = \bbX(v) \cap U(v),
\]
is called the \emph{core} \parencite{shapley1955markets}. The core is a bounded convex polyhedral set, i.e. a \emph{polytope}, lying in the hyperplane $\bbX(v)$. 

\medskip

A game with a nonempty core, i.e., a game for which there exists a preimputation benefiting every coalition, is called a \emph{balanced game} for a reason that will be discussed in Section \ref{sec: the core} (see Theorem \ref{th: BS-first}). In particular, there exists no \emph{coalition structure} (see \textcite{aumann1974cooperative}), i.e., partition of $N$, such that the sum of the worths of the element of the coalition structure is strictly greater than the worth of $N$. For illustration, let $v$ be a game $v: N = \{a, b, c\} \to \bbR$ defined by:
\begin{center}
\begin{tabular}{c||c|c|c|c|c|c|c}
$S$ & $\{a\}$ & $\{b\}$ & $\{c\}$ & $\{a, b\}$ & $\{a, c\}$ & $\{b, c\}$ & $N$ \\
\midrule
$v(S)$ & 5 & 0 & 0 & 8 & 8 & 8 & 10 \\
\end{tabular}.
\end{center}
Let $x \in \bbX(v)$ be a preimputation. If $x$ satisfies $x(\{b, c\}) \geq 8$, then $x(\{a\}) \leq 2$. Therefore, the inequalities $x(\{b, c\}) \geq 8$ and $x(\{a\}) \geq 5$ are incompatible, and $C(v)$ is empty. Dually, we see that the players prefer to be organized according to the coalition structure $\{\{a\}, \{b, c\}\}$, to obtain $13$, rather than form the grand coalition $N$ and get only $10$. But it is not sufficient to look at coalition structures to know whether the core is nonempty. Let $(N,v')$ be another game $v':N = \{a, b, c\} \to \bbR$ defined by:
\begin{equation}
\label{table: game}
\begin{tabular}{c||c|c|c|c|c|c|c}
$S$ & $\{a\}$ & $\{b\}$ & $\{c\}$ & $\{a, b\}$ & $\{a, c\}$ & $\{b, c\}$ & $N$ \\
\midrule
$v'(S)$ & 0 & 0 & 0 & 8 & 8 & 8 & 10 \\
\end{tabular}.
\end{equation}
The only difference between these games is $v(\{a\})$, but it is sufficient for the players to prefer the grand coalition over the coalition structure $\{\{a\}, \{b, c\}\}$. Nevertheless, the core is still empty. Indeed, let $x \in U(v)$ be an upper vector. Then $x(\{a, b\}), x(\{a, c\}), x(\{b, c\}) \geq 8$, hence $2x(N) = x(\{a, b\}) + x(\{a, c\}) + x(\{b, c\}) \geq 24$, so that $x(N) \geq 12$. Then, no upper vector can be a preimputation, and $C(v) = \emptyset$. Let us relax the notion of coalition structure, and consider the three $2$-player coalitions: $\{\{a, b\}, \{a, c\}, \{b, c\}\}$. Each player is included in two of these coalitions. We can imagine that players are able of spending half of their time in each coalition. Because each coalition exists only for half of the time, the worth of it is also halved. Then, the players get a total of 
\[
\frac{1}{2} v(\{a, b\}) + \frac{1}{2} v(\{a, c\}) + \frac{1}{2} v(\{b, c\}) = \frac{3 \cdot 8}{2} = 12, 
\]
that is still greater than $10$, which they can get if they form the grand coalition. The collection $\calB = \{\{a, b\}, \{a, c\}, \{b, c\}\}$ together with the systems of weights $\lambda^\calB = (\lambda_{\{a, b\}} = \frac{1}{2}, \lambda_{\{a, c\}} = \frac{1}{2}, \lambda_{\{b, c\}} = \frac{1}{2})$, as well as the collection $\calB' = \{\{a\}, \{b, c\}\}$ with the weights $\lambda^{\calB'} = ( \lambda_{\{a\}} = 1, \lambda_{\{b, c\}} = 1)$ are examples of \emph{balanced collections} of coalitions.

\begin{definition}
Let $\calB \subseteq \calN$ be a collection of coalitions. We say that $\calB$ is a \emph{balanced collection} if there exists, for each $S \in \calB$, a positive weight $\lambda_S$ such that, for each player $i \in N$, we have $\sum_{S \in \calB_i} \lambda_S = 1$, with $\calB_i = \{S \in \calB \mid i \in S\}$. The quantities $\{\lambda_S \mid S \in \calB\}$ form a \emph{system of balancing weights}. 
\end{definition}

Balanced collections were introduced by \textcite{bondareva1963some} under the name of `$(q, \theta)$-covering', renamed later by \textcite{shapley1967balanced}. They both independently gave a characterization of the set of balanced games based on the balanced collections. The balanced games are exactly those games for which the weighted sum of the worths of the coalitions of any balanced collection does not exceed the worth of the grand coalition, i.e., for any balanced collection $\calB$ with a system of weights $\lambda^\calB$, we have $\sum_{S \in \calB} \lambda^\calB_S v(S) \leq v(N)$. In the same paper, Shapley identified a subset of the set of balanced collections, on which the characterization of balanced games remains valid. This characterization is now known as the Bondareva-Shapley Theorem, and is undoubtedly a fundamental theorem for the study of cooperative behavior. The Bondareva-Shapley Theorem will be discussed in Section \ref{sec: the core}. 

\medskip

Now, let us focus on a given coalition $S \varsubsetneq N$. If the proposed preimputation $x \in \bbX(v)$ satisfies $x(S) < v(S)$, the coalition $S$ can get a higher payment by leaving the grand coalition and getting its worth $v(S)$. Let $y \in \bbX(v)$ be a second preimputation. 

\begin{definition}
\label{def: domination}
We say that $y$ dominates $x$ via $S$, denoted $y \domS x$, if 
\begin{itemize}
\item $y$ is \emph{affordable}, i.e., $y(S) \leq v(S)$, 
\item $y$ \emph{improves $x$ on $S$}, i.e., for all $i \in S$, we have $y_i > x_i$, . 
\end{itemize}
We say that $y$ dominates $x$, noted $y \dom x$, if a coalition $S$ exists such that $y \domS x$. 
\end{definition}

For illustration, consider the game described in Table \eqref{table: game}. Then the payment vectors $x = (5, 3, 2)$, $y = (3, 2, 5)$ and $z = (2, 5, 3)$ are preimputations. We have that $x \ {\rm dom}_{\{a,b\}} \ y$, also $y \ {\rm dom}_{\{a,c\}} \ z$ and $z \ {\rm dom}_{\{b, c\}} \ x$. We see in particular that the domination relation is not transitive. However, the relation of being dominated via a specific coalition is transitive. 

\medskip

According to Gillies \cite{gillies1959solutions}, a coalition $S$ is called \emph{vital} is there exists two preimputation $x$ and $y$ such that $y \domS x$, and this domination is not achieved via any subset of $S$. He also proved that, when studying stability, we can ignore all nonvital coalitions. We will discuss more about vital coalition in Chapter \ref{ChapterB}. 

\medskip

The first condition for domination, affordability of $y$, expresses that the members of coalition $S$ are able to realize $y_{|S}$ when forming a coalition. The second condition, $y$ improves $x$ on $S$, requires that everyone in the coalition $S$ benefits from his payment according to $y$ compared to his payment according to $x$. According to our interpretation of the coalition function, this is what $v(S)$ represents: the amount of money the players of $S$ require to accept the cooperation. 

\medskip

In the more recent literature, a \emph{solution} is a map associating to each game a subset of $\bbX^\leq(v)$. But \textcite{von1944theory}, when they made the first proposal for a solution concept in the broad sense, defined a solution as being a subset of $\bbX^\leq(v)$ that we today call \emph{stable set}, or \emph{von Neumann-Morgenstern stable set}. 

\begin{definition}[\textcite{von1944theory}, \textcite{RM-817}] \leavevmode \newline
A subset $K$ of $V \subseteq \bbX^\leq(v)$ is a $V$-\emph{stable set} if it is
\begin{itemize}
\item \emph{internally stable}, i.e., for all $x \in K$, there is no $y \in K$ such that $y \dom x$,
\item \emph{externally stable}, i.e., for all $x \in V \setminus K$, there exists $y \in K$ such that $y \dom x$.
\end{itemize}
We simply call \emph{stable sets} the $\bbX^\leq(v)$-stable sets.
\end{definition}

This definition is the one of \textcite{RM-817}, who extended the definition of \textcite{von1944theory}. Originally, von Neumann and Morgenstern defined their solutions of a game on a smaller set of payment vectors. Let us denote by $I(v)$ the set of \emph{individually rational} preimputations, i.e, 
\[
I(v) \coloneqq \{x \in \bbX(v) \mid x_i \geq v(\{i\}), \; \text{for all } i \in N\}. 
\]
The elements of $I(v)$ are called \emph{imputations}. The solutions of von Neumann and Morgenstern were defined as $I(v)$-stable sets. Conveniently, an $I(v)$-stable set $K$ such that, for each $i \in N$, there exists an element $x \in K$ satisfying $x_i = v(\{i\})$ is a stable set. Indeed, a preimputation $y$ satisfying $y_i < v(\{i\})$ is dominated by $x$ via $\{i\}$, as already detected by \textcite{RM-817}.  

\medskip 

Notice that no preimputation can be dominated via $N$, no imputation can be dominated by a singleton, and that the core elements can not be dominated at all. 

\medskip

The definition of stable sets is very appealing, however there may be no stable sets, several stable sets, or even a continuum of them, and it is difficult to identify them \parencite{lucas1969proof, lucas1992neumann}. Thus, the concept of stable sets was not very successful till now. Therefore, the main solution has become the core. 

\medskip

Indeed, when nonempty, the core is unique, it consists of all coalitionally rational preimputations and is a polytope defined by a finite number of inequalities. Also, because no core element is undominated, the core is always internally stable and is contained in all the stable sets. However, the core is not necessarily externally stable: there can exist preimputations that are dominated solely by preimputations not belonging to the core. Nevertheless, if the core is externally stable, it is the unique stable set \parencite{driessen2013cooperative}. Studying core stability is the initial motivation of this thesis and will be discussed in Chapter \ref{ChapterD}. 

\medskip

\label{sym: excess} \label{sym: region}
By definition, a preimputation $x$ can be dominated via a coalition $S$ only if the coalition can \emph{improve upon} $x$ by leaving the grand coalition, i.e., $x(S) < v(S)$. In the sequel, we denote by $e(S, x)$ the additional quantity of money that coalition $S$ can acquire by itself, i.e., $e(S, x) \coloneqq v(S) - x(S)$, that we call the \emph{excess of $S$ at $x$}. Let $\calS \subseteq \calN$ be a collection of coalitions. We denote by $X_\calS(v)$ the set of preimputations upon which a coalition can improve if and only if it belongs to $\calS$, i.e., 
\[
X_\calS(v) \coloneqq \{x \in \bbX(v) \mid x(S) < v(S) \text{ if and only if } S \in \calS\}. 
\]
We say that the collection $\calS$ is \emph{feasible} if its associated \emph{region} $X_\calS(v)$ is nonempty. Note that the core is the region associated with the empty collection of coalitions $\calS = \emptyset$, hence the nonempty regions form a partition of $\bbX(v)$. 


\medskip

\label{sym: scalar-product} \label{sym: norm}
Throughout this manuscript, we equip the Euclidean vector space $\bbR^N$, and therefore $\bbX(v)$, with the usual scalar product, denoted by $\langle \cdot, \cdot \rangle$ and the associated norm, denoted by $\lVert \cdot \rVert$ defined, for all $x, y \in \bbR^N$, by 
\[
\langle x, y \rangle = \sum_{i \in N} x_i y_i, \quad \text{and} \quad \lVert x \rVert = \sqrt{\langle x, x \rangle}. 
\]
We say that a vector $x \in \bbR^N$ is a \emph{normal} of a subset $Y \subseteq \bbR^N$ if, for every element $y \in Y$, we have $\langle x, y \rangle = 0$. 

\begin{theorem}[Hilbert projection theorem] \label{th: hilbert-proj} \leavevmode \newline
For every $x \in \bbR^N$ and every nonempty closed convex $K \subseteq \bbR^N$, there exists a unique element $z \in K$ for which $\lVert x - z \rVert$ is equal to $\inf_{y \in K} \lVert x - y \rVert$. 
\end{theorem}

The element $z$ is called the \emph{projection of $x$ onto $K$}. We denote by $\pi_K: \bbR^N \mapsto K$ the map that assigns to $x \in \bbR^N$ its projection $\pi_K(x)$ onto $K$. The map $\pi_K$ is called the \emph{projector} onto $K$, or the \emph{projection operator} onto $K$. The following result is formulated by \textcite{bauschke2011convex}.

\begin{theorem}[Projection Theorem] \leavevmode \newline
Let $K$ be a nonempty closed convex subset of $\bbR^N$. Then, for all $x$ and $y$ in $\bbR^N$, 
\[
y = \pi_K(x) \quad \text{if and only if} \quad \big[ y \in K \quad \text{and for all } z \in K, \ \langle z - y, x - y \rangle \leq 0 \big].
\]
If $K$ is an nonempty affine subspace of $\bbR^N$, for all $x$ and $y$ in $\bbR^N$, 
\[
y = \pi_K(x) \quad \text{if and only if} \quad \big[y \in K \quad \text{and for all } z \in K, \langle z - y, x - y \rangle = 0 \big]. 
\]
\end{theorem}

\section{The affine geometry of preimputations} \label{sec: affine}





\label{sym: sigma}
This subsection focuses on the affine subspace $\bbX(v) \subseteq \bbR^N$. The linear subspace of $\bbR^N$ parallel to $\bbX(v)$ is denoted by $\Sigma$, and its elements are called \emph{side payments}. A side payment represents a translation in the space of preimputations, which is a redistribution of money between the players. Indeed, a preimputation $x$ must satisfy $x(N) = v(N)$, hence a side payment $\sigma$ satisfies $\sigma(N) = 0$. Recall that the affine subspace $\bbA_S$ and the affine halfsubspace $\bbA_S^\geq$ are defined by 
\[
\bbA_S = \{x \in \bbX(v) \mid x(S) = v(S)\}, \quad \text{and} \quad \bbA^\geq_S = \{x \in \bbX(v) \mid x(S) \geq v(S)\}.
\] 

\begin{proposition}
\label{prop: eta}
Let $S$ be a coalition. The vector $\eta^S$, defined by 
\[
\eta^S = \bfone^S - \frac{\lvert S \rvert}{n} \bfone^N, \qquad \text{with} \quad \bfone^S_i = \left\{ \begin{aligned}
& 1, & \text{if } i \in S, \\
& 0, & \text{otherwise}.
\end{aligned} \right.
\]
is a side payment and a normal of $\bbA_S$.
\end{proposition}

\proof
For any coalition $S$, we have $\eta^S(N) = \bfone^S(N) - \frac{\lvert S \rvert}{n} \bfone^N (N) = \lvert S \rvert - \lvert S \rvert = 0$, hence $\eta^S$ is a side payment. Let $x$ and $y$ be two elements of $\bbA_S$. We have 
\[
\langle \eta^S, x - y \rangle = \langle \eta^S, x \rangle - \langle \eta^S, y \rangle = x(S) - \frac{\lvert S \rvert}{n} x(N) - y(S) + \frac{\lvert S \rvert}{n} y(N). 
\]
As $x(S) = y(S) = v(S)$ and $x(N) = y(N) = v(N)$, $\langle \eta^S, x - y \rangle = 0$.
\endproof

We extend the definition above by setting $\eta^\emptyset = \vec{o}$, which coincides with $\eta^N$. The set of vectors $\{\eta^S \mid S \in \calN\}$ and the subspace $\Sigma$ do not depend on the game we are considering, and the coalition function only defines the position of $\bbA_S$ along the line generated by $\eta^S$. We now define the projectors used in this manuscript. 

\begin{proposition}
\label{prop: first-proj}
Let $S \in \calN \setminus \{N\}$ be a coalition and $x$ be a preimputation. Then
\[
\pi_{\bbA_S}(x) = x + \gamma_S(x) \eta^S,
\]
where $\gamma_S(x) \coloneqq e(S,x) / \lVert \eta^S \rVert^2$. Moreover, if $x(S) < v(S)$, we have $\pi_{\bbA_S}(x) \domS x$. 
\end{proposition}

\proof
First, notice that $\lVert \eta^S \rVert^2 = \langle \eta^S, \eta^S \rangle = \eta^S(S)$. Following the Projection Theorem, we first prove that $\pi_{\bbA_S}(x)$ belongs to $\bbA_S$. 
\[
(x + \gamma_S(x)\eta^S)(S) = x(S) + \gamma_S(x) \eta^S(S) = x(S) + \frac{e(S,x)}{\lVert \eta^S \rVert^2} \lVert \eta^S \rVert^2 = v(S). 
\]
Let $z$ be an element of $\bbA_S$. We have
\[ \begin{aligned}
\langle z - (x + \gamma_S(x)\eta^S), x - (x + \gamma_S(x)\eta^S) \rangle & = \langle z - x - \gamma_S(x) \eta^S, - \gamma_S(x) \eta^S \rangle \\
& = - \gamma_S(x) \left( \langle z, \eta^S \rangle - \langle x, \eta^S \rangle - \gamma_S(x) \langle \eta^S, \eta^S \rangle \right) \\
& = - \gamma_S(x) \left( z(S) - x(S) - \gamma_S(x) \lVert \eta^S \rVert^2 \right) \\
& = -\gamma_S(x) \left( e(S,x) - e(S,x) \right) = 0. 
\end{aligned} \]
We use the Projection Theorem to conclude the construction of the projector. To end this proof, notice that, if $e(S, x) > 0$, for all $i \in S$, we have $\eta^S_i = 1 - \frac{\lvert S \rvert}{n} > 0$.  
\endproof

\begin{figure}[ht]
\begin{center}
\begin{tikzpicture}[scale=0.2]
\draw (1.774, 10) node[above right]{$\bbA_S$} -- (11.774, -10);

\filldraw[black] (-16,-6) circle (5pt) node[above left]{$x$};
\filldraw[black] (4.6192, 4.3096) circle (5pt) node[above right]{$\pi_{\bbA_S}(x)$};

\draw[dashed] (-16, -6) -- (4.6192, 4.3096);
\draw (2.7178, 4.9434) -- (3.9854, 5.5772);
\draw (2.7178, 4.9434) -- (3.3516, 3.6758);

\draw[{Stealth[scale=1.5]}-{Stealth[scale=1.5]}] (-15.3763, -7.2474) -- (5.2429, 3.0622); 

\draw[-{Stealth[scale=1.2]}] (9.8192, -6.0904) -- (14, -4) node[below right]{$\eta^S$};

\path (-3.3, -2.7) node[below]{$\dfrac{e(S, x)}{\lVert \eta^S \rVert}$};
\end{tikzpicture}
\caption{Projection of $x$ onto $\bbA_S$.}
\label{fig: proj-1}
\end{center}
\end{figure}

If a coalition $S$ can improve upon a preimputation $x \in \bbX(v)$, the projection $\pi_{\bbA_S}(x)$ shares the excess of $S$ at $x$ equally among the players in $S$, and therefore the projection dominates $x$ via $S$. This fact motivates the use of these projectors in this work to study the domination relations between payment vectors and, more specifically, between preimputations and core elements. Moreover, $\gamma_S(x) \eta^S$ is the shortest side payment able to map the preimputation $x$ onto a preimputation that benefits the coalition $S$ (see Figure \ref{fig: proj-1}). It is the solution to the problem consisting of satisfying coalition $S$ with the shortest side payment. 

\medskip

The idea of projections between preimputations was already existing in the transfer scheme defined by \textcite{cesco1998convergent}. A transfer scheme, first defined and used by \textcite{stearns1968convergent}, is a sequence of preimputations defined by a sequence of transfers, i.e., translation by a side payment. The sequence of preimputations defined by Cesco is defined recursively by projecting onto $\bbA_S$ the last preimputation produced. His projector is defined, for every preimputation $x$, by 
\[
\pi_{\bbA_S}(x) = x + e(S, x) \left( \frac{\bfone^S}{\lvert S \rvert} - \frac{\bfone^{N \setminus S}}{\lvert N \setminus S \rvert} \right). 
\]
Cesco's projectors are identical to ours, but the choice of normals is different. He chose to have a normal of a different norm to have a formulation depending explicitly on the excess $e(S,x)$. Our choice of normals is motivated by the following theorem. 

\begin{theorem} 
Let $S$ and $T$ be coalitions. We have 
\[
\eta^S + \eta^T = \eta^{S \cup T} + \eta^{S \cap T} \qquad \text{and} \qquad \langle \eta^S, \eta^T \rangle = \eta^S(T). 
\]
\end{theorem} 

\proof Notice that if $S$ and $T$ are disjoint, we have 
\[
\eta^S + \eta^T = \bfone^S - \frac{\lvert S \rvert}{n} \bfone^N + \bfone^T - \frac{\lvert T \rvert}{n} \bfone^N = \bfone^{S \cup T} - \frac{\lvert S \rvert + \lvert T \rvert}{n} \bfone^N = \eta^{S \cup T}. 
\]
Otherwise we decompose $\eta^S = \eta^{S \setminus T} + \eta^{S \cap T}$, similarly for $\eta^T$, and we have
\[
\eta^S + \eta^T = \eta^{S \setminus T} + \eta^{S \cap T} + \eta^{T \setminus S} + \eta^{S \cap T} = \eta^{S \cup T} + \eta^{S \cap T}. 
\]
For the second property, the fact that $\eta^S$ is a side payment leads to
\[
\langle \eta^S, \eta^T \rangle = \langle \eta^S, \bfone^T \rangle - \frac{\lvert T \rvert}{n} \langle \eta^S, \bfone^N \rangle = \eta^S(T) - \frac{\lvert T \rvert}{n} \eta^S(N) = \eta^S(T).
\] 
\endproof

From this theorem, we can derive several properties on the set $\{\eta^S \mid S \in \calN\}$, in particular, the repartition of the $\eta^S$ is balanced around the origin of $\Sigma$.  


\begin{theorem}
A collection of coalitions $\mathcal{C} \subseteq \mathcal{N}$ is balanced if and only if there exists a set of positive numbers $\{\theta_S \mid S \in \mathcal{C}\}$ such that 
\[
\sum_{S \in \mathcal{C}} \theta_S \eta^S = 0
\]
\end{theorem}

\proof
Let $\mathcal{C}$ be a balanced collection. Then, there exists positive numbers $\{\lambda_S \mid S \in \mathcal{C}\}$ such that 
\[
\sum_{S \in \mathcal{C}} \lambda_S \bfone^S = \bfone^N. 
\]
Taking the sum of the coefficients of both sides gives that following identity:
\[
\sum_{S \in \mathcal{C}} \lambda_S \lvert S \rvert = n. 
\]
Then we have 
\[
\sum_{S \in \mathcal{C}} \lambda_S \eta^S = \sum_{S \in \mathcal{C}} \lambda_S \bfone^S - \frac{1}{n} \sum_{S \in \mathcal{C}} \lambda_S \lvert S \rvert \bfone^N = \bfone^N - \frac{1}{n} n \bfone^N = 0_{\bbR^N}. 
\]
Let assume now assume that there exists positive numbers $\{\theta_S \mid S \in \mathcal{C}\}$ such that 
\[
\sum_{S \in \mathcal{C}} \theta_S \eta^S = 0_{\bbR^N},
\]
and set $\alpha = \frac{1}{n} \sum_{S \in \mathcal{C}} \theta_S \lvert S \rvert > 0$. It follows that 
\[
\sum_{S \in \mathcal{C}} \theta_S \left( \bfone^S - \frac{\lvert S \rvert}{\lvert N \rvert} \bfone^N \right) = 0_{\bbR^N} \iff \sum_{S \in \mathcal{C}} \theta_S \bfone^S = \alpha \bfone^N \iff \sum_{S \in \mathcal{C}} \left( \alpha^{-1} \theta_S \right) \bfone^S = \bfone^N, 
\]
and then $\mathcal{C}$ is a balanced collection with weights $\{\alpha^{-1} \theta_S \mid S \in \mathcal{C}\}$. 
\endproof


Another surprising result following from the theorem is that, for any coalition $S$ and $T$ both distinct from $N$, if $n$ is a prime number, $\eta^S$ and $\eta^T$ cannot be orthogonal, because 
\[
\langle \eta^S, \eta^T \rangle = \eta^S(T) = \bfone^S(T) - \frac{\lvert S \rvert}{n} \bfone^N(T) = \lvert S \cap T \rvert - \frac{\lvert S \rvert \cdot \lvert T \rvert}{n},
\] 
and $\lvert S \rvert \cdot \lvert T \rvert$ cannot be a multiple of $n$. The combinatorial properties of the normals $\eta^S$ and their associated affine subspaces will be studied in greater detail in Chapter \ref{ChapterB}. 

\medskip

By Proposition \ref{prop: first-proj}, we know that for a preimputation $x$, if $x(S) < v(S)$, we have $\pi_{\bbA_S}(x) \domS x$. But we have no control on where the projection will be. For any preimputation $x$, the relative position of $\pi_{\bbA_S}(x)$ with respect to $\bbA_T$ depends on the excesses $e(S,x)$ and $e(T,x)$, but also on the scalar product $\eta^S(T) = \langle \eta^S, \eta^T \rangle$.

\medskip 

The value and the sign of $\langle \eta^S, \eta^T \rangle$ indicate how correlated the payments of coalitions $S$ and $T$ are, how much their interest overlap. Adding the side payment $\eta^S$ to a preimputation $x$ necessarily benefits coalition $S$, and the value of $\langle \eta^S, \eta^T \rangle$ indicates how much the translation by $\eta^S$ benefits coalition $T$. Indeed, 
\[
\left( x + \eta^S \right)(T) = x(T) + \eta^S(T) = x(T) + \langle \eta^S, \eta^T \rangle. 
\]
To summarize all this information, we define 
\[
\chi_S(T, x) \coloneqq e(S,x) \langle \eta^S, \eta^T \rangle - e(T, x) \lVert \eta^S \rVert^2.
\]

\begin{proposition} \label{prop: chi}
Let $S$ and $T$ be two coalitions, and let $x \in \bbX(v)$. Then
\[
\pi_{\bbA_S}(x) \in \bbA^\geq_T \quad \text{if and only if} \quad \chi_S(T,x) \geq 0.
\]
\end{proposition}

\proof
First, we study the excess of $T$ at the projection onto $\bbA_S$:
\[ \begin{aligned}
e(T, \pi_{\bbA_S}(x)) & = v(T) - x(T) - \gamma_S(x) \eta^S(T) \\
& = e(T,x) - \frac{e(S, x)}{\lVert \eta^S \rVert^2} \langle \eta^S, \eta^T \rangle \\
& = \frac{1}{\lVert \eta^S \rVert^2} \left( e(T, x) \lVert \eta^S \rVert^2 - e(S,x) \langle \eta^S, \eta^T \rangle \right) = \frac{-1}{\lVert \eta^S \rVert^2} \chi_S(T, x). 
\end{aligned} \]
The projection lies into $\bbA^\geq_T$ if and only if $e(T, \pi_{\bbA_S}(x))$ is non-positive and therefore if and only if $\chi_S(T, x)$ is non-negative. 
\endproof

\begin{figure}[ht]
\begin{center}
\begin{subfigure}{0.49\textwidth}
\begin{center}
\begin{tikzpicture}[scale=0.2]
\draw (1.774, 10) node[above right]{$\bbA_S$} -- (11.774, -10);
\draw (-12, -6) node[below left]{$\bbA_T$} -- (14, -6);

\filldraw[black] (-4,0) circle (5pt) node[above left]{$x$};
\filldraw[black] (4.6192, 4.3096) circle (5pt) node[above right]{$\pi_{\bbA_S}(x)$};
\filldraw[black] (-4, -6) circle (5pt) node[below]{$\pi_{\bbA_T}(x)$};

\draw[dashed] (-4, -0) -- (4.6192, 4.3096);
\draw[dashed] (-4, 0) -- (-4, -6);

\draw[-{Stealth[scale=1.2]}] (9.0192, -4.4904) -- (14, -2) node[below right]{$\eta^S$};
\draw[-{Stealth[scale=1.2]}] (8, -6) -- (8, -10) node[below left]{$\eta^T$};
\end{tikzpicture}
\caption{Case $\chi_S(T, x) < 0$.}
\end{center}
\end{subfigure}
\begin{subfigure}{0.49\textwidth}
\begin{center}
\begin{tikzpicture}[scale=0.2]
\draw (1.774, 10) node[above right]{$\bbA_S$} -- (11.774, -10);
\draw (-12, 2) node[below left]{$\bbA_T$} -- (12, 2);

\filldraw[black] (-10, -4) circle (5pt) node[below left]{$x$};
\filldraw[black] (5.0192, 3.5096) circle (5pt) node[above right]{$\pi_{\bbA_S}(x)$};

\draw[dashed] (-10, -4) -- (5.0192, 3.5096);

\draw[-{Stealth[scale=1.2]}] (9.0192, -4.4904) -- (14, -2) node[below right]{$\eta^S$};
\draw[-{Stealth[scale=1.2]}] (-10, 2) -- (-10, 6) node[above left]{$\eta^T$};
\end{tikzpicture}
\caption{Case $\chi_S(T, x) > 0$.}
\end{center}
\end{subfigure}
\caption{Illustrative example of Proposition \ref{prop: chi}.}
\label{fig: chi}
\end{center}
\end{figure}
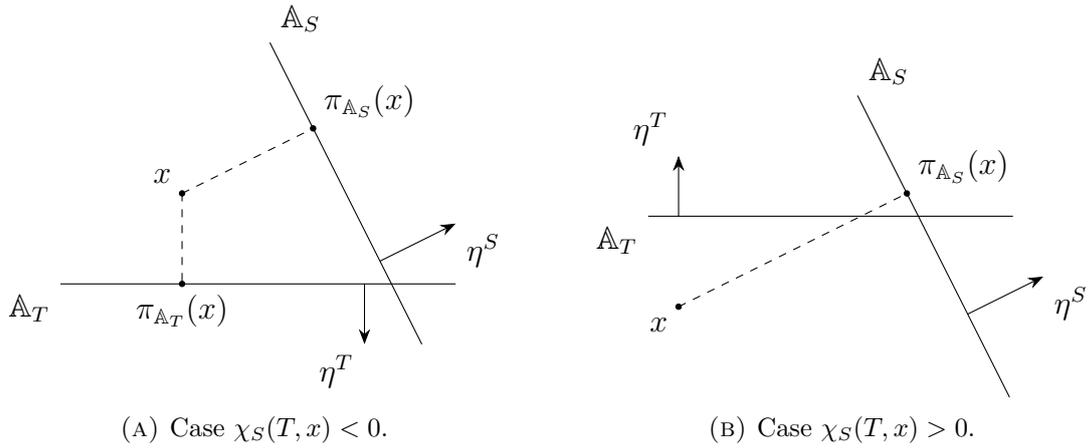

There are intuitive implications for this result. If $\langle \eta^S, \eta^T \rangle \geq 0$, increasing the payment of coalition $S$ by translating a given preimputation along $\eta^S$ also increases the payment of coalition $T$. Therefore, assuming that $e(S, x)$ is sufficiently large, the side payment $\sigma$ between $x$ and $\pi_{\bbA_S}(x)$ increases the payment of $S$ as well as the payment of $T$, and we have $\pi_{\bbA_S}(x)(T) = x(T) + \sigma(T) \geq v(T)$, as we can see on Figure \ref{fig: chi}. 

\medskip

For most of the preimputations, projecting onto an affine subspace $\bbA_S$ is not sufficient to be projected on the core. To do so, we need to be able to define projectors on any intersection of affine subspaces, to simultaneously improve the payment of all the coalitions that can improve upon the considered preimputation. This investigation continues in Section \ref{sec: projection-core} of Chapter \ref{ChapterD}.

\section{Games, balanced collections, and certain polyhedra} \label{sec: the core}

In this section, we discuss the core, its properties, what it represents in economics or any cooperative environment, and how this particular object can be found somewhere else in mathematics, in particular in theory of set functions and combinatorial optimization. The modern definition and usage of the core comes from \textcite{shapley1955markets} (see \cite{shubik1992game, zhao2018three}). For a game $(N, v)$, the \emph{core}, denoted by $C(v)$, is the intersection between the set of preimputations and the set of upper vectors, i.e., the set of preimputations benefiting every coalition:
\[
C(v) = \bbX(v) \cap U(v) = \bigcap_{S \in \calN} \bbA^\geq_S. 
\]
Let $v$ and $w$ be two set functions on the same domain $2^N$. We write $w \geq v$ if, for every $S \in 2^N$, we have $w(S) \geq v(S)$. Then, the core of $v$ can be seen as the set of additive set functions $w$ such that $w \geq v$ under the constraint $w(N) = v(N)$.

\begin{figure}[ht]
\centering
\includegraphics[scale=0.3]{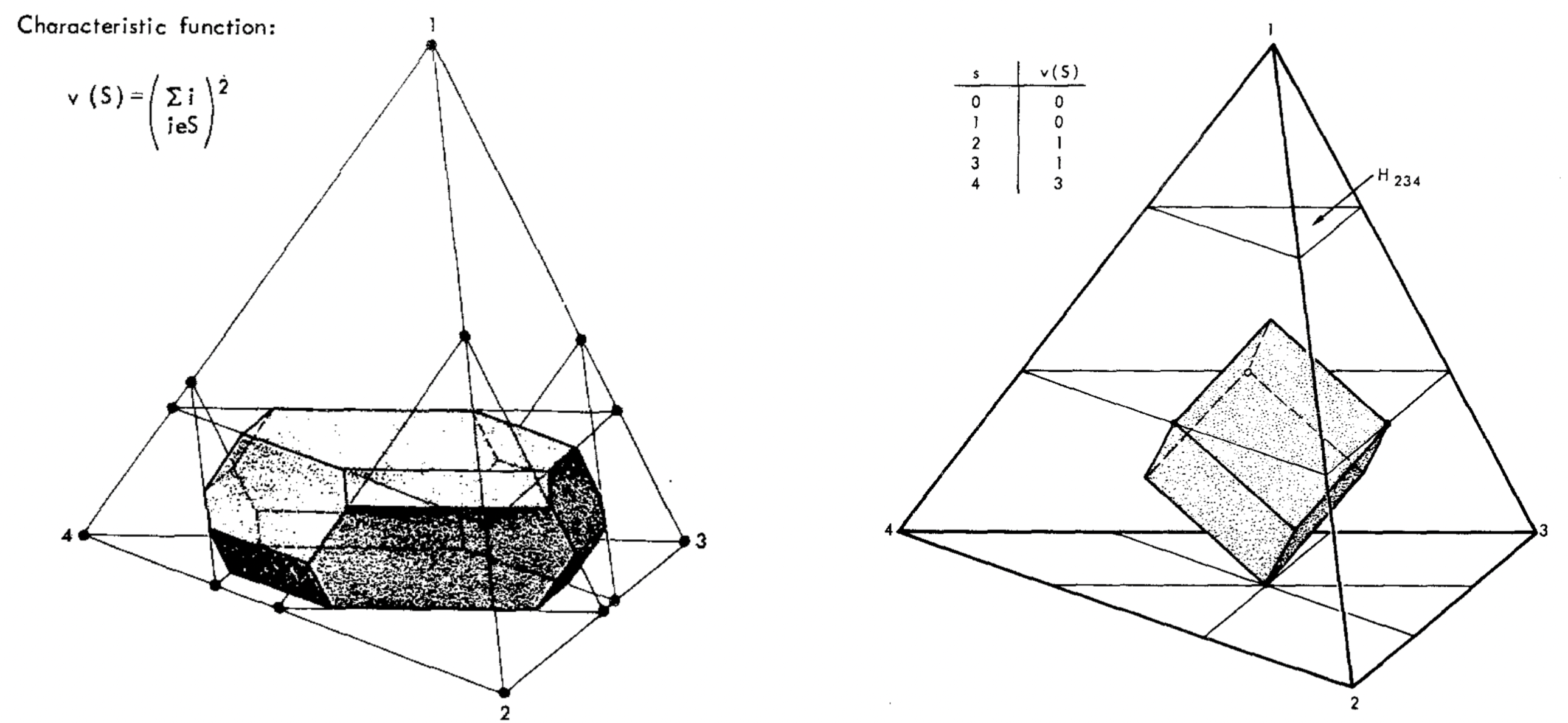}
\caption{Graphical illustrations of cores from \textcite{shapley1971cores}.}
\label{fig: core-Shapley}
\end{figure}

\medskip

In this thesis, we interpret the nonemptiness of the core as a necessary condition for the grand coalition $N$ to form, but not as a sufficient condition. It is necessary because if the core is empty, there exists no preimputation that benefits every coalition, and then some players will prefer to form other coalitions than $N$. However, even if the core is nonempty, there may exist no mechanism for the players to reach these coalitionally rational preimputations, as nothing indicates that all the players and coalitions will succeed to defend their interests. This mechanism can be the result of external intervention, for instance, a planner that has the power to choose which preimputations will be used to share the worth $v(N)$ among the players, in which case the nonemptiness of the core is a sufficient condition. 

\medskip

Nevertheless, if there is no external mediation between the players, a coalition $S$ that could improve upon the current preimputation can threaten to leave the grand coalition $N$. Due to the balancedness of the game, there is no balanced collection containing coalition $S$ which induces a total worth greater than $v(N)$, and then all players outside $S$ will be harmed by the departure of players in $S$. Hence, the players in $N$ never agree on a dominated preimputation. If the core is stable, every preimputation outside the core is dominated by a core element, and then the bargaining process eventually converges to a core element, and cooperative behaviour will emerge.

\subsection{Polyhedra and polytopes}

Before going into the details of the properties of the core, let us recall some notation and basic results about convex polytopes. Most of the notation in this subsection comes from the well-known textbook of \textcite{ziegler2012lectures}. First, in this manuscript, we consider a polytope $P$ in $\bbR^N$ as a convex bounded polyhedron, and a polyhedron is defined as a set of elements of $\bbR^N$ satisfying a finite set of linear inequalities. 

\medskip 

A polyhedron defined by a single linear inequality is called a \emph{half-space}, and a polyhedron defined by a set of linear inequalities which can all become tight (i.e., equalities) with an element $x \in \bbR^N$ is called a \emph{(affine) cone}, and $x$ is called an \emph{apex} of this cone. A polyhedron $P'$ obtained from $P$ by tightening a possibly empty subset of the set of inequalities defining $P$ is called a \emph{face} of $P$. 

\medskip 

\label{sym: aff}
Alternatively, a polytope is characterized by a set of \emph{extreme points}, each element of the polytope being a convex combination of these points. Because we do not assume that the linear inequalities are weak or strict, a polytope is not necessarily topologically closed, nor open, and some extreme points may not belong to the polytope. For a polytope $P \subseteq \bbR^N$, its \emph{affine hull} ${\rm aff}(P)$ is defined as the smallest affine subspace of $\bbR^N$ in which $P$ is included, and the dimension ${\rm dim}(P)$ of $P$ is defined as the dimension of its affine hull ${\rm aff}(P)$. 

\medskip

Polytopes are very popular in combinatorial optimization, especially in linear programming where they represent the set of \emph{feasible solutions}, i.e.,  the set of elements satisfying all the constraints of the optimization problem. A face $F$ of a polytope $P$ is a subset of $P$ where some linear form $L$ on $\bbR^N$ achieves its minimum on $P$, i.e., 
\[
F = \left\{x \in P \; \left| \; L(x) = \min_{y \in P} L(y) \right. \right\}. 
\]
We consider the empty set as a face of any polytope, which we call the empty face. Faces that consists of a single point are called \emph{vertices} of $P$, and correspond with the extreme points of the polytope. The $1$-dimensional faces are called \emph{edges} of $P$, and the $(d-1)$-faces are called \emph{facets} of $P$, where $d$ is the dimension of $P$. 

\medskip

\label{sym: calL(P)}
We call \emph{poset} a finite partially ordered set, that is, a finite set equipped with a binary relation which is reflexive, transitive and antisymmetric. A poset is a \emph{lattice} if every pair of elements of the poset has a supremum in the poset, called the \emph{join} and an infimum, called the \emph{meet}. The set of faces of a polytope $P \subseteq \bbR^N$, ordered by inclusion, defines a lattice called the \emph{face lattice} and is denoted by $\calL(P)$. We say that two polytopes are \emph{combinatorially equivalent} if their face lattices are isomorphic. We use interchangeably the notation $\calL(P)$ to denote the set of faces or the face lattice. 

\medskip

\label{sym: calK_P}
Let $F$ be a face of a polytope $P \subseteq \bbR^N$. We denote by $K_F$ the \emph{normal cone} of $F$ defined as the set of elements $y \in \bbR^N$ such that $\min_{x \in P} \langle y, x \rangle$ is achieved on $F$, i.e., 
\[
K_F \coloneqq \left\{y \in \bbR^N \; \left| \; F \subseteq \arg\min_{x \in P} \; \langle y, x \rangle \right. \right\}. 
\]
The set $\calK_P \coloneqq \{K_F \mid F \in \calL(P) \}$, is called the \emph{normal fan} of $P$. 

\begin{figure}[ht]
\begin{center}
\begin{subfigure}{0.48\textwidth}
\begin{center}
\begin{tikzpicture}[scale=0.5]
\filldraw[thick, fill=gray!15] (-2, 4) -- (2, 4) -- (4, -1) -- (0, -4) -- (-4, -1) -- cycle;

\draw[ultra thick] (-4, -1) -- node[above left]{$F$} (-2, 4);

\draw[dashed] (-2, 4) -- (-2, 6);
\draw[dashed] (2, 4) -- (2, 6);
\draw[dashed] (2, 4) -- (6, 5.6);
\draw[dashed] (4, -1) -- (6, -0.2);
\draw[dashed] (4, -1) -- (6, -3.667);
\draw[dashed] (0, -4) -- (1.5, -6);
\draw[dashed] (0, -4) -- (-1.5, -6);
\draw[dashed] (-4, -1) -- (-6, -3.667);
\draw[dashed] (-4, -1) -- (-6, -0.2);
\draw[dashed] (-2, 4) -- (-6, 5.6);

\filldraw[black] (4, -1) circle (4pt);
\path (2.5, -0.9) node[right] {$G$};
\end{tikzpicture}
\caption{A polytope $P$. The dashed rays go through the vertices and are orthogonal to the facets of $P$.}
\end{center}
\end{subfigure}
\hspace{0.1cm}
\begin{subfigure}{0.48\textwidth}
\begin{center}
\begin{tikzpicture}[scale=0.5]
\filldraw[gray!15] (0, 0) -- (6, 2.4) -- (6, -6) -- (4.5, -6) -- cycle;

\draw (0, 0) -- (0, 6);
\draw (0, 0) -- (6, 2.4);
\draw (0, 0) -- (4.5, -6);
\draw (0, 0) -- (-4.5, -6);
\draw[very thick] (0, 0) -- node[above right]{$K_F$} (-6, 2.4);

\path (4, -1) node[below] {$K_G$};
\end{tikzpicture}
\caption{The normal fan $\calK_P$.}
\end{center}
\end{subfigure}
\caption{A polytope $P$ and its normal fan $\calK_P$.}
\label{fig: normal-fan}
\end{center}
\end{figure}
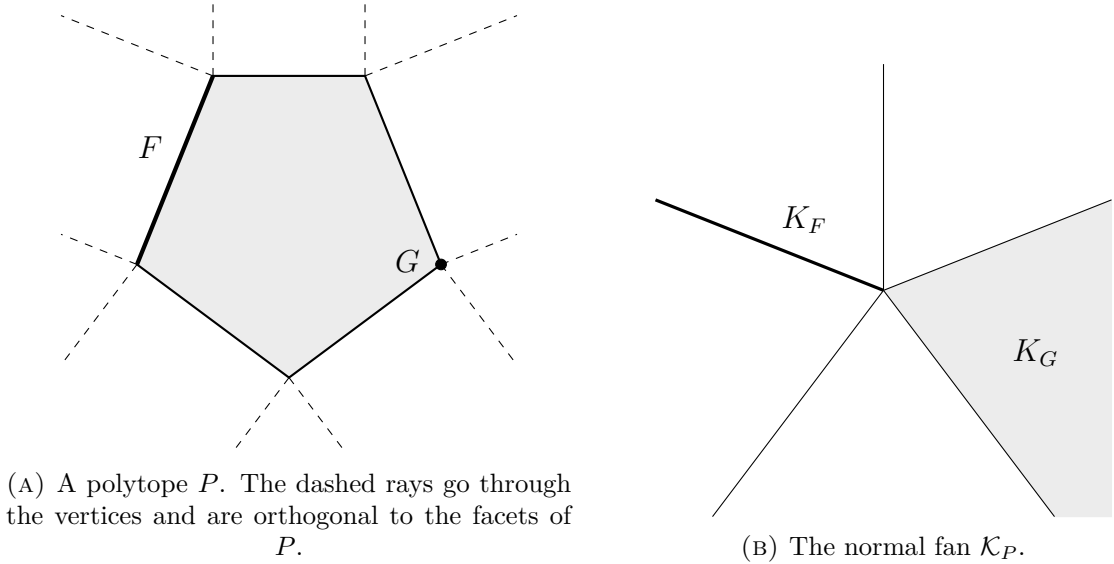

\noindent Let $P$ be a polytope, let $F, G \in \calL(P)$, and denote by $\calK_P$ its normal fan. Then:
\begin{itemize}
\item every nonempty face of a cone in $\calK_P$ is also a cone in $\calK_P$, 
\item the intersection of any two cones in $\calK_P$ is a face of both,
\item if $F \subseteq G$, then $K_F \supseteq K_G$, 
\item the cone $K_F \cap K_G$ is the normal cone of the supremum of $F$ and $G$ in $\calL(P)$. 
\end{itemize}  

\subsection{The minimal balanced collections}

Let $(N, v)$ be a game. Recall that a balanced collection on $N$ is a collection $\calB$ of coalitions of $N$ together with a set of positive weights $\{\lambda_S\}_{S \in \calB}$, called the \emph{balancing weights}, such that, for every player $i \in N$, the sum $\sum_{S \in \calB^i} \lambda_S$ is equal to one, with $\calB^i$ being the collection of coalitions of $\calB$ containing player $i$. In other words, we have
\[
\sum_{S \in \calB} \lambda_S \bfone^S = \bfone^N. 
\]
From a geometrical point of view, the set $\calB$ is balanced if $\bfone^N$ lies in the relative interior of the conical hull of $\{\bfone^S \mid S \in \calB\}$. Using the balanced collections, \textcite{bondareva1963some}, and then \textcite{shapley1967balanced} have found a characterization of games with nonempty cores. 

\begin{theorem}[Bondareva-Shapley Theorem, first version] \label{th: BS-first} \leavevmode \newline
A game $(N, v)$ has a nonempty core if and only if for every balanced collection $\calB$ on $N$ together with any balancing weights $\{\lambda_S\}_{S \in \calB}$, we have 
\begin{equation} 
\label{eq: BS-first}
\sum_{S \in \calB} \lambda_S v(S) \leq v(N). 
\end{equation}
\end{theorem}

This well-known theorem provides the adjective \emph{balanced} to name games having a nonempty core. The interpretation of the result is very natural: a balanced collection is a possible organization for the players, possibly spending fractions of their time among different coalitions. The weighted sum is the total amount of money gathered by the players in $N$ at the end of the game. Then, if the inequality \eqref{eq: BS-first} is not satisfied, there exists a balanced collection which is a more profitable organization for the players than $\{N\}$, therefore $N$ will not form. Notice that the nonexistence of a preimputation which benefits all coalitions is equivalent to the existence of a better organization for the players than $\{N\}$. This fact is a game-theoretical interpretation of the duality theorem of linear programming, which can be used to prove this theorem. 

\medskip

Indeed, consider the linear program $\min_{x \in U(v)} x(N)$. Replacing $U(v)$ by its definition leads to the following formulation:
\[
(P) \qquad \left\{ \begin{aligned}
& \min \, x(N) \\
& \text{ s.t. } x(S) \geq v(S), \quad \text{for all } S \in \calN.
\end{aligned} \right.
\]
Remark that this program is always feasible and that the core of the game is nonempty if and only if the value of the program is $v(N)$. The dual program is:
\[
(D) \qquad 
\left\{ \begin{aligned}
& \max \sum_{S \in \calN} \lambda_S v(S) \\
& \text{ s.t. } \left\{\begin{aligned}
& \sum_{S \in \calN} \lambda_S \bfone^S = \bfone^N, \text{ and} \\
& \lambda_S \geq 0, \text{ for all } S \in \calN.
\end{aligned} \right.
\end{aligned} \right. 
\]
The constraints of the dual program define the aforementioned balanced collections. Because the vector $\lambda^*$ defined by $\lambda^*_S = 0$ for all $S \in \calN$ and $\lambda^*_N = 1$ satisfies the constraints, the program has a solution. Furthermore, the value of the objective function for $\lambda^*$ is $v(N)$. It follows from the duality theorem that the core of the game is nonempty if and only if the value of $(D)$ is $v(N)$, i.e., for all balanced collections $\calB$ on $N$ with balancing weights $\{\lambda_S\}_{S \in \calB}$,
\[
\sum_{S \in \calB} \lambda_S v(S) \leq v(N). 
\]

In practice, this characterization cannot be used for algorithmic purposes because most of the balanced collections can have an infinity of balancing weights. Notice that the balancing weights form a polytope in $\bbR^\calN$, described by
\begin{equation}
\label{eq: polytope-weights}
F = \left\{ \lambda \in \bbR^\calN_+ \ \left| \ \sum_{S \in \calN} \lambda_S \bfone^S = \bfone^N \right. \right\}.
\end{equation}
Each point $\lambda \in F$ represents a balanced collection $\calB$ corresponding to the support of $\lambda$, and the balancing weights of $\calB$ are the corresponding positive coefficients of $\lambda$. Similarly, we can identify each coalition function $v$ with an $(2^n-1)$-dimensional vector; therefore, the vector space $\bbR^\calN$ can be seen as the space of games. Then, the ambient space of the polytope of balancing weights and the vector space of games, together with the usual scalar product is a \emph{dual pairing}, studied in functional analysis. 

\medskip

Because the balanced collections and their weights are associated with a polytope, they are determined by convex combinations of its vertices. 

\begin{definition}
A balanced collection is \emph{minimal} if it does not contain a proper subcollection that is balanced. 
\end{definition}

Shapley proved that the set of minimal balanced collections corresponds to the set of extremal points of $F$, therefore a minimal balanced collection $\calB$ has a unique set of balancing weights $\lambda^\calB$. From this, he stated this sharp version of the Bondareva-Shapley Theorem. For more details about this, see the monograph of \textcite{grabisch2016set}. 

\begin{theorem}[Bondareva-Shapley, sharp version]
\label{th: BS-sharp} \leavevmode \newline
The core of a game $(N, v)$ is nonempty if and only if for every minimal balanced collection $\calB$, we have 
\[
\sum_{S \in \calB} \lambda^\calB_S v(S) \leq v(N). 
\]
Furthermore, none of these inequalities is redundant except for $\calB = \{N\}$. 
\end{theorem}

The sharpness of the result comes from the fact that, if we choose a minimal balanced collection $\calB$, we can find a game that satisfies all the inequalities for minimal balanced collections different from $\calB$, but not the one corresponding to $\calB$. This new characterization requires a much smaller number of inequalities to be checked, provided that we know the minimal balanced collections on $N$. 

\medskip

In chapter \ref{ChapterB}, we present an algorithm checking the nonemptiness of the core, and show that it is significantly faster than the usual algorithms used in linear programming. The only requirement to use a Bondareva-Shapley-like algorithm is to know the set of minimal balanced collections. To do so, we completed and slightly improved the inductive method developed by \textcite{peleg1965inductive}, and implemented it into a working computer program. We are able to compute the set of minimal balanced collections on $\lvert N \rvert = 7$ and any subsets of $N$. This generation is discussed in Section \ref{sec: algorithm}. 

\subsection{Cores of convex games}

So far, we have studied games in a very general framework, coming from arbitrary grounded set functions. In this thesis, we are mainly focused on the core and the domination relation between the preimputations. Moreover, a lot of economic problems modeled with a game-theoretic framework as described here have some common and handy properties to use.

\begin{definition}
Let $\xi: 2^N \to \bbR$ be a set function. We say that $\xi$ is 
\begin{itemize}
\item \emph{grounded} if $\xi(\emptyset) = 0$, 
\item \emph{supermodular} if, for all $S$, $T \in \calN$, we have $\xi(S) + \xi(T) \leq \xi(S \cup T) + \xi(S \cap T)$,
\item \emph{submodular} if, for all $S$, $T \in \calN$, we have $\xi(S) + \xi(T) \geq \xi(S \cup T) + \xi(S \cap T)$. 
\end{itemize}
\end{definition}

Games with supermodular coalition functions are called \emph{convex} \parencite{shapley1971cores}. Many interaction situations can be modeled by these, for instance: production economy with landowners \parencite{shapley1967ownership, driessen2013cooperative}, bankruptcy games \parencite{aumann1985game, driessen2013cooperative}, common pool games with linear cost functions \parencite{o1982problem, driessen2013cooperative}, etc. \textcite{shapley1971cores} studied in great depth the properties of convex games, thanks to the extensive study of submodular set functions, and more specifically polymatroids, by \textcite{edmonds1970submodular}. In particular, convex games have nonempty cores, and their cores are (externally) stable. The result of the nonemptiness of the core comes from the theory of submodular set functions. For an exposition of this theory, see the monograph of \textcite{fujishige2005submodular}. 

\medskip

Let $\xi: 2^N \to \bbR$ be a grounded submodular set function. We denote by ${\rm P}(\xi)$ the \emph{submodular polyhedron} of $\xi$, defined by 
\[
{\rm P}(\xi) \coloneqq \{ x \in \bbR^N \mid x(A) \leq \xi(A), \text{ for all } A \in \calN \},
\]
and by ${\rm B}(\xi)$ the \emph{base polyhedron} of $\xi$, defined by ${\rm B}(\xi) \coloneqq \{x \in {\rm P}(\xi) \mid x(N) = \xi(N)\}$. Denote by $\xi^\#$ the \emph{conjugate set function} of $\xi$, defined, for all $S \subseteq N$, by 
\[
\xi^\#(N \setminus S) = \xi(N) - \xi(S). 
\]

\begin{figure}[ht]
\begin{center}
\begin{tikzpicture}[scale=0.7]
\draw[dashed, ->] (0, -2) -- (0, 5) node[above left]{\footnotesize $x_2$};
\draw[dashed, ->] (-2, 0) -- (7, 0) node[below right]{\footnotesize $x_1$};  

\path[font=\footnotesize] (0, 0) node[below left]{$\vec{o}$};

\draw (1, 3) -- (5, -1);
\draw (1, 3) -- (-2, 3);
\draw (1, 3) -- (1, 5);
\draw (5, -1) -- (5, -2);
\draw (5, -1) -- (7, -1);

\path[font=\footnotesize] (2, 3) node[right]{${\rm U}(v)$};
\path[font=\footnotesize] (0.5, 1.5) node[right]{${\rm P}(\xi)$};

\path[font=\footnotesize] (2.5, -0.5) node[below]{${\rm B(\xi)}$};
\path[font=\footnotesize] (5.5, 0.5) node[above]{${\rm C}(v)$};

\draw[->] (5.5, 0.5) to [out=-90, in=0] (4.5,-0.5);
\draw[->] (2.5, -0.5) to [out=90, in=180] (3.5, 0.5);

\end{tikzpicture}
\caption{Example of the submodular and base polyhedra of a set function $\xi$ and of the core and the set of upper vectors of $v = \xi^\#$.}
\end{center}
\end{figure}
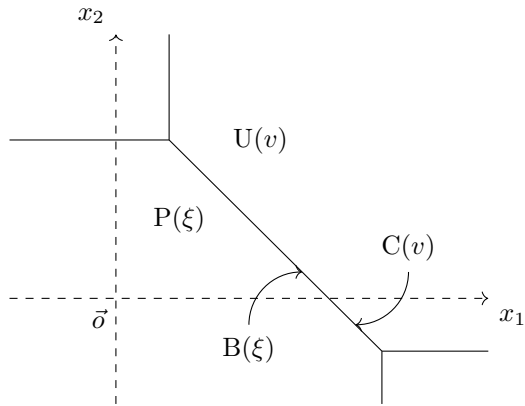

\begin{remark} \label{remark: B=C}
It is known that the conjugate set function $\xi^\#$ of a submodular set function $\xi$ is supermodular, and that ${\rm B}(\xi) = C(v)$ where $v = \xi^\#$ (see \parencite[][Lemma 2.4]{fujishige2005submodular}, \parencite{grabisch2016set}). The last identity does not require submodularity nor supermodularity to hold. 
\end{remark}

Therefore, the theory of submodular functions \parencite[][etc]{edmonds1970submodular, fujishige2005submodular} is, to some extent, dual to Shapley's theory of convex games. 
There remains an important difference between these two theories: the concept of external stability of the core/base polyhedron. There is no analogue in the set function theory. 

\medskip

Originally, Edmonds studied what he called a polymatroid, which is a polytope. His definition differs from the one of Fujishige, who defined a submodular set function under the same name of polymatroid. These two definitions are however deeply connected. For two vectors $x, y \in \bbR^N$, we write $x \leq y$ if, for all $i \in N$, we have $x_i \leq y_i$.

\begin{definition}[\textcite{edmonds1970submodular},  \textcite{schrijver2003combinatorial}] \leavevmode \newline
A \emph{polymatroid} $P$ in $\bbR^N_+$ is a compact nonempty subset of $\bbR^N_+$ such that
\begin{enumerate}
\item if $\vec{o} \leq y \leq x \in P$, then $y \in P$, 
\item for each $z \in \bbR^N_+$ there exists a number $\rho(z)$ such that each maximal vector $x$ of $P \cap \{x \mid x \leq z\}$ satisfies $x(N) = \rho(z)$. 
\end{enumerate}
\end{definition}

This definition resembles a polytopal equivalent of the following definition of a matroid: a \emph{matroid} is a pair $M = (N, B)$ with $N$ a finite set and $B$ a nonempty set of so-called \emph{independent} subsets of $N$ such that 
\begin{enumerate}
\item every subset of an independent set is an independent set, 
\item for every $A \subseteq B$, every maximal independent subset of $A$ has the same cardinality, called the \emph{rank}, $r(A)$, of $A$ (w.r.t. $M$).
\end{enumerate}

\begin{definition}[\textcite{fujishige2005submodular}] \leavevmode \newline
Let $\rho: 2^N \to \bbR$ be a grounded set function satisfying
\begin{enumerate}
\item $A \subseteq B \subseteq N$ implies $\rho(A) \leq \rho(B)$, 
\item for all $A, B \subseteq N$, we have $\rho(A) + \rho(B) \geq \rho(A \cup B) + \rho(A \cap B)$. 
\end{enumerate}
The pair $(N, \rho)$ is called a \emph{polymatroid} and $\rho$ is called \emph{rank function} of the polymatroid. 
\end{definition}

The two definitions are connected thanks to the forthcoming theorem. 

\begin{theorem}[\textcite{edmonds1970submodular}] \label{polymatroid} \leavevmode \newline
Let $\xi: 2^N \to \bbR$ be a grounded, non-decreasing, submodular set function. Then
$P(\xi) = \{x \in \bbR^N_+ \mid x(A) \leq \xi(A), \text{ for all } A \in \calN\}$ is a polymatroid. 
\end{theorem}

Another interesting polytope in the polyhedral combinatorics literature linked to the core is the \emph{(generalized) permutohedron}, derived from the \emph{permutohedron}.  

\label{sym: permutohedron} \label{sym: symmetric-group}
\begin{definition}[\textcite{postnikov2009permutohedra}] \leavevmode \newline
Let $x \in \bbR^N$ such that, for all $i, j \in N$, $x_i \neq x_j$. The \emph{permutohedron} $\Pi_N(x)$ is the convex polytope in $\bbR^N$ defined as the convex hull of all vectors obtained from $x$ by permutation of the coordinates: $\Pi_N(x) \coloneqq {\rm conv} \left\{ x^\sigma \mid \sigma \in \frakS_N \right\}$, where $\frakS_N$ is the group of permutations of $N$ and $x^\sigma = (x_{\sigma(1)}, \ldots, x_{\sigma(n)})$. 
\end{definition}




However, for some authors, the standard permutohedron $\Pi_N$ is defined as 
\[
\Pi_N = \Pi_N((n, n-1, \ldots, 2, 1)).
\]

\begin{example}[\textcite{lancia2018compact}] \label{ex: standard} \leavevmode \newline 
The standard permutohedron is the core of the strictly convex game $(N, v)$ defined, for all $S \in \calN$, by $v(S) = \frac{\lvert S \rvert \left( \lvert S \rvert + 1 \right)}{2}$. Figure \ref{fig: lancia-serafini} illustrates the case $n = 3$. 

\begin{figure}[ht]
\begin{center}
\begin{tikzpicture}[scale=0.8]
\draw[gray!30, dashed] (0, 0, 0) -- (0, 0, 6);
\draw[gray, dashed, ->] (0, 0, 6) -- (0, 0, 8);
\draw[gray!30, dashed] (0, 0, 0) -- (0, 6, 0);
\draw[gray, dashed, ->] (0, 6, 0) -- (0, 7, 0);
\draw[gray!30, dashed] (0, 0, 0) -- (6, 0, 0);
\draw[gray, dashed, ->] (6, 0, 0) -- (7, 0, 0);

\filldraw[black] (0, 0, 6) circle (1pt) node[above left]{\scriptsize $(0, 0, 6) \hspace{3pt}$};
\filldraw[black] (0, 6, 0) circle (1pt) node[left]{\scriptsize $(0, 6, 0) \hspace{3pt}$};
\filldraw[black] (6, 0, 0) circle (1pt) node[above right]{\scriptsize $(6, 0, 0)$};

\draw[black] (0, 0, 6) -- (0, 6, 0);
\draw[black] (0, 6, 0) -- (6, 0, 0);
\draw[black] (6, 0, 0) -- (0, 0, 6);

\filldraw[black] (1, 1, 4) circle (1pt) node[below]{\scriptsize $(1, 1, 4)$};
\filldraw[black] (1, 4, 1) circle (1pt) node[above]{\scriptsize $(1, 4, 1)$};
\filldraw[black] (4, 1, 1) circle (1pt) node[right]{\scriptsize $(4, 1, 1)$};

\fill[gray!15] (1, 1, 4) -- (1, 4, 1) -- (4, 1, 1) -- cycle;

\draw[black] (1, 1, 4) -- (1, 4, 1);
\draw[black] (1, 4, 1) -- (4, 1, 1);
\draw[black] (4, 1, 1) -- (1, 1, 4);

\fill[gray!30] (1, 2, 3) -- (2, 1, 3) -- (3, 1, 2) -- (3, 2, 1) -- (2, 3, 1) -- (1, 3, 2) -- cycle;

\filldraw[black] (1, 2, 3) circle (2pt) node[above left]{\scriptsize $(1, 2, 3)$};
\filldraw[black] (2, 1, 3) circle (2pt) node[below right]{\scriptsize $(2, 1, 3)$};
\filldraw[black] (3, 1, 2) circle (2pt) node[below right]{\scriptsize $(3, 1, 2)$};
\filldraw[black] (3, 2, 1) circle (2pt) node[above right]{\scriptsize $(3, 2, 1)$};
\filldraw[black] (2, 3, 1) circle (2pt) node[above right]{\scriptsize $(2, 3, 1)$};
\filldraw[black] (1, 3, 2) circle (2pt) node[above left]{\scriptsize $(1, 3, 2)$};

\draw[black] (1, 2, 3) -- (2, 1, 3) -- (3, 1, 2) -- (3, 2, 1) -- (2, 3, 1) -- (1, 3, 2) -- cycle;
\end{tikzpicture}
\caption{Drawing of the core (gray) of $(N, v)$. In white is the set $\bbX(v) \cap \bbR^N_+$, and in light gray is the set of imputations $I(v)$.}
\label{fig: lancia-serafini}
\end{center}
\end{figure}
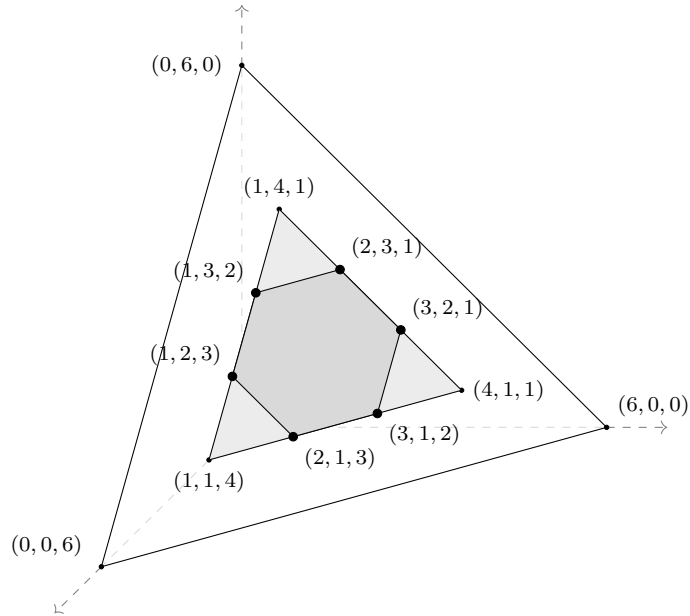

\end{example}

In polyhedral combinatorics, since \textcite{postnikov2009permutohedra}, people are studying deformations of permutohedra to deal with a more general type of polytopes. A \emph{generalized permutohedron} is a polytope obtained from a permutohedron by translating facets so that the directions of edges are preserved (i.e., are parallel), while some of the edges may accidentally degenerate into a single point. 

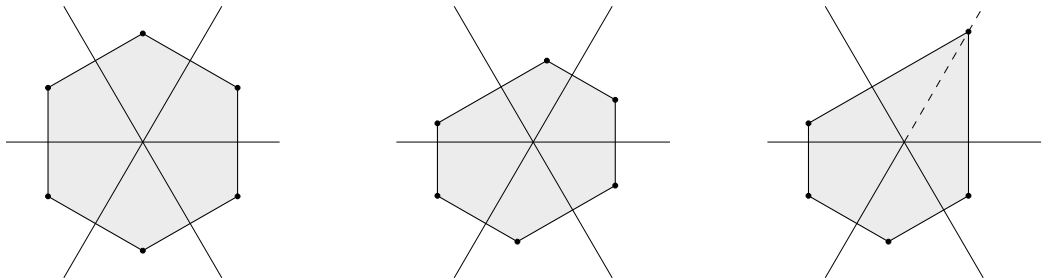
\begin{figure}[ht]
\begin{center}
\begin{subfigure}{0.32\textwidth}
\begin{center}
\begin{tikzpicture}[scale=0.18]
\filldraw[fill=gray!15] (0, 8) -- (6.9285, 4) -- (6.9285, -4) -- (0, -8) -- (-6.9285, -4) -- (-6.9285, 4) -- cycle;

\draw[very thin] (-10, 0) -- (10, 0);
\draw[very thin] (-5.774, -10) -- (5.774, 10);
\draw[very thin] (5.774, -10) -- (-5.774, 10);

\filldraw[black] (0,8) circle (5pt);
\filldraw[black] (6.9285, 4) circle (5pt);
\filldraw[black] (6.9285, -4) circle (5pt);
\filldraw[black] (0, -8) circle (5pt);
\filldraw[black] (-6.9285, -4) circle (5pt);
\filldraw[black] (-6.9285, 4) circle (5pt);
\end{tikzpicture}
\caption{Standard permutohedron.}
\end{center}
\end{subfigure}
\begin{subfigure}{0.64\textwidth}
\begin{center}
\begin{tikzpicture}[scale=0.18]
\filldraw[fill=gray!15] (1, 6) -- (6, 3.113) -- (6, -3.2048) -- (-1.1528, -7.3348) -- (-7, -3.9586) -- (-7, 1.3808) -- cycle;

\draw[very thin] (-10, 0) -- (10, 0);
\draw[very thin] (-5.774, -10) -- (5.774, 10);
\draw[very thin] (5.774, -10) -- (-5.774, 10);

\filldraw[black] (1, 6) circle (5pt);
\filldraw[black] (6, 3.113) circle (5pt);
\filldraw[black] (6, -3.2048) circle (5pt);
\filldraw[black] (-1.1528, -7.3348) circle (5pt);
\filldraw[black] (-7, -3.9586) circle (5pt);
\filldraw[black] (-7, 1.3808) circle (5pt);
\end{tikzpicture}
\hspace{1cm}
\begin{tikzpicture}[scale=0.18]
\filldraw[fill=gray!15] (4.6969, 8.1346) -- (4.6969, -3.9572) -- (-1.1528, -7.3348) -- (-7, -3.9586) -- (-7, 1.3808) -- cycle;

\draw[very thin] (-10, 0) -- (10, 0);
\draw[very thin] (-5.774, -10) -- (0, 0); 
\draw[very thin, dashed] (0, 0) -- (5.774, 10);
\draw[very thin] (5.774, -10) -- (-5.774, 10);

\filldraw[black] (4.6969, 8.1346) circle (5pt);
\filldraw[black] (4.6969, -3.9572) circle (5pt);
\filldraw[black] (-1.1528, -7.3348) circle (5pt);
\filldraw[black] (-7, -3.9586) circle (5pt);
\filldraw[black] (-7, 1.3808) circle (5pt);
\end{tikzpicture}
\caption{Generalized permutohedra.}
\end{center}
\end{subfigure}
\caption{Three generalized permutohedra. The lines represent the symmetry axis of coordinate transpositions.}
\end{center}
\end{figure}

More formally, we have the following definition.

\begin{definition}[\textcite{postnikov2009permutohedra}] \leavevmode \newline
A \emph{generalized permutohedron} is the convex hull of $n!$ points $x^w$ labeled by $w \in \frakS_N$ such that, for any $w \in \frakS_N$ and any adjacent transposition $\tau_i = (i, i+1)$, we have 
\[
x^w - x^{\tau_i(w)} = \alpha^{w, i} \left( \bfone^{\{w(i)\}} - \bfone^{\{w(i+1)\}} \right) \qquad \text{where } \alpha_{w, i} \geq 0. 
\]
\end{definition} 


Postnikov proved that all permutohedra are combinatorially equivalent (i.e., have isomorphic face lattices), and reformulated the inequality of \textcite{rado1952inequality} to show that there exists a bijection between the facets of any permutohedron $\Pi_N$ and $\calN$. Then, generalized permutohedra are parameterized by sets of $2^n - 1$ coordinates $\{z_S\}_{S \in \calN}$, one for each coalition. However, not all the sets of $2^n - 1$ numbers can generate a generalized permutohedron. Each generalized permutohedron is of the form
\[
\left\{ x \in \bbR^N \mid x(N) = z_N, \text{ and } x(S) \geq z_S \text{ for all } S \in \calN \right\}. 
\]
Here, we recognize the definition of the core, where the parameters $\{z_S \mid S \in \calN\}$ are the worths $v(S)$. \textcite{castillo2022deformation} called these parameters \emph{deforming vectors}. 

\medskip

A convex game induces a deformation of the standard permutohedron. Moreover, generalized permutohedra are deeply connected to submodular functions by the Submodularity Theorem \parencite{castillo2022deformation}. 

\begin{theorem}[Submodularity Theorem] \label{th: submodularity} \leavevmode \newline
There exists a bijection between generalized permutohedra $P$ with $\dim(P) \leq n-1$ and grounded submodular set functions on $2^N$. 
\end{theorem}

Another proof of the Submodularity Theorem can be found in \textcite{rehberg2022pruned}. The relations between these polytopes are summarized in the following table. 

\vspace{0.2cm}

\begin{figure}[ht]
\begin{center}
\begin{tabular}{c|c|c|c}
\begin{minipage}{3cm} \centering Set functions \end{minipage} & Polymatroids & Permutohedra & Convex games \\
\midrule
\midrule
\begin{minipage}{3cm} \centering Submodular polyhedron \end{minipage} & \begin{minipage}{3cm} \centering Extended polymatroid \end{minipage} & - & Upper vectors \\
\midrule
- & Polymatroid & - & - \\
\midrule
Base polyhedron & Base polytope & \begin{minipage}{3cm} \centering Generalized permutohedron \end{minipage} & Core \\
\midrule 
- & - & Permutohedron & \begin{minipage}{3cm} \centering $C(v)$ with $v: S \mapsto  \frac{\lvert S \rvert ( \lvert S \rvert + 1)}{2}$ \end{minipage} \\
\midrule
\begin{minipage}{3cm} \centering \textcite{schrijver2003combinatorial}, \textcite{fujishige2005submodular} \end{minipage} & \begin{minipage}{3.1cm} \centering \textcite{edmonds1970submodular}, \textcite{schrijver2003combinatorial} \end{minipage} & \begin{minipage}{4.2cm} \centering \textcite{postnikov2009permutohedra},\\ \textcite{castillo2022deformation} \end{minipage} & \begin{minipage}{2.9cm} \centering \textcite{shapley1971cores}, \textcite{grabisch2016set} \end{minipage} \\
\end{tabular}
\caption{Similarly defined polyhedra in different theories.}
\label{table: convex-polytopes}
\end{center}
\end{figure}

To conclude this part on convex games, let us give some properties of their cores. We denote by $F_S$ the face of the core defined by $F_S \coloneqq {\rm C}(v) \cap \bbA_S$. The game is said to be \emph{exact} (Schmeidler, \cite{schmeidler1972cores}) if $F_S$ is nonempty for all coalition $S \in \calN$. We say the core of a balanced game $(N, v)$ has a \emph{regular configuration} if, for all $S, T \in \calN$, we have $F_S \cap F_T \subseteq F_{S \cap T} \cap F_{S \cup T}$. 

\begin{proposition}[\textcite{shapley1971cores}] \leavevmode \newline
A balanced game is convex if and only if its core has a regular configuration. Moreover, a convex game is exact. 
\end{proposition}

To any balanced game, we can associate a unique exact game with the same core. 

\begin{definition}
Let $(N, v)$ be a balanced game. The \emph{lower envelope}, or \emph{exact cover}, of $(N, v)$ is the game $(N, v_*)$ defined, for all $S \in \calN$, by $v_*(S) = \min_{x \in {\rm C}(v)} x(S)$. 
\end{definition}

Geometrically, the lower envelope of a game is constructed by ``pushing'' the hyperplanes $\bbA_S$ towards the core until they touch it. Indeed, the core of a game and the core of its lower envelope coincide \parencite{grabisch2016set}. 

\begin{proposition}
The lower envelope of a game $(N, v)$ is convex if and only if $C(v)$ is a generalized permutohedron. 
\end{proposition}

\proof It is a corollary of the Submodularity Theorem. Because $C(v)$ is a generalized permutohedron, there exists a unique submodular function $\xi$ for which ${\rm B}(\xi) = C(v)$. Using Remark \ref{remark: B=C}, we convert this set function $\xi$ into the coalition function $v' \equiv \xi^\#$ of a convex game. Then $C(v') = C(v)$ and because $(N, v')$ is convex, it is exact. Yet, the core being a compact polyhedron, the lower envelope of a game is unique, thus $(N, v')$ is the lower envelope of $(N, v)$. 
\endproof

As a conclusion, the theory of convex games and their cores is well developed. But a supermodular coalition function is a restrictive condition, and many social interactions or economic situations can not be modeled by convex games. To model these problems, we need a more general theory. 

\subsection{Totally balanced games}

The theoretical model of a game, and the mathematical concept of a coalition function without any specific structure, are hardly applicable in practice due to the tedious enumeration of the worths of all coalitions. However, the theory of cooperative games encompasses several more precise theories, with explicitly defined coalition functions, arising from concrete situations. These more precise theories makes the study of cooperative games relevant, because each of these inherits the results and the properties from a larger, abstract theory we study here.

\medskip 

In this subsection, we present the \emph{linear production games}, the \emph{market games} and the \emph{flow games}. More details about these games can be found in the survey of Tijs about combinatorial optimization games \cite{tijs1992lp}. Subsequently, we recall that each balanced game is ``equivalent'' to a game in each of these classes, while explaining what we mean by equivalent. Any balanced game showing no \textit{a priori} structure can therefore be converted into a game belonging to one of these three classes. There are many more classes of games that are equivalent to these three, which we can't study in depth here, for instance the glove-market games defined by Apartsin and Holzman \cite{apartsin2003core}, or risk allocation games defined by Cs{\'o}ka, Herings and K{\'o}czy \cite{csoka2009stable}.

\begin{definition}
Let $(N, v)$ be a game, and let $S \in \calN$ be a coalition. The subgame $(S, v_{|S})$ is the game on $S$ whose coalition function $v_{|S}$ coincides with $v$ on the subcoalitions of $S$. A game is \emph{totally balanced} if all of its subgames are balanced. 
\end{definition}

Kroupa and Studen{\'y} \cite{kroupa2019facets} have studied the cone formed by the totally balanced games, and identified the ones forming the facets of it. 


\paragraph{Linear production games.} The definition of a linear production games is due to \textcite{owen1975core}. Let $N$ be a finite nonempty set of players, and let $b^i \in \bbR^m$ be a nonnegative vector of resources assigned to player $i \in N$. From these resources, players can produce $p$ different goods which can be sold at a given market price. For any suitable $(j, k)$, the $j$-th good requires $a_{kj} > 0$ units of the $k$-th resource, and can be sold at a price $c_j \geq 0$. We denote by $A$ the matrix formed by the coefficients $a_{kj}$. The players included in $S \in \calN$ pool their resources, represented by $b^S \coloneqq \sum_{i \in S} b^i$, to produce finished goods in order to maximize their profit. Then the linear production game $(N, v)$ associated with $(A, b = \{b^i\}_{i \in N}, c)$ is defined, for all $S \in \calN$, by $v(S) = \max_{x \in \bbR^p} \ \big\{ \langle c, x \rangle \mid A x \leq b^S \big\}$.

\begin{theorem}[\textcite{rosenmuller2013game}] \leavevmode \newline
A nonnegative game is a linear production game if and only if it is totally balanced. 
\end{theorem}

\paragraph{Market games.} The definition of a market game comes from \textcite{shapley1969market}. Let $N$ be a finite set of players, or \emph{traders}, let $G$ be the non-negative orthant of a finite-dimensional vector space, called the \emph{commodity space}, let $A = \{a^i \mid i \in N\} \in G^N$ be an indexed set of elements in $G$, called the \emph{initial endowments}, and let $U = \{u^i \mid i \in N\}$ be an indexed set of continuous, concave functions $u^i: G \to \bbR$ called the \emph{utility functions}. For any coalition $S \in \calN$, a set of endowments $\{x^i\}_{i \in S} \subseteq G$ such that $\sum_{i \in S} x^i = \sum_{i \in S} a^i$ is called a \emph{feasible $S$-allocation} of the market $(N, G, A, U)$, and we denote their set by $X^S$. The \emph{market game} generated by this market is a game $(N, v)$ such that, for all $S \in \calN$, we have $v(S) = \max_{x \in X^S} \sum_{i \in S} u^i(x^i)$. 

\begin{theorem}[\textcite{shapley1969market}]  \leavevmode \newline
A game is a market game if and only if it is totally balanced. 
\end{theorem}

For a specific subclass of market games, called \emph{assignment games}, \textcite{solymosi2001assignment} found the set of games with a stable core. 

\paragraph{Flow games.} Another class of games with a wide range of applications is the class of flow games. These games are described by \textcite{kalai1982totally} as being useful for modeling problems of profit sharing in an integrated production system with alternative production routes. Let $G = (V, E)$ be a directed graph, with $V$ being the set of vertices containing an \emph{initial node} $s$ and a \emph{terminal node} $t$, $E$ being the set of edges, and let $N$ be a finite set of players. Let $u: E \mapsto \bbR$ associate each edge with its \emph{capacity} and let $p: E \mapsto N$ associate each edge with a player which owns it. 

\begin{figure}[ht]
\begin{center}
\begin{tikzpicture}[node distance={23mm}, thick, main/.style = {draw, circle}] 
\node[main] (s) {$s$}; 
\node[main] (1) [right of=s] {$1$}; 
\node[main] (2) [below right of=s] {$2$}; 
\node[main] (3) [right of=1] {$3$};
\node[main] (4) [below right of=1] {$4$};
\node[main] (t) [below right of=3] {$t$};

\draw[->] (s) to node[midway, above] {\small $3, a$} (1);
\draw[->] (s) to node[midway, below left] {\small $3, b$} (2);
\draw[->] (1) to node[midway, above] {\small $3, b$} (3);
\draw[->] (1) to node[midway, right] {\small $\; 2, a$} (2);
\draw[->] (2) to node[midway, below] {\small $2, c$} (4); 
\draw[->] (3) to node[midway, above right] {\small $2, a$} (t);
\draw[->] (3) to node[midway, left] {\small $4, c \;$} (4);
\draw[->] (4) to node[midway, below] {\small $3, a$} (t);
\end{tikzpicture} 
\end{center}
\caption{Directed graph describing a flow game with $N = \{a, b, c\}$.}
\label{fig: flow}
\end{figure}
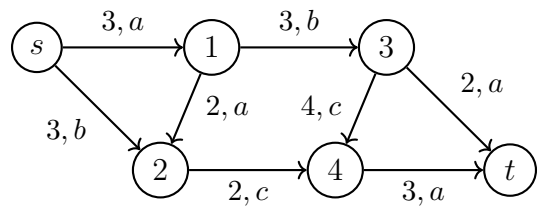

For a coalition $S \in \calN$, let $G^S$ be the subgraph restricted to the edges owned by a player in $S$. Then the coalition function $v$ of the flow game $(N, v)$ associated with $G$ maps the coalitions $S$ to the maximal amount of flow carried throughout $G^S$ between the node $s$ and the node $t$. Following the example depicted in Figure \ref{fig: flow}, the coalition function of the associated flow game is

\begin{center}
\begin{tabular}{c||c|c|c|c|c|c|c}
$S$ & $\{a\}$ & $\{b\}$ & $\{c\}$ & $\{a, b\}$ & $\{a, c\}$ & $\{b, c\}$ & $N$ \\
\midrule
$v(S)$ & 0 & 0 & 0 & 2 & 2 & 0 & 5 \\
\end{tabular}.
\end{center}

\noindent The flow corresponding to the coalition $N$ is 

\begin{center}
\begin{tikzpicture}[node distance={23mm}, thick, main/.style = {draw, circle}] 
\node[main] (s) {$s$}; 
\node[main] (1) [right of=s] {$1$}; 
\node[main] (2) [below right of=s] {$2$}; 
\node[main] (3) [right of=1] {$3$};
\node[main] (4) [below right of=1] {$4$};
\node[main] (t) [below right of=3] {$t$};

\draw[->] (s) to node[midway, above] {\small $3$} (1);
\draw[->] (s) to node[midway, below left] {\small $2$} (2);
\draw[->] (1) to node[midway, above] {\small $3$} (3);
\draw[dashed, ->] (1) to (2);
\draw[->] (2) to node[midway, below] {\small $2$} (4); 
\draw[->] (3) to node[midway, above right] {\small $2$} (t);
\draw[->] (3) to node[midway, left] {\small $1$} (4);
\draw[->] (4) to node[midway, below] {\small $3$} (t);
\end{tikzpicture} 
\end{center}

\noindent with the dashed link not carrying any flow. The number associated to each edge is the flow transiting through the edge, and respects the capacity constraint of each of them. 

\begin{theorem}[\textcite{kalai1982totally}] \leavevmode \newline
A nonnegative game is a flow game if and only if it is totally balanced. 
\end{theorem}

As these examples have shown, totally balanced games represent a much wider class of economic situations than convex games and are solely defined as games having nonempty cores independently of the choice of the grand coalition. There are at least two other strong assets for totally balanced games which we present below. 

\medskip 

%

\textcite{shapley1969market} say that two games are \emph{domination-equivalent} if they have the same imputation sets and the same domination relations on them. Two domination-equivalent games have the same core, and they have precisely the stable sets. 

\medskip 

In the same paper, they defined the totally balanced cover of a game. 

\begin{definition}[\textcite{shapley1969market}] \leavevmode \newline
Let $(N, v)$ be a game, and denote by $\bbB(S)$ the set of minimal balanced collections on $S$. The \emph{(totally balanced) cover} of $(N, v)$ is the game $(N, \overline{v})$ defined, for all $S \in \calN$, by 
\[
\overline{v}(S) = \max_{\calB \in \bbB(S)} \sum_{T \in \calB} \lambda^\calB_T v(T).
\] 
\end{definition} 

\begin{proposition}[\textcite{shapley1969market}] \leavevmode \newline
A balanced game is domination-equivalent to its totally balanced cover. 
\end{proposition}

This result reduces the study of the external stability of cores of balanced games to the study of the external stability of cores of totally balanced games. Therefore, studying totally balanced games allows us to study a very broad class of games, but also through the d-equivalence, all the balanced games. This is why in this thesis we always consider if needed, totally balanced games. Unlike the convex games, it is not clear which type of set functions characterizes the totally balanced games, however \textcite{kalai1982totally} found a very interesting and useful characterization while studying flow games. Let us call those games that have an additive coalition function \emph{inessential}. A game $(N, v)$ is said to have a \emph{finite representation} \parencite{rosenmuller2013game} if there exists a finite set of inessential games $\{(N, w^i) \mid i \in I\}$ such that $v = \bigwedge_{i \in I} w^i$, i.e., for all $S \in \calN$, we have
\[
v(S) = \min_{i \in I} w^i(S). 
\]

\begin{theorem}[\textcite{kalai1982totally}] \label{theorem: finite-inessential} \leavevmode \newline
A game is totally balanced if and only if it has a finite representation.
\end{theorem}

This characterization links totally balanced games with tropical polynomials. 

\begin{definition}
A \emph{tropical polynomial} is a map $\psi: \bbR^N \to \bbR$ defined as the minimum of finitely many affine maps, whose linear parts have rational coefficients, i.e., 
\[
\psi: x \in \bbR^N \mapsto \min_{i \in I} \{c^i + \langle z^i, x \rangle \} \in \bbR, 
\]
where the coefficients of all $z^i$ are rational. 
\end{definition}

In the standard definition of a tropical polynomial, the coefficients $c^i$ are usually allowed to take the value $+ \infty$, provided that there is at least one finite coefficient. Because the term $c^i + l^i(x)$ is not taken into account by the $\min$ operator if $c^i$ is infinity, it plays no role in the map $\psi$. 

\medskip

We say that a game $(N, v)$ is \emph{rational} if its coalition function only takes rational values. Assuming that a game is rational should not limit much its range of economic applications, because a quantity of money is rarely expressed with irrational numbers, or can be approximated by rational numbers. 

\begin{proposition}
For every rational, totally balanced game $(N, v)$, there exists a tropical polynomial $\psi: \bbR^N \to \bbR$ such that, for all $S \in \calN$, we have $\psi \left( \bfone^S \right) = v(S)$. If $\psi$ is a tropical polynomial such that, for all $i \in I$, we have $c^i=0$, then there exists a totally balanced game which can be extended to it. 
\end{proposition}

\proof
It is a corollary of Theorem \ref{theorem: finite-inessential} and the Riesz-Fr{\'e}chet representation theorem. Let $v$ be the coalition function of a totally balanced game. By Theorem \ref{theorem: finite-inessential}, there exists a finite set of additive coalition functions $\{a^i \mid i \in I\}$ such that $v(S) = \min_{i \in I} a^i(S)$. Then we extend each $a^i$ using the representation theorem, and we obtain a tropical polynomial. Let now $\psi$ be a tropical polynomial with all the coefficients $c^i$ being zero. Then it is defined as the minimum of linear forms, which are associated with $n$-dimensional vectors, themselves being associated with inessential games forming a representation of a totally balanced game. 
\endproof

In the game-theoretical literature, different concepts of game extensions already exist. Back in 1972, \textcite{owen1972multilinear} defined a \emph{multilinear extension} for any game $(N, v)$. \textcite{algaba2004lovasz}, studied the Choquet integral (also called the Lovász extension, see \textcite{grabisch2016set}) in the context of cooperative game theory. The Choquet and the Sugeno integrals are also used as extensions of capacities in decision theory \parencite{grabisch2016set}. But this extension into a tropical polynomial aims to build deeper connections with combinatorial optimization and discrete geometry. 

\medskip

To summarize, in this chapter, we have presented the foundations on which this thesis is built. First, we have introduced the main objects that we will investigate: the preimputations, the stable sets, the core and the (minimal) balanced collections. Next, we provided the notation we use to study the Euclidean space of payment vectors and the affine subspace of preimputations, and we showed that these are relevant in the study of projections, which will be a substantial part of our contribution to cooperative game theory. Subsequently, we presented a few results about cores of convex games, as well as some similarities they share with other objects studied in other fields of mathematics, in particular in combinatorial optimization and discrete geometry. The results about cores of convex games are the starting point of the investigation of cores of totally balanced games, in particular their coincidence with the von Neumann-Morgenstern stable sets, which is the primary goal of this thesis.


\chapter{Balancedness for nonemptiness} 

\label{ChapterB} 




The aim of this chapter is three-fold: to describe the generation of minimal balanced collections, to implement the Bondareva-Shapley Theorem as an algorithm and compare this method with well-known linear programming algorithms, and finally to extend the range of applications of these algorithms to a larger class of polyhedra. A significant part of these results comes from \textcite{laplace2023minimal}.

\medskip 

Our algorithm generating the minimal balanced collection is based on Peleg's inductive method \cite{peleg1965inductive}. From the set of minimal balanced collections on a set of players $N$, we can construct the minimal balanced collections on $N \cup \{p\}$, with $p$ being a player not included in $N$, by means of four construction procedures. Three of these procedures consist of finding a specific set of coalitions in the collection, in which we can add the new player, possibly adding a singleton $\{p\}$ to the new collection. The fourth procedure generates minimal balanced collections on $N \cup \{p\}$ from pairs of collections on $N$, and ensures to have built all the collections. Thanks to this algorithm, we are able to give the numbers of minimal balanced collections till $n=7$, which coincide for $n \leq 4$ with the numbers given by Shapley \cite{shapley1967balanced}. The list of these numbers is accessible under the number A355042 of the Online Encyclopedia of Integer Sequences \cite{oeis}. 

\medskip 

In a second part, we show that our algorithm is relevant because it outspeeds the previous methods to generate the minimal balanced collections. Indeed, this generation is equivalent to the vertex enumeration of a specific polytope. We use the Avis-Fukuda algorithm to compare its performance with our algorithm. Using the minimal balanced collections, we implement the Bondareva-Shapley Theorem as a computer program, and compare it to classical linear programming algorithms. It turns out that our implementation of Bondareva-Shapley Theorem is more than seven times faster than the revised simplex method natively implemented in Python. 

\medskip 

Finally, we define a new class of polyhedra, called \emph{basic polyhedra}, on which it is possible to apply our generalization of the Bondareva-Shapley Theorem (Theorem \ref{th: basic}). Roughly speaking, a basic polyhedron is a polyhedron the adherence of which is the core of a well-chosen coalition function. These polyhedra are very common in cooperative game theory, as they are usually either defined as a set of solutions with specific properties, like the core, or their nonemptiness is equivalent to the satisfaction of a property. For the latter, exactness is a good example, because the nonemptiness of the basic polyhedron consisting of core elements giving a payment of $v(S)$ to a coalition $S$ is equivalent to the exactness of $S$. Moreover, all the properties equivalent to the nonemptiness of a basic polyhedron are prosperity properties.





\section{The algorithm} \label{sec: algorithm}

So far, there are two known methods for generating minimal balanced collections. The first one, due to \textcite{peleg1965inductive}, is specifically devoted to the generation of minimal balanced collections and proceeds by induction on the number of players $n$. The second one uses any vertex enumeration method for convex polyhedra, applied on the specific polytope described in equation \eqref{eq: polytope-weights}. One popular algorithm for vertex enumeration is the one of \textcite{avis1991pivoting}, which we will use for performance comparisons. 

\medskip

Peleg's method constructs, from the minimal balanced collections defined on a set $N$, all those that are defined on the set $N' = N \cup \{p\}$, with $p$ a new player that was not included in $N$. As far as we know, Peleg's inductive method has never been implemented as an algorithm, perhaps due to the rather abstract way it is described, far from any algorithmic considerations. For this reason, we translate Peleg's methods and results from an algorithmic point of view, reproving his results in our new formalism for the sake of clarity and completeness. In the following, the main result is divided into four cases where the fourth one is slightly generalized compared to Peleg's method. 

\medskip

Let $\calC = \{S_1, \ldots, S_k\}$ be a balanced collection of $k$ coalitions on $N$ and $p$ be a new player not in $N$. Denote by $[k]$ the set $\{1, \ldots, k\}$ for any positive integer $k$. If $\lambda^\calC$ is a system of balancing weights for $\calC$ and $I \subseteq [k]$ is a subset of indices, denote by $\lambda^\calC_I$ the sum $\sum_{i \in I} \lambda^\calC_{S_i}$. Also, denote by $A^\calC$ the incidence $(n \times k)$-matrix formed by the $k$ column vectors $\bfone^{S_1}, \ldots, \bfone^{S_k}$. Denote by $\rk \left(A^\calC \right)$ the rank of the matrix $A^\calC$, i.e., the dimension of the Euclidean space spanned by its columns viewed as $n$-dimensional vectors. 

\paragraph{First case.} Let $\calC$ be a minimal balanced collection on $N$. Take $I \subseteq [k]$ such that $\lambda^\calC_I = 1$. Denote by $\calC'$ the new collection in which the coalitions $\{S_i \mid i \in I\}$ contain the new player $p$ as an additional member and the other coalitions $\{S_j \mid j \in [k] \setminus I\}$ are kept unchanged. 

\begin{lemma} \label{lemma: first}
$\calC'$ is a minimal balanced collection on $N'$. 
\end{lemma}

\proof
Because $\calC$ is a minimal balanced collection, the equalities $\sum_{S \in \calC', S \ni i} \lambda^\calC_S = 1$ are already satisfied for any player $i \in N$. By definition of $I$, we also have that $\sum_{S \in \calC', S \ni p} \lambda^\calC_S = \lambda^\calC_I = 1$. Then, $\calC$ is balanced. Moreover, the minimality of $\calC'$ implies the minimality of $\calC'$. 
\endproof

\paragraph{Second case.} Let $\calC$ be a minimal balanced collection on $N$. Take $I \subseteq [k]$ such that $\lambda^\calC_I < 1$. Denote by $\calC'$ the new collection in which the coalitions $\{S_i \mid i \in I\}$ contain the new player $p$ as an additional member and the other coalitions $\{S_j \mid j \in [k] \setminus I\}$ are kept unchanged, and to which the coalition $\{p\}$ is added: 
\[
\calC' = \{S_i \cup \{p\} \mid i \in I\} \cup \{S_j \mid j \in [k] \setminus I\} \cup \{\{p\}\}. 
\]

\begin{lemma} \label{lemma: second}
$\calC'$ is a minimal balanced collection on $N'$. 
\end{lemma}

\proof
Because $\calC$ is a minimal balanced collection, the equalities $\sum_{S \in \calC', S \ni i} \lambda^\calC_S = 1$ are already satisfied for any player $i \in N$. Define $\lambda^{\calC'}$ such that $\lambda^{\calC'}_S = \lambda^\calC_S$ for $S \in \calC$ and $\lambda^{\calC'}_{\{p\}} = 1 - \lambda^\calC_I$. Therefore,
\[
\sum_{\substack{S \in \calC' \\ S \ni p}} \lambda^{\calC'}_S = \lambda^{\calC'}_{\{p\}} + \lambda^\calC_I = 1 - \lambda^\calC_I + \lambda^\calC_I = 1. 
\]
Then $\calC'$ is balanced. We cannot obtain a balanced subcollection of $\calC'$ by discarding one element of $\{S_i \mid i \in I\}$ because $\calC$ is minimal, nor can we discard coalition $\{p\}$ because $\lambda^\calC_I < 1$. So $\calC'$ is minimal. 
\endproof

\paragraph{Third case.} Let $\calC$ be a minimal balanced collection on $N$. Take a subset $I \subseteq [k]$ and an index $\delta \in [k] \setminus I$ such that $1 > \lambda^\calC_I > 1 - \lambda^\calC_{S_\delta}$. Denote by $\calC'$ the new collection in which the coalitions $\{S_i \mid i \in I\}$ contain the new player $p$ as an additional member, the other coalitions $\{S_j \mid j \in [k] \setminus I\}$ are kept unchanged, and to which the coalition $S_\delta \cup \{p\}$ is added: 
\[
\calC' = \{S_i \cup \{p\} \mid i \in I\} \cup \{S_j \mid j \in [k] \setminus I\} \cup \{S_\delta \cup \{p\}\}. 
\]

\begin{lemma} \label{lemma: third}
$\calC'$ is a minimal balanced collection on $N'$. 
\end{lemma}

\proof
Define $\lambda^{\calC'}$ by $\lambda^{\calC'}_S = \lambda^\calC_S$ for $S \in \calC \setminus \{S_\delta\}$, and
\[
\lambda^{\calC'}_{S_\delta \cup \{p\}} = 1 - \lambda^\calC_I \qquad \text{ and } \qquad \lambda^{\calC'}_{S_\delta} = \lambda^\calC_{S_\delta} - \lambda^{\calC'}_{S_\delta \cup \{p\}} > 0 \text{ (by assumption)}. 
\]
Let $i \in N$ be a player. If $i \not \in S_\delta$, by balancedness of $\calC$, $\sum_{S \in \calC', S \ni i}\lambda^{\calC'}_S = 1$. If $i \in S_\delta$, then
\[
\sum_{\substack{S \in \calC' \\ S \ni i}} \lambda^{\calC'}_S =  \lambda^{\calC'}_{S_\delta \cup \{p\}} + \lambda^{\calC'}_{S_\delta} + \sum_{\substack{S \in \calC \setminus \{S_\delta\} \\ S \ni i}} \lambda^{\calC'}_S = \lambda^\calC_{S_\delta} + \sum_{\substack{S \in \calC \setminus \{S_\delta\} \\ S \ni i}} \lambda^\calC_S = \sum_{\substack{S \in \calC \\ S \ni i}} \lambda^\calC_S,
\]
that is equal to $1$ by balancedness of $\calC$. Concerning player $p$, 
\[
\sum_{\substack{S \in \calC' \\ S \ni p}} \lambda^{\calC'}_{S_\delta \cup \{p\}} = \lambda^{\calC'}_{S_\delta \cup \{p\}} + \lambda^{\calC'}_I = 1 - \lambda^\calC_I + \lambda^\calC_I = 1. 
\]
Then $\calC$ is balanced. Because none of the coalitions $S \in \calC$ or $S_\delta \cup \{p\}$ can be discarded to obtain a balanced subcollection, the proof is finished. 
\endproof

\paragraph{Fourth case.} Let $\calC^1$ and $\calC^2$ be two distinct minimal balanced collections on $N$ such that $\calC = \calC^1 \cup \calC^2$ satisfies $\lvert \calC \rvert = k$ and $\rk \left(A^\calC\right) = k-1$. Define two systems of balancing weights for $\calC$, defined, for all $S \in \calC$, by 
\[
\mu_S = \left\{ \begin{aligned}
& \lambda^{\calC^1} \text{ if } S \in \calC^1, \\
& 0 \text{ otherwise}, 
\end{aligned} \right. \qquad 
\nu_S = \left\{ \begin{aligned}
& \lambda^{\calC^2} \text{ if } S \in \calC^2, \\
& 0 \text{ otherwise}. 
\end{aligned} \right. 
\]
Assume that there exists $I \subseteq [k]$ such that $\mu_I = \sum_{i \in I} \mu_{S_i} \neq \nu_I = \sum_{i \in I} \nu_{S_i}$ and 
\[
t^I = \frac{1 - \mu_I}{\nu_I - \mu_I} \in (0, 1). 
\]
Denote by $\calC'$ the new collection in which the coalitions $\{S_i \mid i \in I\}$ contain the new player $p$ as an additional member and the other coalitions $\{S_j \mid j \in [k] \setminus I\}$ are kept unchanged. 

\begin{lemma} \label{lemma: fourth}
$\calC'$ is a minimal balanced collection on $N'$. 
\end{lemma}

\proof
Define $\lambda^{\calC'}$, for all $S \in \calC'$, by $\lambda^{\calC'} = (1-t^I) \mu_S + t^I \nu_S$. Because $\lambda^{\calC'}$ is a convex combination of two systems of balancing weights of $\calC$, for all players $i \in N$ we have $\sum_{S \in \calC', S \ni i} \lambda^{\calC'}_S = 1$. Concerning player $p$, 
\[
\sum_{\substack{S \in \calC' \\ S \ni p}} \lambda^{\calC'}_S = \lambda^{\calC'}_I = \left( 1 - t^I \right) \mu_I + t^I \nu_I = \mu_I + t^I \left( \nu_I - \mu_I \right) = \mu_I + 1 - \mu_I = 1. 
\]
Then $\calC'$ is a balanced collection. Now, we prove the minimality of $\calC'$ as a balanced collection, i.e., the uniqueness of $\lambda^{\calC'}$ as a system of balancing weights for $\calC'$. The system of balancing weights for $\calC$, denoted by $\Lambda(\calC)$, is the convex set $\Lambda(\calC) = \conv(\mu, \nu)$, and then $\Lambda(\calC') \subseteq \Lambda(\calC)$. More precisely, to satisfy the balancedness for player $p$ we have $\Lambda(\calC') \subseteq \{\lambda \in \bbR^{k+1} \mid \lambda_I = 1\} \cap \conv(\mu, \nu) $. Because it is the intersection between two non-parallel one-dimensional sets, $\lambda(\calC')$ contains at most one solution, which is $\lambda^{\calC'}$. 
\endproof

\paragraph{Final algorithm.} It is now possible to construct, from the set of minimal balanced collections on a set $N$, the set of minimal balanced collections on another set $N' = N \cup \{p\}$ (see Algorithm \ref{construction}: \textsc{AddNewPlayer}). 

\begin{breakablealgorithm}
\caption{AddNewPlayer} \label{construction}
\begin{algorithmic}[1]
\Require A set of minimal balanced collections $\bbB(N)$ on a set $N$
\Ensure A set of minimal balanced collections $\bbB(N')$ on a set $N' = N \cup \{p\}$
\Procedure{AddNewPlayer}{$\bbB_N, p$}
\For{$(\calC^1, \calC^2) \in \bbB_N \times \bbB_N$}
\State $\calC \gets \calC^1 \cup \calC^2$ and $k \gets \lvert \calC \rvert$
\If{$\rk(A^\calC) = k-1$}
\For{$I \subseteq [k]$ \textbf{such that} $t^I \in \ ]0, 1[$}
\State \textbf{for} $i \in I$ \textbf{do} add $S_i \cup \{p\}$ with weights $(1-t^I)\mu_{S_i} - t^I \nu_{S_i}$ to $\calC'$
\State \textbf{for} $i \not \in I$ \textbf{do} add $S_i$ with weights $(1-t^I)\mu_{S_i} - t^I \nu_{S_i}$ to $\calC'$
\State add $\calC'$ to $\bbB(N')$
\EndFor
\EndIf
\EndFor
\For{$\calC \in \bbB_N$}
\State $k \gets \lvert \calC \rvert$
\For{$I \subseteq [k]$ \textbf{such that} $\lambda^\calC_I \leq 1$}
\State $C' \gets \varnothing$
\State \textbf{for} $i \in I$ \textbf{do} add $S_i \cup \{p\}$ with weights $\lambda^\calC_{S_i}$ to $\calC'$
\State \textbf{for} $i \not \in I$ \textbf{do} add $S_i$ with weights $\lambda^\calC_{S_i}$ to $\calC'$
\State \textbf{if} $\lambda^\calC_I < 1$ \textbf{then} add $\{p\}$ with weight $1 - \lambda^\calC_I$ to $\calC'$
\State add $\calC'$ to $\bbB(N')$
\For{$\delta \in [k] \setminus I$ \textbf{such that} $\lambda_{S_\delta} > 1 - \lambda^\calC_I$}
\State $C' \gets \varnothing$
\State \textbf{for} $i \in I \setminus \{\delta\}$ \textbf{do} add $S_i \cup \{p\}$ with weights $\lambda^\calC_{S_i}$ to $\calC'$
\State \textbf{for} $i \not \in I \cup \{\delta\}$ \textbf{do} add $S_i$ with weights $\lambda^\calC_{S_i}$ to $\calC'$
\State add $S_\delta \cup \{p\}$ with weight $1 - \lambda^\calC_I$ to $\calC'$
\State add $S_\delta$ with weight $\lambda^\calC_{S_\delta} + \lambda^\calC_I - 1$ to $\calC'$
\State add $\overline{\calC}$ to $\bbB(N')$
\EndFor
\EndFor
\EndFor
\Return $\bbB(N')$
\EndProcedure
\end{algorithmic}
\end{breakablealgorithm}

\begin{theorem}
Let $N$ be a finite set. Algorithm \ref{construction} {\normalfont \textsc{AddNewPlayer}}, which takes as an input the set of all minimal balanced collections on $N$, generates all the minimal balanced collections on $N' = N \cup \{p\}$. 
\end{theorem}

\proof
Thanks to the four previous lemmas, the algorithm generates only minimal balanced collections on $N'$. It remains to prove that every minimal balanced collection is generated by this algorithm. Let $\calB$ be a minimal balanced collection on $N'$. If the player $p$ is removed from each coalition of $\calB$, the collection is still balanced. Denote by $\calB_{-p}$ this new collection. 

\medskip

If $\{p\} \not \in \calB$ and $\calB_{-p}$ is a minimal balanced collection, then $\calB$ is generated by the first case (Lemma \ref{lemma: first})

\medskip

If $\{p\} \in \calB$: since $\calB$ has a unique system of balancing weights, $\calB_{-p}$ has only one system of balancing weights, and so it is a minimal balanced collection, and $\calB$ is generated by the second case (Lemma \ref{lemma: second}). 

\medskip 

If $\{p\} \not \in \calB$ and there are two identical coalitions in $\calB_{-p}$: the minimality of $\calB$ implies the minimality of $\calB_{-p}$ when the two identical coalitions are merged and their weights added. Then $\calB$ is generated by the third case (Lemma \ref{lemma: third}). 

\medskip 

Assume that $\{p\} \not \in \calB$, no coalitions in $\calB_{-p}$ are identical, and $\calB_{-p}$ is not a minimal balanced collection. Because $\calB$ is a minimal balanced collection of $k$ coalitions, $\rk(A^\calB) = k$, and then $\rk(A^{\calB_{-p}}) = k-1$. Consequently, the set of solutions of the following system of inequalities 
\begin{equation}
\label{proof-algo}
A^{\calB_{-p}} \lambda = \bfone^N, \quad \lambda \geq \vec{o},
\end{equation}
is one-dimensional and has the form $\lambda = \lambda_0 + t \lambda_1$, where $\lambda_0$ is a system of balancing weights for $\calB_{-p}$, where $t$ is a real number and $\lambda_1$ is a non-zero solution of the homogeneous system
\[
A^{\calB_{-p}} \lambda = \vec{o}, \quad \lambda \geq \vec{o}. 
\]
The set of solutions of \eqref{proof-algo} being bounded and one-dimensional, it is a nondegenerate segment $[\alpha, \beta]$. Let $U_\alpha = \{i \mid \alpha_i > 0\}$ and $U_\beta = \{i \mid \beta_i > 0\}$. Clearly, $U_\alpha$ and $U_\beta$ are subsets of $[k]$ and $U_\alpha \cup U_\beta = [k]$. Let $\calB^\alpha = \{S_i \in \calB \mid i \in U_\alpha\}$ and $\calB^\beta = \{S_i \in \calB \mid i \in U_\beta\}$. $\alpha^*$, the restriction of $\alpha$ to $U_\alpha$ is a system of balancing weights for $\calB^\alpha$, and $\beta^*$, the restriction of $\beta$ to $U_\beta$, is a system of balancing weights for $\calB^\beta$. Since $\alpha$ and $\beta$ are extremal solutions of the system \eqref{proof-algo}, $\calB^\alpha$ and $\calB^\beta$ are minimal balanced collections. Then $\calB$ is the union of $\calB^\alpha$ and $\calB^\beta$ and is generated by the fourth case (Lemma \ref{lemma: fourth}). 
\endproof

With the procedure \textsc{AddNewPlayer} used recursively, all the minimal balanced collections on $N$ are generated from the ones of $\{1, 2\}$. 

\begin{example}
Let $N = \{a, b, c, d\}$ and $N' = N \cup \{e\}$. Let $S_1 = \{a, b\}$, $S_2 = \{a, c\}$, $S_3 = \{a, d\}$ and $S_4 = \{b, c, d\}$. Then, $\calC = \{S_1, S_2, S_3, S_4\}$ is a minimal balanced collection with the system of weights $\lambda = \left(\frac{1}{3}, \frac{1}{3}, \frac{1}{3}, \frac{2}{3} \right)$. 

\paragraph{First case.} Remark that the set $I = \{1, 4\}$ satisfies the equation $\lambda_I = 1$. Therefore, we construct the minimal balanced collection: $\calC_1 = \{\{a, b, e\}, \{a, c\}, \{a, d\}, \{b, c, d, e\}\}$, with the system of weights $\lambda^{\calC_1} = \left( \frac{1}{3}, \frac{1}{3}, \frac{1}{3}, \frac{2}{3} \right)$. 

\paragraph{Second case.} Let $I = \{4\}$. We have $\lambda_I = \frac{2}{3} < 1$, therefore, we construct the minimal balanced collection: $\calC_2 = \{\{a, b\}, \{a, c\}, \{a, d\}, \{b, c, d, e\}, \{e\}\}$, with the system of weights $\lambda^{\calC_2} = \left( \frac{1}{3}, \frac{1}{3}, \frac{1}{3}, \frac{2}{3}, \frac{1}{3} \right)$.

\paragraph{Third case.} Let $I = \{1, 2\}$ and $\delta = 4$. Then $\lambda_I = \frac{2}{3}$ and $1 - \lambda_{S_\delta} = \frac{1}{3}$. Therefore, $1 > \lambda_I > 1 - \lambda_{S_\delta}$ and the following minimal balanced collection can be constructed: $\calC_3 = \{\{a, b, e\}, \{a, c, e\}, \{a, d\}, \{b, c, d\}, \{b, c, d, e\}\}$, with the system of weights $\lambda^{\calC_3} = \left( \frac{1}{3}, \frac{1}{3}, \frac{1}{3}, \frac{1}{3}, \frac{1}{3} \right)$. 

\paragraph{Fourth case.} For the last case, we consider another framework. Let $N = \{a, b\}$. $\calC_1 = \{\{a\}, \{b\}\}$, $\calC_2 = \{\{a, b\}\}$ are the two minimal balanced collections on $N$. Set $\calC \coloneqq \calC_1 \cup \calC_2 = \{\{a\}, \{b\}, \{a, b\}\}$. Let $\mu = (1, 1, 0)$ and $\nu = (0, 0, 1)$. We have
\[
\rk \left( A^\calC \right) = \rk \left( \begin{pmatrix} 1 & 0 & 1 \\ 0 & 1 & 1 \end{pmatrix} \right) = 2 = k-1. 
\]
Finally, let $I = \{1, 2\}$. Then, $\mu_I = 2$, $\nu_I = 0$, and $t^I = \frac{1 - \mu_I}{\nu_I - \mu_I} = \frac{1}{2} \in (0, 1)$. The following collection can be constructed: $\calC = \{\{a, b\}, \{b, c\}, \{a, c\}\}$, with the system of weights $\lambda^{\calC} = \frac{1}{2} \mu + \frac{1}{2} \nu = \left( \frac{1}{2}, \frac{1}{2}, \frac{1}{2} \right)$. 
\end{example}

\begin{remark} \label{remark: systems}
It is possible to adapt Algorithm \ref{construction} to compute the minimal balanced collections on every set system $\calF \subseteq 2^N$ on which the game is defined. We start again at $N = \{a, b\}$ and $\bbB(N) = \{\{\{a, b\}\}, \{\{a\}, \{b\}\}\}$, and for each minimal balanced collection generated, we check if this is a subset of a coalition in $\calF$. 
\end{remark}

\paragraph{Results and performance.} We implemented the above algorithm in Python\footnote{Computing device: Intel Xeon W-1250, CPU 3.30 GHz, 32 GB RAM.}, and found the following results and performance, given in Table \ref{table: perf}. 

\begin{table}[ht]
\begin{center}
\begin{tabular}{lcccccccc}
\toprule
$n$ \hspace{0.5cm} & \hspace{3pt} 2 \hspace{3pt} & \hspace{3pt} 3 \hspace{3pt} & \hspace{3pt} 4 \hspace{3pt} & \hspace{3pt} 5 \hspace{3pt} & \hspace{18pt} 6 \hspace{18pt} & \hspace{3pt} 7 \hspace{3pt} \\ 
\midrule
$k$ & 2 & 6 & 42 & 1,292 & 200,214 & 132,422,036 \\
$t$ & 0s & 0s & 0s & 1s & 244s & 63 hours \\
\bottomrule
\end{tabular}
\caption{Number $k$ of minimal balanced collections and CPU time $t$ according to the number $n$ of players.}
\label{table: perf}
\end{center}
\end{table}

The four first numbers correspond to the ones found by \textcite{shapley1967balanced}. None of the sequences already known in the OEIS \parencite{oeis} shared the same first numbers, so this one was added to the Encyclopaedia, and it can be accessed under the number A355042\footnote{see \href{https://oeis.org/A355042}{https://oeis.org/A355042}.}. Moreover, the lists of minimal balanced collections\footnote{Available on request from the author.} have been stored till $n = 7$. 

\medskip

The second method to compute the minimal balanced collections is by finding the vertices of a specific polytope. Let $F$ be the polytope defined by 
\[
F = \left\{ \lambda \in \bbR^\calN_+ \ \left| \ \sum_{S \in \calN} \lambda_S \bfone^S = \bfone^N \right. \right\}. 
\]
It is easy to check that the vertices of $F$ are in bijection with the minimal balanced collections on $N$. Indeed, an element $\lambda \in F$ is a vertex if and only if its support is a minimal balanced collection with the corresponding balancing weights \parencite[see, e.g.,][Corollary 3.1.9]{peleg2007introduction}. The reason is the following. Consider $\lambda$ an element of $F$. By definition, $\calB = \{S \in \calN \mid \lambda_S > 0\}$ is a balanced collection with balancing weight system $\lambda$. If $\lambda$ is a vertex, it cannot be obtained as a convex combination of other vectors in $F$, hence the balancing weight system is unique and the corresponding balanced collection is minimal. 

\medskip 

Consequently, generating all minimal balanced collections on $N$ amounts to finding all vertices of $F$. We have used the well-known Avis-Fukuda algorithm \parencite{avis1991pivoting} for vertex enumeration, available in the pycddlib package in Python. Running it for $n = 6$ yields the following performance (the one of our algorithm is recalled), see Table \ref{table: comparison-avis-fukuda}. The comparison indicates that our algorithm outperforms the Avis-Fukuda algorithm. 

\begin{table}[ht]
\begin{center}
\begin{tabular}{lr}
\toprule
Algorithm used & CPU time \\
\midrule
Our algorithm & 244 seconds \\
Avis-Fukuda algorithm \hspace{1cm} & 1764 seconds \\
\bottomrule
\end{tabular}
\end{center}
\caption{Comparison of the CPU times with $n = 6$.}
\label{table: comparison-avis-fukuda}
\end{table}

Now that we know the minimal balanced collections, we can use an algorithmic implementation of the Bondareva-Shapley Theorem for games defined with at most seven players. An important general remark for this implementation and the subsequent applications is that the minimal balanced collections do not depend on the game under consideration, but only on $n$. Therefore, there is no need to generate them for each application, but just to export them from some storage device. Until $n=7$, this gives a computational advantage compared to other methods based on linear programming and polyhedra, as it is shown in Table \ref{table: comparison-lp}. 

\medskip

Let $(N, v)$ be a game. The question is whether the core $C(v)$ of this game is nonempty. Consider the following linear program:
\[ \left\{ \begin{aligned}
& \min \, x(N) \\
& \text{ s.t. } x(S) \geq v(S), \quad \forall S \in \calN.
\end{aligned} \right. \]
Clearly, $C(v)$ is nonempty if and only if the optimal value of this program is $x(N) = v(N)$ (see the discussion following Theorem \ref{th: BS-first}). Therefore, one simple way to check the nonemptiness of the core is to solve this program and compute its optimal value. 

\medskip 

Another way is to take the dual program of the previous one. This was done by Bondareva and Shapley independently, as we have discussed in Section \ref{sec: the core}. It directly leads to minimal balanced collections and the Bondareva-Shapley Theorem (Theorem \ref{th: BS-sharp}), which states that $C(v)$ is nonempty if and only if, for any minimal balanced collection $\calB$ on $N$, we have 
\begin{equation}
\label{eq: BS}
\sum_{S \in \calB} \lambda^\calB_S v(S) \leq v(N). 
\end{equation}
This result shows that the nonemptiness of the core can be checked by a simple algorithm inspecting inequality \eqref{eq: BS} for each minimal balanced collection. The test terminates once we have found a minimal balanced collection not satisfying \eqref{eq: BS}. 

\medskip 

In order to compare both approaches, we fixed $n = 6$ and generated 5,000 different games in the following way: for all coalitions $S \in \calN \setminus \{N\}$, the worths $v(S)$ are drawn at random in the interval $[0, 5]$, while $v(N)$ is fixed to $50$. Doing so, each generated game has a nonempty core, as $\left( \frac{50}{6}, \ldots, \frac{50}{6} \right)$ is a core element for any generated game. 

\medskip 

Therefore, in the algorithm based on the Bondareva-Shapley Theorem, all inequalities have to be checked in order to conclude the nonemptiness of the core (which is the most unfavorable case). To solve the linear program, we have used the revised simplex method, already programmed in Python. Similar to the comparison with the Avis-Fukuda algorithm, both algorithms are implemented in the same language, so the comparison is fair. The results are given in Table \ref{table: comparison-lp}. 

\begin{table}[ht]
\begin{center}
\begin{tabular}{lr}
\toprule 
Algorithm used & Accumulated CPU time \\
\midrule 
Our algorithm & 0.96 seconds \\
Revised simplex method & 24.85 seconds \\
\bottomrule
\end{tabular}
\end{center}
\caption{Comparison of the CPU time for checking the balancedness of 5,000 games with $n = 6$, for both methods.}
\label{table: comparison-lp}
\end{table}

We conclude that the algorithm based on minimal balanced collections (provided that they are available) is much faster than an approach based on linear programming. 

\medskip 

\label{sym: calE(v)}
Additionally, the minimal balanced collections permit us to check whether the core is full-dimensional, i.e., if its affine space is $\bbX(v)$. A coalition $S$ is \emph{effective} (for $(N, v)$) if, for all $x \in C(v)$, we have $x(S) = v(S)$. We denote by $\calE(v)$ the collection of coalitions which are effective for $(N, v)$. Equivalently, $S \in \calE(v)$ if and only if $C(v) \subseteq \bbA_S(v)$. Therefore, if $\calE(v)$ is different from $\{N\}$, the core is not full-dimensional. The following result allows for obtaining all effective coalitions of a game. 

\begin{proposition} \label{prop: effectiveness}
Let $(N, v)$ be a balanced game. $\calE(v)$ is the union of all minimal balanced collections $\calB$ satisfying
\begin{equation} \label{eq: effectiveness}
\sum_{S \in \calB} \lambda^\calB_S v(S) = v(N). 
\end{equation}
\end{proposition}

\proof
Let $\calB$ be a minimal balanced collection satisfying \eqref{eq: effectiveness} and let $x \in C(v)$. Then, 
\[
v(N) = x(N) = \sum_{S \in \calB} \lambda^\calB_S x(S) \geq \sum_{S \in \calB} \lambda^\calB_S v(S) = v(N). 
\]
For all $S \in \calB$, we have $\lambda^\calB_S > 0$, then $x(S) = v(S)$, and $\calB \subseteq \calE(v)$. \\
For the reverse inclusion, let $S \in \calE(v)$. As $\{N\}$ is a minimal balanced collection, assume that $S \neq N$. It remains to show that $S$ is contained in some minimal balanced collection satisfying \eqref{eq: effectiveness}. Assume the contrary. Then, by the Bondareva-Shapley Theorem, there exists $\varepsilon > 0$ such that the game $(N, v^\varepsilon)$ that differs from $(N, v)$ only inasmuch as $v^\varepsilon(S) = v(S) + \varepsilon$ is still balanced. Hence, for $x \in C(N, v^\varepsilon)$, it follows that $x(S) > v(S)$ and $x \in C(v)$, then the desired contradiction has been obtained. 
\endproof

Notice that, because $\calE(v)$ is a union of minimal balanced collections, it is itself a balanced collection. To compute it, we simply need to run over all the minimal balanced collections and store the ones that satisfy \eqref{eq: effectiveness}. This procedure can therefore be used at the same time we check the balancedness of a game, with a negligible additional time. Table \ref{table: comparison-lp} certifies that this method is considerably faster than any other linear programming method to check whether the core is full-dimensional. 

\medskip 

The output of this computation also defines the affine span of the core, that is 
\[
\text{aff} \left( C(v) \right) = \bigcap_{S \in \calE(v)} \bbA_S(v). 
\]
However, there may be some redundant hyperplanes in the definition of the affine span. 

\section{Nonempty polyhedra and set functions}

The Bondareva-Shapley Theorem and its algorithmic implementation can be used in a wider range of applications than cooperative game theory. Any polytope satisfying a few conditions, quite natural in mathematical economics and combinatorics, can be studied using minimal balanced collections. 

\begin{definition} \label{def: basic}
A \emph{basic polyhedron} $P \subseteq \bbR^N$ is defined as the set of solutions of a system of linear inequalities of the form
\[ \left\{ \begin{tabular}{lcll}
$\sum_{j \in N} x_j$ & $=$ & $b_N$, & \\
$\sum_{j \in N} a_{ij} x_j$ & $\geq$ & $b_i$, & \quad for all $i \in \calU_1$, \\
$\sum_{j \in N} a_{ij} x_j$ & $>$ & $b_i$, & \quad for all $i \in \calU_2$, \\
$\sum_{j \in N} a_{ij} x_j$ & $\leq$ & $b_i$, & \quad for all $i \in \calD_1$, \\
$\sum_{j \in N} a_{ij} x_j$ & $<$ & $b_i$, & \quad for all $i \in \calD_2$.
\end{tabular} \right. \]
where all the coefficients $a_{ij}$ belong to $\{0, 1\}$. 
\end{definition}

For convenience, we denote the following indices sets $\calU = \calU_1 \cup \calU_2$, $\calD = \calD_1 \cup \calD_2$ and $\calI = \calU \cup \calD$. Because each coefficient $a_{ij}$ is either $0$ or $1$, we can identify any index with a coalition $S \in \calN$, via the map $i \mapsto S = \{j \in N \mid a_{ij} = 1\}$. The coalition associated with each index being unique, we keep the same notation $\calU$ and $\calD$ to denote the sets of coalitions and the sets of indices. 

\medskip 

Let $S \in \calD_1$. We have that $\sum_{j \in S} x_j = x(S) \leq b_S$. By multiplying both sides by $-1$ and adding $b_N$, we obtain $x(N) - x(S) \geq b_N - b_S$, which can be rewritten as $x(N \setminus S) \geq b_N - b_S$. The same procedure can be performed for indices in $\calD_2$. We obtain the following system of linear inequalities:
\[ \left\{ \begin{tabular}{rcll}
$x(N)$ & $=$ & $b_N$, & \\
$x(S)$ & $\geq$ & $b_S$, & \quad $\forall S \in \calU_1$, \\
$x(S)$ & $>$ & $b_S$, & \quad $\forall S \in \calU_2$, \\
$x(N \setminus S)$ & $\geq$ & $b'_{N \setminus S} \coloneqq b_N - b_S$, & \quad $\forall S \in \calD_1$, \\
$x(N \setminus S)$ & $>$ & $b'_{N \setminus S} \coloneqq b_N - b_S$, & \quad $\forall S \in \calD_2$.
\end{tabular} \right. \]
Denote by $\calD^c$ the set of complements of elements of $\calD$, i.e., $\calD^c = \{N \setminus S \mid S \in \calD\}$. For any basic polyhedron $P \subseteq \bbR^N$, we denote $\calF_P \coloneqq \{\emptyset, N\} \cup \calU \cup \calD^c$ and
\begin{equation} \label{eq: basic-game}
\beta_S = \left\{ \begin{tabular}{ll}
$0$ & \quad if $S = \emptyset$, \\
$\max \{b_S, b'_S\}$ & \quad if $S \in \calU \cap \calD^c$, \\
$b_S$ & \quad if $S \in (\calU \setminus \calD^c) \cup \{N\}$, \\
$b'_S$ & \quad if $S \in \calD^c \setminus \calU$. 
\end{tabular} \right. 
\end{equation}
Becuase there is one $\beta_S$ for each $S$ in $\calF_P$, we can define a set function on $\calF_P$ by $v_P(S) = \beta_S$. The pair $(\calF_P, v_P)$ \label{sym: calF_P-v_P} defines a \emph{game associated with $P$}, with $v_P$ only defined on the coalitions included in the set system $\calF_P$. We can interpret a game defined on a set system as modeling a situation where the formation of coalitions that are not $\calF_P$ is impossible. To read more about games on set systems, we refer to the survey of \textcite{grabisch2013core}. See also the core of \emph{incomplete games} \parencite{vcerny2022approximations}. 

\medskip 

The Bondareva-Shapley Theorem on games defined on set systems still operates, we only have to consider the minimal balanced collections included in the given set system. As previously discussed in Remark \ref{remark: systems}, our algorithm also functions to compute these minimal balanced collections. However, if we already know all the minimal balanced collections on $N$, it is more efficient to proceed to an inclusion check when running through all the minimal balanced collections on $N$. By Remark \ref{remark: B=C}, we know that the base polyhedron of the set function $\xi_P \equiv v_P^\#$ defined on the domain $\calF^c_P$ is the same set as the core of $(\calF_P, v_P)$, which is the adherence of $P$ in $\bbR^N$. 

\medskip

We are now able to present the main result of Chapter \ref{ChapterB}.  Denote by $\calS_P$ the collection $\calS_P = \calU_2 \cup \calD^c_2$ of all coalitions implying strict inequalities. 

\begin{theorem} \label{th: basic}
Let $P$ be a basic polyhedron. $P$ is nonempty if and only if, for all minimal balanced collections $\calB \subseteq \calF_P$, we have 
\[
\sum_{S \in \calB} \lambda^\calB_S \beta_S \leq \beta_N \qquad \text{and} \qquad \sum_{S \in \calB} \lambda^\calB_S \beta_S \neq \beta_N \quad \text{if } \calB \cap \calS_P \neq \emptyset. 
\]
Equivalently, $P$ is nonempty if and only $(\calF_P, v_P)$ is balanced and $\calS_P \cap \calE(\calF_P, v_P) = \emptyset$.
\end{theorem}

\proof
By construction of $v_P$, we have that 
\[
P = C(v_P) \setminus \bigcup_{S \in \calS_P} \bbA_S(v_P). 
\]
Assume that $(\calF_P, v_P)$ is balanced and that $\calS_P \cap \calE(\calF_P, v_P) = \emptyset$. Then, $C(\calF_P, v_P)$ is nonempty and, for all $S \in \calS_P$, there exists a core element $x^S \in C(v_P)$ such that $x^S(S) > v_P(S)$. Let $x^{\calS_P}$ denote the convex mid-point of these points, i.e., 
\[
x^{\calS_P} = \frac{1}{\lvert \calS_P \rvert} \sum_{S \in \calS_P} x^S.
\] 
Because the core is convex, we have $x^{\calS_P} \in C(v_P)$, and because all the $x^S$ are in the core, we have, for all $S \in \calS_P$, that $x^{\calS_P}(S) > v_P(S)$. Then $x^{\calS_P} \in P$. 

\medskip 

Assume now that $P \neq \emptyset$. Let $x \in P$. We have $x \in C(v_P)$, and therefore $(\calF_P, v_P)$ is balanced by the Bondareva-Shapley Theorem. Also, for all $S \in \calS_P$, we have $x(S) > v_P(S)$ because $x$ is a core element and does not belong to any of the $\bbA_S(v_P)$. Therefore, none of the $S \in \calS_P$ can belong to $\calE(v_P)$, and then $\calS_P \cap \calE(v_P) = \emptyset$. 
\endproof

This result is relevant because many properties or solutions in cooperative game theory depend on the existence of one, or many, specific preimputations forming a basic polyhedron. The simplest example is the basic polyhedron of coalitionally rational preimputations, i.e., the core. Let us consider a second example. 

\begin{example}
Let $x$ be a preimputation, let $S$ be a coalition and denote $\calI_S = \{\{i\} \mid i \in S\}$. Then, the set of preimputations dominating $x$ via $S$ is the basic polyhedron 
\[
\Delta_S(x) \coloneqq \{y \in \bbX(v) \mid y(S) \leq v(S) \text{ and, for all } i \in S, y_i > x_i\}. 
\] 
We have $\lvert S \rvert + 1$ inequalities indexed by $\calU = \calU_2 = \calI_S$, and $\calD = \calD_1 = \{S\}$ using the same notation as in Definition \ref{def: basic}. Then 
\[
\calF_{\Delta_S(x)} = \{\emptyset, N, N \setminus S\} \cup \calI_S.
\] 
The only non-trivial minimal balanced collection included in $\calF_{\Delta_S(x)}$ is the partition $\calB = \calI_S \cup \{N \setminus S\}$. The set function $v_{\Delta_S(x)}$ is defined on $\calF_{\Delta_S(x)}$ by 
\[
v_{\Delta_S(x)}(T) = \left\{ \begin{tabular}{ll}
$v(N) - v(S)$ & \quad if $T = N \setminus S$, \\
$x_i$ & \quad if $T \in \calI_S$, \\
$0$ & \quad if $T = \emptyset$.
\end{tabular} \right.
\]
The conditions of Theorem \ref{th: basic} become $\sum_{i \in S} v_{\Delta_S(x)}(\{i\}) + v_{\Delta_S(x)}(N \setminus S) < v(N)$, which is equivalent to $x(S) + v(N) - v(S) < v(N)$ and we retrieve that $x$ is dominated via $S$ if and only if $x(S) < v(S)$. 
\end{example}

\begin{definition}
Let $(N, v)$ be a game and denote by $v^0$ the restriction of $v$ to $2^N \setminus \{N\}$. A property $\calP$ on games is a \emph{prosperity property} if for each $v^0$ there exists a real number $\alpha_0 \geq \sum_{i \in N} v(\{i\})$ such that $(N, v)$ has property $\calP$ if and only if $v(N) \geq \alpha_0$.
\end{definition}

\begin{proposition}
All properties on games which are equivalent to the nonemptiness of a closed basic polyhedron are prosperity properties. 
\end{proposition}

\proof
Denote by $P$ the closed basic polyhedron whose nonemptiness is equivalent to the property we study. Denote by $(\calF_P, v_P)$ the game associated with $P$, as defined in Equation \ref{eq: basic-game}. Then, $P$ is nonempty if and only if $v(N) \geq \max_{\calB \in \bbB(N)} \sum_{S \in \calB} \lambda^\calB_S v(S)$ because $P$ is closed, and $\alpha_0 = \max_{\calB \in \bbB(N)} \sum_{S \in \calB} \lambda^\calB_S v(S)$. 
\endproof

In the next section, we introduce a few properties of coalitions and collections of coalitions, that are equivalent to the nonemptiness of a basic polyhedron. We show now that basic polyhedra are a generalization of generalized permutohedra. For a polytope $P \subseteq \bbR^N$, we denote by $A^P$ and $b^P$ a matrix $A^P \in \bbR^{k \times n}$ and a vector $b^P \in \bbR^k$ such that $P = \{x \in \bbR^N \mid A^P x \leq b^P\}$.   

\begin{definition}
A polytope $Q \subseteq \bbR^N$ is a \emph{deformation} of another polytope 
\[
P = \{x \in \bbR^N \mid A^P x \leq b^P\}
\] 
if there exists a vector $b^Q \in \bbR^\calN$ such that the following two conditions are satisfied: 
\begin{itemize}
\item The polytope $Q$ can be written as $Q = \{x \in \bbR^N \mid A^P x \leq b^Q\}$,
\item For any vertex $x$ of $P$, if $\{F_i\}_{i \in I}$ are the facets of $P$ containing $x$, then
\[
\bigcap_{i \in I} \ \left\{y \in \bbR^N \ \left| \ \sum_{j \in N} a^P_{ij} y_j = b^Q_i \right. \right\} \text{ is a vertex of } Q. 
\]
\end{itemize}
We call $b^Q$ a \emph{deforming} vector of $Q$. If $Q'$ is a polytope only satisfying the first condition, we call it a \emph{distortion} of polytope $P$, and $b^{Q'}$ is called a \emph{distorting} vector of $Q'$. 
\end{definition}

A distortion $Q$ of a polytope $P$ is a transformation consisting of translations of its facets along their normals. During this process, some faces can disappear, and the set of the normals of the facets of $Q$ is a subset of the set of normals of the facets of $P$. Notice that the combinatorial structure of the polytope, as described by its face lattice, can be completely different. On the contrary, a deformation is a distortion that preserves the direction of the edges of the initial polytope. 

\medskip 

\begin{figure}[ht]
\begin{center}
\begin{tabular}{>{\centering\arraybackslash}m{4cm}>{\centering\arraybackslash}m{3cm}>{\centering\arraybackslash}m{4cm}}
\begin{tikzpicture}[scale=0.65]
\coordinate (A) at (1, 2, 3);
\coordinate (B) at (1, 3, 2);
\coordinate (C) at (2, 1, 3);
\coordinate (D) at (2, 3, 1);
\coordinate (E) at (3, 1, 2);
\coordinate (F) at (3, 2, 1);
\coordinate (G) at (4, 1, 2);
\coordinate (H) at (4, 2, 1);
\coordinate (I) at (2, 4, 1);
\coordinate (J) at (2, 1, 4);
\coordinate (K) at (1, 4, 2);
\coordinate (L) at (1, 2, 4);
\coordinate (M) at (1, 4, 3);
\coordinate (N) at (1, 3, 4);
\coordinate (O) at (3, 1, 4);
\coordinate (P) at (3, 4, 1);
\coordinate (Q) at (4, 3, 1);
\coordinate (R) at (4, 1, 3);
\coordinate (S) at (4, 2, 3);
\coordinate (T) at (4, 3, 2);
\coordinate (U) at (3, 2, 4);
\coordinate (V) at (3, 4, 2);
\coordinate (W) at (2, 4, 3);
\coordinate (Z) at (2, 3, 4);

\fill[gray!15] (E) -- (G) -- (H) -- (F) -- cycle;
\fill[gray!15] (B) -- (D) -- (I) -- (K) -- cycle;

\filldraw (A) circle (2pt);
\filldraw (B) circle (2pt);
\filldraw (C) circle (2pt);
\filldraw (D) circle (2pt);
\filldraw (E) circle (2pt);
\filldraw (F) circle (2pt);
\filldraw (G) circle (2pt);
\filldraw (H) circle (2pt);
\filldraw (I) circle (2pt);
\filldraw (J) circle (2pt);
\filldraw (K) circle (2pt);
\filldraw (L) circle (2pt);
\filldraw (M) circle (2pt);
\filldraw (N) circle (2pt);
\filldraw (O) circle (2pt);
\filldraw (P) circle (2pt);
\filldraw (Q) circle (2pt);
\filldraw (R) circle (2pt);
\filldraw (S) circle (2pt);
\filldraw (T) circle (2pt);
\filldraw (U) circle (2pt);
\filldraw (V) circle (2pt);
\filldraw (W) circle (2pt);
\filldraw (Z) circle (2pt);

\draw[dashed, gray] (S) -- (T);
\draw[dashed, gray] (T) -- (V);
\draw[dashed, gray] (U) -- (S);
\draw[dashed, gray] (V) -- (W);
\draw[dashed, gray] (W) -- (Z);
\draw[dashed, gray] (Z) -- (U);

\draw (A) -- (C);
\draw (B) -- (A);
\draw (C) -- (E);
\draw (D) -- (B);
\draw (E) -- (F);
\draw[ultra thick] (F) -- (D);

\draw (J) -- (L);
\draw (L) -- (N);
\draw[dashed, gray] (N) -- (Z);
\draw (O) -- (J);
\draw[dashed, gray] (U) -- (O);
\draw[dashed, gray] (Z) -- (U);

\draw (D) -- (I);
\draw (F) -- (D);
\draw (H) -- (F);
\draw (I) -- (P);
\draw (P) -- (Q);
\draw (Q) -- (H);

\draw (G) -- (H);
\draw[dashed, gray] (Q) -- (T);
\draw (R) -- (G);
\draw[dashed, gray] (S) -- (R);

\draw (K) -- (I);
\draw (M) -- (K);
\draw[dashed, gray] (P) -- (V);
\draw[dashed, gray] (W) -- (M);

\draw (A) -- (L);
\draw (K) -- (B);
\draw (N) -- (M);

\draw (E) -- (G);
\draw (J) -- (C);
\draw (R) -- (O);
\end{tikzpicture} &  &
Core of $(N, v)$ with 
\[ \begin{aligned}
& N = \{1, 2, 3, 4\}, \text{ and } \\
& v(S) = \frac{\lvert S \rvert (\lvert S \rvert + 1)}{2}.
\end{aligned}
\]
\end{tabular}
\caption{The standard permutohedron.}
\label{fig: standard}
\end{center}
\end{figure}
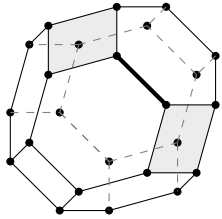

The core of the game defined in Figure \ref{fig: standard} is the standard permutohedron.

\begin{figure}[ht]
\begin{center}
\begin{tabular}{>{\centering\arraybackslash}m{4cm}>{\centering\arraybackslash}m{3cm}>{\centering\arraybackslash}m{4cm}}
\begin{tikzpicture}[scale=0.63]
\coordinate (A) at (0.1, 0.8, 2.7);
\coordinate (B) at (0.1, 2.0, 1.5);
\coordinate (C) at (0.5, 0.4, 2.7);
\coordinate (D) at (1.1, 0.4, 2.1);
\coordinate (E) at (0.7, 2.0, 0.9);
\coordinate (F) at (1.1, 1.6, 0.9);
\coordinate (G) at (0.9, 2.4, 5.1);
\coordinate (H) at (1.3, 2.0, 5.1);
\coordinate (I) at (1.3, 0.4, 5.1);
\coordinate (J) at (0.5, 0.4, 5.1);
\coordinate (K) at (0.1, 2.4, 5.1);
\coordinate (L) at (0.1, 0.8, 5.1);
\coordinate (M) at (0.1, 3.6, 3.9);
\coordinate (N) at (0.1, 3.6, 1.5);
\coordinate (O) at (0.7, 3.6, 0.9);
\coordinate (P) at (1.5, 3.6, 0.9);
\coordinate (Q) at (1.5, 3.6, 3.3);
\coordinate (R) at (0.9, 3.6, 3.9);
\coordinate (S) at (1.9, 2.0, 4.5);
\coordinate (T) at (1.9, 3.2, 3.3);
\coordinate (U) at (1.9, 1.6, 0.9);
\coordinate (V) at (1.9, 3.2, 0.9);
\coordinate (W) at (1.9, 0.4, 4.5);
\coordinate (Z) at (1.9, 0.4, 2.1);

\fill[gray!15] (B) -- (N) -- (O) -- (E) -- cycle;
\fill[gray!15] (D) -- (Z) -- (U) -- (F) -- cycle;

\filldraw (A) circle (2pt);
\filldraw (B) circle (2pt);
\filldraw (C) circle (2pt);
\filldraw (D) circle (2pt);
\filldraw (E) circle (2pt);
\filldraw (F) circle (2pt);
\filldraw (G) circle (2pt);
\filldraw (H) circle (2pt);
\filldraw (I) circle (2pt);
\filldraw (J) circle (2pt);
\filldraw (K) circle (2pt);
\filldraw (L) circle (2pt);
\filldraw (M) circle (2pt);
\filldraw (N) circle (2pt);
\filldraw (O) circle (2pt);
\filldraw (P) circle (2pt);
\filldraw (Q) circle (2pt);
\filldraw (R) circle (2pt);
\filldraw (S) circle (2pt);
\filldraw (T) circle (2pt);
\filldraw (U) circle (2pt);
\filldraw (V) circle (2pt);
\filldraw (W) circle (2pt);
\filldraw (Z) circle (2pt);

\draw (A) -- (B);
\draw (B) -- (E);
\draw (C) -- (A);
\draw (D) -- (C);
\draw[ultra thick] (E) -- (F);
\draw (F) -- (D);

\draw[dashed, gray] (G) -- (H);
\draw[dashed, gray] (H) -- (I);
\draw (I) -- (J);
\draw (J) -- (L);
\draw[dashed, gray] (K) -- (G);
\draw (L) -- (K);

\draw[dashed, gray] (G) -- (R);
\draw[dashed, gray] (H) -- (G);
\draw[dashed, gray] (Q) -- (T);
\draw[dashed, gray] (R) -- (Q);
\draw[dashed, gray] (S) -- (H);
\draw[dashed, gray] (T) -- (S);

\draw (M) -- (N);
\draw (N) -- (O);
\draw (O) -- (P);
\draw[dashed, gray] (P) -- (Q);
\draw[dashed, gray] (Q) -- (R);
\draw[dashed, gray] (R) -- (M);

\draw[dashed, gray] (T) -- (V);
\draw (U) -- (Z);
\draw (V) -- (U);
\draw[dashed, gray] (W) -- (S);
\draw (Z) -- (W);

\draw (F) -- (U);
\draw (O) -- (E);
\draw (V) -- (P);

\draw (B) -- (N);
\draw (L) -- (A);
\draw (M) -- (K);

\draw (C) -- (J);
\draw (I) -- (W);
\draw (Z) -- (D);
\end{tikzpicture} & &  
Core of $(N, w)$ with 
\[ \begin{aligned}
& N = \{1, 2, 3, 4\}, \text{ and } \\
& w(S) = \left( \, \sum_{i \in S} i \right)^2.
\end{aligned}
\]
\end{tabular}
\caption{A generalized permutohedron.}
\label{fig: generalized}
\end{center}
\end{figure}
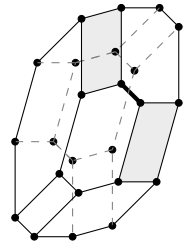

In Figure \ref{fig: generalized}, a convex game is defined and its core is drawn. We remark that any facet is parallel to its corresponding facet on the standard permutohedron, and that the directions of the edges are preserved. Therefore the game describes a deformation of the standard permutohedron, and its core is a generalized permutohedron. 

\begin{figure}[ht]
\begin{center}
\begin{tabular}{>{\centering\arraybackslash}m{4cm}>{\centering\arraybackslash}m{1cm}>{\centering\arraybackslash}m{6.3cm}}
\begin{tikzpicture}[scale=0.63]
\coordinate (A) at (0.1, 0.8, 2.7);
\coordinate (D1) at (0.1, 1.6, 1.9);
\coordinate (B) at (0.1, 2.0, 1.5);
\coordinate (C) at (0.5, 0.4, 2.7);
\coordinate (C1) at (0.6, 0.4, 2.6);
\coordinate (D) at (1.1, 0.4, 2.1);
\coordinate (E) at (0.7, 2.0, 0.9);
\coordinate (F) at (1.1, 1.6, 0.9);
\coordinate (G) at (0.9, 2.4, 5.1);
\coordinate (H) at (1.3, 2.0, 5.1);
\coordinate (I) at (1.3, 0.4, 5.1);
\coordinate (J) at (0.5, 0.4, 5.1);
\coordinate (K) at (0.1, 2.4, 5.1);
\coordinate (L) at (0.1, 0.8, 5.1);
\coordinate (M) at (0.1, 3.6, 3.9);
\coordinate (H1) at (0.1, 3.6, 1.9);
\coordinate (N) at (0.1, 3.6, 1.5);
\coordinate (G1) at (1.1, 3.6, 0.9);
\coordinate (O) at (0.7, 3.6, 0.9);
\coordinate (P) at (1.5, 3.6, 0.9);
\coordinate (Q) at (1.5, 3.6, 3.3);
\coordinate (R) at (0.9, 3.6, 3.9);
\coordinate (S) at (1.9, 2.0, 4.5);
\coordinate (T) at (1.9, 3.2, 3.3);
\coordinate (B1) at (1.9, 2.1, 0.9);
\coordinate (U) at (1.9, 1.6, 0.9);
\coordinate (V) at (1.9, 3.2, 0.9);
\coordinate (W) at (1.9, 0.4, 4.5);
\coordinate (A1) at (1.9, 0.4, 2.6);
\coordinate (Z) at (1.9, 0.4, 2.1);

\coordinate (E1) at (0.6, 1.6, 1.4);
\coordinate (F1) at (1.1, 2.1, 0.9);

\fill[gray!15] (D1) -- (H1) -- (G1) -- (F1) -- (E1) -- cycle;
\fill[gray!15] (C1) -- (A1) -- (B1) -- (F1) -- (E1) -- cycle;

\filldraw (A) circle (2pt);
\filldraw (D1) circle (2pt);
\filldraw (C) circle (2pt);
\filldraw (C1) circle (2pt);
\filldraw (G) circle (2pt);
\filldraw (H) circle (2pt);
\filldraw (I) circle (2pt);
\filldraw (J) circle (2pt);
\filldraw (K) circle (2pt);
\filldraw (L) circle (2pt);
\filldraw (M) circle (2pt);
\filldraw (H1) circle (2pt);
\filldraw (G1) circle (2pt);
\filldraw (P) circle (2pt);
\filldraw (Q) circle (2pt);
\filldraw (R) circle (2pt);
\filldraw (S) circle (2pt);
\filldraw (T) circle (2pt);
\filldraw (B1) circle (2pt);
\filldraw (V) circle (2pt);
\filldraw (W) circle (2pt);
\filldraw (A1) circle (2pt);

\filldraw (E1) circle (2pt);
\filldraw (F1) circle (2pt);

\draw (A) -- (D1);
\draw (D1) -- (E1);
\draw (C) -- (A);
\draw (C1) -- (C);
\draw[ultra thick] (E1) -- (F1);
\draw (E1) -- (C1);

\draw[dashed, gray] (G) -- (H);
\draw[dashed, gray] (H) -- (I);
\draw (I) -- (J);
\draw (J) -- (L);
\draw[dashed, gray] (K) -- (G);
\draw (L) -- (K);

\draw[dashed, gray] (G) -- (R);
\draw[dashed, gray] (H) -- (G);
\draw[dashed, gray] (Q) -- (T);
\draw[dashed, gray] (R) -- (Q);
\draw[dashed, gray] (S) -- (H);
\draw[dashed, gray] (T) -- (S);

\draw (M) -- (H1);
\draw (H1) -- (G1);
\draw (G1) -- (P);
\draw[dashed, gray] (P) -- (Q);
\draw[dashed, gray] (Q) -- (R);
\draw[dashed, gray] (R) -- (M);

\draw[dashed, gray] (T) -- (V);
\draw (B1) -- (A1);
\draw (V) -- (B1);
\draw[dashed, gray] (W) -- (S);
\draw (A1) -- (W);

\draw (F1) -- (B1);
\draw (G1) -- (F1);
\draw (V) -- (P);

\draw (D1) -- (H1);
\draw (L) -- (A);
\draw (M) -- (K);

\draw (C) -- (J);
\draw (I) -- (W);
\draw (A1) -- (C1);
\end{tikzpicture} & & 
Core of $(N, w')$ with 
\[ \begin{aligned}
& N = \{1, 2, 3, 4\}, \text{ and } \\
& w'(S) = \left\{ \begin{tabular}{ll}
$30$, & if $S = \{2, 3\}$, \\
$20$, & if $S = \{1, 3\}$, \\
$w(S)$, & otherwise.
\end{tabular} \right. 
\end{aligned}
\]
\end{tabular}
\caption{A general basic polyhedron.}
\label{fig: basic}
\end{center}
\end{figure}

In Figure \ref{fig: basic}, the game only describes a distortion of the standard permutohedron, because the bold edge does not have the same direction as in the standard permutohedron. Indeed, the worths of the coalitions corresponding to the gray facets increased such that the facets are pushed too much inwards. We see that the coalition function is no longer supermodular: 
\[
w'(\{2, 3\}) + w'(\{1, 3\}) = 30 + 20 = 50 > 45 = 36 + 9 = w'(\{1, 2, 3\}) + w'(\{3\}). 
\]
The core is therefore a general basic polyhedron and not a generalized permutohedron. 

\begin{table}[ht]
\begin{center}
\begin{tabular}{lr}
\toprule
Polytopes & Games \\
\midrule
\small{Standard permutohedron} & \small{$v: S \mapsto \lvert S \rvert (\lvert S \rvert + 1) /2$} \\
\small{Generalized permutohedron} & \small{convex games} \\
\small{Nonempty basic polyhedron} \hspace{1cm} & \small{(totally) balanced games} \\
Empty set & non-balanced games \\
\bottomrule
\end{tabular}
\caption{Typology of the basic polyhedron.}
\end{center}
\end{table}

The Submodularity Theorem together with Remark \ref{remark: B=C} state that generalized permutohedra are deformations of the standard permutohedron induced by convex games. If the game is not convex, but only balanced, the generated polytope is a general basic polyhedron. If the game is not even balanced, the polytope is no longer nonempty. Then, basic polyhedra are a generalization of generalized permutohedra, much more adequate to study cooperative games as we will see in the next subsection. 

\section{Various applications in cooperative game theory} \label{sec: properties}

In this section, we are looking at specific aspects of cooperative games, such as coalitions, or collections of coalitions, which carry important information about the game, and at a transitive sub-relation of domination called \emph{outvoting}. The properties satisfied by these coalitions, or collections of coalitions, are called \emph{exactness}, \emph{feasibility}, \emph{vitality}, \emph{strict vital-exactness} and \emph{extendability}. They are equivalent to the existence of specific preimputations, that, for each property, define a basic polyhedron.  

\medskip

Throughout the section let $(N, v)$ be a balanced game. For a coalition $S$, denote by $(S, v_{|S})$ the subgame on $S$, in which only the subcoalitions of $S$ are considered, and by $(N, v^S)$ the game that may differ from $(N, v)$ only inasmuch as $v^S(N \setminus S) = v(N) - v(S)$. This definition can be extended to a collection of coalitions $\calS$ such that, for all $S \in \calS$, we have $v^\calS(N \setminus S) = v(N) - v(S)$ and $v^\calS(T) = v(T)$ otherwise. 

\paragraph{Exactness.} A coalition $S$ is \emph{exact} (for $(N, v)$) if there exists a core element $x \in C(v)$ such that $x(S) = v(S)$. Hence, a coalition $S$ is exact if and only if $\bbA_S$ intersects the core. The following result permits us to build an algorithm that checks exactness. 

\begin{proposition} \label{prop: exactness}
Let $(N, v)$ be a balanced game. A coalition $S$ is exact if and only if $(N, v^S)$ is balanced. 
\end{proposition}

\proof
By definition, $S$ is exact if there exists a core element $x \in C(v)$ such that $x(S) = v(S)$. Because the inequality $x(S) \geq v(S)$ is already involved in the definition of the core, by adding the inequality $x(S) \leq v(S)$ we have defined the basic polyhedron of preimputations that we are looking for, which we denote by $P$. By balancedness, we have that $v(S) + v(N \setminus S) \leq v(N)$ and thus $v(N \setminus S) \leq v(N) - v(S) = v^S(N \setminus S)$. Therefore, the inequality $x(N \setminus S) \geq v(N \setminus S)$ is redundant in the definition of $P$, and then the set function associated with $P$ is $v^S$. Theorem \ref{th: basic} finishes the proof, its second condition being superfluous in this case because there is no strict inequality. 
\endproof


The proof of Proposition \ref{prop: exactness} resembles the methods used in the paper of Lohmann, Borm and Herings \cite{lohmann2012minimal}, of Cs{\'o}ka, Herings and K{\'o}czy \cite{csoka2011balancedness} and of Studen{\'y} and Kratochv{\'i}l \cite{studeny2022facets}. Indeed, the negative weight that is used in the definition of so-called \emph{min-semi-balanced} collections, and the \emph{exceptional} coalition associated to it, corresponds to the coalition of which we are cheking the exactness. Flipping the inequality $x(S) \geq v(S)$ to $x(S) \leq v(S)$ leads to the  same computations than taking a negative weight. Then, using min-semi-balanced collections, or checking the balancedness of $(N, v^S)$ for each $S \in \calN$ is completely equivalent. However, the main idea in this thesis is to always use the same set of minimal balanced collections, which are actually computed, to deal with any situation. That is why we prefer to alter the game rather than the collections. 

\medskip 

The exact coalitions represent the coalitions that are critical in the situation we model. First, they form a \emph{core-describing} collection of coalitions, i.e., 
\[
C(v) = \{x \in \bbX(v) \mid x(S) \geq v(S), \; \text{for all } S \text{ exact}\}. 
\]
Moreover, they are the coalitions via which a preimputation can be dominated by a core element. Indeed, if a coalition $S$ is exact, then there exists a core element $x \in C(v)$ such that $x(S) = v(S)$. Therefore, this element $x$ is affordable for the coalition $S$ and is in the core. Algorithm \ref{algo: exactness} checks whether a coalition is exact for a given game. 

\begin{breakablealgorithm}
\caption{Exactness checking subroutine} \label{algo: exactness}
\begin{algorithmic}[1]
\Require A coalition $S$, a balanced coalition function $v$, the set $\bbB(N)$
\Ensure The Boolean value: `$S$ is exact'
\Procedure{IsExact}{$S$, $v$, $\bbB(N)$}
\State Define $v^S$
\For{$\calB \in \bbB(N)$}
\If{$\sum_{T \in \calB} \lambda^\calB_T v^S(T) > v(N)$}
\Return \textbf{False}
\EndIf
\EndFor 
\Return \textbf{True}
\EndProcedure
\end{algorithmic}
\end{breakablealgorithm}

Furthermore, Gillies found a necessary condition for the core to be a stable set, based on exactness. 

\begin{proposition}[\textcite{gillies1959solutions}] \label{prop: necessary-exact} \leavevmode \newline
A balanced game has a stable core only if each singleton is exact. 
\end{proposition}

This condition is also a sufficient conditions on different classes of games, for example: matching games, minimum coloring game \cite{shellshear2009core}. 

\medskip

As the set of exact coalitions can be computed (see Proposition \ref{prop: exactness} and Algorithm \ref{algo: exactness}), the necessary condition of Gillies can be easily checked. Another interesting consequence of this result is the expansion of the space in which the core is externally stable (see discussion following Definition \ref{def: domination}). 

\paragraph{Feasibility.} Let $(N, v)$ be a balanced game and $\calF \subseteq \calN$ be a core-describing collection of coalitions, i.e., 
\[
C(v) = \{x \in \bbX(v) \mid x(S) \geq v(S), \forall S \in \calF\}. 
\]
Let us consider a subcollection $\calS \subseteq \calF$, and consider the following subset of $\bbX(v)$:
\[
X_\calS = X_\calS^\calF(v) \coloneqq \{ x \in \bbX(v) \mid x(S) < v(S) \iff S \in \calS\}. 
\]
We call $X_\calS^\calF(v)$ the \emph{region} associated with $\calS$, w.r.t.\ $\calF$ and $v$. 

\begin{definition} \label{def: feasibility}
The collection $\calS$ is $\calF$-feasible if the region $X_\calS^\calF(v)$ is nonempty. 
\end{definition}

The regions form a partition of $\bbX(v)$, with $X^\calF_\emptyset(v) = C(v)$. If no confusion occurs, the collection is simply said to be \emph{feasible} and the region is simply denoted by $X_\calS$. Next, we provide some properties of the feasible collections. 

\begin{definition}
A collection of coalitions $\calS \subseteq \calN$ is said to be \emph{unbalanced} if it does not contain a balanced collection. 
\end{definition}

\begin{lemma}[\textcite{grabisch2021characterization}] \label{lemma: blocking} \leavevmode \newline
Let $(N, v)$ be a balanced game and let $\calS \subseteq \calF$. The following holds. 
\begin{enumerate}
\item If $\calS$ is feasible, then it is an unbalanced collection. 
\item For $S, S' \in \calS$ such that $S \cup S' = N$, no $x \in X_\calS$ is dominated via $S$ or $S'$. 
\end{enumerate}
\end{lemma}

We postpone to Chapter \ref{ChapterC} the discussions about unbalanced collections. A feasible collection $\calS$ that only contains coalitions satisfying the second condition above is called a \emph{blocking feasible collection}. A characterization that can be translated into an algorithm is needed to compute the feasible collections. Recall that we denote $\calS^c = \{N \setminus S \mid S \in \calS\}$ and $\calF_\calS = \left( \calF \setminus \calS \right) \cup \calS^c$. 

\begin{lemma} \label{lemma: feasibility}
A collection $\calS \subseteq \calF$ is feasible (w.r.t.\ $\calF$) if and only if $(\calF_\calS, v^\calS)$ is balanced and $\calS^c \cap \calE(v^\calS) = \emptyset$. 
\end{lemma}

\proof
The proof is straightforward because the region $X_\calS$ is a basic polyhedron. Let $S \in \calS$. For any element $x \in X_\calS$, if any, we have that $x(S) < v(S)$. Multiplying both sides by $-1$ and adding $v(N)$ leads to the strict inequality $x(N \setminus S) > v(N) - v(S) = v^\calS(N \setminus S)$. Then, for all elements $x \in X_\calS$, if any, and for all $S \in \calF_\calS$, we have $x(S) \geq v^\calS(S)$, with strict inequalities for coalitions in $\calS^c$. We apply Theorem \ref{th: basic} to conclude the proof. 
\endproof

\begin{breakablealgorithm}
\caption{Feasibility checking algorithm} \label{algo: feasibility}
\begin{algorithmic}[1]
\Require A balanced coalition function $v$, a set system $\calF$, a set $\calS \subseteq \calF$, the set $\bbB(N)$
\Ensure The Boolean value: `$\calS$ is feasible'
\Procedure{IsFeasible}{$\calS$, $\calF$, $v$, $\bbB(N)$}
\For{$\calB \in \bbB(N)$ \textbf{such that} $\calB \subseteq 
    (\calF\setminus \calS)\cup\calS^c$} {\bf do}
\If{$\calB \cap \calS^c \neq \varnothing$ \textbf{and} $\sum_{S \in \calB}\lambda_S^\calB v^\calS(S) \geq v(N)$}
\Return \textbf{False}
\ElsIf{$\calB \cap \calS^c = \varnothing$ \textbf{and} $\sum_{S \in \calB}\lambda_S^\calB v^\calS(S) > v(N)$}
\Return \textbf{False}
\EndIf
\EndFor
\Return \textbf{True}
\EndProcedure
\end{algorithmic}
\end{breakablealgorithm}


\paragraph{Extendability.} A coalition $S$ is called \emph{extendable} (w.r.t.\ $(N, v)$) if, for any $x \in C(v_{|S})$, there exists $y \in C(v)$ such that $x = y_S$, where $y_S$ is the restriction of $y$ to coordinates in $S$.  Extendability was defined by \textcite{kikuta1986core} to study the stability of the core. Indeed, they prove that, if a game is \emph{extendable}, i.e., if all coalitions are extendable, then its core is a stable set. It comes from the fact that any preimputation that is dominated via an extendable coalition, is dominated by a core element. 

\medskip 

To check whether a coalition is extendable, by convexity of the core, it is sufficient to check if each vertex of $C(v_{|S})$ can be extended to a core element. Let $z \in C(v_{|S})$. Recall that $\calI_S = \{\{i\} \mid i \in S\}$ and denote by $v_{S, z}$ the set function defined by 
\[
v_{S, z} (T) = \left\{ \begin{tabular}{ll}
$z_i$ & \quad if $T \in \calI_S$, $T = \{i\}$, \\
$\max \{v(N) - z(N \setminus T), v(T)\}$ & \quad if $T \in \{\{N \setminus \{i\}\} \mid i \in S\}$, \\
$v(T)$ & \quad otherwise. 
\end{tabular} \right. 
\]

\begin{proposition} \label{prop: extendability}
Let $(N, v)$ be a balanced game, let $S$ be a coalition and let $z \in C(v_{|S})$. There exists $x \in C(v)$ such that $x_S = z$ if and only if $(N, v_{S, z})$ is balanced. 
\end{proposition}

\proof
The question is whether the polytope, denoted by $P$, and defined by 
\[
P = \{x \in C(v) \mid \text{for all } i \in S, x_i = z_i\}, 
\]
is nonempty. For all $i \in S$, the equality $x_i = z_i$ can be rewritten as a set of two inequalities, $x_i \geq z_i$ and $x_i \leq z_i$. In the last inequality, we can multiply both sides by $-1$, and add $v(N)$ to obtain $x(N \setminus \{i\}) \geq v(N) - z_i$. Moreover, because $z \in C(v_{|S})$, we have that $z_i \geq v(\{i\})$, implying that the inequalities $x_i \geq v(\{i\})$ are redundant in the definition of $P$. Therefore, the polyhedron $P$ is the set of elements satisfying
\[ \left\{ \begin{tabular}{rcll}
$x_i$ & $\geq$ & $z_i$, & \quad for all $i \in S$, \\
$x(N \setminus \{i\})$ & $\geq$ & $v(N) - z_i$, & \quad for all $i \in S$, \\
$x(T)$ & $\geq$ & $v(T)$, & \quad for all $T \in \calN \setminus \calI_S$. 
\end{tabular} \right.\]
By Theorem \ref{th: basic}, $P$ is nonempty if and only if $(N, v_{S, z})$ is balanced. The second condition of the theorem is superfluous since no strict inequality is involved. 
\endproof

Proposition \ref{prop: extendability} gives us a necessary and sufficient condition for the existence of an extension of a vertex of $C(v_{|S})$ to an element of $C(v)$, based upon a balancedness check. If there exists an extension for each extreme point of $C(v_{|S})$, by convexity of the core, any element of $C(v_{|S})$ can be extended. Algorithm \ref{algo: extendability} checks if each vertex of $C(v_{|S})$ can be extended to an element of $C(v)$. 

\begin{breakablealgorithm}
\caption{Extendability checking algorithm} \label{algo: extendability}
\begin{algorithmic}[1]
\Require A coalition $S$, a balanced coalition function $v$
\Ensure The Boolean value: `$S$ is extendable'
\Procedure{IsExtendable}{$S$, $v$, $\bbB(N)$}
\For{$\xi \in {\rm ext}(C(v_{|S}))$}
\State define $v_{S, \xi}$
\For{$\calB \in \bbB(N)$}
\If{$\sum_{T \in \calB} \lambda^\calB_T v_{S, \xi}(T) > v(N)$}
\Return \textbf{False}
\EndIf
\EndFor
\EndFor
\Return \textbf{True}
\EndProcedure
\end{algorithmic}
\end{breakablealgorithm}

\textcite{kikuta1986core} have provided a sufficient condition for a game to have a stable core via extendability. 

\begin{theorem}[\textcite{kikuta1986core}] \leavevmode \newline
An extendable game has a nonempty and stable core. 
\end{theorem}

The extendability of a game can be checked by using Algorithm \ref{algo: extendability}, but it is time-consuming. However, this property can be considerably weakened as follows. Say that a game $(N, v)$ is \emph{$\calF$-weakly extendable} if each $\calF$-feasible collection of coalitions contains a minimal (w.r.t.\ inclusion) coalition that is extendable. 

\begin{proposition} \label{prop: weak-extandability}
A $\calF$-weakly extendable game has a nonempty and stable core. 
\end{proposition}

\proof
Let $\calS$ be a $\calF$-feasible collection and $S$ be extendable and minimal w.r.t.\ inclusion in $\calS$. Take $y \in X_\calS$. Then $y(S) < v(S)$ and, for all $T \in \calF \cap \left( 2^S \setminus \{S\} \right)$, we have $y(T) \geq v(T)$. Define $z_S \in \bbR^S$ by 
\[
z_S = y_S + \frac{v(S) - y(S)}{\lvert S \rvert} \bfone^S. 
\]
Notice that $z_S$ is the projection of $y_S$ in $\bbR^S$ onto $\bbX(v_{|S})$. Clearly, $z_S \in C(v_{|S})$ and for all $i \in S$ we have $(z_S)_i > y_i$. As $S$ is extendable, there exists $x \in C(v)$ such that $x_S = z_S$. Then $x \domS y$. 
\endproof

\paragraph{Vitality.} A coalition $S$ is \emph{vital} (for $(N, v)$) if there exists $x \in C(v_{|S})$ such that $x(T) > v(T)$, for all $T \in 2^S \setminus \{\emptyset, S\}$. Originally, \textcite{gillies1959solutions} defined a vital coalition as a minimal coalition via which domination between two given preimputations can be achieved, and showed that domination between two given preimputations can always be achieved via a vital coalition. Moreover, setting the worth $v(S)$ of all non-vital coalitions to $0$ generates a new game that is d-equivalent to the original one. He also provided a characterization of vital coalitions in terms of minimal balanced collections. 

\begin{lemma}[\textcite{gillies1959solutions}] \label{lemma: vitality} \leavevmode \newline
A coalition $S$ is vital if and only if, for any minimal balanced collection $\calB \neq \{S\}$ on $S$ together with its system of balancing weights $\lambda^\calB$, we have
\[
\sum_{T \in \calB} \lambda^\calB_T v(T) < v(S).
\]
\end{lemma}

Equivalently, a coalition $S$ is vital if and only if the core $C(v_{|S})$ is full-dimensional in $\bbR^S$, i.e., if and only if $(S, v)$ is balanced and $\calE(v_{|S}) = \{S\}$, a formulation that reminds us of Theorem \ref{th: basic}. 

\begin{remark}
The \emph{essential} coalitions \parencite{branzei2005strongly, huberman1980nucleolus} play a similar role to the one of vital coalitions, in the sense that nonessential coalitions are redundant when studying domination. We can use Lemma \ref{lemma: vitality} to find the essential coalitions as well, by replacing the minimal balanced collections by the partitions. Moreover, the essential coalitions are core-defining, and characterize the nucleolus \cite{huberman1980nucleolus}. 
\end{remark}

\paragraph{Strict vital-exactness.} \label{sym: sve} A coalition $S$ is \emph{strictly vital-exact} (for $(N, v)$) if there exists a core element $x \in C(v)$ such that $x(S) = v(S)$ and, for all $T \varsubsetneq S$, we have $x(T) > v(T)$. Strict vital-exactness, defined by \textcite{grabisch2021characterization}, implies vitality and exactness. In particular, an exact singleton is strictly vital-exact. Denote the collection of strictly vital-exact coalitions by $\calVE(v)$. 

\medskip 

From the definition, we see that a coalition $S$ is strictly vital-exact if and only if a certain basic polyhedron is nonempty. 

\begin{proposition} \label{prop: sve}
A coalition $S$ is strictly vital-exact if and only if it is exact and
\[
\calE(v^S) \cap \left(2^S \setminus \{S\}\right) = \emptyset. 
\]
\end{proposition}

\proof
The coalition $S$ is strictly vital-exact if and only if the following polyhedron  
\[
P = \{x \in C(v) \mid x(S) = v(S) \text{ and, for all } T \varsubsetneq S, x(T) > v(T)\}. 
\]
is nonempty. As $P$ is a basic polyhedron, we can apply Theorem \ref{th: basic} to have that $P$ is nonempty if and only if $(N, v^S)$ is balanced and $\calE(v^S) \cap \left( 2^S \setminus \{S\} \right) = \emptyset$. By Proposition \ref{prop: exactness}, we reformulate: $P$ is nonempty if and only if $S$ is exact and 
\[
\calE(v^S) \cap \left( 2^S \setminus \{S\} \right) = \emptyset.
\] 
\endproof

Algorithm \ref{algo: sve} checks whether a coalition is strictly vital-exact for a given game. 

\begin{breakablealgorithm}
\caption{Strict vital-exactness checking algorithm} \label{algo: sve}
\begin{algorithmic}[1]
\Require A coalition $S$, a balanced coalition function $v$, the set $\bbB(N)$
\Ensure The Boolean value: `$S$ is strictly vital-exact'
\Procedure{IsStrictlyVitalExact}{$S$, $v$, $\bbB(N)$}
\For{$\calB \in \bbB(N)$}
\If{$\sum_{T \in \calB} \lambda^\calB_T v^S(T) > v(N)$}
\Return \textbf{False}
\ElsIf{$\sum_{T \in \calB} \lambda^\calB_T v^S(T) = v(N)$}
\For{$T \in \calB$}
\If{$T \cap S^c = \varnothing$}
\Return \textbf{False}
\EndIf
\EndFor
\EndIf
\EndFor
\Return \textbf{True}
\EndProcedure
\end{algorithmic}
\end{breakablealgorithm}

\begin{example}[Biswas, Parthasarathy, Potters and Voorneveld \cite{biswas1999large}] \leavevmode \newline 
Let $N = \{a, b, c, d, e\}$ and $v(S) = \min \{x(S), y(S)\}$ with 
\[
x = (2, 1, 0, 0, 0) \qquad y = (0, 0, 1, 1, 1). 
\]
The game is exact, and the set of strictly vital-exact coalitions is 
\[
\calVE(v) = \{\{i\} \mid i \in N\} \cup \{\{b, c\}, \{b, d\}, \{b, e\}, \{a, c, d\}, \{a, c, e\}, \{a, d, e\}\}. 
\]
Let $S = \{a, c\}$. It is exact because the game is exact. It is also vital, becauce $v(S) = 1 > 0 = v(\{a\}) + v(\{c\})$ and $\{\{a\}, \{c\}\}$ is the only minimal balanced collection on $\{a, c\}$ which differs from $\{\{a, c\}\}$. But $S$ is not strictly vital-exact. 
\end{example}

The following result presents a necessary condition for the core to be a stable set based on strictly vital-exact coalitions, as well as an opportunity to reduce the algorithmic complexity of the study of core stability. 

\begin{proposition} \label{prop: sve-describing}
Let $(N, v)$ be a balanced game. The core is a stable set only if $\calVE(v)$ is core-describing, i.e., 
\[
C(v) = \{x \in \bbX(v) \mid x(S) \geq v(S), \text{ for all } S \in \calVE(v)\}. 
\]
\end{proposition}

\proof
Assume that the core is stable, and suppose by contradiction that there exists $y \in \bbX(v) \setminus C(v)$ such that, for all $S \in \calVE(v)$, we have $y(S) \geq v(S)$. Because the core is stable, there exists $x \in C(v)$ such that $x \dom y$. Choose a minimal (w.r.t.\ inclusion) coalition $S$ such that $x \domS y$. Then, $v(S) = x(S) > y(S)$, and for all $T \in 2^S \setminus \{\emptyset, S\}$, we have $x(T) > v(T)$. Therefore, $S$ is strictly vital-exact, which is a contradiction. 
\endproof

This important result shows that checking core stability should begin by finding all strictly vital-exact coalitions (using Algorithm \ref{algo: sve}), and check whether these coalitions determine the core. If they do, one should work on $\calF = \calVE(v)$ instead of $2^N$, as this considerably reduces the complexity.  

\paragraph{Outvoting.} We now introduce the definition of \emph{outvoting}, a transitive sub-relation of domination defined by \textcite{grabisch2021characterization}, that was inspired by a definition given by \textcite{kulakovskaja1971necessary}. In view of Proposition \ref{prop: sve-describing}, let $(N, v)$ be a balanced game for which the collection $\calVE(v)$ of strictly vital-exact coalitions is core-describing. Therefore, we work on the game $(\calVE(v), v)$. 

\begin{definition}
A preimputation $y$ \emph{outvotes} another preimputation $x$ via $S \in \calVE(v)$, written $y \succ_S x$, if $y \text{ dom}_S \ x$ and, for all $T \in \calVE(v) \setminus 2^S$, we have $y(T) \geq v(T)$. Also, $y$ outvotes $x$ ($y \succ x$) if there exists a coalition $S \in \calVE(v)$ such that $y \succ_S x$. 
\end{definition}

Denote by $M(v) = \{x \in \bbX(v) \mid y \not \succ x, \, \forall y \in \bbX(v)\}$ the set of preimputations that are maximal w.r.t.\ outvoting. 

\begin{proposition}[\textcite{kulakovskaja1971necessary}, \textcite{grabisch2021characterization}] \label{prop: C=M} \leavevmode \newline
Let $(N, v)$ be a balanced game. Then $C(v) = M(v)$ if and only if $C(v)$ is a stable set. 
\end{proposition}

This new characterization of the stability of the core using the two binary relations is very useful. We already know that the core is internally stable, and we have that $M(v)$ is externally stable \parencite{grabisch2021characterization}. Then, the core $C(v)$ is a stable set if and only if all preimputations $x \in \bbX(v) \setminus C(v)$ are outvoted. By definition of outvoting, the set of preimputations outvoting $x$ via a given coalition $S$ is a basic polyhedron, whose nonemptiness can be determined using Theorem \ref{th: basic}. The following result has been already proved by \textcite{grabisch2021characterization} in a different way. 

\medskip

Let $x$ be a preimputation and let $S \in \calVE(v)$. Let $v^S_x$ be a set function defined on 
\[ \begin{aligned}
& \calF_S \coloneqq \left( \calVE(v) \setminus 2^S \right) \cup \{N \setminus S\} \cup \calI_S \\
& \text{by} \quad v^S_x (T) = \left\{ \begin{tabular}{ll}
$x_i$ & \quad if $T \in \calI_S$, $T = \{i\}$, \\
$v^S(T)$ & \quad if $T \in \calF_S \setminus \calI_S$,
\end{tabular} \right.
\end{aligned} \]
when $\calI_S$ denotes the collection $\calI_S = \{\{i\} \mid i \in S\}$. 

\begin{proposition}
Let $x$ be a preimputation, and let $S \in \calF_v$. Then $x$ is outvoted by some preimputation via $S$ if and only if $(\calF_S, v^S_x)$ is balanced and $\calI_S \cap \calE(\calF_S, v^S_x) = \emptyset$. 
\end{proposition}

\proof
Let $O_S(x)$ denote the polyhedron of preimputations which outvote $x$ via $S$. By flipping the inequality $z(S) \leq v(S)$ into $z(N \setminus S) \geq v(N) - v(S)$ as we usually do, a preimputation $z \in O_S(x)$ satisfies 
\[ \left\{ \begin{tabular}{rcll}
$z_i$ & $>$ & $x_i$, & \quad for all $i \in S$, \\
$z(T)$ & $\geq$ & $v^S(T)$, & \quad for all $T \in \calF_v \setminus 2^S \cup \{N \setminus S\}$. 
\end{tabular} \right. \]
We see that $O_S(x)$ is a basic polyhedron. By Theorem \ref{th: basic}, $O_S(x)$ is nonempty if and only if $(\calF_S, v^S_x)$ is balanced and $\calI_S \cap \calE(\calF_S, v^S_x) = \emptyset$.
\endproof



To summarize, in this chapter, we have presented an algorithm to compute the set of minimal balanced collections on any finite set of players, and on any set system. The Python implementation of the algorithm and the minimal balanced collections up to 7 players can be provided upon request. Next, a new family of polyhedra is introduced, the \emph{basic polyhedra}, as well as a characterization of their nonemptiness. These polyhedra are ubiquitous in mathematical economics, especially in cooperative game theory, as our many examples have revealed.


\chapter{The combinatorial structure of a game} 

\label{ChapterC} 




In this chapter, we discuss some interactions between combinatorics and cooperative game theory which have been relatively unexplored. First, we study the similarities between the balanced collections and some hypergraphs, regular or uniform. Secondly, we investigate how a hyperplane arrangement, called the resonance arrangement, can be a powerful tool in the study of cooperative games.

\medskip 

A hypergraph is a generalization of a graph, where edges can contain more than two nodes. In the first connection we make between balanced collections and hypergraphs, we will interpret the edges as coalitions, and then show that a balanced collection is a \emph{regular} hypergraph, i.e., a hypergraph for which each node is contained in the same number of edges. After, we reverse the incidence between edges and nodes, to study hypergraphs the edges of which have all the same cardinality. These hypergraphs are said to be \emph{uniform} and are in a duality relation with the regular hypergraphs. We therefore explain how the uniform hypergraphs, which are rather simple to study and generate, can help us to generate (minimal) balanced collections. To study them, we use the theory of \emph{combinatorial species} developed by Joyal \cite{joyal1981theorie}, which is an abstract and systematic method for deriving generating functions of discrete structures. We construct the species of $k$-uniform hypergraphs of size $p$, as an intermediary step to construct the species of (minimal) balanced collections. 

\medskip 

In the second part, we study how hyperplane arrangements can be useful in the study of cooperative games. A hyperplane arrangement is a set of hyperplanes in a specific vector space. The one we are interested in is the \emph{resonance arrangement} of the vector space of side payments, where each hyperplane is the set of side payments leaving the payment of a specific coalition fixed. We notice that the connected components of the complement of the union of the hyperplanes, called the \emph{chambers} of the arrangement, are in bijection with the maximal unbalanced collections, where a collection is said to be unbalanced if it does not contain a balanced collection. Later, we show that a coalition function can be seen as a \emph{distortion} of the resonance arrangement. The combinatorial properties of this distorted arrangement are closely related to the set of feasible regions and the facial structure of the core, and a few results about these connections are presented.





\section{The combinatorics of balanced collections} \label{sec: mbc}

Balanced collections are known in other scientific areas under different names, especially in combinatorics and mathematical physics. We start with combinatorics. 





\subsection{Hypergraph theory}

Most of the notation and the definitions of this subsection are due to \textcite{berge1984hypergraphs}. A \emph{(undirected) hypergraph} $\calH$ is a pair $\calH = (N, E)$ where $N$ is the set of elements called \emph{nodes} and $E$ is a spanning collection of nonempty subsets of $N$, called \emph{hyperedges} or simply \emph{edges}, i.e., the union of all edges coincides with the set of nodes.

\medskip

We call $\lvert N \rvert$ the \emph{order} of $\calH$ and $\lvert E \rvert$ the \emph{size} of $\calH$. The degree of a node $x \in N$, denoted by $\delta(x)$, is the number of edges that contain $x$. If each node has the same degree $d$, the hypergraph is said to be \emph{$d$-regular}. If each edge has cardinality $k$, the hypergraph is said to be \emph{$k$-uniform}. For a hypergraph $\calH$, we denote by $A^\calH$ its incidence matrix, defined by 
\[
A^\calH_{ij} = \left\{ \begin{aligned}
1, & \quad \text{if the $i$-th node lies in the $j$-th edge}, \\
0, & \quad \text{otherwise}. 
\end{aligned} \right.
\]
The hypergraph $\calH'$ which has an incidence matrix $A^{\calH'}$ being the transpose matrix of $A^\calH$ is the \emph{dual} hypergraph of $\calH$. Counting the non-zero entries of each row and column shows that the dual hypergraph of a $k$-uniform hypergraph is $k$-regular, and the dual hypergraph of a $d$-regular hypergraph is $d$-uniform. 

\medskip

In the sequel, we denote with Greek letters the nodes of hypergraphs, and by Roman letters the players in balanced collections. Because a hypergraph can have more than one occurence of the same edges, the collection of edges is a \emph{multiset}, and not a set in the usual sense. However, we will denote multisets like sets, meaning with braces, as we can see in the following example. 

\begin{example} Let us consider two dual hypergraphs 
\[
\calH = \left( \{\alpha, \beta, \gamma\}, \Big\{\{\beta\}, \{\alpha, \gamma\}\Big\} \right) \qquad \calH' = \left( \{\alpha, \beta\}, \Big\{\{\beta\}, \{\alpha\}, \{\beta\}\Big\} \right).
\]
The adjacency matrices are 
\[
A^\calH = \begin{bmatrix} 0 & 1 \\ 1 & 0 \\ 0 & 1 \end{bmatrix} \qquad A^{\calH'} = \begin{bmatrix} 0 & 1 & 0 \\ 1 & 0 & 1 \end{bmatrix}
\]
We see that $\calH$ is simple, i.e., no two edges are identical, and regular, while $\calH'$ is uniform and not simple. Also, they are not defined on the same node set. 
\end{example}

\textcite{shapley1967balanced} defined the \emph{depth} of a minimal balanced collection $\calB$, denoted by $\text{depth}(\calB)$, as the least common multiple of the denominators of the elements of the system of balancing weights associated with $\calB$. We extend here his definition to pairs $(\calB, \lambda)$ of balanced collections together with one compatible system of balancing weights, as the smallest least common multiple of the denominators of the elements of $\lambda$. 

\begin{example} Consider the following balanced collection on $N = \{a, b, c, d, e, f, g\}$.  
\[
\begin{tabular}{ccccccc}
\toprule
$\{a, b\}$ & $\{a, c\}$ & $\{a, d\}$ & $\{b, c, d\}$ & $\{e, f\}$ & $\{e, g\}$ & $\{f, g\}$\\
\midrule
\sfrac{1}{3} & \sfrac{1}{3} & \sfrac{1}{3} & \sfrac{2}{3} & \sfrac{1}{2} & \sfrac{1}{2} & \sfrac{1}{2} \\
\bottomrule
\end{tabular}
\]
The denominators of the weights are $2$ and $3$, hence its depth is $6$. Consider now the balanced collection on $N = \{a, b, c\}$ defined by $\calB = \{abc, ab, c\}$. Two possible systems of balancing weights are 
\[
\lambda = \left( \frac{1}{2}, \frac{1}{2}, \frac{1}{2} \right) \qquad \text{and} \qquad \lambda' = \left( \frac{1}{3}, \frac{2}{3}, \frac{2}{3} \right). 
\]
The depth of $\left( \calB, \lambda \right)$ is $2$, and the depth of $\left( \calB, \lambda' \right)$ is $3$. 
\end{example}

\begin{proposition}
The set of $d$-regular hypergraphs of order $\lvert N \rvert = n$ is in bijection with the set of pairs $(\calB, \lambda)$ with $\calB$ a balanced collection of depth $d$ with $n$ players and $\lambda$ a system of balancing weights of $\calB$. 
\end{proposition}

\proof
Let $\calH = (N, E)$ be a $d$-regular hypergraph of order $n$. Denote by $\calB$ the set of edges of $E$, with only one occurrence of each element. Define a function $\lambda$ that assigns to each element $S$ of $\calB$ the number of occurrences of $S$ in $E$ divided by $d$. Then, $\calB$ forms a balanced collection with the system of balanced weights defined by $\lambda$. \\
Now, let $\calB$ be a balanced collection with a system of balancing weights $\{\lambda_S\}_{S \in \calB}$ of depth $d$. Denote by $E$ the multiset constructed by including $d \lambda_S$ occurrences of each coalitions $S$ in $\calB$. By definition of the depth, $d \lambda_S$ is always an integer. Set $\calH = (N, E)$. For all $x \in N$, the sum of the occurrences of the edges containing $x$ is 
\[
\sum_{\substack{S \in E \\ S \ni x}} d \lambda_S = d \sum_{\substack{S \in \calB \\ S \ni x}} \lambda_S = d. 
\]
Therefore, $\calH$ is a $d$-regular hypergraph. 
\endproof



Let $\calH$ be a uniform hypergraph of order $n$. For its associated balanced collection, an edge represents the collection of coalitions a given player belongs to. Then, because each player belongs to the same number of coalitions, counted with multiplicities, the collection is balanced.

\begin{table}[ht]
\begin{center}
\begin{tabular}{lrr}
\toprule
\multirow{2}{*}{\begin{minipage}{3cm} balanced \\ collection \end{minipage}}  & \multicolumn{2}{r}{hypergraph} \\
\cmidrule{2-3}
 & regular & uniform \\
\midrule
depth ($d$) & $d$-regularity & $k$-uniformity \\
\midrule
players & nodes & edges \\
\midrule
$d \cdot \sum_{S \in \calB} \lambda_S$ & size & order \\
\bottomrule
\end{tabular}
\caption{Equivalences between hypergraphs and balanced collections.}
\end{center}
\end{table}

\begin{example} 
Let $\calH$ be the $2$-uniform hypergraph on $N = \{\alpha, \beta, \gamma, \delta, \varepsilon\}$ with the collection of edges: $\{\{\alpha, \beta\}, \{\beta, \gamma\}, \{\alpha, \gamma\}, \{\delta, \varepsilon\}\}$. Because $\calH$ is uniform, we know that its dual is a balanced collection, which we construct now. The size of $\calH$ is $4$ and its order is $5$. Then the balanced collection involves $4$ players, and the sum of the integer multiplicities of the coalitions is $5$. Because $\calH$ is $2$-uniform, the depth of the balanced collection is $2$. To know the coalitions and their \emph{integer weights}, i.e., their weights multiplied by the depth of the collection, we must know the dual of $\calH$, which is 
\[
\calH' = \left( \{\alpha', \beta', \gamma', \delta'\}, \Big\{ \{\alpha', \gamma'\}, \{\alpha', \beta'\}, \{\beta', \gamma'\}, \{\delta'\}, \{\delta'\} \Big\} \right). 
\]
Therefore, the pair $(\calB, \lambda)$ associated with $\calH$ is 
\[
\calB = \big\{\{a, c\}, \{a, b\}, \{b, c\}, \{d\}\big\}, \qquad \lambda = \left( \frac{1}{2}, \frac{1}{2}, \frac{1}{2}, \frac{2}{2} = 1 \right). 
\]
\end{example}

Writing a program generating uniform hypergraphs of a given size is extremely easy: it suffices to take arbitrary sets of same cardinality, and relabel their elements to fit in the notation $N = \{1, \ldots, n\}$. This may be a lead for a more efficient way to generate minimal balanced collections. The main drawback of this approach is the difficulty to anticipate the minimality of the collection, which makes the apparent ease of the method vanish. 

\begin{definition}
Let $\calH = (N, E)$ be a hypergraph, let $A \subseteq N$ and $X \subseteq E$. The hypergraph denoted by $\calH_A$ and defined by
\[
\calH_A = \left(A, \Big\{ S \cap A \ \Big| \ S \in E \text{ and } S \cap A \neq \emptyset  \Big\} \right). 
\]
is the \emph{subhypergraph} of $\calH$ induced by $A$. The hypergraph $\calH^X = (N, X)$ is the \emph{partial hypergraph} of $\calH$ generated by $X$. 
\end{definition}

Notice that the subhypergraph of a hypergraph corresponds to a partial hypergraph of its dual. Let $\calH = (N, E)$ be a uniform hypergraph. If there exists $A \varsubsetneq N$ such that $\calH_A$ is uniform, then $\calH_{N \setminus A}$ is also a uniform hypergraph. We say that a uniform hypergraph is \emph{minimal} if no proper subhypergraph is uniform, and we say that a regular hypergraph is \emph{minimal} if no proper partial hypergraph is regular.


\begin{theorem} \label{th: minimal-balanced=minimal-uniform}
The balanced collection associated with a minimal uniform hypergraph is minimal.
\end{theorem}

\proof
Let $\calH = (N, E)$ be a minimal $k$-uniform hypergraph of size $n$. First, let us prove that $\calH'$, the dual hypergraph of $\calH$, is minimal $k$-regular of order $n$. Consider $A^\calH$ the incidence matrix of $\calH$. We have that the sum of the entries of each of the $n$ columns is $k$. By minimality, it is impossible to remove a row of $A^\calH$ and still have all the columns having the same sum of their entries. Then, it is impossible to remove a column to its transpose matrix such that the sum of the entries of its $n$ rows remains constant, so its dual is minimal $k$-regular of order $n$. We have already shown that from each regular hypergraph $\calH'$ we have a balanced collection $\calB$ together with a system of balancing weights, and the minimality of $\calH'$ as a regular hypergraph implies the minimality of $\calB$ as a balanced collection. 
\endproof

\begin{example} Consider the hypergraph $\calH$ defined on $N = \{\alpha, \beta, \gamma, \delta, \varepsilon, \zeta, \eta, \iota\}$ with the collection of edges: 
\[
E = \{\{\alpha, \beta, \gamma, \eta, \iota\}, \{\alpha, \delta, \varepsilon, \zeta, \iota\}, \{\beta, \delta, \varepsilon, \zeta, \eta\}, \{\gamma, \delta, \varepsilon, \eta, \iota\}\}. 
\]
It is a $5$-uniform hypergraph of size $4$ and order $8$. Its associated balanced collection is defined on $\{a, b, c, d\}$, has $8$ coalitions counted with multiplicities, and is of depth $5$: 

\[
\begin{tabular}{ccccccc}
\toprule
$\{a, b\}$ & $\{a, c\}$ & $\{a, d\}$ & $\{b, c, d\}$ & $\{b, c\}$ & $\{a, c, d\}$ & $\{a, b, d\}$\\
\midrule
\sfrac{1}{5} & \sfrac{1}{5} & \sfrac{1}{5} & \sfrac{2}{5} & \sfrac{1}{5} & \sfrac{1}{5} & \sfrac{1}{5} \\
\bottomrule
\end{tabular}
\]

\medskip

\noindent Let $A = \{\alpha, \beta, \gamma, \delta, \varepsilon\}$. Then, the subhypergraph $\calH_A$ is still uniform. We have
\[ \begin{aligned}
& \calH_A = \left( A, \big\{\{\alpha, \beta, \gamma\}, \{\alpha, \delta, \varepsilon\}, \{\beta, \delta, \varepsilon\}, \{\gamma, \delta, \varepsilon\} \big\} \right), \\
\text{and} \quad & \calH_{N \setminus A} = \left( N \setminus A, \big\{\{\eta, \iota\}, \{\zeta, \iota\}, \{\zeta, \eta\}, \{\eta, \iota\} \big\} \right). 
\end{aligned} \]
None of these two hypergraphs has a subhypergraph which is uniform, therefore they are minimal uniform hypergraphs. Their associated minimal balanced collections are
\[
\big\{ \{a, b\}, \{a, c\}, \{a, d\}, \{b, c, d\} \big\} \quad \text{and} \quad \big\{ \{b, c\}, \{a, c, d\}, \{a, b, d\} \big\}. 
\]
Notice that the union of these two minimal balanced collections is the balanced collection $\calB$ associated with the initial hypergraph $\calH$. 
\end{example}

\medskip

The systems of balancing weights of a balanced collection can also be seen as a matching. As in graph theory, a \emph{matching} is a collection of disjoint edges spanning $N$. There also exists a fractional analog to a matching. 

\begin{definition}
A \emph{fractional matching} in a hypergraph $\calH = (N, E)$ is a function $\mu: E \to [0, 1]$ such that, for every node $x \in N$, we have 
\[
\sum_{\substack{S \in E \\ S \ni x}} \mu(S) \leq 1. 
\]
A fractional matching is said to be \emph{perfect} if, for every node $x \in N$, we have 
\[
\sum_{\substack{S \in E \\ S \ni x}} \mu(S) = 1. 
\]
\end{definition}

It is easy to see that a simple hypergraph on $N$ admitting a perfect fractional matching $\mu$ induces a balanced collection on $N$, which is $\{S \in E \mid \mu(S) > 0\}$. The converse holds as well. 

\subsection{Enumeration of uniform hypergraphs} 

Let us now study the uniform hypergraphs. To do so, we use the \emph{species of structures}, and the corresponding operations on formal power series developed by \textcite{joyal1981theorie}. Most of the following definitions come from \textcite{bergeron1998combinatorial}. 

\medskip

Informally, a species of structures is a rule, $\text{F}$, associating with each finite set $U$, a finite set $\text{F}[U]$ which is ``independent of the nature'' of the elements of $U$. The members of $\text{F}[U]$, called \emph{${\rm F}$-structures}, are interpreted as combinatorial structures on the set $U$ given by the rule $\text{F}$. The fact that the rule is independent of the nature of the elements of $U$ is expressed by invariance under relabeling. More precisely, to any bijection $\sigma: U \to V$, the rule $\text{F}$ associates a bijection $\text{F}[\sigma]: \text{F}[U] \to \text{F}[V]$, which transforms each $\text{F}$-structures on $U$ into an (isomorphic) $\text{F}$-structure on $V$. 

\begin{example}
Let $\textsc{Gr}$ be the rule associating to each finite set $U$ the set of graphs with $U$ as a set of nodes. For any set $V$ such that there exists a bijection $\sigma: U \to V$ between them, the structures obtained from the graphs on $U$ by relabeling the nodes with $\sigma$ is a graph on $V$. The important combinatorial information carried by a graph is the number of nodes and the set of edges, not the nodes themselves. The same remark applies to hypergraphs, trees, balanced collections, etc. 
\end{example}

Let us now give the formal definition of a combinatorial species of structures.

\begin{definition}[\textcite{joyal1981theorie}] \leavevmode \newline
A \emph{species (of structures)} is a rule $\text{F}$ which
\begin{itemize}
\item produces, for each finite set $U$, a finite set $\text{F}[U]$, 
\item produces, for each bijection $\sigma: U \to V$, a function $\text{F}[\sigma]: \text{F}[U] \to \text{F}[V]$. 
\end{itemize}
The functions $\text{F}[\sigma]$ should further satisfy the following functorial properties:
\begin{itemize}
\item for all bijections $\sigma: U \to V$ and $\tau: V \to W$, we have $\text{F}[\tau \circ \sigma] = \text{F}[\tau] \circ \text{F}[\sigma]$, 
\item for the identity map $\text{Id}_U: U \to U$, we have $\text{F}[\text{Id}_U] = \text{Id}_{\text{F}[U]}$. 
\end{itemize}
An element $s \in \text{F}[U]$ is called an \emph{$\text{F}$-structure on $U$}. The function $\text{F}[\sigma]$ is called the \emph{transport} of $\text{F}$-structures \emph{along} $\sigma$. 
\end{definition}

Define the rule $\textsc{Sh}$ which assigns to a set $U$ the set of minimal balanced collections on $U$, and which assigns to a bijection $\sigma: U \to V$, a function $\textsc{Sh}[\sigma]: \textsc{Sh}[U] \to \textsc{Sh}[V]$, acting on a minimal balanced collection $\calB \in \textsc{Sh}[U]$ as follows:
\[
\textsc{Sh}[\sigma](\calB) = \big\{\{\sigma(u) \mid u \in S\} \mid S \in \calB\big\}. 
\]

\begin{theorem}
The rule {\normalfont \textsc{Sh}} defines a species of structures. 
\end{theorem}

\proof
First, for any finite set $U$, the number of $\textsc{Sh}$-structures is finite because it is bounded by $2^{2^n}$. Let $\sigma$ be a bijection $\sigma: U \to V$, and let $\calB \in \textsc{Sh}[U]$ be a minimal balanced collection on $U$. Let $u$ be an element of $U$, and denote by $\calB_u$ the collection $\calB_u = \{S \in \calB \mid u \in S\}$. Then, we have $\sum_{S \in \calB_u} \lambda_S = 1$, which implies $\sum_{S \in \textsc{Sh}[\sigma](\calB_v)} \lambda_S = 1$ with $v = \sigma(u)$. Then $\textsc{Sh}[\sigma](\calB) \in \textsc{Sh}[V]$. Now, we prove the functoriality of the functions $\textsc{Sh}[\sigma]$. Let $\sigma: U \to V$ and $\tau: V \to W$, and let $\calB \in \textsc{Sh}[U]$. 
\[ \begin{aligned}
\textsc{Sh}[\sigma \circ \tau](\calB) & = \{\{\tau(\sigma(u)) \mid u \in S\} \mid S \in \calB\}, \\
& = \{\{\tau(v)\mid v = \sigma(u), \ u \in S\} \mid S \in \calB\}, \\
& = \{\{\tau(v) \mid v \in T\} \mid T \in \textsc{Sh}[\sigma](\calB)\}, \\
& = \left( \textsc{Sh}[\tau] \circ \textsc{Sh}[\sigma] \right) (\calB). 
\end{aligned} \]
Finally, let $\text{Id}_U$ be the identity map on $U$. Then, for all $\calB \in \textsc{Sh}[U]$, 
\[
\textsc{Sh}[\text{Id}_U](\calB) = \{\{\text{Id}_U(u) \mid u \in S \} \mid S \in \calB\} = \calB,
\]
and $\textsc{Sh}[\text{Id}_U] = \text{Id}_{\textsc{Sh}[U]}$. 
\endproof

To each species of structures $\text{F}$ is associated a formal power series related to the enumeration of $\text{F}$-structures. They are enumerated by the \emph{generating series} of $\text{F}$, denoted $\text{F}(x)$. For all sets $U$ of cardinality $n$, the number of $\text{F}$-structures on $U$ only depends on $n$. This property follows from the functoriality of $\text{F}$ because all the $\text{F}[\sigma]$ are bijections. Hence, the cardinalities $\lvert \text{F}[U] \rvert$ are completely characterized by the sequence of numbers $\{f_n = \lvert \text{F}[\{1, \ldots, n\}] \rvert \mid n \in \bbN\}$. 

\begin{definition}
The \emph{generating series} of a species of structures $\text{F}$ is the formal power series
\[
\text{F}(x) = \sum_{n \geq 0} f_n \frac{x^n}{n!}. 
\]
\end{definition}

With formal series we are not concerned by the convergence rules, as they only are a handy notation to denote a sequence of integers. For instance, the sequence $(1, 2, 4, 8, \ldots)$ is denoted by $e^{2x}$. 

\begin{example} \label{sym: varsigma} Let $\varsigma^{[p]}$ be the species of $p$-multisets, associating to each finite set $U$, the set of multisets of cardinality $p$ on $U$. By definition of the multisets coefficients, we have 
\[
\varsigma^{[p]}(x) = \sum_{n \geq 0} \frac{n^{\overline{p}}}{p!} \frac{x^n}{n!}, \qquad \text{with} \quad n^{\overline{p}} = \prod_{i = 0}^{p-1}(n+i) = n(n+1) \ldots (n + p -1). 
\]
\end{example}
With the correspondence between species and formal power series, it is possible to perform combinatorial constructions on species, and to report them into power series, and vice versa. Let $\wp^{[2]}$ denote the species of $2$-subsets, associating to each finite set $U$ the set of subsets of $2$ elements of $U$, and let $\wp$ denote the species of subsets, associating to each finite set $U$ the collection of subsets of $U$. Then, we can compose the two species in the following way: we first apply $\wp^{[2]}$ to $U$, to obtain the set of all subsets of cardinality $2$ of $U$, and in a second step we apply $\wp$, to take a subset of the set of $2$-subsets of $U$. This operation, called the \emph{functorial composition} of species of structures, permits the construction of the species of simple graphs $\textsc{Gr}$ from the species of $2$-subsets and the species of subsets. 

\begin{example}
The generating series of the species $\wp^{[2]}$ of $2$-subsets and the species $\wp$ of subsets are, respectively, 
\[
\wp^{[2]}(x) = \sum_{n \geq 0} \binom{n}{2} \frac{x^n}{n!} \qquad \text{and} \qquad \wp(x) = \sum_{n \geq 0} 2^n \frac{x^n}{n!} = e^{2x}, 
\]
we obtain the generating series of simple graph by composing the two previous ones according to the combinatorial construction we performed: 
\[
\textsc{Gr}(x) = \sum_{n \geq 0} 2^{\binom{n}{2}} \frac{x^n}{n!}. 
\]
\end{example}

The calculus of formal power series counts several `natural' combinatorial operations. We present formally in the next result the operations we need in our construction. 

\begin{definition}[\textcite{bergeron1998combinatorial}] \leavevmode \newline
Let $\text{F}$ and $\text{G}$ be two species of structures. 
\begin{itemize}
\item The species $\text{F} \cdot \text{G}$, called the \emph{product} of $\text{F}$ and $\text{G}$, is defined as follows: an $(\text{F} \cdot \text{G})$-structure is an ordered pair $s = (f, g)$, where
\begin{itemize}
\item $f$ is an $\text{F}$-structure on a subset $U_1 \subseteq U$, 
\item $g$ is a $\text{G}$-structure on a subset $U_2 \subseteq U$, 
\item $\{U_1, U_2\}$ is a partition of $U$. 
\end{itemize} 
The transport along a bijection $\sigma: U \to V$ is carried out by setting, for each $(\text{F} \cdot \text{G})$-structure $s = (f, g)$ on $U$, 
\[
(\text{F} \cdot \text{G})[\sigma](s) = \left( \text{F}[\sigma_{|U_1}](f), \text{G}[\sigma_{|U_2}](g) \right). 
\]
\item The species $\text{F} \ssq \text{G}$, called the \emph{functorial composite} of $\text{F}$ and $\text{G}$, is defined as follows: an $(\text{F} \ssq \text{G})$-structure on $U$ is an $\text{F}$-structure on the set $\text{G}[U]$ of all the $\text{G}$-structures on $U$, i.e., for any finite set $U$, $(\text{F} \ssq \text{G})[U] = \text{F}[\text{G}[U]]$. The transport along a bijection $\sigma: U \to V$ is carried out by setting $(\text{F} \ssq \text{G})[\sigma] = \text{F}[\text{G}[\sigma]]$. 
\end{itemize}
\end{definition}

The generating series of the newly constructed species of structures are given, from the generating series of F and G, by the following result. 

\begin{proposition}[\textcite{bergeron1998combinatorial}] \leavevmode \newline
Let F and G be two species of structures. 
\begin{itemize}
\item The generating series of the species $\text{F} \cdot \text{G}$ satisfies the equality
\[
(\text{F} \cdot \text{G})(x) = \text{F}(x) \text{G}(x) = \sum_{n \geq 0} \left( \sum_{k = 0}^n \binom{n}{k} f_k g_{n-k} \right) \frac{x^n}{n!}. 
\]
\item The generating series of the species $\text{F} \ssq \text{G}$ satisfies the equality
\[
(\text{F} \ssq \text{G})(x) = \text{F}(x) \ssq \text{G}(x) \coloneqq \sum_{n \geq 0} f_{g_n} \frac{x^n}{n!}. 
\]
\end{itemize}
\end{proposition}

Now, we need to define the species to which we want to apply the operations described above. Let $\text{E}$ be the species of sets defined by $\text{E}[U] = {U}$. For each finite set $U$, there is a unique $\text{E}$-structure, namely the set $U$ itself. Let $\text{E}_k$ be the species of sets of cardinality $k$, i.e., $\text{E}_k[U] = {U}$ if $\lvert U \rvert = k$, and $\text{E}_k[U] = \emptyset$ otherwise. We define the species of nonempty sets $\text{E}_+$ by $\text{E}_+ = \sum_{k \geq 1} \text{E}_k$. Their generating series are
\[
\text{E}(x) = \sum_{n \geq 0} \frac{x^n}{n!} = e^x, \qquad \text{E}_k(x) = \frac{x^k}{k!}, \qquad \text{and} \qquad \text{E}_+(x) = \sum_{n \geq 1} \frac{x^n}{n!} = e^x - 1. 
\]
From these, we construct the species of subsets $\wp$ \label{sym: wp} which satisfies the combinatorial equation $\wp \coloneqq \text{E} \cdot \text{E}$. In a similar manner, we construct the species of $k$-subsets by $\wp^{[k]} \coloneqq \text{E}_k \cdot \text{E}$.  Their generating series are therefore $\wp(x) = e^{2x} = \sum_{n \geq 0} 2^n \frac{x^n}{n!}$ and
\[ \begin{aligned} 
\wp^{[k]}(x) = \frac{x^k}{k!} \sum_{n \geq 0} \frac{x^n}{n!} = \sum_{n \geq 0} \frac{x^{n+k}}{k!n!} = \sum_{n \geq k} \frac{x^n}{k!(n-k)!} = \sum_{n \geq 0} \binom{n}{k} \frac{x^n}{n!}, 
\end{aligned} \]
giving the well-known combinatorial interpretation of binomial coefficients. 

\begin{proposition}
\label{prop: comb-equation}
The species of $k$-uniform hypergraphs of size $p$, which we denote by ${\normalfont \textsc{Hyp}}_{k,p}$, satisfies the following combinatorial equation:
\[
{\normalfont \textsc{E}} \cdot {\normalfont \textsc{Hyp}}_{k,p} = \varsigma^{[p]} \ssq \wp^{[k]}. 
\]
\end{proposition}

\proof
The right-hand side of the equation defines the species of $p$-multisets of $k$-subsets. On the left-hand side, we have the product between the species of sets $\text{E}$ and the species of $k$-uniform hypergraphs of size $p$. The reason from the presence of $\text{E}$ in the equation is because a hypergraph must be \emph{spanning}, i.e., every node must be included in an edge. The construction $\varsigma^{[p]} \ssq \wp^{[k]}$ does not ensure that, so for every $(\varsigma^{[p]} \ssq \wp^{[k]})$-structure, we can partition the set of nodes into nodes which are included in one $k$-subset, and the others. Equivalently, for any $(\text{E} \cdot \textsc{Hyp}_{p, k})$-structure, there is a partition $\{U_1, U_2\}$, on which the structure is a $\textsc{Hyp}_{p, k}$-structure on $U_2$, and just a set of nodes on $U_1$. 
\endproof

To solve this equation, and be able to compute the generating series, we need a species ``$\text{E}^{-1}$''. It is an example of \emph{virtual species}, a solution to a combinatorial equation which does not necessarily correspond to an actual species. Its purpose is to solve combinatorial equations like the one of Proposition \ref{prop: comb-equation}, or more simply, $\text{F} = \text{E} \cdot \text{G}$, when we want to find G knowing F. In this case, a G-structure is a spanning F-structure. In the literature, the conversion from F to G, or G to F, is called a \emph{binomial transform}, or an \emph{Euler transform}, and is handled using `$\text{E}^{-1}$' in combinatorial equations. By decomposing $\text{E} = \text{E}_0 + \text{E}_+$, we have 
\[
\text{E}^{-1} = (\text{E}_0 + \text{E}_+)^{-1} = \sum_{n \geq 0} (-1)^n \left( E_+ \right)^n, 
\]
and therefore, 
\[
\text{E}^{-1}(x) = \sum_{n \geq 0} (-1)^n (\text{E}_+)^n(x) = \sum_{n \geq 0}(-1)^n (e^x - 1)^n = \sum_{n \geq 0} (1 - e^x)^n  = e^{-x}. 
\]

Recall that $n^{\overline{p}} = n (n+1) \ldots (n + p - 1)$. 

\begin{theorem}
The generating series of the species ${\normalfont \textsc{Hyp}}_{k,p}$ is
\[
{\normalfont \textsc{Hyp}}_{k,p}(x) = \sum_{n \geq 0} \left( \sum_{i = 0}^n (-1)^{n-i} \binom{n}{i} \frac{\binom{i}{k}^{\overline{p}}}{p!} \right) \frac{x^n}{n!}. 
\]
\end{theorem}

\proof
The generating series of the species of $k$-uniform hypergraphs of size $p$ is 
\[
\textsc{Hyp}_{k,p}(x) = \textsc{E}^{-1}(x) \cdot \left( \varsigma^{[p]}(x) \ssq \wp^{[k]}(x) \right). 
\]
Let us first compute $\varsigma^{[p]}(x) \ssq \wp^{[k]}(x)$.
\[
\varsigma^{[p]}(x) \ssq \wp^{[k]}(x) = \left( \sum_{n \geq 0} \frac{n^{\overline{p}}}{p!} \frac{x^n}{n!} \right) \ssq \left( \sum_{n \geq 0} \binom{n}{k} \frac{x^n}{n!} \right) = \sum_{n \geq 0} \frac{\binom{n}{k}^{\overline{p}}}{p!} \frac{x^n}{n!}. 
\] 
Then, we have 
\[
\textsc{Hyp}_{k,p}(x) = \left( \sum_{n \geq 0} (-1)^n \frac{x^n}{n!} \right) \cdot \left( \sum_{n \geq 0} \frac{\binom{n}{k}^{\overline{p}}}{p!} \frac{x^n}{n!} \right) = \sum_{n \geq 0} \left( \sum_{i = 0}^n (-1)^{n-i} \binom{n}{i} \frac{\binom{i}{k}^{\overline{p}}}{p!} \right) \frac{x^n}{n!}. 
\]
\endproof

Finding a way to express `minimality' into combinatorial equations may lead to a new way to generate and count the minimal balanced collections. 




%
%
%

\begin{example}
Let us count the spanning $2$-uniform hypergraphs of size $3$, with no more than three nodes. Then, because the hypergraphs is $2$-uniform, $n$ only goes from $2$ to $3$. Then the number we are looking for is 
\[ \begin{aligned} 
\sum_{n = 2}^3 \left( \sum_{i = 2}^n (-1)^{n-i} \binom{n}{i} \frac{\binom{i}{2}^{\overline{p}}}{p!} \right) & = (-1)^0 \binom{2}{2} \frac{\binom{2}{2}^{\overline{3}}}{3!} + (-1)^1 \binom{3}{2} \frac{\binom{2}{2}^{\overline{3}}}{3!} + (-1)^0 \binom{3}{3} \frac{\binom{3}{2}^{\overline{3}}}{3!} \\
& = 1 - 3 + 10 = 8.
\end{aligned} \]

\begin{table}[ht]
\begin{center}
\begin{tabular}{lcccc}
\toprule
nodes & \hspace{1cm} & hypergraphs & \hspace{1cm} & balanced collections \\
\midrule
$\{\alpha, \beta\}$ & & $\big\{\{\alpha, \beta\}, \{\alpha, \beta\}, \{\alpha, \beta\}\big\}$ & & $\big\{\{a, b, c\}\big\}$ \\
\midrule
\multirow{7}{*}{$\{\alpha, \beta, \gamma\}$} & & $\big\{\{\alpha, \beta\}, \{\alpha, \beta\}, \{\alpha, \gamma\}\big\}$ & & $\big\{\{a, b, c\}, \{a, b\}, \{c\} \big\}$ \\
& & $\big\{\{\alpha, \beta\}, \{\alpha, \beta\}, \{\beta, \gamma\}\big\}$ & & $\big\{\{a, b\}, \{a, b, c\}, \{c\} \big\}$ \\
& & $\big\{\{\alpha, \gamma\}, \{\beta, \gamma\}, \{\alpha, \gamma\}\big\}$ & & $\big\{\{a, c\}, \{b\}, \{a, b, c\} \big\}$ \\
& & $\big\{\{\alpha, \gamma\}, \{\alpha, \beta\}, \{\alpha, \gamma\}\big\}$ & & $\big\{\{a, b, c\}, \{b\}, \{a, c\} \big\}$ \\
& & $\big\{\{\alpha, \beta\}, \{\beta, \gamma\}, \{\beta, \gamma\}\big\}$ & & $\big\{\{a\}, \{a, b, c\}, \{b, c\} \big\}$ \\
& & $\big\{\{\alpha, \gamma\}, \{\beta, \gamma\}, \{\beta, \gamma\}\big\}$ & & $\big\{\{a\}, \{b, c\}, \{a, b, c\} \big\}$ \\
& & $\big\{\{\alpha, \beta\}, \{\beta, \gamma\}, \{\alpha, \gamma\}\big\}$ & & $\big\{\{a, c\}, \{a, b\}, \{b, c\} \big\}$ \\
\bottomrule
\end{tabular}
\caption{List of $k$-uniform hypergraphs of size $p$ and order $\leq 3$.}
\label{tab: ex-uni-hypergraph}
\end{center}
\end{table}

The $8$ corresponding hypergraphs are listed in Table \ref{tab: ex-uni-hypergraph}. They only generate $5$ balanced collections: 
\[
\{abc\}, \{ab, ac, bc\}, \{abc, ab, c\}, \{abc, ac, b\}, \{abc, bc, a\}, 
\]
with the last three appearing twice in the table. Remark that $\{abc\}$ is not of depth $2$, but the multiset of coalitions $\{abc, abc\}$ can be seen as an absurd balanced collection of depth $2$. Usually, in a balanced collection we do not consider multiple occurrences of the same coalition, this is why we changed it into $\{abc\}$. 

\medskip 

By carefully inspecting the balanced collections appearing twice in the table, we see that a pair of players is completely included in a coalition, or not at all. We call these pairs \emph{macro-players} \parencite{faigle2016games}. Indeed, $\{a, b\}$ is a macro-player in $\{abc, ab, c\}$, $\{a, c\}$ is a macro-player in $\{abc, ac, b\}$, and finally $\{b, c\}$ is a macro-player in $\{abc, bc, a\}$. 
\end{example}

\paragraph{Thermal quantum field theory and unbalanced collections} \leavevmode \newline

A collection of subsets of $N$ which does not contain a balanced collection is said to be \emph{unbalanced}. It is \emph{maximal} if no proper super-collection of it is unbalanced. Strangely enough, maximal unbalanced collections are also a topic whose importance is rapidly growing in mathematical physics, especially in quantum field theory \parencite{evans1994being, early2017canonical, liu2019adjoint}. 

\medskip

In mathematical physics, \emph{thermal quantum field theory} is a set of methods to calculate expectation values of physical observables of a \emph{quantum field} at finite temperature. Quantum field theory is a theoretical framework that combines classical field theory (for example, Newtonian gravitation or Maxwell's equations of electromagnetic fields), special relativity and quantum mechanics. Quantum field theory treats particles as excited states of their underlying quantum fields, which are more fundamental than the particles. 

\medskip 

Key objects of this theory are the \emph{correlators}, also called \emph{Green functions}, that are used to calculate various \emph{observables}, i.e., self-adjoint operators on the Hilbert \emph{space of states} $\bbH$ that extract some physical properties from a particular state of the studied system. These correlators are all encoded in a generating functional, called the \emph{partition function}, in the same way a sequence of integers is encoded in a generating function. 

\medskip

With the \emph{imaginary time formalism}, the difference between the partition function in thermal quantum field theory and in zero-temperature quantum field theory is a thermal weight $e^{-\beta H}$, which is actually the action of a \emph{time-evolution} $e^{-i H T}$, that operates a shift in time of $-i \beta$. The physicists aim to extract the corresponding correlators from this new partition function which includes this thermal weight. In the computation, a function $\Phi$ appears, that takes as an input a set of complex energies $\{z_i\}_{i \in I}$ satisfying $\sum_{i \in I} z_i = 0$, called the \emph{imaginary Matsubara energies}. Physicists are interested in the analytic continuations of $\Phi$, which exist only where, for all subsets $J \subseteq I$, we have 
\[
\sum_{i \in J} z_i \not \in \bbR.
\] 
We remark that, for $J \subseteq I$, the set $H_J = \{z \in \bbC^{\lvert I \rvert} \mid \sum_{i \in J} z_i \in \bbR\}$ is a hyperplane of the energy space. It has been proven by \textcite{evans1992n} that the analytic continuations of $\Phi$, which produce solutions called \emph{(thermal) generalized retarded functions} \parencite{evans1994being}, and the chambers of the hyperplane arrangement $\{H_J\}_{J \subseteq I}$, called the \emph{resonance arrangement}, or the \emph{all-subset arrangement}, are in bijection. In the sequel, we will see that the chambers of the resonance arrangement are in bijection with the set of maximal unbalanced collections.

\section{The resonance arrangement} \label{sec: resonance}

In this section, we discuss the resonance arrangement, which is a very regular structure in $\bbR^N$, and explore links between this arrangement and cooperative game theory. The term `resonance arrangement' was coined by \textcite{shadrin2008chamber} while studying double Hurwitz numbers in algebraic geometry, but this arrangement is already considered since at least \textcite{evans1994being} in the context of quantum field theory. Surprisingly, resonance arrangements are found in various fields, in psychometrics and economics under the name of \emph{restricted all-subsets arrangements} \parencite{kamiya2011ranking, kamiya2012arrangements}, in mathematical physics and quantum field theory \parencite{early2017canonical, evans1994being, liu2019adjoint}, in representation theory \parencite{billera2018boolean, billey2021boolean}, and mostly in combinatorics \parencite{aguiar2017topics, bjorner2015positive, cavalieri2011wall, gutekunst2021root, kuhne2020universality}. We show in the sequel that the resonance arrangement appears very naturally in cooperative game theory. For more details about the resonance arrangement, see \textcite{kuhne2020universality}. As often as possible, we follow the notation and definitions of \textcite{aguiar2017topics}. 

\begin{definition}
\label{def: resonance}
The \emph{(restricted) resonance arrangement} $\calA_R$ of $\bbR^N$ is defined by 
\[
\calA_R \coloneqq \left\{ H^R_S \mid S \in \calN \setminus \{N\} \right\} \quad \text{where} \quad H^R_S \coloneqq \left\{ x \in \bbR^N \mid x(N) = 0 \text{ and } x(S) = 0\right\}. 
\]
\end{definition}

The definitions vary in the literature. Some authors define the resonance arrangement as simply the set of hyperplanes $\left\{ \{x \in \bbR^N \mid x(S) = 0 \} \mid S \in \calN \right\}$. The next result shows that the difference between the two definitions is negligible. 

\begin{proposition}
\label{prop: restricted}
The restricted resonance arrangement of $\bbR^N$ is isomorphic to the resonance arrangement of an $(n-1)$-dimensional Euclidean space. 
\end{proposition}

\proof
Without loss of generality, let $N = \{1, \ldots, n\}$ and $N' = N \setminus \{n\}$. Denote by $\calA_R$ the resonance arrangement defined as in Definition \ref{def: resonance}, and let $\calA^{N'}$ be 
\[
\calA^{N'} \coloneqq \left\{H_S \mid S \in \calN \setminus \{N\} \right\} \quad \text{where} \quad H_S \coloneqq \left\{x \in \bbR^{N'} \mid x(S) = 0 \right\}. 
\]
Notice that the linear subspace $\{x \in \bbR^N \mid x(N) = 0\}$ is isomorphic to $\bbR^{N'}$. Let $S$ be a coalition of $N'$. We have 
\[
H^R_S = \left\{x \in \bbR^N \mid x(N) = 0 \text{ and } x(S) = 0 \right\} \simeq \left\{x \in \bbR^{N'} \mid x(S) = 0 \right\} = H_S \in \calA^{N'}.
\] 
Then, $\calA^{N'}$ is isomorphic to a subarrangement of $\calA_R$. To complete the proof, let $S$ be a coalition of $N$ such that $\lvert S \rvert = n-1$. Then $S$ is not a proper coalition of $N'$, but $N \setminus S$ is, and because
\[ \begin{aligned}
H^R_S & = \left\{ x \in \bbR^N \mid x(N) = 0 \text{ and } x(S) = 0 \right\} \\ 
& = \left\{x \in \bbR^N \mid x(N) = 0 \text{ and } x(N \setminus S) = 0 \right\} = H^R_{N \setminus S},
\end{aligned} \] 
we have that $H^R_S \simeq H_{N \setminus S}$, and then $\calA_R \simeq \calA^{N'}$. 
\endproof

Then, the restricted resonance arrangement satisfies all the properties of a non-restricted resonance arrangement, we thus call simply them resonance arrangement when no confusion occurs. 

\medskip

Let $\calA$ be a hyperplane arrangement. Each hyperplane divides the ambient space $\bbR^N$ into two closed half-spaces, the intersection of which is the given hyperplane. We call \emph{faces} of the arrangement the subsets of $\bbR^N$ obtained by taking the intersection, for any hyperplane, of itself or one of its closed half-spaces. In other words, for each hyperplane, two elements of the same face belongs to the same side of the hyperplane, or are contained by the hyperplane itself. Denote by $\Sigma[\calA]$ the set of faces of $\calA$, which is a poset under inclusion. A maximal face of $\Sigma[\calA]$ is called a \emph{chamber}, the set of which is denoted by $\Gamma[\calA]$. 

\subsection{The posets of chambers}

The first link between the resonance arrangement and cooperative game theory has been made by \textcite{billera2012maximal}, where the authors show that there exists a bijection between the chambers of $\calA_R$ and the \emph{maximal unbalanced collections}. The link is based on a characterization of unbalanced collections of Billera, Moore, Moraites, Wang and Williams \cite{billera2012maximal}. 

\begin{proposition}[Billera, Moore, Moraites, Wang and Williams \cite{billera2012maximal}] \label{prop: unbalanced} \leavevmode \newline
A collection $\calS \subseteq \calN$ is unbalanced if and only if there exists a side payment $\sigma \in \Sigma$ such that, for all $S \in \calS$, we have $\sigma(S) > 0$. 
\end{proposition}

It is important to notice that unbalancedness is not equivalent to the negation of balancedness. The collection $\{\{a, b\}, \{a, c\}, \{b, c\}, \{a\}\}$ is not balanced on $\{a, b, c\}$, but contains $\{\{a, b\}, \{a, c\}, \{b, c\}\}$, which is balanced. We call the collections that are not balanced but that contain a balanced collection \emph{weakly balanced}. The name is motivated by two aspects: first, by removing some coalitions, a weakly balanced collection can become balanced. Secondly, relaxing the positivity condition on the balancing weights in the definition of balancedness leads to weak balancedness. 

\begin{example}
These two collections are maximal unbalanced collections with $\sigma$ a side payment illustrating the previous result: 
\begin{itemize}
\item for $n = 3$: $\{\{a, b\}, \{a, c\}, \{a\}\}$ and, as a side payment, $\sigma = (2, -1, -1)$;
\item for $n=4$: $\{\{a\}, \{a, b\}, \{a, c\}, \{a, d\}, \{a, b, c\}, \{a, b, d\}, \{a, c, d\}\}$ and, as a side payment, $\sigma = (3, -1, -1, -1)$. 
\end{itemize}
The side payment $\sigma$ represents a direction which improves the payment of all coalitions included in the unbalanced collection. 
\end{example}

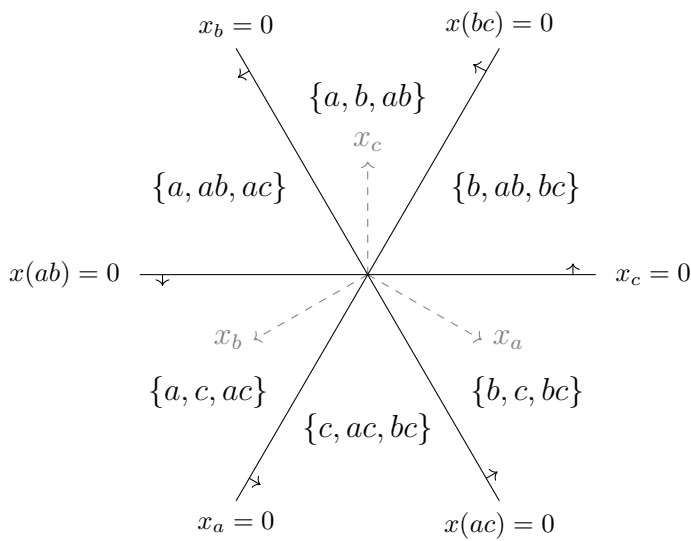
\begin{figure}[ht]
\begin{center}
\begin{tikzpicture}[scale=0.3]
\draw[gray, dashed, thin, ->] (0, 0) -- (0, 5) node[above] {$x_c$};
\draw[gray, dashed, thin, ->] (0, 0) -- (-5, -2.887) node[left] {$x_b$};
\draw[gray, dashed, thin, ->] (0, 0) -- (5, -2.887) node[right] {$x_a$};

\draw (-10, 0) node[left] {\footnotesize $x(ab) = 0 \ $} -- (10, 0) node[right] {\footnotesize $\ x_c = 0$};
\draw (-5.774, -10) node[below] {\footnotesize $x_a = 0$} -- (5.774, 10) node[above] {\footnotesize $x(bc) = 0$};
\draw (5.774, -10) node[below] {\footnotesize $x(ac) = 0$} -- (-5.774, 10) node[above] {\footnotesize $x_b = 0$};

\draw[->] (-9, 0) -- (-9, -0.5);
\draw[->] (9, 0) -- (9, 0.5);
\draw[->] (5.198, 9) -- (4.621, 9.289);
\draw[->] (-5.197, 9) -- (-5.697, 8.711);
\draw[->] (-5.197, -9) -- (-4.697, -9.289);
\draw[->] (5.198, -9) -- (5.698, -8.711);

\path (0, 9) node[below] {$\{a, b, ab\}$};
\path (0, -8) node[above] {$\{c, ac, bc\}$};
\path (-6.5, 5) node[below] {$\{a, ab, ac\}$};
\path (6.5, 5) node[below] {$\{b, ab, bc\}$};
\path (-7, -4) node[below] {$\{a, c, ac\}$};
\path (7, -4) node[below] {$\{b, c, bc\}$};
\end{tikzpicture}
\caption{The restricted resonance arrangement for $n=3$ in the plane $x(N) = 0$. Arrows indicate the normal vector to the hyperplane. The 6 maximal unbalanced collections (subsets are written without commas and braces) correspond to the 6 chambers.}
\label{fig: resonance}
\end{center}
\end{figure}

This very simple characterization permits us to see that maximal unbalanced collections are the same as positive set sum systems defined by \textcite{bjorner2015positive} and that they are related to the resonance arrangement (see Figure \ref{fig: resonance}). 

\medskip

Let $C$ be a chamber of the restricted resonance arrangement, and let $x \in \text{relint}(C)$. Because $x$ does not belong to any hyperplane of $\calA_R$, there is no coalition $S \in \calN$ for which we have $x(S) = 0$. Then, for all coalitions $S \in \calN$, we have that either $x(S) > 0$ or $x(S) < 0$. Also, we see that for any other element $y \in \text{relint}(C)$ of the same chamber, the number $y(S)$ will have the same sign as the one of $x(S)$ because there is no hyperplane of $\calA_R$ separating $x$ and $y$.

\medskip 

Then, we can identify for each chamber of $\calA_R$ a set of coalitions $\calS$ such that, for all $S \in \calS$, and for all $x \in \text{relint}(C)$, we have $x(S) > 0$, and for all $T \not \in \calS$, we have $x(T) < 0$. By Proposition \ref{prop: unbalanced}, we have that each chamber is associated with an unbalanced collection. 

\medskip 

The maximality comes from the fact that $x \in C$ is a side payment and does not belong to any hyperplane of $\calA_R$. For a given coalition $S \in \calS$, if $x(S) > 0$, it implies $x(N \setminus S) < 0$, and therefore all unbalanced collections associated with a chamber have cardinality $2^{n-1}-1$, as they include exactly one coalition from any pair of complement, nonempty and proper coalitions of $N$.

\medskip 

Then, for any unbalanced collection $\calS$ associated with a chamber $C$, and for any coalition $S$ such that $S \not \in \calS$, the collection $\calS \cup \{S\}$ will fail to be unbalanced, because it will contain the balanced collection $\{S, N\setminus S\}$. Then, any unbalanced collection associated with a chamber is maximal. 

\medskip

The number of chambers for resonance arrangements is known for $n \leq 9$, see Table \ref{table: unbalanced} and sequence \href{http://oeis.org/A034997}{A034997} \parencite{oeis}, whereas the number of minimal balanced collections is known only for $n \leq 7$ \parencite{laplace2023minimal}, see sequence \href{http://oeis.org/A355042}{A355042} \parencite{oeis}. Notice that the number of maximal unbalanced collections increases more slowly than the number of minimal balanced collections. 

\medskip 

\begin{table}[ht]
\begin{center}
\begin{tabular}{lcccccccc}
\toprule
$n$ \hspace{0.5cm} & 2 & 3 & 4 & 5 & 6 & 7 & 8 & 9 \\ 
\midrule
$k$ & 2 & 6 & 32 & 370 & 11,292 & 1,066,044 & 347,326,352 & 419,172,756,930 \\
$l$ & 2 & 6 & 42 & 1,292 & 200,214 & 132,422,036 & ? & ? \\
\bottomrule
\end{tabular}
\caption{Number $k$ of maximal unbalanced collections and number $l$ of minimal balanced collections according to the number $n$ of players.}
\label{table: unbalanced}
\end{center}
\end{table}

There are other deep connections between cooperative games and the resonance arrangement. Let $(N, v)$ be a game, and denote by $\calA(v)$ the set defined by 
\[
\calA(v) \coloneqq \left\{ \bbA_S(v) \mid S \in \calN \setminus \{N\} \right\}.
\]
Let $(N, o)$ be the \emph{null game}, defined, for all $S \in \calN$, by $o(S) = 0$. We can easily see that $\calA(o)$ is the restricted resonance arrangement of $\bbR^N$, because for all $S \in \calN$, the hyperplane $\bbA_S(o)$ corresponds to $H^R_S$. In the proof of Prop. \ref{prop: restricted}, we have seen that $\bbA_S(o) = \bbA_{N \setminus S}(o)$, therefore pairs of hyperplanes of the resonance arrangement associated with complement proper coalitions merge. Let $(N, v)$ be a game and let $S$ be a coalition. We define the \emph{slab} of $S$ (or equivalently $N \setminus S$) as the subset of $\bbX(v)$ denoted by $\slab_S(v)$, defined by 
\[
\slab_S(v) \coloneqq \{x \in \bbX(v) \mid x(S) \geq v(S) \text{ and } x(N \setminus S) \geq v(N \setminus S)\}. 
\]
Modifying the worth of $S$ such that $v(S) + v(N \setminus S) \leq v(N)$ will generate a nonempty slab. Then, a game can be seen as a \emph{distortion} of the resonance arrangement, creating a `slab arrangement' where any hyperplane $\bbA_S(o)$ is translated by $v(S)$ along its normal vector relatively to $\bbX(v)$, i.e., along $\eta^S$. 

\begin{figure}[ht]
\begin{center}
\begin{subfigure}{0.49\textwidth}
\begin{center}
\begin{tikzpicture}[scale=0.25]
\draw[cyan] (-10, 0) node[left] {\footnotesize $\bbA_{ab}$} -- (10, 0) node[right] {\footnotesize $\bbA_c$};
\draw[purple] (-5.774, -10) node[left] {\footnotesize $\bbA_a$} -- (5.774, 10) node[right] {\footnotesize $\bbA_{bc}$};
\draw[orange] (5.774, -10) node[right] {\footnotesize $\bbA_b$} -- (-5.774, 10) node[left] {\footnotesize $\bbA_{ac}$};


\end{tikzpicture}
\caption{Arrangement $\calA(o)$.}
\end{center}
\end{subfigure}
\begin{subfigure}{0.49\textwidth}
\begin{center}
\begin{tikzpicture}[scale=0.25]
\fill[cyan!10] (-10, 0) -- (10, 0) -- (10, -2.578) -- (-10, -2.578) -- cycle;
\fill[purple!10] (-7.706, -10) -- (3.841, 10) -- (7.788, 10) -- (-3.760, -10) -- cycle;
\fill[orange!10] (5.774, -10) -- (9.233, -10) -- (-2.314, 10) -- (-5.774, 10) -- cycle;

\fill[gray!10] (0.764, 4.669 ) -- (2.737, 1.251) -- (2.014, 0) -- (0, 0) -- (-0.966, 1.673) -- cycle;
\fill[gray!10] (4.948, -2.578) -- (1.488, -2.578) -- (1.007, 1.744) -- (2.014, 0) -- (3.459, 0) -- cycle;
\fill[gray!10] (-3.420, -2.578) -- (-1.932, 0) -- (0, 0) -- (1.007, 1.744) -- (0.526, -2.578) -- cycle;

\fill[gray!50] (0, 0) -- (2.014, 0) -- (1.007, -1.744) -- cycle;

\draw[cyan] (-10, 0) node[left] {\footnotesize $\bbA_{ab}$} -- (10, 0);
\draw[cyan] (-10, -2.578) -- (10, -2.578) node[right] {\footnotesize $\bbA_c$};
\draw[purple] (-7.706, -10) node[left] {\footnotesize $\bbA_a$} -- (3.841, 10); 
\draw[purple] (-3.760, -10) -- (7.788, 10) node[right] {\footnotesize $\bbA_{bc}$};
\draw[orange] (5.774, -10) -- (-5.774, 10) node[left] {\footnotesize $\bbA_{ac}$};
\draw[orange] (-2.314, 10) -- (9.233, -10) node[right] {\footnotesize $\bbA_b$};


\end{tikzpicture}
\caption{Arrangement $\calA(v)$.}
\end{center}
\end{subfigure}
\caption{Balanced distortion of the resonance arrangement $\calA(o)$ by a game $(N, v)$. The core $C(v)$ is the darkest area.}
\label{fig: slab}
\end{center}
\end{figure}
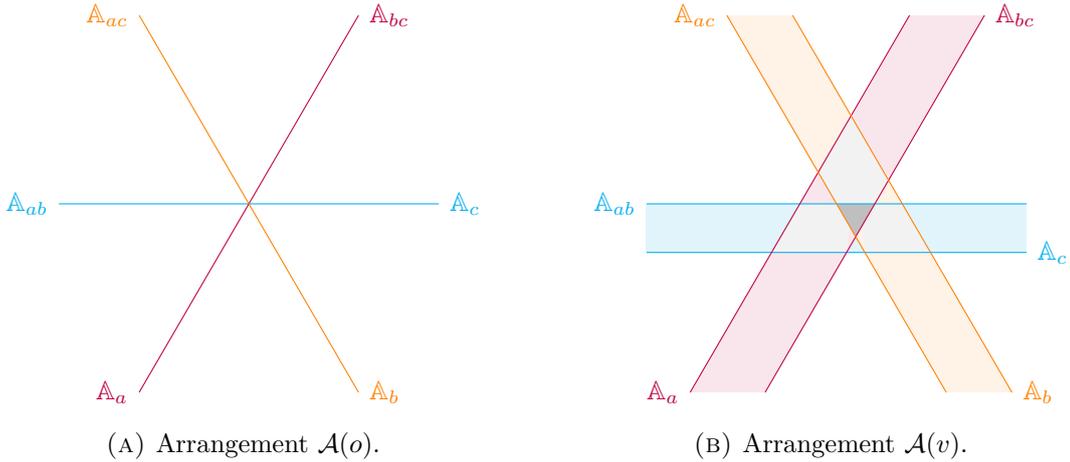

Because every hyperplane $\bbA_S(o)$ passes through the origin $\vec{o}$, the resonance arrangement is said to be \emph{central}. Like the hyperplane arrangements, we say that a slab arrangement is central if the intersection of all the slabs is nonempty. Because the intersection of all the slabs is a basic polyhedron, a slab arrangement defined from the resonance arrangement is central if and only if it is generated by a balanced game. 

\medskip

The set $\Gamma[\calA(v)]$ obtained by the distortion induced by $(N, v)$ may be radically different from the original $\Gamma[\calA_R]$. Previously, all chambers were cones, pointed at $\vec{o}$, in bijection with the set of maximal unbalanced collections. Now, some chambers can be bounded polyhedra, i.e., polytopes, or non-conic unbounded polyhedra. However, all these chambers are closely related to the feasible collections. We denote by $\calR(v)$ the set of regions of $(N, v)$ corresponding to the feasible collections. 

\begin{proposition}
Let $(N, v)$ be a balanced game. $\calR(v)$ is a graded poset. 
\end{proposition}

\proof
We transfer the partial order defined by the inclusion on the feasible collections to the set of regions. The rank of a region in the poset is the cardinality of the associated feasible collection. 
\endproof

\begin{example} \label{ex: posets}
Consider the game drawn in Figure \ref{fig: dist-res}. 
\medskip 

\begin{figure}[ht]
\begin{center}
\begin{tikzpicture}[scale=1]
\fill[black!50] (0.846, 1) -- (-0.309, 3) -- (3-0.309, 3) -- (3, 2.46) -- (2.156, 1) -- cycle;

\fill[black!25] (-0.312, 3) -- (-1.5, 5.055) -- (0, 7.65) -- (2.688, 3) -- cycle;
\fill[black!25] (-0.312, 3) -- (-2.688, 3) -- (-3.844, 1) -- (0.844, 1) -- cycle;
\fill[black!25] (3, 2.46) -- (2.688, 3) -- (3.312, 3) -- cycle;
\fill[black!25] (3, 2.46) -- (2.156, 1) -- (3.844, 1) -- cycle;
\fill[black!25] (2.156, 1) -- (0.844, 1) -- (1.5, -0.135) -- cycle; 

\fill[black!10] (-0.312, 3) -- (-2.688, 3) -- (-1.5, 5.055) -- cycle;
\fill[black!10] (0, 7.65) -- (-1.751, 10.680) -- (-4, 9.38) -- (-1.5, 5.055) -- cycle;
\fill[black!10] (0, 7.65) -- (1, 9.38) -- (5.497, 6.78) -- (5, 5.92) -- (3.312, 3) -- (2.688, 3) -- cycle;
\fill[black!10] (3, 2.46) -- (3.312, 3) -- (6, 3) -- (6, 1) -- (3.844, 1) -- cycle;
\fill[black!10] (2.156, 1) -- (3.844, 1) -- (5, -1) -- (3.27, -2) -- (2.578, -2) -- (1.5, -0.125) -- cycle;
\fill[black!10] (0.844, 1) -- (1.5, -0.135) -- (0.422, -2) -- (-3.27, -2) -- (-5, -1) -- (-3.844, 1) -- cycle; 
\fill[black!10] (-3.844, 1) -- (-6, 1) -- (-6, 3) -- (-2.688, 3) -- cycle;

\draw (-6, 1) node[below] {$\bbA_{c}$} -- (6, 1);
\draw (-5, -1) -- (1, 9.38) node[above right] {$\bbA_{a}$};
\draw (-1, 9.38) node[left] {$\bbA_{b}$} -- (5, -1);

\draw (-6, 3) -- (6, 3) node[above] {$\bbA_{ab}$};
\draw (-4, 9.38) node[left] {$\bbA_{ac}$} -- (2.578, -2);
\draw (0.422, -2) -- (5, 5.92) node[right] {$\bbA_{bc}$};

\draw[->] (4.8, 5.57) -- (4.6, 5.69);
\draw[->] (0.8, 9.034) -- (1, 8.919);
\draw[->] (-0.8, 9.034) -- (-1, 8.919);
\draw[->] (-3.8, 9.034) -- (-3.6, 9.149);
\draw[->] (-5.8, 1) -- (-5.8, 1.2);
\draw[->] (5.8, 3) -- (5.8, 2.8);

\path[white] (1.5, 2.6) node[below] {$C(v)$};
\path (2, 6) node[right] {$X_{\{b, ab\}}$};
\path (-3.5, 5.5) node[below] {$X_{\{a, ab, ac\}}$};
\path (-1.5, 0) node[below] {$X_{\{c, ac\}}$};
\path (-1.5, 2.2) node[below] {$X_{\{ac\}}$};
\path (-1.5, 3.9) node[below] {$X_{\{ab, ac\}}$};
\path (-1.5, 8) node[below] {$X_{\{a, ab\}}$};
\path (0.2, 5) node[below] {$X_{\{ab\}}$};
\path (-5, 2.5) node[below] {$X_{\{a, ac\}}$};
\path (-5.5, 0.3) node[below] {$X_{\{a, c, ac\}}$};
\path (0, 10) node[below] {$X_{\{a, b, ab\}}$};
\path (3, 0.2) node[below] {$X_{\{c, bc\}}$};
\path (5.5, 4.5) node[below] {$X_{\{b, ab, bc\}}$};
\path (5.5, 0.5) node[below] {$X_{\{b, c, bc\}}$};
\path (5, 2.4) node[below] {$X_{\{b, bc\}}$};
\path (3.05, 1.8) node[below] {$X_{\{bc\}}$};
\path (1.5, 0.95) node[below] {$X_{\{c\}}$};
\path (1.55, -1.35) node[below] {\footnotesize $X_{\{c, ac, bc\}}$};
\path (3, 4) node[below] {$X_{\{b\}}$};
\draw (2.9, 3.3) -- (3, 2.75);
\end{tikzpicture}
\caption{Resonance arrangement distorted by a game $(N, v)$ in $\bbX(v)$ for $N = \{a, b, c\}$. The rank of a region increases with its lightness.}
\label{fig: dist-res}
\end{center}
\end{figure}
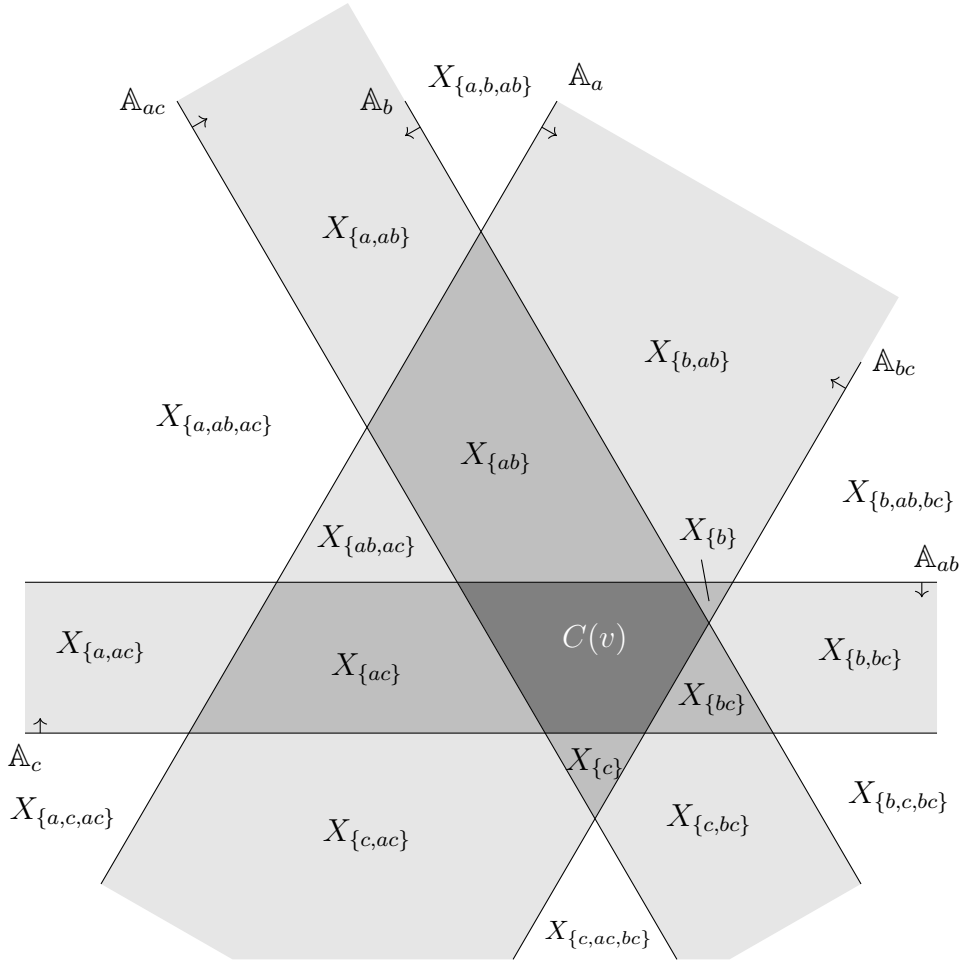
We see that each region has a specific rank, which is the number of hyperplanes separating it from the core. Notice that $X_{\{b, ab\}}$ is neither a polytope, nor a cone. 
\end{example}

The set of chambers $\Gamma[\calA(v)]$ has a very similar structure. 

\begin{proposition}
Let $(N, v)$ be a balanced game. $\Gamma[\calA(v)]$ is a graded poset. 
\end{proposition}

\proof
First, let us define a partial order on $\Gamma[\calA(v)]$. Let $C \in \Gamma[\calA(v)]$ be a chamber. We denote by $\text{sep}(C)$ the set of separating hyperplanes between the core and $C$, i.e., hyperplanes for which the core and $C$ lie on different closed half-spaces generated by the given hyperplane. Let $C_1$ and $C_2$ be two chambers of $\Gamma[\calA(v)]$. We say that $C_1$ is lower than $C_2$, denoted by $C_1 \leq_\Gamma C_2$, if $\text{sep}(C_1) \subset \text{sep}(C_2)$. We end this proof by defining the rank function of the poset. Let $C$ be a chamber of $\Gamma[\calA(v)]$, and let $\rho: \Gamma[\calA(v)] \to \bbN$ be defined by $\rho(C) = \lvert \text{sep}(C) \rvert$. We have that, for all $C_1$, $C_2 \in \Gamma[\calA(v)]$ such that $C_1 <_\Gamma C_2$ we have $\rho(C_1) < \rho(C_2)$, and if $C_2$ covers $C_1$, i.e., if $C_1 \leq_\Gamma C_2$ and only one hyperplane separates the two chambers, we have $\rho(C_1) + 1 = \rho(C_2)$, then $\rho$ is a rank function. 
\endproof

\begin{proposition}
Let $(N, v)$ be a balanced game. Then there exists an order-preserving injection of $\Gamma[\calA(v)]$ into $\calR(v)$. If the core $C(v)$ is full-dimensional, this injection is a bijection. 
\end{proposition}

\proof
Let $C$ be a chamber $C \in \Gamma[\calA(v)]$ and let $x \in \text{int}(C)$. Then, there exists a collection of coalitions $\calS \subseteq \calN$ such that, for all $S \in \calS$ we have $x(S) < v(S)$ and, for all $T \not \in \calS$, we have $x(T) > v(T)$. Then $\text{int}(C)$ is the interior of the region $X_\calS$, therefore $\calS$ is feasible. Assume now that $C(v)$ is full-dimensional, i.e., that $N$ is vital, and let $\calS$ be a feasible collection. By Lemma \ref{lemma: vitality}, for all minimal balanced collections $\calB$ on $N$, we have 
\[
\sum_{S \in \calB} \lambda^\calB_S v(S) < v(N). 
\]
Let us focus on the core $C(v^\calS)$. By Lemma \ref{lemma: feasibility}, we have that $\calS^c \cap \calE(N, v^\calS) = \emptyset$. Because $N$ is vital, we have that $\calE(v) = \{N\}$, and then $\calE(v^\calS) = \{N\}$. Therefore, we have that $C(v^\calS)$ is full-dimensional, which implies that $X_\calS$ is also full-dimensional, and then included in a unique chamber of $\calA(v)$. 
\endproof

In Example \ref{ex: posets}, the core is full-dimensional, hence the regions and the chambers are in bijections. Even if the chambers of $\calA(v)$ differ from the one of $\calA_R$, all the previous chambers remain in the new $\Gamma[\calA(v)]$.

\begin{proposition}
Let $(N, v)$ be a balanced game. Then there exists an injection of $\Gamma[\calA_R]$ into the set of maximal elements of $\Gamma[\calA(v)]$. 
\end{proposition}

\proof
Let $C$ be a chamber of $\Gamma[\calA_R]$. Then, there exists a maximal unbalanced collection $\calS$ such that, for all $x \in \text{relint}(C)$ and for all $S \in \calS$, we have $x(S) < v(S)$ and $x(N \setminus S) > v(N \setminus S)$. This defines a basic polyhedron, whose associated set function is $v^\calS$. The domain of $v^\calS$ is 
\[
\calF_v = (\calN \setminus \calS) \cup \calS^c = \calS^c \cup \{N\}. 
\]
The only minimal balanced collection included in $\calF_v$ is $\{N\}$, then the condition of Theorem \ref{th: basic} is always satisfied for all $(N, v)$. The maximality of the elements of $\Gamma[\calA_R]$ in $\Gamma[\calA(v)]$ is given by the maximality of the unbalanced collection associated with these and Lemma \ref{lemma: blocking}. 
\endproof

\label{sym: lcv}
In Example \ref{ex: posets}, the chambers of $\calA_R$ are the white ones. Let $(N, v)$ be a balanced game. The face lattice of the core, denoted by $\calL_C(v)$ is the poset of the faces of the core (including the empty face), partially ordered by inclusion. Let $F \in \calL_C(v)$ be a face of $C(v)$. Denote by $\calE(F)$ the collection of coalitions that are efficient for all elements of $F$ but not for all elements of the core, i.e., 
\[
\calE(F) \coloneqq \{ S \in \calN \setminus \calE(v) \mid F \subseteq \bbA_S(v)\}.
\] 

\begin{example} 
The diagram \ref{fig: face-lattice-core} represents the face lattice of the core in Figure \ref{fig: dist-res}. 

\medskip

\begin{figure}[h!]
\begin{center}
\begin{tikzcd}[row sep=1.4cm, column sep=1.6cm]
& & F_\emptyset = C(v)  \\
F_{\{ab\}} \arrow[urr] & F_{\{b\}} \arrow[ur] & F_{\{bc\}} \arrow[u] & F_{\{c\}} \arrow[ul] & F_{\{ac\}} \arrow[ull] \\
F_{\{ab, \, ac\}} \arrow[u] \arrow[urrrr] & F_{\{b, \, ab\}} \arrow[ul, crossing over] \arrow[u, crossing over]  & F_{\{b, \, bc\}} \arrow[u, crossing over] \arrow[ul, crossing over] & F_{\{c, \, bc\}} \arrow[u, crossing over] \arrow[ul, crossing over] & F_{\{c, \, ac\}} \arrow[u, crossing over] \arrow[ul, crossing over] \\
& & \emptyset \arrow[ull] \arrow[ul]  \arrow[u] \arrow[ur] \arrow[urr]
\end{tikzcd}
\caption{Face-lattice $\calL_C(v)$ of the core of $(N, v)$ from Table \ref{fig: dist-res}.}
\label{fig: face-lattice-core}
\end{center}
\end{figure}
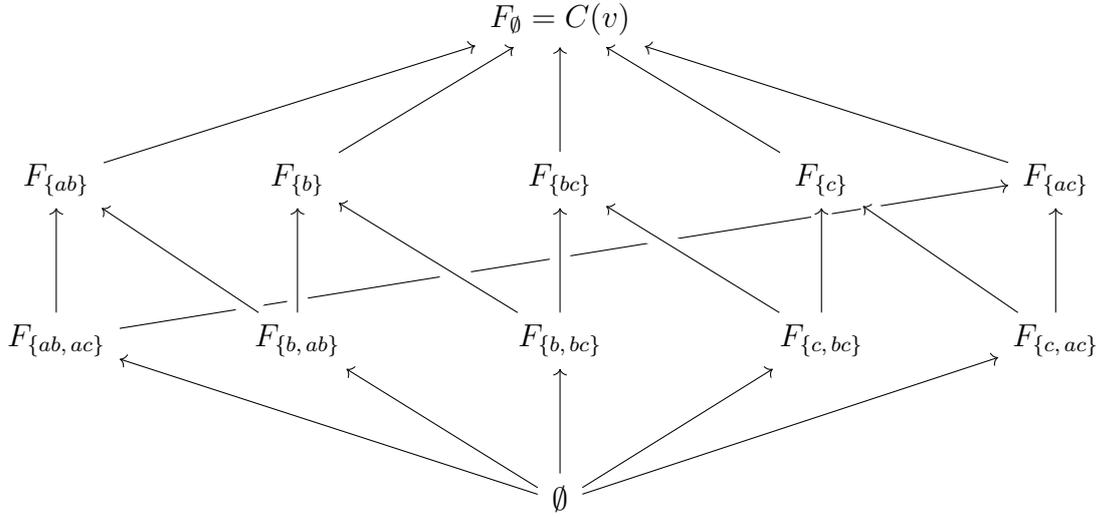
Notice that player $a$ is not alone in any collection $\calE(F)$, as $\{a\}$ is not exact. In particular, this game does not have a stable core. 
\end{example}

\begin{proposition} \label{prop: face-lattice}
Let $(N, v)$ be a balanced game. Then, there exists an order-reversing injection of $\calL_C(v) \setminus \{\emptyset\}$ into $\Gamma[\calA(v)]$. 
\end{proposition}

\proof
Let $F \in \calL_C(v)$ be a proper face of the core. We first show that $\calE(F)$ is a feasible collection. To show that, we use Lemma \ref{lemma: feasibility} relative to feasible collections. 

\medskip 

Let $\calF = \left(2^N \setminus \calE(F)\right) \cup \calE(F)^c$. The game $(\calF, v^{\calE(F)})$ is balanced because $F \subseteq C(\calF, v^{\calE(F)})$. Let us focus now on the effective coalitions for $(\calF, v^{\calE(F)})$. Let $S \in \calE(F)^c$. If $S \in \calE(F) \cap \calE(F)^c$, then $v^{\calE(F)}(S) + v^{\calE(F)}(N \setminus S) = v(S) + v(N) - v(S) = v(N)$, which implies $S$ and $N \setminus S$ are included in $\calE(v)$ and then $S$ can not be in $\calE(F)$. So we assume that $S \in \calE(F)^c$ and $\calE(F) \cap \calE(F)^c = \emptyset$. We prove by contradiction that $S \notin \calE(\calF, v^{\calE(F)})$. Assume the contrary. Then $C(\calF, v^{\calE(F)}) \subseteq \bbA_S(v)$, which implies $F \in \subseteq \bbA_S(v)$. Then, $S \in \calE(F)$, and the desired contradiction is obtained.

\medskip 

The order reversion comes from the fact that a face $F$ included in another face $G$ is included in more hyperplanes than $G$. For any coalition in $\calE(G)$, we have $F \subseteq G \subseteq \bigcap_{S \in \calE(G)} \bbA_S(v)$, and then $\calE(G) \subseteq \calE(F)$. 
\endproof

To summarize, we have the following inclusion relations between the posets, where $\bbU(N)$ denotes the set of unbalanced collections on $N$, $\bbU^{\max}(N)$ the maximal unbalanced collections on $N$, and $\hookrightarrow$ denotes an order-preserving (reversing for $\calL_C(v)$) injection. 
\[ \begin{aligned}
& \bbU^{\max}(N) \equiv \Gamma[\calA_R] \hookrightarrow \Gamma[\calA(v)] \hookrightarrow \calR(v) \hookrightarrow \bbU(N) \\
& \text{and} \qquad \calL_C(v) \hookrightarrow \calR(v) \hookrightarrow \bbU(N).  
\end{aligned} \]

\begin{example}
We see in Figure \ref{fig: poset-regions} that the nonempty faces of the core can be injected with their order reversed in the poset of regions $\calR(v)$. 

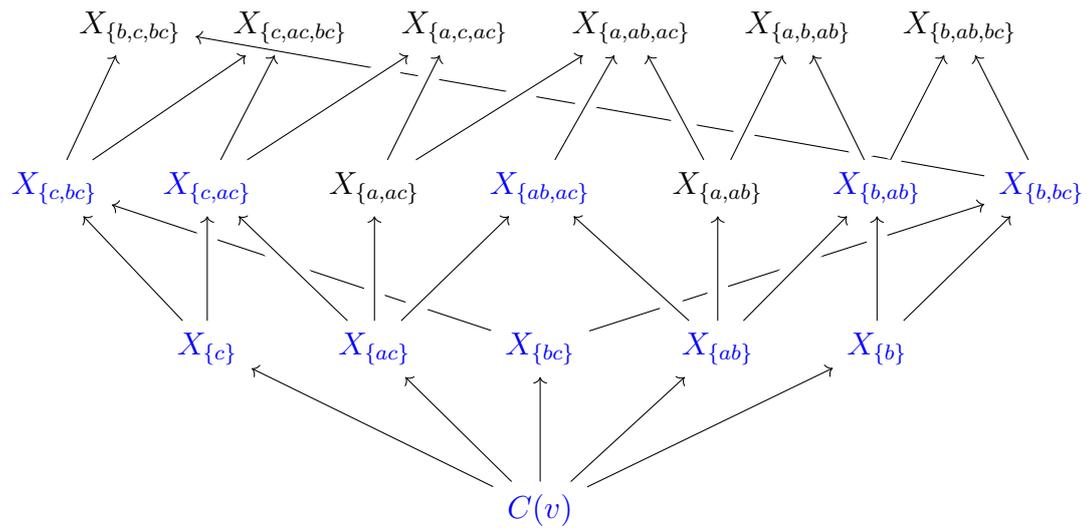
\begin{figure}[ht]
\begin{center}
\begin{tikzcd}[row sep=1.4cm, column sep=-0.6cm]
& X_{\{b, c, bc\}} \arrow[from=drrrrrrrrrrr] & & X_{\{c, ac, bc\}} & & X_{\{a, c, ac\}} & & X_{\{a, ab, ac\}} & & X_{\{a, b, ab\}} & & X_{\{b, ab, bc\}} & \\
\textcolor{blue}{X_{\{c, bc\}}} \arrow[from=drrrrrr] \arrow[ur, crossing over] \arrow[urrr, crossing over] & & \textcolor{blue}{X_{\{c, ac\}}} \arrow[ur, crossing over] \arrow[urrr, crossing over] & & X_{\{a, ac\}} \arrow[ur, crossing over] \arrow[urrr, crossing over] & & \textcolor{blue}{X_{\{ab, ac\}}} \arrow[ur, crossing over] & & X_{\{a, ab\}} \arrow[ul, crossing over] \arrow[ur, crossing over] & & \textcolor{blue}{X_{\{b, ab\}}} \arrow[ul, crossing over] \arrow[ur, crossing over] & & \textcolor{blue}{X_{\{b, bc\}}} \arrow[ul] \\
& & \textcolor{blue}{X_{\{c\}}} \arrow[ull] \arrow[u, crossing over] & & \textcolor{blue}{X_{\{ac\}}} \arrow[ull, crossing over] \arrow[u, crossing over] \arrow[urr, crossing over] & & \textcolor{blue}{X_{\{bc\}}} \arrow[urrrrrr] & & \textcolor{blue}{X_{\{ab\}}} \arrow[ull, crossing over] \arrow[u, crossing over] \arrow[urr, crossing over] & & \textcolor{blue}{X_{\{b\}}} \arrow[u, crossing over] \arrow[urr] & & \\
& & & & & & \textcolor{blue}{C(v)} \arrow[ullll] \arrow[ull] \arrow[u] \arrow[urr] \arrow[urrrr]
\end{tikzcd}
\caption{Poset $\calR(N, v)$ of regions of $(N, v)$ from Figure \ref{fig: dist-res}. In blue we have the proper-face--semi-lattice of the core, in reverse order.}
\label{fig: poset-regions}
\end{center}
\end{figure}
Notice that the adherence of the blue regions, in bijection with the collections defining faces of the core, intersects the core. 
\end{example}





\chapter{Stability, domination and projections} 

\label{ChapterD} 




This chapter is devoted to the study of core stability, i.e., the coincidence between the core and the von Neumann-Morgenstern stable sets. Determining whether a core is stable is a long-standing problem, that has been around for more than 70 years. Before 2021 and the theorem of \textcite{grabisch2021characterization}, we had no characterization of stable cores on the general class of balanced games. Notwithstanding, for specific classes of games, such as \emph{assignment games} \parencite{bednay2014stable, raghavan2006assignment, shapley1971assignment, solymosi2001assignment}, and \emph{chain-component additive games} \parencite{van2008core}, important results exist. Also, convex games, defined by \textcite{shapley1971cores}, have stable cores, as do \emph{extendable} games \parencite{kikuta1986core}, games with \emph{large cores} \parencite{sharkey1982cooperative}, and \emph{strongly vital-exact extendable games} \parencite{shellshear2009core}. Deeper investigations on cores and stable sets of assignment games can be found in \parencite{nunez2015survey, nunez2013neumann}. 

\medskip

Another approach that has been successfully adopted is to modify the notion of stable sets in order to encompass similar intuitive properties of domination with a different set of maximal elements. Usually, this is done by iterating the domination process between preimputations. The \emph{myopic stable set} defined by \textcite{demuynck2019myopic}, which extends the notions developed in \textcite{herings2009farsightedly, herings2019stability}, uses a property called \emph{asymptotic external stability}, which resembles an iterative equivalent of the external stability as defined by \textcite{von1944theory}. A similar approach has been used by \textcite{beal2008farsighted} in their definition of a \emph{farsighted stable set} and by \textcite{ray2015farsighted}. The number of iterations and the `accessibility' of the core is discussed in \textcite{beal2013optimal}. In Chapter \ref{ChapterA}, we already discussed a recursive process of accumulated improvements of preimputations with the \emph{transfer schemes}, defined by \textcite{stearns1968convergent}, and developed by \textcite{cesco1998convergent} using projection operators. We will study in greater details projection operators between preimputations in Section \ref{sec: projection-core}. 

\medskip

In the first section of this chapter, we present the new characterization of \textcite{grabisch2021characterization}, which we translated into a working computer program \parencite{laplace2023minimal}. In the second section, we study two types of cones, one consisting of preimputations improving the payments of a specific collection of coalitions, and the other consisting of preimputations strictly increasing the payment of players in a specific coalition. By studying their intersection we provide a new necessary condition for core stability, which can be implemented to increase the efficiency of the algorithm of the first section of this chapter. In the last two sections, we present a new way to study core stability and domination relations between preimputations by using projection operators onto intersections of affine subspaces and polar regions. 

\section{Nested balancedness}

The characterization uses the outvoting relation we introduced in Chapter \ref{ChapterB}. A preimputation $x$ \emph{outvotes} another preimputation $y$ via a coalition $S$ if $x \domS y$ and, for all $T \not \subseteq S$, we have $x(T) \geq v(T)$. We write it $x \succ_S y$. The set of preimputations which are not outvoted is denoted by $M(v)$, and we have that $M(v) = C(v)$ if and only if $C(v)$ is a stable set \parencite{grabisch2021characterization}. 

\medskip 

Let $(N, v)$ be a balanced game. In the sequel, we only consider strictly vital-exact coalitions, as we discussed in Chapter \ref{ChapterB}, and we assume that $\calVE(v)$, the set of strictly vital-exact coalitions, is core-describing (see Proposition \ref{prop: sve-describing}). To know whether the core is a stable set, we are looking for nonoutvoted preimputations. For a given preimputation $x$, the set of preimputations outvoting it via $S$ is 
\[
P_S^\succ(x) \coloneqq \left\{ y \in \bbX(v) \ \left| \
\begin{tabular}{ll}
$y_i > x_i$ & for all $i \in S$ \\
$y(S) \leq v(S)$ & \\
$y(T) \geq v(T)$ & for all $T \not \subseteq S$
\end{tabular} \right. \right\}. 
\]

The name of this method of \emph{nested balancedness}, developed by Grabisch and Sudh{\"o}lter \cite{grabisch2021characterization}, consists of checking the nonemptiness of a specific polyhedron, composed of preimputations for which another specific polyhedron is nonempty. 

\medskip 

Recall that $v^S$ is the coalition function which may differ from $v$ only inasmuch as $v^S(N \setminus S) = v(N) - v(S)$, and recall that $\calI_S = \{\{i\} \mid i \in S\}$. Using Theorem \ref{th: basic}, a preimputation $x$ is outvoted via a coalition $S$ if and only if $(\calF_{P_S^\succ(x)}, v_{P_S^\succ(x)})$ is balanced and $\calI_S \cap \calE(\calF_{P_S^\succ(x)}, v_{P_S^\succ(x)}) = \emptyset$, where 
\begin{equation} \label{eq: coalition-function-outvoting}
\begin{aligned}
& \calF_S \coloneqq \left( \calVE(v) \setminus 2^S \right) \cup \{N \setminus S\} \cup \calI_S, \quad \text{and} \\
& v_{P_S^\succ(x)} (T) = \left\{ \begin{tabular}{ll}
$x_i$ & \quad if $T \in \calI_S$, $T = \{i\}$, \\
$v^S(T)$ & \quad if $T \in \calF_S \setminus \calI_S$.
\end{tabular} \right.
\end{aligned} 
\end{equation}

Because $S$ is a strictly vital-exact coalition, it is in particular exact, and therefore by Proposition \ref{prop: exactness} the game $(\calF_{P_S^\succ(x)}, v^S)$ is balanced. It implies that many of the inequalities checked for Theorem \ref{th: basic} are already satisfied, and need not to be checked. The only inequalities which need to be checked are the ones involving the new worths $v_{P_S^\succ(x)}(\{i\}) = x_i$ for $i \in S$, which are the inequalities corresponding to the minimal balanced collections $\calB$ such that $\calB \cap \calI_S \neq \emptyset$. 

\medskip

\label{sym: aggrieved}
Let $\phi$ be the map assigning to each preimputation $x$ the set of coalitions defined by 
\[
\phi(x) \coloneqq \{S \in \calVE(v) \mid x(S) < v(S)\}. 
\]
It is sufficient to check whether $x$ is not outvoted via any $S \in \phi(x)$, because $x$ cannot be dominated via a coalition not belonging to $\phi(x)$. Hence, a preimputation $x$ is not ouvoted if for all coalitions $S \in \phi(x)$, there exists a minimal balanced collection $\calB$ satisfying $\calB \cap \calI_S \neq \emptyset$, such that
\begin{equation}
\label{eq: non-basic}
\sum_{\substack{i \in S \\ \{i\} \in \calB}} \lambda^\calB_{\{i\}} x_i \geq v(N) - \sum_{T \in \calB \setminus \calI_S} \lambda^\calB_T v^S(T). 
\end{equation}
In other words, for all feasible collection $\calS \subseteq \calVE(v)$, we have that $M(v) \cap X_\calS(v) \neq \emptyset$ if and only if there exists $S \in \calS$, $x \in X_\calS(v)$ and $\calB$ satisfying $\calB \cap \calI_S \neq \emptyset$ such that equation \eqref{eq: non-basic} holds. 

\medskip

Let $\calS$ be a feasible collection, $S \in \calS$ be a coalition, and $\calB$ be a minimal balanced collection intersecting $\calI_S$. Consider the following linear program. 
\[ \left(P^\calB_S \right) \qquad 
\left\{
\begin{aligned}
& \min x(N) \\
& \text{s.t.} \left\{ \begin{tabular}{ll}
$x(T) \geq v(T)$, & for all $T \not \in \calS$, \\
$x(N \setminus T) > v(N) - v(T)$, & for all $T \in \calS$, \\
$\sum_{\{i\} \in \calB \cap \calI_S} \lambda^\calB_{\{i\}} x_i \geq v(N) - \sum_{T \in \calB \setminus \calI_S} \lambda^\calB_T v^S(T)$.
\end{tabular} \right. 
\end{aligned} \right.
\]
The last constraint of the program corresponds to equation \eqref{eq: non-basic}, the first two constraints ensure that $x$ belongs to the region $X_\calS(v)$. If the optimal value of the program is $v(N)$, then there exists a preimputation $x \in X_\calS(v)$ which is not outvoted via $S$. Then, if for all $S \in \calS$ there exists a minimal balanced collection $\calB$ intersecting $\calI_S$ such that the optimal values of each program $(P^\calB_S)$ is $v(N)$, we have that $M(v) \cap X_\calS(v) \neq \emptyset$.  

\medskip

Because of the presence of the weights $\lambda^\calB_{\{i\}}$ on the left-hand side of \eqref{eq: non-basic} and of the last constraint, we cannot use Theorem \ref{th: basic} about basic polyhedra, but we still use the same idea: we check the existence of the preimputations satisfying the constraints using a linear program, namely $(P^\calB_S)$, and we take the dual program. Because the polyhedron is no longer basic, the constraints of the dual program do not correspond to the balanced collections. 

\medskip

To overcome this problem, we use the \emph{balanced sets} of vectors. 

\begin{definition}[\textcite{grabisch2021characterization}] \leavevmode \newline
Let $Z \subseteq \bbR^N_+ \setminus \{\vec{o} \,\}$. We say that $Z$ is a \emph{balanced set} if there exists a system of positive balancing weights $\lambda = \{\lambda_z \mid z \in Z\}$ such that $\sum_{z \in Z} \lambda_z z = \bfone^N$. The set $Z$ is a \emph{minimal} balanced set if it does not contain a proper subset that is balanced. 
\end{definition}

The geometric interpretation of balancedness still holds: a set $Z \subseteq \bbR^N_+ \setminus \{\vec{o} \,\}$ is balanced if and only if the vector $\bfone^N$ lies in the relative interior of the conic span of $Z$. From a linear algebra point of view, the system $A^Z \lambda = \bfone^N$ must have nonnegative solutions, where $A^Z$ is the matrix whose columns are the vectors in $Z$.

\medskip

Let $\calS$ be a feasible collection and $\calB$ be a minimal balanced collection. For $S \in \calS$, let $z^S \in \bbR^N$ be given by 
\[
z^S_j = \left\{ \begin{tabular}{ll}
$\lambda^\calB_{\{i\}}$, & if $j = i$ and $\{i\} \in \calB \cap \calI_S$, \\
$0$, & otherwise. 
\end{tabular} \right.
\]
Define $\Omega^\calB_\calS = \left\{ \bfone^{N \setminus S} \mid S \in \calS \right\} \cup \left\{ \bfone^T \mid T \not \in \calS \right\} \cup \left\{ z^S \mid S \in \calS \right\}$. Moreover, let $\varepsilon > 0$ and for each $z \in Z$, define $a^\varepsilon_z \coloneqq \max\{A_\varepsilon \cup B \cup C\}$, where
\[ \begin{tabular}{rcl}
$A_\varepsilon$ & $=$ & $\Big\{v(N) - v(S) + \varepsilon \mid \bfone^{N \setminus S} = z, S \in \calS \Big\}$, \\
$B$ & $=$ & $\Big\{v(T) \mid \bfone^T = z, T \not \in \calS \Big\}$, \\
$C$ & $=$ & $\Big\{v(N) - \sum_{T \in \calB \setminus \calI_S} \lambda^\calB_T v^S(T) \mid z = z^S, S \in \calS \Big\}$. 
\end{tabular} \]
Then, the dual program of $( P^\calB_S )$ can be expressed by 
\[ \left( D^\calB_S \right) \quad \left\{ \begin{aligned}
& \max \sum_{z \in Z} \lambda_z a^\varepsilon_z \\
& \text{s.t.} \left\{ \begin{tabular}{l}
$\sum_{z \in Z} \lambda_z z = \bfone^N$ and, \\
$\lambda_z \geq 0$, for all $z \in Z$. 
\end{tabular} \right. 
\end{aligned} \right. \]
The constraints of the program $(D^\calB_S)$ correspond to the definition of balanced sets, more precisely the balanced sets which are subsets of $\Omega^\calB_\calS$. Using once again the Duality Theorem, because both feasible sets are nonempty, if the optimal value of $(D^\calB_\calS)$ is $v(N)$, it is also the case for $(P^\calB_\calS)$, which is equivalent to the existence of preimputations not being outvoted via $S$. Furthermore, the optimal value of $(D^\calB_\calS)$ is $v(N)$ if and only if, for all minimal balanced sets $Z \subseteq \Omega^\calB_\calS$ and for all $\varepsilon > 0$, we have 
\[
\sum_{z \in Z} \lambda_z a^\varepsilon_z \leq v(N). 
\]
To get rid of the $\varepsilon$, we can distinguish two cases, as it is done in Theorem \ref{th: basic}. Denote
$a_z \coloneqq \max \{A \cup B \cup C\}$ where $A \coloneqq \{v(N) - v(S) \mid \bfone^{N \setminus S} = z, S \in \calS\}$. We denote by $\bbB(\Omega^\calB_\calS)$ the set of minimal balanced sets $Z \subseteq \Omega^\calB_\calS$, and by $\bbB_0(\Omega^\calB_\calS)$ the set of minimal balanced sets $Z \subseteq \Omega^\calB_\calS$ such that $a_z = v(N) - v(S)$ for some $z \in Z$. Then, the optimal value of $(D^\calB_\calS)$ is $v(N)$ if and only if, for all $Z \in \bbB(\Omega^\calB_\calS) \setminus \bbB_0(\Omega^\calB_\calS)$, we have $\sum_{z \in Z} \lambda_z a_z \leq v(N)$, and for all $Z \in \bbB_0(\Omega^\calB_\calS)$, we have $\sum_{z \in Z} \lambda_z a_z < v(N)$.

\medskip 

We are now able to give the main result of \textcite{grabisch2021characterization}.

\begin{theorem}[\textcite{grabisch2021characterization}] \label{th: GS-main} \leavevmode \newline
Let $(N, v)$ be a balanced game. Then $(N, v)$ has a stable core if and only if for all feasible collection $\calS$, for all coalitions $S \in \calS$, for all minimal balanced collections $\calB$ satisfying $\calB \cap \calI_S \neq \emptyset$, there exists a minimal balanced set $Z \in \bbB \left( \Omega^\calB_\calS \right)$ such that 
\begin{equation} \label{eq: GS-main}
\left\{ \begin{aligned}
& \sum_{z \in Z} \lambda_z a_z \geq v(N), \quad \text{if } Z \in \bbB_0 \left( \Omega^\calB_\calS \right), \\ 
& \sum_{z \in Z} \lambda_z a_z > v(N), \quad \text{otherwise}.
\end{aligned} \right. 
\end{equation}
\end{theorem}

All the quantifiers in the characterization of Theorem \ref{th: GS-main} are taken over finite sets, so it can be tested algorithmically within in a finite number of computations. In order to build the most efficient algorithm, it is required to take into account and check the necessary conditions presented in Chapter \ref{ChapterB}: 
\begin{itemize}
\item The game must be balanced (Theorem \ref{th: BS-sharp}), 
\item The singletons must be exact (Proposition \ref{prop: necessary-exact}), 
\item The strictly vital-exact coalitions must be core-describing (Proposition \ref{prop: sve-describing}), 
\item No feasible collection can be a blocking feasible collection (Lemma \ref{lemma: blocking}),
\end{itemize}
and take advantages of the known sufficient conditions: 
\begin{itemize}
\item A weakly extendable game has a stable core (Proposition \ref{prop: weak-extandability}), 
\item A vital-exact extendable game has a stable core (\textcite{shellshear2009core}). 
\end{itemize}
To do so, it is necessary to 
\begin{itemize}
\item check whether a coalition is exact (Algorithm \ref{algo: exactness}), 
\item compute the set of strictly vital-exact coalitions (Algorithm \ref{algo: sve}), 
\item compute the set of extendable coalitions (Algorithm \ref{algo: extendability}), 
\item compute the set of feasible collections (Algorithm \ref{algo: feasibility}). 
\end{itemize}
The complete algorithm is presented below in pseudocode. Denote by $\text{val}(P)$ the optimal value of a linear program $(P)$, and, for a coalition $S \in \calN$, denote by $\bbB_S(\calF_S)$ the set of minimal balanced collections $\calB \subseteq \calF_S$ such that $\calB \cap \calI_S \neq \emptyset$ (see \eqref{eq: coalition-function-outvoting}). 

\begin{breakablealgorithm}
\caption{Core stability checking algorithm} \label{algo: core-stability}
\begin{algorithmic}[1]
\Require A coalition function $v$, the set of minimal balanced collections $\bbB(N)$
\Ensure The Boolean value: `$(N,v)$ has a stable core'
\Procedure{IsCoreStable}{$v$, $\bbB(N)$}
\For{$\calB \in \bbB(N)$} \algorithmiccomment{Checking balancedness}
\If{$\sum_{S \in \calB} \lambda^\calB_S v(S) > v(N)$}
\Return \textbf{False}
\EndIf
\For{$i \in N$} \algorithmiccomment{Checking exactness of the singletons}
\If{\textbf{not} \Call{IsExact}{$\{i\}$, $v$, $\bbB(N)$}}
\Return \textbf{False}
\EndIf
\EndFor
\EndFor
\State $\calVE(v) \gets \emptyset$, $\mathcal{E}xt(v) \gets \emptyset$
\For{$S \in 2^N \setminus \{\emptyset\}$}
\If{\Call{IsStrictlyVitalExact}{$S$, $v$, $\bbB(N)$}}
\State Add $S$ to $\calVE(v)$
\EndIf
\If{\Call{IsExtendable}{$S$, $v$, $\bbB(N)$}}
\State Add $S$ to $\mathcal{E}xt(v)$ 
\EndIf
\EndFor
\If{\textbf{not} \Call{IsCoreDescribing}{$\calVE(v)$, $v$}} \algorithmiccomment{see Proposition \ref{prop: sve-describing}}
\Return \textbf{False}
\EndIf
\For{$\calS \subseteq \calVE(v)$ such that $\calS \not \subseteq \bbB(N)$}
\If{\Call{IsFeasible}{$\calS$, $\calVE$, $v$, $\bbB(N)$}}
\If{$\calS = \{S_1, S_2\}$ such that $S_1 \cup S_2 = N$} \algorithmiccomment{see Lemma \ref{lemma: blocking}}
\Return \textbf{False}
\Else
\For{$S \in \mathcal{E}xt(v) \cap \calS$} \algorithmiccomment{see Proposition \ref{prop: extendability}}
\If{$S$ minimal (w.r.t. inclusion) in $\calS$}
\State Go to the next feasible collection $\calS$
\EndIf
\EndFor
\If{$\max_{S \in \calS} \min_{\calB \in \bbB_S(\calF_S)} \text{val}(P^\calB_\calS) = v(N)$}
\Return \textbf{False}
\EndIf
\EndIf
\EndIf
\EndFor
\Return \textbf{True}
\EndProcedure
\end{algorithmic}
\end{breakablealgorithm}

Let us illustrate this algorithm with some examples, executed on the following computing device: Apple M1 chip, CPU 3.2 GHz, 16 GB RAM. We solved the linear program using the minimal balanced sets. 

\begin{example}[4-person game]
Let $(N, v)$ be the game defined by $N = \{a, b, c, d\}$ and $v(S) = 0.6$ if $\lvert S \rvert = 3$, $v(N) = 1$ and $v(T) = 0$ otherwise. The algorithm returns that the set $\calE(v)$ only contains $N$. The set of strictly vital-exact coalitions is 
\[
\calVE(v) = \{\{i\} \mid i \in N\} \cup \{N \setminus \{i\} \mid i \in N\}. 
\]
The collection $\{\{a, c, d\}, \{a, b, c\}\}$ is a blocking feasible collection, so by Lemma \ref{lemma: blocking} the core is not stable. The CPU time for this example is 0.1 second.  
\end{example}

\begin{example}[5-person game] \label{ex: biswas}
Let $(N, v)$ be the exact game defined by Biswas, Parthasarathy, Potters and Voorneveld \cite{biswas1999large}, defined on $N = \{a, b, c, d, e\}$ by $v(S) = \min \{x(S), y(S)\}$, with $x = (2, 1, 0, 0, 0)$ and $y = (0, 0, 1, 1, 1)$. The core of this game is the convex hull of $x$ and $y$. For this game, the set of effective proper coalitions is
\[
\calE(v) \setminus \{N\} = \{\{b, c\}, \{b, d\}, \{b, e\}, \{a, c, d\}, \{a, c, e\}, \{a, d, e\}\}.
\]
The set of strictly vital-exact coalitions is $\calVE(v) = \calE(v) \cup \{\{i\} \mid i \in N\}$. The feasible collections which do not contain a minimal extendable coalition are the nonempty subsets of $\{\{a, c, d\}, \{a, c, e\}, \{a, d, e\}\}$. The collection $\{\{a, c, e\}, \{a, d, e\}\}$ does not satisfy the condition of Theorem \ref{th: GS-main}, therefore the core is not a stable set. The CPU time for this example is 1.5 seconds. 

\medskip

Let $(N, v')$ be the same game, except that we shift the affine subspace of preimputations by setting $v'(N) = 3.1$. The set $\calE(v')$ becomes $\{N\}$, and the core is the convex hull of the following points:
\[ \begin{aligned}
C(v') = {\rm conv} \left\{ \vphantom{\begin{pmatrix} 0 \\ 0 \\ 1.1 \\ 1 \\ 1 \end{pmatrix}} \right. & 
\begin{pmatrix} 0.1 \\ 0 \\ 1 \\ 1 \\ 1 \end{pmatrix},
\begin{pmatrix} 0 \\ 0.1 \\ 1 \\ 1 \\ 1 \end{pmatrix},
\begin{pmatrix} 0 \\ 0 \\ 1.1 \\ 1 \\ 1 \end{pmatrix},
\begin{pmatrix} 0 \\ 0 \\ 1 \\ 1.1 \\ 1 \end{pmatrix},
\begin{pmatrix} 0 \\ 0 \\ 1 \\ 1 \\ 1.1 \end{pmatrix},
\begin{pmatrix} 1.9 \\ 1 \\ 0 \\ 0.1 \\ 0.1 \end{pmatrix},
\begin{pmatrix} 1.9 \\ 1 \\ 0.1 \\ 0 \\ 0.1 \end{pmatrix},
\begin{pmatrix} 1.9 \\ 1 \\ 0.1 \\ 0.1 \\ 0 \end{pmatrix}, \\
&
\begin{pmatrix} 2.1 \\ 1 \\ 0 \\ 0 \\ 0 \end{pmatrix},
\begin{pmatrix} 2 \\ 1.1 \\ 0 \\ 0 \\ 0 \end{pmatrix},
\begin{pmatrix} 2 \\ 1 \\ 0.1 \\ 0 \\ 0 \end{pmatrix},
\begin{pmatrix} 2 \\ 1 \\ 0 \\ 0.1 \\ 0 \end{pmatrix},
\begin{pmatrix} 2 \\ 1 \\ 0 \\ 0 \\ 0.1 \end{pmatrix},
\begin{pmatrix} 0.1 \\ 0.1 \\ 0.9 \\ 1 \\ 1 \end{pmatrix}, 
\begin{pmatrix} 0.1 \\ 0.1 \\ 1 \\ 0.9 \\ 1 \end{pmatrix},
\begin{pmatrix} 0.1 \\ 0.1 \\ 1 \\ 1 \\ 0.9 \end{pmatrix} 
\left. \vphantom{\begin{pmatrix} 0.1 \\ 0.1 \\ 1 \\ 1 \\ 0.9 \end{pmatrix}} \right\}
\end{aligned} \]

The set of strictly vital-exact coalitions now contains $14$ coalitions, while the previous game had $11$ strictly vital-exact coalitions. The additional ones are $\{a, c\}$, $\{a, d\}$, $\{a, e\}$. The number of feasible collections which do not contain a minimal extendable coalition considerably increases to $51$, but none of them is a blocking feasible collection. The largest feasible collection contains $6$ strictly vital-exact coalitions. The estimated time for the algorithm to check if this specific collection satisfies the condition of Theorem \ref{th: GS-main} is greater that $200$ hours. 
\end{example}

\begin{example}[6-person game]
Let $(N, v)$ be the game defined by \textcite{studeny2022facets} on $N = \{a, b, c, d, e, f\}$ by $v(N) = 10$ and 
\[
\begin{tabular}{lcr}
\toprule
$v(S)$ & \hspace{1cm} & for all $S$ in \\
\midrule \midrule
$v(S) = 2$ & & \makecell[r]{$\{b, e\}, \{c, e\}, \{a, b, e\}, \{b, c, e\}, \{b, d, e\}, \{b, e, f\}, \{a, b, d, e\}$, \\ $\{a, b, d, f\}, \{a, b, e, f\}, \{b, d, e, f\}, \{a, b, d, e, f\}$} \\
\midrule
$v(S) = 3$ & & $\{c, d, e\}$ \\
\midrule
$v(S) = 4$ & & \makecell[r]{$\{a, c, e\}, \{a, c, f\}, \{c, d, f\}, \{c, e, f\}, \{a, b, c, e\}, \{a, c, d, e\}$, \\ $\{c, f\}, \{a, c, d, f\}, \{a, c, e, f\}, \{b, c, d, e\}, \{a, b, c, d, e\}$} \\ 
\midrule
$v(S) = 6$ & & \makecell[r]{$\{b, c, f\}, \{a, b, c, f\}, \{b, c, d, f\}, \{b, c, e, f\}$, \\ $\{a, b, c, d, f\}, \{a, b, c, e, f\}$} \\
\midrule
$v(S) = 8$ & & $\{c, d, e, f\}, \{a, c, d, e, f\}, \{b, c, d, e, f\}$ \\
\midrule
$v(S) = 0$ & & otherwise. \\
\bottomrule
\end{tabular}
\]
The set $\calE(v)$ is only composed of $N$. The set of strictly vital-exact coalitions is
\[
\{\{i\} \mid i \in N\} \cup \{\{b, e\}, \{c, f\}, \{a, c, e\}, \{b, c, f\}, \{a, b, d, f\}, \{b, c, d, e\}, \{c, d, e, f\}\},
\]
and the feasible collections which do not contain a minimal extendable coalition are the nonempty subsets of $\{\{a, c, e\}, \{c, d, e, f\}, \{b, c, d, e\}\}$. The core of the game is not a stable set because the feasible collection $\{\{a, c, e\}, \{c, d, e, f\}\}$ does not satisfy the condition of Theorem \ref{th: GS-main}. The CPU time for this example is 18 minutes and 12 seconds, among which 43 seconds for computing the set of minimal balanced collections on a set of $6$ players. 
\end{example}

\section{Domination and augmentation cones}

Throughout this section, let $(N, v)$ be a balanced game. Let $x$ be a preimputation outside of the core. Then, there exists a coalition $S \in \calN$ such that $x(S) < v(S)$. By redistributing a small enough amount of money from the players outside $S$ among the players in $S$, we may be able to construct a preimputation dominating $x$ via $S$. Then, any preimputation not in the core is dominated, and the set of dominating preimputations form a polytope. Moreover, if $x$ is a preimputation belonging to $X_\calS$, then there exists a polytope of dominating preimputations for each coalition $S$ in $\calS$. If one of these polytopes intersects the core, then any element in this intersection is a core element dominating $x$. But if none of these polytopes intersects the core, then the core cannot be externally stable. This observation briefly presents the characterization and the algorithm we discuss in this section. 

\medskip

The main objects we are studying in the sequel are the domination cones and the augmentation cones. When we say \emph{cone}, we usually mean \emph{affine cone}, but we omit the adjective \emph{affine} when an apex is clearly identifiable from the definition. If needed, to emphasize that a cone has $\vec{o}$ has an apex, we say that it is a \emph{linear} cone. Let $\phi: \bbX(v) \to 2^\calN$ be the map associating to each preimputation $x$, the set of coalitions 
\[
\phi(x) \coloneqq \{S \in \calN \mid x(S) < v(S)\}. 
\]
In other words, $\phi(x)$ is the feasible collection associated with the region to which $x$ belongs, i.e., for all $x \in \bbX(v)$, we have $x \in X_{\phi(x)}$. 

\begin{definition} \label{sym: dom-cone}
Let $x$ be a preimputation and $S$ a coalition. We denote by $\delta_S(x)$ the \emph{domination cone} of $x$ with respect to $S$, defined by 
\[
\delta_S(x) \coloneqq \{y \in \bbX(v) \mid y_i > x_i, \; \text{for all } i \in S\}. 
\]
\end{definition}

The domination cone $\delta_S(x)$ does not only include preimputations dominating $x$ because some of them are not affordable for $S$, i.e., there exist some $y \in \delta_S(x)$ such that $y(S) > v(S)$. In particular, the set $\delta_S(x)$ contains no preimputation dominating $x$ whenever $S \not \in \phi(x)$. 

\begin{definition} \label{sym: aug}
Let $x$ be a preimputation. We denote by $\text{Aug}(x)$ the \emph{augmentation cone} of $x$, defined by 
\[
\text{Aug}(x) \coloneqq \{y \in \bbX(v) \mid y(S) \geq x(S), \; \text{for all } S \in \phi(x)\}.
\]
\end{definition}

The augmentation cone $\text{Aug}(x)$ represents all the directions along which we can translate $x$ to simultaneously increase or preserve the payment of all coalitions in $\phi(x)$. Because a feasible collection is unbalanced, the nonemptiness of ${\rm Aug}(x)$ is guaranteed by Proposition \ref{prop: unbalanced}. 

\begin{figure}[ht]
\begin{center}
\begin{tikzpicture}[scale=0.3]
\fill[gray!15] (0, 0) -- (5.774, 10) -- (14, 10) -- (14, 0) -- cycle;
\fill[gray!15] (0, 0) -- (5, -10) -- (-5.774, -10) -- cycle;
\fill[pattern color=black!30, pattern={Lines[angle=60, distance=5pt]}] (0, 0) -- (14, 0) -- (14, -10) -- (5, -10) -- cycle;

\draw (1.774, 10) node[above left]{$\bbA_{ab}$} -- (11.774, -10);
\draw (-7.774, -10) -- (3.774, 10) node[above right]{$\bbA_a$};
\draw (-10, -6) node[left]{$\bbA_{ac}$} -- (14, -6);

\filldraw[black] (0,0) circle (5pt) node[above left]{$x$};

\draw[dashed] (0, 0) -- (5.774, 10);
\draw[dashed] (0, 0) -- (14, 0);
\draw[dashed] (0, 0) -- (-5.774, -10);
\draw[dashed] (0, 0) -- (5, -10);

\path (8, 5) node[below] {$\delta_{ab}(x)$};
\path (0, -7) node[below] {$\delta_{ac}(x)$};
\path (6.7, -3.8) node[below] {$\text{Aug}(x)$};
\end{tikzpicture}
\caption{Preimputation $x \in X_{\{ab, ac\}}$ with its two domination cones (in gray) and its augmentation cone (dashed).}
\label{fig: dom-aug}
\end{center}
\end{figure}
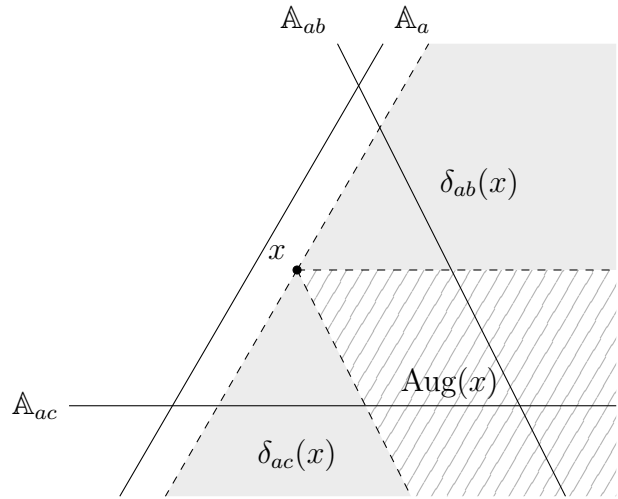

In the case of Figure \ref{fig: dom-aug}, the augmentation cone intersects none of the domination cones, as they are open sets. Then, preimputation $x$ cannot be dominated by a core element, and the core is not stable. Moreover, as we can see on Figure \ref{fig: dom-aug}, applying a translation to $x$ such that it stays in $X_{\{ab, ac\}}$ will not make the cones intersect. Then, no preimputation of $X_{\{ab, ac\}}$ will be dominated by a core element. Let $\zeta(x)$ be the cone defined by 
\[
\zeta(x) \coloneqq {\rm Aug}(x) \cap \left( \bigcup_{S \in \phi(x)} \delta_S(x) \right). 
\]

\begin{definition}
Let $\calS$ be a feasible collection. We say that $X_\calS$ is a \emph{blind spot} if, for all $x \in X_\calS$, we have $\zeta(x) = \emptyset$. 
\end{definition}

The feasible collections corresponding to the blind spots regions are called \emph{blocking feasible collections} (see Lemma \ref{lemma: blocking}). 

\begin{lemma} \label{lemma: necessary-blind-spot}
Let $(N, v)$ be a balanced game. The core $C(v)$ is a stable set only if, for all preimputations $x \in \bbX(v)$, we have $\zeta(x) \neq \emptyset$. 
\end{lemma}

\proof
Assume that there exists $x \in \bbX(v)$ such that $\zeta(x) = \emptyset$. Then, because $C(v) \subseteq {\rm Aug}(x)$, $x$ is not dominated by a core element, and the core cannot be stable. 
\endproof

Let $x$ be a preimputation outside of the core. Then, $\phi(x) \neq \emptyset$. If $\zeta(x) \neq \emptyset$, there exists a preimputation ensuring a better payment for all coalitions in $\phi(x)$, which is `accessible' for some $S \in \phi(x)$, through a domination process. It is not always possible to have a preimputation $y$ dominating $x$ which increases the payment of all coalitions in $\phi(x)$, as we can see in Figure \ref{fig: dom-aug}. 

\begin{proposition}
Let $\calS$ be a feasible collection, and let $y \in X_\calS$ such that $\zeta(y) \neq \emptyset$. Then, for all $x \in X_\calS$, we have $\zeta(x) \neq \emptyset$. 
\end{proposition}

\proof
Assume that there exists $y \in X_\calS$, $z \in \bbX(v)$ and a coalition $S \in \calS$ such that $z \in \text{Aug}(y) \cap \delta_S(y) \subseteq \zeta(y)$. Let $x \in X_\calS$. Denote by $\sigma = x - y$ the side payment from $y$ to $x$. For all $i \in S$, we have
\[
(z + \sigma)_i = z_i + \sigma_i > y_i + \sigma_i = (y + \sigma)_i = x_i. 
\]
Hence, $z + \sigma \in \delta_S(x)$. Moreover, for all $T \in \calS$, we have
\[
(z + \sigma)(T) = z(T) + \sigma(T) \geq y(T) + \sigma(T) = (y + \sigma)(T) = x(T). 
\]
Therefore, $z + \sigma \in \text{Aug}(x) \cap \delta_S(x) \subseteq \zeta(x)$. 
\endproof

We therefore know that the impossibility, for a given preimputation $x$, to be translated into another preimputation $y$ giving a better payment to all coalitions $S \in \phi(x)$ while dominating $x$ only depends on the set $\phi(x)$. In other words, the possibility to content a collection of coalitions with their payment does not depend on the values of the previous payments, but on the set system formed by the collection of coalitions. However, the feasibility of a collection depends on the coalition function $v$. 

\begin{theorem} \label{th: blind-spot}
Let $\calS$ be a feasible collection. Then $X_\calS$ is not a blind spot if and only if there exists a coalition $S \in \calS$ such that $\calS \cup \calI_S$ is unbalanced on $N$. 
\end{theorem}

\proof
Let $x \in X_\calS$, and denote by $P$ the polyhedron $\text{Aug}(x) \cap \delta_S(x)$. $P$ being a basic polyhedron, we denote by $(\calF_P, v_P)$ the game associated to $P$. Notice that, for all $T \in \calF_P$, we have $v_P(T) = x(T)$. 

\medskip 

We assume that $P \neq \emptyset$. By Theorem \ref{th: basic}, we have that $(\calF_P, v_P)$ is balanced and $\calI_S \cap \calE( \calF_P, v_P ) = \emptyset$. Because $\calS$ is feasible, it is unbalanced by Lemma \ref{lemma: blocking}. Therefore, if there exists a minimal balanced collection $\calB \subseteq \calF_P$, it must intersect $\calI_S$. Moreover, 
\[
\sum_{T \in \calB} \lambda_T v_P(T) = \sum_{T \in \calB} \lambda_T x(T) = x(N) = v(N).  
\]
Then, by Proposition \ref{prop: effectiveness}, we have $\calB \subseteq \calE(\calF_P, v_P)$, but this contradicts Theorem \ref{th: basic}. Hence, $\calF_P$ is unbalanced. 

\medskip

Assume now that there is no balanced collection $\calB \subseteq \calF_P$. Then by Theorem \ref{th: basic} the set $P = \text{Aug}(x) \cap \delta_S(x)$ is nonempty for all preimputations, and for all $x \in X_\calS$, we have $\zeta(x) \neq \emptyset$.
\endproof

\begin{example}
Let us consider the game depicted in Figure \ref{fig: dom-aug}. The feasible collection is $\calS = \{ab, ac\}$. Take $S = ab$. Then, the set system is $\calF_S = \{ab, ac, a, b\}$, which contains the balanced collection $\{ac, b\}$. For $T = ac$, the set system is $\calF_T = \{ab, ac, a, c\}$, which contains $\{ab, c\}$. Then $X_{\{ab, ac\}}$ is a blind spot. 
\end{example}

This result is useful to solve real-world problems of reallocation of payments without any external intervention. Usually, trying to satisfy a collection of overlapping coalitions complaining about a preimputation can be complicated because of the complexity of the computation. Also, knowing if there is a way to satisfy everyone is already a difficult task. If we assume an external intervention, which can force players to agree on a preimputation without involving domination, there always exists a preimputation improving the payment of any unbalanced collection of unsatisfied coalitions. 

\medskip 

However, if there is no external intervention, we need a coalition $S$ for which the new preimputation, which increases the payment for all coalitions in $\calS$, dominates the previous one. Theorem \ref{th: blind-spot} gives a characterization, easily implementable as a computer program, provided that we know the set of minimal balanced collections. 

\medskip

Furthermore, this algorithmic procedure can be incorporated into Algorithm \ref{algo: core-stability}, to discard all the blocking feasible collections. Then, the algorithm focuses on the regions which only include preimputations $x$ satisfying $\zeta(x) \neq \emptyset$. 

\medskip

Finally, the converse of Lemma \ref{lemma: necessary-blind-spot} does not hold. Indeed, consider the game $(N, v)$ defined in Example \ref{ex: biswas} on $N = \{a, b, c, d, e\}$ by $v(S) = \min \{x(S), y(S)\}$ with $x = (2, 1, 0, 0, 0)$ and $y = (0, 0, 1, 1, 1)$. The collection $\calS = \{ace, ade\}$ is feasible because $(1, 1, \text{\sfrac{1}{2}}, \text{\sfrac{1}{2}}, 0 ) \in X_\calS \neq \emptyset$, and not blocking because player $b$ is contained in none of these coalitions, but does not satisfy the condition of Theorem \ref{th: GS-main}. Therefore, there exists $x \in X_\calS$ which is not dominated by a core element, even if $\zeta(x) \neq \emptyset$. 

\section{Projection onto the core} \label{sec: projection-core}

In the previous section, we described a characterization which allows us to know whether there exists a side payment simultaneously improving the payment of a collection of coalitions, and if this side payment could be supported by a domination process involving one of the coalitions in the collection. In this section, we are looking for the smallest (according to the Euclidean norm) side payment between an affine subspace of $\bbX(v)$ and a given preimputation. 

\medskip

Let us consider a preimputation $x$, an element $y$ of a given affine subspace $V \subseteq \bbX(v)$, and the side payment $\sigma$ such that $x + \sigma = y$. Finding the side payment with the minimal norm is equivalent to, for the given preimputation $x$, solving this minimization problem $\min_{y \in V} \lVert y - x \rVert$. By definition of the Euclidean norm, we have 
\[
\arg \min_{y \in V} \; \lVert y - x \rVert = \arg \min_{y \in V} \sqrt{ \sum_{i \in N} \left( y_i - x_i \right)^2} = \arg \min_{y \in V} \sum_{i \in N} \left( y_i - x_i \right)^2. 
\]
By the Hilbert projection theorem (Theorem \ref{th: hilbert-proj}), the element $y$ is the (orthogonal) projection of $x$ onto $V$. The terms $(y_i - x_i)$ in the formula above are the adjustments of the individual players' payments induced by the side payment $\sigma = y - x$. Then, $\sigma$ is the side payment reorganizing the payments of individual players which minimize the sum of the square of the adjustments, while preserving the total sum of payments. Moreover, by the Hilbert projection theorem, it is the only side payment with these properties. In order to identify these side payments, we study projectors on $\bbX(v)$. 

\begin{definition}
A \emph{projector} $P$ is an idempotent linear map, i.e., $P \circ P = P$.
\end{definition}

The image of $P: \bbR^N \to \bbR^N$, denoted by $\text{im}(P)$ and called the \emph{column space} in the context of linear algebra, is the subspace of $\bbR^N$ spanned by the column of $P$. Then, $P$ takes as an input any element $x$ of the space $\bbR^N$ and associates to it an element $z \in \text{im}(P)$, which is the closest element of $\text{im}(P)$ from $x$ according to the Hilbert projection theorem (Theorem \ref{th: hilbert-proj}). 

\begin{remark}[Four fundamental subspaces]  \label{remark: fundamental-subspaces}
Each matrix $A$ a size $(n \times k)$ defines four subspaces, called the \emph{four fundamental subspaces} of $A$, which are 
\begin{itemize}
\item the \emph{column space} of $A$: ${\rm im}(A) \coloneqq \{Ay \mid y \in \bbR^k\} \subseteq \bbR^n$, 
\item the \emph{row space} of $A$: ${\rm im}(A^\top) \coloneqq \{A^\top x \mid x \in \bbR^n\} \subseteq \bbR^k$, 
\item the \emph{kernel} of $A$: ${\rm ker}(A) \coloneqq \{y \in \bbR^k \mid Ay = 0\}$, 
\item the \emph{left kernel} of $A$: ${\rm ker}(A^\top) \coloneqq \{ x \in \bbR^n \mid A^\top x = 0\}$. 
\end{itemize}
These four subspaces are related in the following way (see Fact 2.25, \parencite{bauschke2011convex}):
\[
\bbR^n = {\rm im}(A) \oplus {\rm ker}\left( A^\top \right) \qquad \text{and} \qquad \bbR^k = {\rm im} \left( A^\top \right) \oplus {\rm ker}(A). 
\]
\end{remark}

In our case, we are interested in subspaces defined as subsets of $\bbX(v)$ for which a collection $\calS$ of coalitions is effective. By Proposition \ref{prop: sve-describing}, we always assume in this section collections of strictly vital-exact coalitions. Recall that $\eta^S = \bfone^S - \frac{\lvert S \rvert}{n} \bfone^N$ and denote by $\langle \calS \rangle$ \label{sym: span} the subspace of $\Sigma \subseteq \bbR^N$ spanned by the set of vectors $\{\eta^S \mid S \in \calS\}$, and simply write $\langle S \rangle$ whenever $\calS = \{S\}$. Recall that $\bbA_\calS$ is the subspace of $\bbX(v) \subseteq \bbR^N$ defined by $\bbA_\calS = \{x \in \bbX(v) \mid x(S) = v(S), \forall S \in \calS\}$. 

\medskip

Each projection in $\bbX(v)$ is naturally decomposed as the sum of two elements: a preimputation and a side payment. Recall that the set of side payments $\Sigma$ is the linear subspace of $\bbR^N$ which is parallel to $\bbX(v)$. For each coalition $S \in \calS$, the affine subspace $\bbA_S$ has a corresponding linear subspace $\bbH_S$ included in $\Sigma$ with the same dimension as $\bbA_S$, which was defined by 
\[
\bbH_S \coloneqq \{\sigma \in \Sigma \mid \sigma(S) = 0\}. 
\]
The set $\{\bbH_S \mid S \in \calN\}$ forms the restricted resonance arrangement of $\bbR^N$, and does not depend on the game $(N, v)$. Notice also that $\eta^S$ is a normal to the subspace $\bbH_S$. For any subspace $V$ of a vector space $E$, the \emph{orthogonal complement} of $V$, denoted by $V^\perp$, is defined by 
\[
V^\perp \coloneqq \{x \in E \mid \langle x, y \rangle = 0, \forall y \in V\}. 
\]
Any subspace $V$ of a finite-dimensional vector space $E$ induces an orthogonal decomposition of $E$ into $E = V \oplus V^\perp$, i.e., for each element $x \in E$, we can write in a unique way $x = y + z$ with $y \in V$ and $z \in V^\perp$. 

\begin{remark} \label{remark: orth-decomp}
Let $P$ be a projector. Because the projection of an element is unique, any element $x \in E$ can be decomposed in a unique way as $x = y + z$ with $y = Px$ and $z = x - Px = (I_n - P)x$, with $y \in \text{im}(P)$ and $z \in {\rm im}(P)^\perp = {\rm ker}(P^\top)$ by Remark \ref{remark: fundamental-subspaces}. Therefore, we have that $I_n - P$ is the projector onto ${\rm ker}(P^\top)$. 
\end{remark}

Let $V$ be a linear subspace of $\Sigma$. Even if it is also a linear subspace of $\bbR^N$, in the sequel, we always consider the orthogonal complement $V^\perp$ as a subset of $\Sigma$, i.e., 
\[
V^\perp = \{\sigma \in \Sigma \mid \langle \sigma, y \rangle = 0, \, \forall y \in V\}. 
\]
We do not have a well-defined equivalent of the orthogonal complement for affine subspace, but it is still possible to have an orthogonal decomposition of it. 

\begin{remark} \label{remark: decomposition}
Let $S$ be a coalition. Using Proposition \ref{prop: first-proj}, any preimputation $x$ can be decomposed as $x = \pi_{\bbA_S}(x) - \gamma_S(x) \eta^S$. Naturally, $\pi_{\bbA_S}(x)$ belongs to $\bbA_S$ and $- \gamma_S(x) \eta^S$ belongs to $\langle S \rangle$, hence we have the decomposition $\bbX(v) = \bbA_S \oplus \langle S \rangle$. 
\end{remark}

Our main objective in this section is to compute, for any preimputation not included in the core, its projection onto the core. According to the Hilbert projection theorem, the projection of $x \in \bbX(v) \setminus C(v)$ onto the core is the closest preimputation $y \in C(v)$ from $x$. Then, $y$ belongs to a face $F$ of the core, for which there exists a collection of coalition $\calS$ such that $F \subseteq \bbA_\calS$. Then, the first part of this section is devoted to the study of the nonemptiness of $\bbA_\calS$. 

\medskip 

In the second part, we construct the projector $\pi_{\langle \calS \rangle}: \Sigma \to \langle \calS \rangle$, which associates to each side payment $\sigma$ its projection onto $\langle \calS \rangle$. The problem amounts to finding the linear combination of the $\eta^S$, for $S \in \calS$, which is the closest from $\sigma$. 

\medskip 

From this projector $\pi_{\langle \calS \rangle}$, we build the projector $\pi_{\bbA_\calS}: \bbX(v) \to \bbA_\calS$, which associates to each preimputation its projection onto $\bbA_\calS$, by using the fact that $\langle \calS \rangle^\perp$ and $\bbA_\calS$ are parallel. The projector is expressed in terms of the $\eta^S$ for $S \in \calS$, with specific coefficients for which an exact formula is provided. 

\medskip 

In another part, we give an alternative formulation of the projector $\pi_{\bbA_\calS}: \bbX(v) \to \bbA_\calS$, in terms of the excesses of the given preimputation at the coalitions in $\calS$, and a set of vectors uniquely defined from the set $\{\eta^S \mid S \in \calS\}$. 

\medskip 

Moreover, we give an algorithm which associate, to any preimputation $x$, a collection $\calS$ of coalitions for which $\pi_{\bbA_\calS}(x)$ belongs to the core.  

\medskip 

Finally, we use these formulas and algorithms to compute the distance between a preimputation and the core of a game, and propose a map $\bbX(v) \to \bbR$, which associate to each preimputation $x$ a real value measuring the failure of $x$ to be in the core, which we apply to market games. 

\subsection{Nonemptiness of $\bbA_\calS$}

In general, the nonemptiness of $\bbA_\calS$ depends only on $\calS$. 

\begin{definition}
Let $\calS \subseteq \calN$ be a collection of coalitions. We say that $\calS$ is an \emph{independent} collection if $\{\eta^S \mid S \in \calS\}$ forms a linearly independent set of vectors. 
\end{definition}

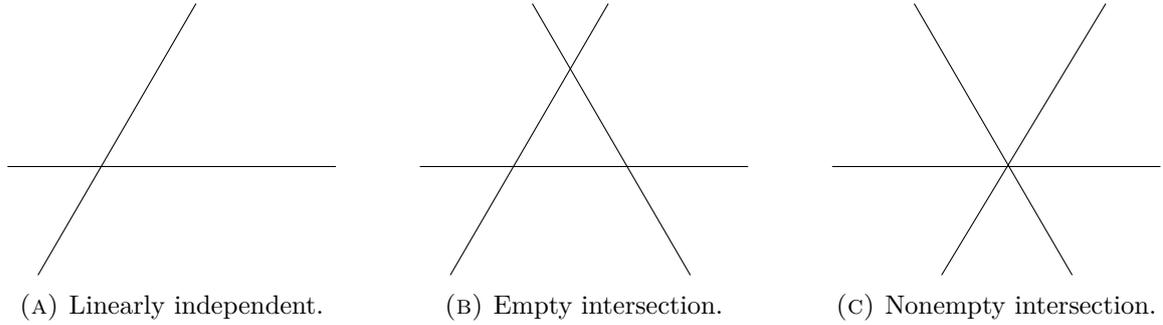
\begin{figure}[ht]
\begin{center}
\begin{subfigure}{0.3\textwidth}
\begin{center}
\begin{tikzpicture}[scale=0.18]
\draw (-7.774, -10) -- (3.774, 10);
\draw (-10, -2) -- (14, -2);
\end{tikzpicture}
\caption{Linearly independent.}
\end{center}
\end{subfigure}
\hspace{0.5cm}
\begin{subfigure}{0.3\textwidth}
\begin{center}
\begin{tikzpicture}[scale=0.18]
\draw (-1.774, 10) -- (9.774, -10);
\draw (-7.774, -10) -- (3.774, 10);
\draw (-10, -2) -- (14, -2);
\end{tikzpicture}
\caption{Empty intersection.}
\end{center}
\end{subfigure}
\hspace{0.5cm}
\begin{subfigure}{0.3\textwidth}
\begin{center}
\begin{tikzpicture}[scale=0.18]
\draw (-6, 10) -- (5.55, -10);
\draw (-4, -10) -- (8, 10);
\draw (-12, -2) -- (12, -2);
\end{tikzpicture}
\caption{Nonempty intersection.}
\end{center}
\end{subfigure}
\caption{Different configurations of hyperplanes in the plane.}
\label{fig: independent}
\end{center}
\end{figure}

A collection which is not independent is said to be \emph{dependent}. We see that $\bbA_\calS$ is nonempty if $\calS$ is independent. However, it is not sufficient, as depicted in Figure \ref{fig: independent}. 

\medskip

\begin{definition} 
We say that a coalition function $v$ is \emph{nonsingular} on a collection $\calS$ if, for all nonempty dependent subcollections $\calT \subseteq \calS$, we have $\bbA_\calT = \emptyset$. 
\end{definition}

\label{sym: gram}
Let $\calS = \{S_1, \ldots, S_k\} \subseteq \calN$ be a collection of coalitions. We denote by $H$ and $G$ respectively the matrices defined by 
\[
H = \begin{bmatrix} \eta^{S_1} & \ldots & \eta^{S_k} \end{bmatrix} \qquad \text{and} \qquad G = H^\top H.
\]
The matrix $G$ is called the \emph{Gram matrix} of the collection $\calS$, and its general term satisfies $g_{ij} = \langle \eta^{S_i}, \eta^{S_j} \rangle$. Then, $G$ captures all the information about the interdependence and the correlations between the payments of coalitions in $\calS$. If needed, we can specify the considered collection in the notation by writing $H_\calS$ and $G_\calS$. 

\medskip 

The symmetry of the scalar product implies the symmetry of the Gram matrix. Furthermore, for all $x \in \bbR^\calS$, we have 
\[
x^\top G x = x^\top H^\top H x = \langle Hx, Hx \rangle = \lVert Hx \rVert^2 \geq 0, 
\]
hence $G$ is positive semidefinite. We can use the Gram matrix to know whether a collection is independent. 

\begin{proposition}
\label{prop: non-singular}
A collection $\calS$ is independent if and only if $G$ is nonsingular. 
\end{proposition}

\proof
We have for all $x \in \bbR^\calS$, $x^\top G x \geq 0$. Then $G$ is positive semidefinite, i.e., all its eigenvalues are nonnegative. For $G$ to be nonsingular, we need to have only positive eigenvalues, i.e., to have $G$ positive definite. Then $G$ is nonsingular if and only if, for all $x \in \bbR^\calS \setminus \{\vec{o} \, \}$, 
\[
x^\top G x = \lVert Hx \rVert^2 = \left\lVert \sum_{S \in \calS} x_S \eta^S \right\rVert^2 > 0, 
\]
i.e., if and only if, for all $x \in \bbR^\calS \setminus \{\vec{o} \, \}$, we have $\sum_{S \in \calS} x_S \eta^S \neq 0$, which is the definition of $\{\eta^S \mid S \in \calS\}$ being a linearly independent set of vectors. 
\endproof

We now have a sharp characterization of the nonemptiness of $\bbA_\calS$ when $v$ is nonsingular on the set system on which it is defined. It is consistent with the fact that our formula for the projector onto $\bbA_\calS$ involves the inverse of $G$, as we will see shortly. 



\medskip




\subsection{Construction of the projector $\pi_{\langle \calS \rangle} : \Sigma \to \langle \calS \rangle$}

Because the columns of $H$ are the normal vectors $\eta^S$, any element of $\langle \calS \rangle$ can be written as $Hy$, with $y \in \bbR^\calS$. Intuitively, the side payment between a preimputation $x$ and its projection onto $\bbA_\calS$ is a linear combination of the normal vectors $\eta^S$, for $S \in \calS$. Now, we are looking for the coefficients of this linear combination. For a given side payment $\sigma$, we want to find the vector of coefficients $y \in \bbR^\calS$ which minimizes
\[
\varphi_H(y) = \lVert \sigma - Hy \rVert^2 = \left( \sigma - Hy \right)^\top \left( \sigma - Hy \right) = \sigma^\top \sigma - y^\top H^\top \sigma - \sigma^\top H y + y^\top G y.
\]
The critical points of $\varphi_H$ are the elements $y \in \bbR^\calS$ satisfying $\nabla \varphi_H(y) = 0$. We have the three following gradients: 
\[
\nabla \left( y^\top H^\top \sigma \right) = H^\top \sigma, \qquad \nabla \left( \sigma^\top Hy \right) = H^\top \sigma, \qquad \nabla \left (y^\top G y \right) = 2 G y, 
\]
which lead, by linearity of $\nabla$, to
\[
\nabla \varphi_H(y) = 2 G y - 2 H^\top \sigma. 
\]
Then, the critical point of $\varphi_H$ satisfy the so-called \emph{normal equation}, i.e., 
\begin{equation} \label{eq: normal-equation}
Gy = H^\top \sigma. 
\end{equation}
The Hessian of $\varphi_H$ is ${\rm Hess}(\varphi_H) = 2G$, which is positive definite when $\calS$ is independent. Then there is a unique critical point, which is a global minimum. The unique solution $y$ of the normal equation determines the projection of $\sigma$ onto $\langle \calS \rangle$, which is $Hy$.

\medskip

To manipulate matrix equations such as the normal equations, because not all matrices have an inverse, we need a generalization of usual inverses of matrices.  

\begin{theorem}[\textcite{penrose1955generalized}] \label{th: penrose} \leavevmode \newline
For any matrix $A$, there exists a unique matrix $X$ satisfying
\[
AXA = A, \qquad XAX = X, \qquad \left( AX \right)^\top = AX, \qquad \text{and} \qquad \left( XA \right)^\top = XA. 
\]
\end{theorem}

These equations are called the \emph{Penrose equations}, and their solution is called the \emph{Moore-Penrose inverse}, or \emph{generalized inverse}, of $A$ and is denoted by $A^\dag$. We notice that if $A$ is nonsingular, we have that $A^\dag = A^{-1}$. Moreover, if the columns of $A$ are linearly independent, we have 
\[
A^\dag = \left( A^\top A \right)^{-1} A^\top, 
\]
and if the rows of $A$ are linearly independent, we have
\[
A^\dag = A^\top \left( A A^\top \right)^{-1}. 
\]
When the columns of $A$ are linearly independent, the matrix $A^\dag$ is a \emph{left inverse} for $A$, because $A^\dag A = \left( A^\top A \right)^{-1} A^\top A = I_n$. Similarly, when the rows of $A$ are linearly independent, $A^\dag$ is a \emph{right inverse}. Then, if $\calS$ is independent, the Moore-Penrose inverse of $H$ is the matrix 
\[
H^\dag = \left( H^\top H \right)^{-1} H^\top = G^{-1} H^\top. 
\]
Then, we can solve the normal equation \eqref{eq: normal-equation}, and find $\pi_{\langle \calS \rangle}(\sigma)$:
\[
y = G^{-1} H^\top \sigma = H^\dag \sigma \qquad \text{and} \qquad \pi_{\langle \calS \rangle}(\sigma) = H y = H H^\dag \sigma. 
\]

\subsection{Construction of the projector $\pi_{\bbA_\calS}: \bbX(v) \to \bbA_\calS$}

We have built the projector onto ${\rm im}(H) = \langle \calS \rangle$, from the matrix $H$ and its Moore-Penrose inverse $H^\dag$. Our initial goal is to build a projector onto the affine subspace $\bbA_\calS$ for all preimputations. We introduce $b^\calS \in \bbR^\calS$ defined, for all $S \in \calS$, by 
\[
b^\calS_S = v(S) - \lvert S \rvert \frac{v(N)}{n}. 
\]
Let $L = H^\top$. We can describe $\bbA_\calS$ using the matrix $L$ and $b^\calS$, by
\[
\bbA_\calS = \left\{x \in \bbX(v) \mid Lx = b^\calS\right\}. 
\]
The linear subspace of $\Sigma$ parallel to $\bbA_\calS$ is therefore $\{\sigma \in \Sigma \mid L\sigma = 0\} = {\rm ker}(L) \cap \Sigma$. From Remark \ref{remark: fundamental-subspaces}, we know that $\bbR^N = \langle \calS \rangle \oplus {\rm ker}(L)$. Then, $\pi_{{\rm ker}(L)} = I_n - \pi_{\langle \calS \rangle}$. To deduce $\pi_{\bbA_\calS}$ from $\pi_{{\rm ker}(L)}$, we use the following lemma. 

\begin{lemma}[\textcite{bauschke2011convex}] \label{lemma: proj-affine} \leavevmode \newline
Let $K$ be a nonempty closed convex subset of $\bbR^N$, and let $x, y \in \bbR^N$. Then
\[
\pi_{y + K}(x) = y + \pi_K (x-y). 
\]
\end{lemma}

Using Lemma \ref{lemma: proj-affine}, we have, for any $y \in \bbA_\calS$, that 
\[
\pi_{\bbA_\calS}(x) = y + \pi_{{\rm ker}(L)}(x - y) = y + \left(I_n - H H^\dag \right) \left( x - y \right) = x - H G^{-1} L \left( x - y \right). 
\]
Indeed, any element of $\bbA_\calS$ can be written as $y + z$ with $z \in {\rm ker}(L)$ because the element of ${\rm ker}(L)$ do not change the payment of coalitions in $\calS$. Because $y \in \bbA_\calS$, we have $Ly = b^\calS$. Denote by $e(\calS, x) \in \bbR^\calS$ the vector defined, for all $S \in \calS$, by $e(\calS, x)_S = e(S, x)$.

\begin{theorem} \label{th: main-formula-proj}
Let $\calS$ be an independent collection. For all $x \in \bbX(v)$, we have
\[
\pi_{\bbA_\calS}(x) = x + \sum_{S \in \calS} \gamma^\calS_S(x) \eta^S, 
\]
where $\gamma^\calS_S(x) = \left( G^{-1} e(\calS, x) \right)_S$. 
\end{theorem}

\proof
To prove this theorem, we use the Projection Theorem for affine subspaces. Let $x$ be a preimputation. First, let us show that $\pi_{\bbA_\calS}(x) \in \bbX(v)$. We have that $\pi_{\bbA_\calS}(x)(N) = x(N) + \sum_{S \in \calS} \gamma^\calS_S(x) \eta^S(N) = v(N)$. Also, notice that 
\[
L(x - y) = Lx - b^\calS = - e(\calS, x). 
\]
Now, let us prove that $\pi_{\bbA_\calS}(x)$ belongs to $\bbA_\calS$, i.e., that $L \pi_{\bbA_\calS}(x) = b^\calS$:
\[
L \pi_{\bbA_\calS}(x) = L \left( x + H G^{-1} e(\calS, x) \right) = Lx - L H G^{-1} \left( Lx - b^\calS \right). 
\]
Because $L= H^T$, we have that $LH = G$, and it follows that
\[
L \pi_{\bbA_\calS}(x) = Lx - \left( Lx - b^\calS \right) = b^\calS, 
\]
therefore $\pi_{\bbA_\calS}(x) \in \bbA_\calS$. Let us prove now that $x - \pi_{\bbA_\calS}(x)$ is orthogonal to $\bbA_\calS$. Denote by $\gamma^\calS_S(x)$ the coordinate $\left( G^{-1} e(\calS, x) \right)_S$. For all $y \in \bbA_\calS$, we have 
\[
- \langle y - \pi_{\bbA_\calS}(x), x - \pi_{\bbA_\calS}(x) \rangle = \sum_{S \in \calS} \gamma^\calS_S(x) \langle y - \pi_{\bbA_\calS}(x), \eta^S \rangle.
\]
Because both $y$ and $\pi_{\bbA_\calS}(x)$ belong to $\bbA_\calS$, we have that $y(N) = v(N) = \pi_{\bbA_\calS}(x)(N)$ and, for all $S \in \calS$, that $y(S) = \pi_{\bbA_\calS}(x)(S)$. Then, $\langle y, \eta^S \rangle = \langle \pi_{\bbA_\calS}(x), \eta^S \rangle$, and finally 
\[
- \langle y - \pi_{\bbA_\calS}(x), x - \pi_{\bbA_\calS}(x) \rangle = \sum_{S \in \calS} \gamma^\calS_S(x) \left( \langle y, \eta^S \rangle - \langle \pi_{\bbA_\calS}(x), \eta^S \rangle \right) = 0, 
\]
which concludes the proof by the Projection Theorem. 
\endproof

We have now found a closed-form formula for the projection of any given preimputation $x$ onto an affine subspace $\bbA_\calS$, provided that $\calS$ is an independent collection. It requires no iterative computations, and can be implemented within a few lines of codes. The main algorithmic effort is to compute the inverse the Gram matrix $G$. One way to avoid that is to use an existing solver of linear systems to find $\gamma^\calS(x)$ satisfying 
\begin{equation} \label{eq: cramer}
G \gamma^\calS(x) = e(\calS, x). 
\end{equation}

\begin{remark}[Cramer's rule \parencite{cramer1750introduction}] \label{remark: cramer-rule}
We can use Cramer's rule to solve the linear system \eqref{eq: cramer}, which express the coordinates of $\gamma^S(x)$ in terms of the determinant of $G$, called the \emph{Gramian} of $\calS$. Let $G^S_x$ denote the matrix form where we replace the column composed of all the scalar products involving $\eta^S$ with the vector $e(\calS, x)$, i.e., 
\[
G^{S_i}_x = \begin{bmatrix} \lVert \eta^{S_1} \rVert^2 & \ldots & \langle \eta^{S_1}, \eta^{S_{i-1}} \rangle & e(S_1, x) & \langle \eta^{S_1}, \eta^{S_{i+1}} \rangle & \ldots & \langle \eta^{S_1}, \eta^{S_k} \rangle \\
\langle \eta^{S_2}, \eta^{S_1} \rangle & \ldots & \langle \eta^{S_2}, \eta^{S_{i-1}} \rangle & e(S_2, x) & \langle \eta^{S_2}, \eta^{S_{i+1}} \rangle & \ldots & \langle \eta^{S_2}, \eta^{S_k} \rangle \\
\vdots & & & \vdots & & & \vdots \\
\langle \eta^{S_j}, \eta^{S_1} \rangle & \ldots & \langle \eta^{S_j}, \eta^{S_{i-1}} \rangle & e(S_j, x) & \langle \eta^{S_j}, \eta^{S_{i+1}} \rangle & \ldots & \langle \eta^{S_j}, \eta^{S_k} \rangle \\
\vdots & & & \vdots & & & \vdots \\
\langle \eta^{S_k}, \eta^{S_1} \rangle & \ldots & \langle \eta^{S_k}, \eta^{S_{i-1}} \rangle & e(S_k, x) & \langle \eta^{S_k}, \eta^{S_{i+1}} \rangle & \ldots & \lVert \eta^{S_k} \rVert^2 \end{bmatrix} 
\]
Cramer's rule gives that 
\[
\gamma^\calS_S(x) = \frac{\det G^S_x}{\det G}, 
\]
therefore the formula of the projector for any preimputation $x$ is 
\[
\pi_{\bbA_\calS}(x) = x + \frac{1}{\det G} \sum_{S \in \calS} \det G^S_x \cdot \eta^S. 
\]
The relevance of Cramer's rule here is mainly for theoretical considerations. Even if the computational complexity of Cramer's rule has greatly reduced in the past years (see \textcite{habgood2012condensation}), the Cholesky decomposition is preferred to solve linear systems of equations with a symmetric positive definite matrix of coefficients. 
\end{remark}

\begin{example}
Let us apply Theorem \ref{th: main-formula-proj} and Cramer's rule to find the projector of $x \in \bbX(v)$ onto $\bbA_S$, for a coalition $S$. We have that $G = \begin{bmatrix} \lVert \eta^S \rVert^2 \end{bmatrix}$, and $G^S_x = \begin{bmatrix} e(S, x) \end{bmatrix}$. It follows that 
\[
\pi_{\bbA_S}(x) = x + \frac{1}{\det G} \sum_{S \in \{S\}} \det G^S_x \cdot \eta^S = x + \frac{e(S, x)}{\lVert \eta^S \rVert^2} \eta^S,
\]
which coincides with the formula of Proposition \ref{prop: first-proj}.
\end{example}

\subsection{Alternative formula for $\pi_{\bbA_\calS}: \bbX(v) \to \bbA_\calS$}

It is possible to express the projection of a given preimputation in a simpler way, using a new set of vectors, uniquely determined from $\{\eta^S \mid S \in \calS\}$. Denote by $\delta_{ij}$ the Kronecker delta, defined by $\delta_{ij} = 1$ if $i = j$ and $\delta_{ij} = 0$ otherwise. 

\begin{definition}
Let $\{e_1, \ldots, e_k\}$ and $\{f_1, \ldots, f_k\}$ be two sets of vectors in $\bbR^N$. We say that they form a \emph{biorthogonal system} if they satisfy $\langle e_i, f_j \rangle = \delta_{ij}$. 
\end{definition}

A common example of a biorthogonal system is the pair formed by a basis of a Euclidean space together with its dual basis. Then, for simplicity, if the sets are linearly independent, we say that $\{f_1, \ldots, f_k\}$ is a dual basis of $\{e_1, \ldots, e_k\}$, as it is in ${\rm span}(\{e_1, \ldots, e_k\})$. 

\begin{proposition} \label{prop: dual-basis}
Let $\{e_1, \ldots, e_k\}$ be a set of linearly independent vectors in $\bbR^N$, and $\{f_1, \ldots, f_k\}$ be a dual basis of $\{e_1, \ldots, e_k\}$. Set $E = \begin{bmatrix} e_1 & \ldots & e_k \end{bmatrix}$ and $F = \begin{bmatrix} f_1 & \ldots & f_k \end{bmatrix}$, as well as $G^E = E^\top E$ and $G^F = F^\top F$. The following equalities hold. 
\[
E^\top F = F^\top E = I_k, \qquad \text{and} \qquad \left( G^E \right)^{-1} = G^F.
\]
\end{proposition}

\proof
Let us look at the entries of $E^\top F$. From the definition of biorthogonal systems, we deduce $\left( E^\top F \right)_{ij} = \langle e_i, f_j \rangle = \delta_{ij}$, and $\left( F^\top E \right)_{ij} = \langle f_i, e_j \rangle = \delta_{ij}$, then $E^\top F = I_k = F^\top E$. Let us prove now that $\left( G^E \right)^{-1} = G^F$. Denote $K = \{1, \ldots, k\}$ and let $x \in {\rm span}(\{e_1, \ldots, e_k\})$. We write the coordinates of $x$ in the bases $\{e_1, \ldots, e_k\}$ and $\{f_1, \ldots, f_k\}$ respectively by $x = \sum_{i \in K} \alpha_i e_i$ and $x = \sum_{i \in K} \beta_i f_i$. We denote by $g^E_{ij}$ and $g^F_{ij}$ the general terms of $G^E$ and $G^F$ respectively. We can express the coordinates in one basis in terms of the coordinates in the other basis by 
\[ \begin{aligned}
\alpha_p & = \sum_{i \in K} \alpha_i \langle e_i, f_p \rangle = \langle x, f_p \rangle = \sum_{i \in K} \beta_i \langle f_i, f_p \rangle = \sum_{i \in K} g^F_{ip} \beta_i, \qquad \forall p \in K, \\
\text{and} \qquad \beta_q & = \sum_{j \in K} \beta_j \langle f_j, e_q \rangle = \langle x, e_q \rangle = \sum_{j \in K} \alpha_j \langle e_j, e_q \rangle = \sum_{j \in K} g^E_{jq} \alpha_j, \qquad \forall q \in K.
\end{aligned} \]
Combining these two expressions yields 
\[
\alpha_p = \sum_{i \in K} g^F_{ip} \left( \sum_{j \in K} g^E_{ji} \alpha_j \right) = \sum_{j \in K} \left( \sum_{i \in K} g^E_{ji} g^F_{ip} \right) \alpha_j = \sum_{j \in K} \left( G^E G^F \right)_{jp} \alpha_j, \qquad \forall p \in K. 
\]
Then, we have that $\left( G^E G^F \right)_{ij} = \delta_{ij}$, and doing the same calculations on the $\beta$'s gives $\left( G^F G^E \right)_{ij} = \delta_{ij}$, hence $G^F$ is the inverse of $G^E$. 
\endproof

Let $\calS = \{S_1, \ldots, S_k\}$ be an independent collection. We denote by $\{h^{S_1}, \ldots, h^{S_k}\}$ the dual basis of $\{\eta^{S_1}, \ldots, \eta^{S_k}\}$. Set $H^\circ = \begin{bmatrix} h^{S_1} & \ldots & h^{S_k} \end{bmatrix}$. 

\begin{lemma} \label{lemma: dual-basis}
Let $\calS$ be an independent collection of coalitions. Then 
\[
h^{S_j} = \sum_{i \in K} g^{[-1]}_{ij} \eta^{S_i} \qquad \text{and} \qquad H^\circ = H G^{-1} = L^\dag, 
\]
where $g^{[-1]}_{ij}$ denotes the general term of $G^{-1}$. 
\end{lemma}

\proof
The second fact is simply the translation of the first one in terms of matrices, we will therefore only prove the first fact. Let $x \in \langle \calS \rangle$, and we write the coordinates of $x$ in basis $\{\eta^S \mid S \in \calS\}$ and $\{h^S \mid S \in \calS\}$ respectively by $x = \sum_{S \in \calS} \alpha_S \eta^S$ and $x = \sum_{S \in \calS} \beta_S h^S$. Using Proposition \ref{prop: dual-basis} gives
\[
\alpha_{S_p} = \sum_{i \in K} \alpha_{S_i} \langle \eta^{S_i}, h^{S_p} \rangle = \langle x, h^{S_p} \rangle = \sum_{i \in K} \beta_{S_i} \langle h^{S_i}, h^{S_p} \rangle = \sum_{i \in K} g^{[-1]}_{ip} \beta_{S_i}, \qquad \forall p \in K.
\]
By setting all the coefficients $\beta_S$ to $0$ except for $\beta_{S_j} = 1$, we get $\alpha_{S_p} = g^{[-1]}_{jp}$ for all $p \in K$. Then, $h^{S_j} = x = \sum_{i \in K} g^{[-1]}_{ij} \eta^{S_i}$. We can rewrite it as 
\[
h^{S_j} = \eta^{S_1} g^{[-1]}_{1j} + \eta^{S_2} g^{[-1]}_{2j} + \ldots + \eta^{S_k} g^{[-1]}_{kj} = \begin{bmatrix} \eta^{S_1} & \ldots & \eta^{S_k} \end{bmatrix} \left( G^{-1} \right)^{\text{col}}_j,
\]
with $\left( G^{-1} \right)^{\text{col}}_j$ denoting the $j$-th column of $G^{-1}$, and then we have $H^\circ = H G^{-1}$. Because the rows of $L$ are linearly independent, we have 
\[
L^\dag = L^\top \left( L L^\top \right)^{-1} = H \left( H^\top H \right)^{-1} = H G^{-1} = H^\circ, 
\]
which concludes the proof. 
\endproof

Using Lemma \ref{lemma: dual-basis}, we can rewrite the formula of the projector. 

\begin{theorem} \label{th: proj-h}
Let $\calS$ be an independent collection. For all $x \in \bbX(v)$, we have 
\[
\pi_{\bbA_\calS}(x) = x + \sum_{S \in \calS} e(S, x) h^S.
\]
\end{theorem}

\proof
From Theorem \ref{th: main-formula-proj}, we already have that $\pi_{\bbA_\calS}(x) = x + H G^{-1} e(\calS, x)$. By Lemma \ref{lemma: dual-basis}, we have $H G^{-1} = H^\circ$, and therefore, 
\[
\pi_{\bbA_\calS}(x) = x + H^\circ e(\calS, x) = x + \sum_{S \in \calS} e(S, x) h^S. 
\]
\endproof


\begin{figure}[ht]
\begin{center}
\begin{subfigure}{0.49\textwidth}
\begin{center}
\begin{tikzpicture}[scale=0.2]
\draw[->] (-8, 6) -- (-8, 8) node[above]{$\eta^{ac}$};
\draw[->] (2.25, 9.034) -- (4.25, 10.184) node[above right]{$\eta^{ab}$};

\draw (1.774, 10) node[above left]{$\bbA_{ab}$} -- (11.774, -10);
\draw (-10, 6) node[left]{$\bbA_{ac}$} -- (14, 6);

\filldraw[black] (-5,-5) circle (5pt) node[below left]{$x$};

\draw[dashed] (-5, -5) -- (3.774, 6);

\draw[thick, blue, ->] (-5, -5) -- node [midway, below right]{$\gamma^\calS_{ab}(x) \eta^{ab}$} (3.774, -0.613);

\draw[color=white!50, fill=white!50] (-2.3, 1.5) rectangle (3.5, 3.8);
\draw[thick, blue, ->] (3.774, -0.613) -- node[midway, left]{$\gamma^\calS_{ac}(x) \eta^{ac}$} (3.774, 6);
\end{tikzpicture}
\end{center}
\end{subfigure}
\begin{subfigure}{0.49\textwidth}
\begin{center}
\begin{tikzpicture}[scale=0.2]
\draw[->] (-8, 6) -- (-8, 8) node[above]{$\eta^{ac}$};
\draw[->] (2.25, 9.034) -- (4.25, 10.184) node[above right]{$\eta^{ab}$};

\draw (1.774, 10) node[above left]{$\bbA_{ab}$} -- (11.774, -10);
\draw (-10, 6) node[left]{$\bbA_{ac}$} -- (14, 6);

\filldraw[black] (-5,-5) circle (5pt) node[below left]{$x$};

\draw[dashed] (-5, -5) -- (3.774, 6);

\draw[thick, blue, ->] (-5, -5) -- node [midway, below]{\footnotesize $e(ac, x) h^{ac}$} (9.274, -5);

\draw[thick, blue, ->] (9.274, -5) -- node[midway, above right]{\footnotesize $e(ab, x) h^{ab}$} (3.774, 6);
\end{tikzpicture}
\end{center}
\end{subfigure}
\caption{Decompositions of the side payment between $x$ and its projection onto $\bbA_\calS$ with $\calS = \{ab, ac\}$.}
\label{fig: two-proj}
\end{center}
\end{figure}

\medskip

Let $\calS \subseteq \calN$ be independent. We know that the side payment between the projection onto $\bbA_\calS$ and a given preimputation is a linear combination of the normal $\{\eta^S \mid S \in \calS\}$ or of $\{h^S \mid S \in \calS\}$, as we can see in Figure \ref{fig: two-proj}. 

\subsection{Using the appropriate projector on a given preimputation}

Assume now that $\calS$ is feasible, but not independent. Let $(N, v)$ be a balanced game nonsingular on $\calS$. Then, the affine subspace $\bbA_\calS$ is empty, and there is no obvious choice of independent subcollection $\calT \subseteq \calS$ such that, for all $x \in X_\calS$, we have $\pi_{\bbA_\calT}(x) \in C(v)$. Moreover, there may be several independent subcollections satisfying this property, and they might be nonmaximal with respect to inclusion. 

\medskip

The next result is the first step to the construction of an algorithm finding a proper independent subcollection $\calT \subseteq \calS$, and generalize Proposition \ref{prop: chi} to the general case. Let $\calS$ be an independent collection, $T \not \in \calS$ be a coalition and $x \in \bbX(v)$. We define
\[
\chi_{\calS}(T, x) = \sum_{S \in \calS} \det G^S_x \langle \eta^S, \eta^T \rangle - e(T,x) \, \det G.
\]

\begin{proposition}
Let $\calS \subseteq \calN$ be independent, $T \in \calN \setminus \calS$ and $x \in \bbX(v)$. Then 
\[
\pi_{\bbA_\calS}(x) \in \bbA^\geq_T \quad \text{ if and only if } \quad \chi_\calS(T, x) \geq 0. 
\]
\end{proposition}

\proof
First, we study the excess of $T$ at the projection onto $\bbA_\calS$:
\[ 
e(T, \pi_{\bbA_\calS}(x)) = v(T) - x(T) - \sum_{S \in \calS} \gamma^\calS_S(x) \eta^S(T).
\]
Using Cramer's rule (see Remark \ref{remark: cramer-rule}) gives
\[
e(T, \pi_{\bbA_\calS}(x)) = e(T, x) - \sum_{S \in \calS} \frac{\det G^S_x}{\det G} \langle \eta^S, \eta^T \rangle = -\left( \det G \right)^{-1} \chi_\calS(T, x). 
\]
The projection lies into $\bbA^\geq_T$ if and only if $e(T, \pi_{\bbA_\calS}(x))$ is nonpositive, and therefore if and only if $\chi_\calS(T, x)$ is nonnegative. 
\endproof

\begin{figure}[ht]
\begin{center}
\begin{tikzpicture}[scale=0.35]
\fill[gray!15] (5.1548, -2) -- (1, 5.1957) -- (3.774, 10) -- (11.3677, 10) -- (14, 5.4411) -- (14, -2) -- cycle;

\draw (-1.774, 10) node[above left]{$\bbA_{ab}$} -- (9.774, -10);
\draw (-7.774, -10) -- (3.774, 10) node[above right]{$\bbA_a$};
\draw (-10, -2) -- (14, -2) node[right]{$\bbA_{ac}$};

\path (7, 4) node[below] {$C(v)$};

\filldraw[black] (-6,-5) circle (5pt) node[below left]{$x$};

\draw[dashed] (-6, -5) -- (3.665, 0.5804);

\draw[thick, blue, ->] (-6, -5) -- (1, 5.196);
\draw[thick, blue, ->] (-6, -5) -- (-3.155, -2);
\draw[thick, blue, ->] (-6, -5) -- (5.155, -2);
\end{tikzpicture}
\caption{Projections on $\bbA_\calT$ for all maximal independent collection $\calT \subseteq \{ab, ac, a\}$, with the projection on the core.}
\label{fig: multi-proj}
\end{center}
\end{figure}
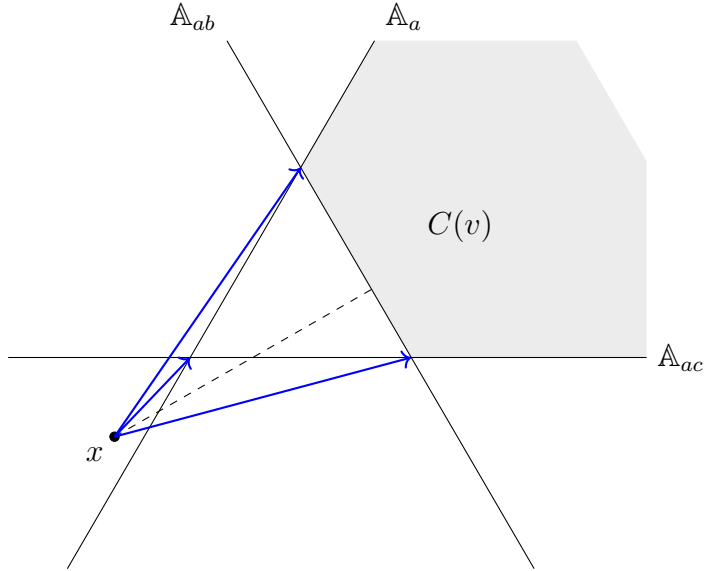

We now have a characterization to check whether the projection onto an affine subspace $\bbA_\calT$ with $\calT \subseteq \calS$ cross or lies into all the affine hyperplanes $\bbA_S$, for $S \in \calS$. 

\medskip 

One drawback of this characterization is the frequent use of determinants. But the computation of $\det G^S_x$ can be done in terms of Gramians using Laplace's expansion formula. To compute determinant of these Gramians, we use the following result. 

\begin{proposition} \label{prop: gramian}
Let $\calS$ be an independent collection, and $\calV = \{v_S \mid S \in \calS\}$ be an orthonormal basis of $\langle \calS \rangle$. Then, 
\[
\det G = \left( \prod_{S \in \calS} v_S(S) \right)^2. 
\]
\end{proposition}

\proof
The proof is based on the QR decomposition of $H$ by the Gram-Schmidt process. Indeed, the QR decomposition of $H = QR$ gives an orthogonal matrix $Q$ whose columns form an orthonormal basis $\calV = \{v_{S_1}, \ldots, v_{S_k}\}$ of $\langle \calS \rangle$, and an upper triangular matrix $R$ defined by 
\[
R = \begin{bmatrix} \langle v_1, \eta^{S_1} \rangle & \langle v_1, \eta^{S_2} \rangle & \langle v_1, \eta^{S_3} \rangle & \ldots & \langle v_1, \eta^{S_k} \rangle \\
0 & \langle v_2, \eta^{S_2} \rangle & \langle v_2, \eta^{S_3} & \ldots & \langle v_2, \eta^{S_k} \rangle \\
0 & 0 & \langle v_3, \eta^{S_3} \rangle & \ldots & \langle v_3, \eta^{S_k} \rangle \\
\vdots & \vdots & \vdots & \ddots & \vdots \\
0 & 0 & 0 & \ldots & \langle v_k, \eta^{S_k} \rangle \end{bmatrix}.
\]
Therefore, the determinant of $R$ is $\det R = \prod_{S \in \calS} \langle v_S, \eta^S \rangle = \prod_{S \in \calS} v_S(S)$ because, for all $S \in \calS$, $v_S \in \langle \calS \rangle \subseteq \Sigma$. By the multiplicativity of the  determinant, we have that 
\[
\det G = \det \left( H^\top H \right) = \det \left( R^\top Q^\top Q R \right) = \left( \det R \right)^2, 
\]
which leads to $\det G = \left( \prod_{S \in \calS} v_S(S) \right)^2$. 
\endproof

Combining this result with the Gram-Schmidt process leads to Algorithm \ref{algo: gramian}, which has a complexity of $\mathcal{O}(n k^2)$, which is slightly slower than the usual ones used to compute determinants, which have a complexity of $\mathcal{O}(k^3)$. However, the idea is not to compute directly a Gramian with this algorithm, but to derive the Gramian of a collection $\calS$ by updating the one of the collection $\calS \setminus S$ with $S \in \calS$. In the next result, we present a projector onto lines in $\Sigma$ that we use in the Algorithm computing the Gramian. 

\begin{lemma} \label{lemma: proj-gram-schmidt}
Let $S$ be a coalition, and let $\sigma \in \Sigma$ be a side payment satisfying $\lVert \sigma \rVert = 1$. Then the projection of $\eta^S$ onto ${\rm span}(\sigma)$ is 
\[
\pi_{{\rm span}(\sigma)}(\eta^S) = \sigma(S) \cdot \sigma. 
\]
\end{lemma}

\proof
The usual projector onto lines gives
\[
\pi_{{\rm span}(\sigma)}(\eta^S) = \frac{\langle \eta^S, \sigma \rangle}{\langle \sigma, \sigma \rangle} \sigma = \langle \eta^S, \sigma \rangle \sigma. 
\]
Because $\sigma$ is a side payment, we have $\langle \bfone^N, \sigma \rangle = \sigma(N) = 0$. Then, it remains 
\[
\langle \eta^S, \sigma \rangle \sigma = \left( \langle \bfone^S, \sigma \rangle - \frac{\lvert S \rvert}{n} \langle \bfone^N, \sigma \rangle \right) \sigma = \langle \bfone^S, \sigma \rangle \sigma = \sigma(S) \cdot \sigma.
\] 
\endproof

\begin{breakablealgorithm}
\caption{Gramian $\det G$ computation} \label{algo: gramian}
\begin{algorithmic}[1]
\Require The Gramian $\det G$ of $\calS$, an orthonormal basis $\calV$ of $\langle \calS \rangle$, a coalition $S \not \in \calS$
\Ensure The Gramian $\det G_*$ of $\calS \cup \{S\}$, an orthonormal basis $\calV'$ of $\calS \cup \{S\}$
\Procedure{UpdateGramian}{$\calS$, $S$}
\State Set $v_S \gets \eta^S$
\For{$v_T \in \calV$}
\State $v_S \gets v_S - v_T(S) \cdot v_T$ \Comment{see Lemma \ref{lemma: proj-gram-schmidt}}
\EndFor
\If{$\lVert v_S \rVert = 0$}
\Return $\det G_* = 0$
\Else
\State $v_S \gets v_S / \lVert v_S \rVert$
\State $\det G_* = \left(v_S (S)\right)^2 \cdot \det G$ \Comment{see Proposition \ref{prop: gramian}}
\State Add $v_S$ to $\calV$
\EndIf
\Return $\calV$, $\det G_*$
\EndProcedure
\end{algorithmic}
\end{breakablealgorithm}

Because the size of the Gram matrices depends on the size of the collections and not on the number of players, we need to work with the smallest collections. Moreover, we see on Figure \ref{fig: multi-proj} that the projection onto the core is not necessarily the projection associated with a large independent collection. The following result states that the projection on the core is the projection onto one of these smallest collections. 

\begin{definition}
Let $x$ be a preimputation outside of the core. We say that a collection $\calS$ is a \emph{reaching collection} for $x$ if it is independent and $\pi_{\bbA_\calS}(x) \in C(v)$. 
\end{definition}

We denote by $\rho(\calS, x)$ the set of reaching subcollections of $\calS$ for $x$. If $\calS = \phi(x)$, we simply write $\rho(\phi(x), x) \eqqcolon \rho(x)$.

\begin{proposition}
Let $x \in \bbX(v) \setminus C(v)$. The projection of $x$ onto the core coincide with the projection onto $\bbA_\calT$, with $\calT$ being minimal in $\rho(x)$.
\end{proposition}

\proof
Let us first recall the notation we used in the previous chapter. We denote by $\calL_C(v)$ the face lattice of the core. For any face $F \in \calL_C(v)$ of the core, we denote by $\calE(F)$ the set of coalitions $\calE(F) \coloneqq \{S \in \calN \setminus \calE(v) \mid F \subseteq \bbA_S(v)\}$. 

\medskip 

Let $x \in X_\calS$. The projection of $x$ onto the core is necessarily achieved via a reaching collection. Let $\calT_0$ be one of them. Then $\pi_{\bbA_\calT}(x)$ belongs to a face of the core, denoted by $F_0$. If there does not exist another face $F_1$ containing $F_0$ such that $\calE(F_1)$ belongs to $\rho(x)$, then, because the injection of $\calL_C(v)$ into the set of unbalanced collections is order-reversing by Proposition \ref{prop: face-lattice}, $\calT$ is minimal in $\rho(x)$. Then, we assume that there exists a face $F_1 \supseteq F_0$ such that $\calE(F_1) \eqqcolon \calT_1 \in \rho(x)$, and repeat the same process with $\calT_1$. Because the number of players is finite, the number of collection of coalitions is also finite, therefore there exists a face $F_p$ containing $F_0$ and a collection $\calT_p$ contained in $\calT_0$ such that $\calT_p$ is minimal in $\rho(x)$. Because $\pi_{\bbA_{\calT_0}}(x)$, which is the projection onto the core, belongs to $F_0$, it also belongs to $F_p$. 
\endproof

Because the set of independent subcollections of any collection is a matroid, they form a poset under inclusion of their spanned space. Therefore, to explore this poset and find a minimal reaching collection for any given preimputation, we use a breadth-first algorithm, detailed in Algorithm \ref{algo: find-reach-coll}.  

\begin{breakablealgorithm}
\caption{Finding minimal reaching collections} \label{algo: find-reach-coll}
\begin{algorithmic}[1]
\Require A balanced coalition function $v$, the set of strictly vital-exact coalitions $\calVE(v)$, a preimputation $x$
\Ensure The set of minimal reaching collections for $x$
\Procedure{FindReachingCollection}{$v$, $\calVE(v)$, $x$}
\State Set $\calS \gets \emptyset$ 
\For{$S \in \calVE(v)$}
\If{$x(S) < v(S)$}
\State Add $S$ to $\calS$
\EndIf
\EndFor
\State $\mathfrak{K} \gets \{\{S\} \mid S \in \calS\}$ \Comment{Set of collections to check}
\State $\mathbb{O} \gets \emptyset$ \Comment{Set of outputs}
\For{$\calT \in \mathfrak{K}$}
\If{$\max_{S \in \calVE(v)} e(S, \pi_{\bbA_\calT}(x)) \leq 0$}
\State Add $\calT$ to $\mathbb{O}$
\Else
\For{$S \in \calS$}
\If{$\calT \cup \{S\} \not \in \mathfrak{K}$}
\If{\Call{UpdateGramian}{$\calT$, $S$} $\neq 0$}
\State Add $\calT \cup \{S\}$ to $\mathfrak{K}$
\State Store new Gramian
\EndIf
\EndIf
\EndFor
\State Remove $\calT$ from $\mathfrak{K}$
\EndIf
\EndFor
\Return $\mathbb{O}$
\EndProcedure
\end{algorithmic}
\end{breakablealgorithm}

We know that Algorithm \ref{algo: find-reach-coll} gives at least one output when the coalition function $v$ is balanced, because the core has at least a nonempty face. The face corresponding to the union of all the outputs of the algorithm is the smallest face where lies the projection onto the core. It is the intersection of the faces corresponding to the collections in $\mathbb{O}$. 

\medskip

During this section, the goal was to project a preimputation onto the core. Because the objects involved in the formulas are simply the (possible) normals of the facets of the core, we can apply any of these algorithms and formulas to any basic polyhedron. 

\subsection{Application: measuring and correcting market failures.}

In the sequel, we propose a quantification of the \emph{failure} of a market, for a given payment vector, using the model of \emph{market games} defined by \textcite{shapley1969market}. Recall that a \emph{market} is a mathematical model, denoted by $\calM = (N, G, A, U)$, where 
\begin{itemize}
\item $N$ is a finite set of \emph{players}, or \emph{traders}, 
\item $G$ is the nonnegative orthant of a finite-dimensional vector space, called the \emph{commodity space}, 
\item $A = \{a^i \mid i \in N\} \subseteq G^N$ is an indexed set of elements in $G$, called the \emph{initial endowments}, 
\item $U = \{u^i \mid i \in N \}$ is an indexed set of continuous, concave function $u^i: G \to \bbR$, called the \emph{utility functions}. Notice that there is no assumption concerning the monotonicity of the utility functions. 
\end{itemize}

The initial endowments can be interpreted as the belongings of each player at the initial state of the economy, and they will trade these commodities with other players to maximize their utility. Let $S$ be a coalition. If the players in $S$ form a market, the whole quantity of commodities is $\sum_{i \in S} a^i$. Then, a set of endowments $\{x^i \mid i \in S\} \subseteq G$ such that $\sum_{i \in S} x^i = \sum_{i \in S} a^i$ is called a \emph{$S$-feasible allocation} of the market $\calM$, and we denote their set by $X^S$. The \emph{market game} generated by this market is a game $(N, v)$ whose coalition function is defined, for all $S \in \calN$, by 
\[
v(S) = \max_{x \in X^S} \sum_{i \in S} u^i(x^i). 
\]
As we have seen in Chapter \ref{ChapterA}, market games are totally balanced, and their core is referred to as the core of the associated market. The core of a market is then the set of payment vectors that coalitions cannot achieve by forming a market on their own. 

\medskip

It is well-known that the competitive equilibria of an economy lie in its core. Consistently, \textcite{shapley1975competitive} proved that the set of competitive outcomes of a market is included in the core of its associated market game. Therefore, they satisfy very desirable properties, such as Pareto optimality and coalitional rationality, but the players of the market do not necessarily reach one of these points as a payment vector. Following the interpretation we adopted throughout this manuscript, a positive outcome for the interaction between players leading to the existence of the game must be a payment vector from the core, to avoid the departure of some players from the grand coalition. This leads to the following definition. 

\begin{definition}
Let $\calM$ be a market, and $(N, v)$ its associated game. Let $x$ be a preimputation which is not included in the core. We define the \emph{failure} of the market $\calM$, or equivalently of the game $(N, v)$, at $x$, denoted by $\mu_v(x)$, as the following quantity
\[
\mu_v(x) \coloneqq \min_{y \in \partial C(v)} \lVert x - y \rVert, 
\]
where $\partial C(v)$ denotes the frontier of the core $C(v)$. 
\end{definition}

The failure of the market at a specific preimputation $x$ is the distance from $x$ to the core. It represents the volume of the smallest reallocation of money, i.e., the Euclidean norm of a shortest side payment, needed to go from $x$ to its closest element in the core. 

\medskip 

When players join a market, they put in common all their initial endowments, that they will share later. Players in $S$, with the sum of their initial endowments, are able to redistribute them in a way that the sum of their utility after the redistribution is $v(S)$. Because market games are totally balanced, the larger the grand coalition is, the better should be the payment of every coalition. 

\medskip 

Then, if the allocation of commodities leads to a preimputation not included in the core, for at least one coalition $S$ we have $x(S) < v(S)$, which is absurd. If there is a broader range of commodities to distribute, the payments of all the coalitions must be greater than the payment then can have by themselves. The failure at a specific preimputation $x$ quantify how much $x$ misses the core and fails to distribute correctly the payments. The players in $S$ may want to leave this market, together with their initial endowments, which decreases the other players' payments by balancedness. 

\medskip

The same interpretation holds for linear production games or flow games. When several players forming a coalition $S$ join the game, they bring raw materials in the case of a linear production games, or new edges in the graph for a flow game. By themselves, they can achieve $v(S)$, but by joining their effort with other players, for instance to have more diversity of raw materials, or a denser graph, with a greater capacity, we can expect that everyone is doing better than before joining, which can be an interpretation of being totally balanced. They cannot accept to have a payment lower than $v(S)$, and the failure at a given preimputation quantifies the cumulative and nested inability of the market to give a proper payment to all the coalitions not satisfied by their payment. 

\begin{proposition}
Let $\calM$ be a market, and $(N, v)$ be its associated game. Let $x$ be a preimputation not included in the core. Then
\[
\mu_v(x) = \min_{\calS \in \rho(x)} \left\lVert \sum_{S \in \calS} \gamma^\calS_S(x) \eta^S \right\rVert = \min_{\calS \in \rho(x)} \left\lVert \sum_{S \in \calS} e(S, x) h^S \right\rVert. 
\]
\end{proposition}

\proof
Because the core is bounded, closed and convex, the projection of $x \in \bbX(v) \setminus C(v)$ lies on the frontier $\partial C(v)$. By the Hilbert projection theorem, the projection is the point minimizing the distance, then 
\[
\mu_v(x) = \lVert x - \pi_{C(v)}(x) \rVert
\]
By the same argument, the projection onto the core coincides with the projection onto a face of the core that is the closest from the projected preimputation. Then, 
\[
\mu_v(x) = \min_{S \in \rho(x)} \lVert x - \pi_{\bbA_\calS}(x) \rVert.
\]
Applying the formula of Theorem \ref{th: main-formula-proj} finishes the proof. 
\endproof

It may be possible to use the orthonormal basis of $\calS$ computed in the iterative procedure computing the Gramian of $\calS$ to let the norm and the sum commute. 

\medskip

Using the projectors to compute a market failure at a given preimputation also gives an explicit side payment to redistribute the global worth without changing it. Moreover, this side payment is the one of smallest norm. We can interpret it as the side payment with the smallest redistribution cost, and therefore the optimal reallocation from $x$ to a core element. Let $\calS$ be the reaching collection defining the core projection. Then the optimal reallocation 
\[
\sigma_{\text{opt}} = \sum_{S \in \calS} \gamma^\calS_S(x) \eta^S
\]
works as follows: for each coalition $S \in \calS$, we collect from each player the worth $\frac{\lvert S \rvert}{n} \gamma^\calS_S(x)$, and give $\gamma^\calS_S(x)$ to each player in $S$.  

\medskip 

All the preimputations belonging to the frontier of the core have a failure of $0$, and we can even extend this function to an element $x$ of the interior of the core by $\mu_v(x) \coloneqq - \min_{y \in \partial C(v)} \lVert x - y \rVert$, and merge these two functions, for all $x \in \bbX(v)$, by 
\[
\mu_v(x) = (-1)^{\bfone_{C(v)}(x)} \min_{y \in \partial C(v)} \lVert x - y \rVert, 
\] 
with $\bfone_{C(v)}$ being the indicator function of the core. These quantities are linked to solutions concepts developed by the same authors, before defining market games. To read more about it, see \textcite[][Chapter 7]{peleg2007introduction}. We denote $\calN_* \coloneqq \calN \setminus \{N\}$. 

\begin{definition}[\textcite{shapley1966quasi}] \leavevmode \newline
Let $\varepsilon$ be a real number. The $\varepsilon$-core of the game $(N, v)$, denoted by $C_\varepsilon(v)$, is defined by 
\[
C_\varepsilon(v) \coloneqq \{x \in \bbX(v) \mid e(S, x) \leq \varepsilon, \hspace{1pt} \forall S \in \calN_* \}. 
\]
\end{definition}

Notice that $C_0(v) = C(v)$. They also defined the smallest of these sets, called the \emph{least-core} and denoted it by $LC(v)$, as the intersection of all nonempty $\varepsilon$-cores of $(N, v)$. It is possible to define the least-core as an $\varepsilon_0$-core, with $\varepsilon_0 = \min_{x \in \bbX(v)} \max_{S \in \calN_*} \, e(S, x)$. Then, we have that 
\[
LC(v) = \arg \min_{x \in \bbX(v)} \max_{S \in \calN_*} \, e(S,x). 
\] 

\begin{definition}
Let $\calM$ be a market, and $(N, v)$ its associated game. We define the \emph{failure core}, denoted by $C_\mu (v)$, by 
\[
C_\mu (v) = \arg \min_{x \in \bbX(v)} \mu_v(x). 
\]
\end{definition}

The projection of an element from the interior of a polytope onto its frontier is the projection onto a facet of the polytope. Indeed, any face of the polytope is the intersection of some facets, and by the triangle inequality, the distance from an intersection of facets is longer than the distance from a facet. Then, the distance between an element $x \in C(v)$ and $\partial C(v)$ is given by $\min_{y \in \partial C(v)} \lVert x - y \rVert = \min_{S \in \calN_*} \, \lVert x - \pi_{\bbA_S}(x) \rVert$. Using the formula of Proposition \ref{prop: first-proj} leads to
\[
\min_{y \in \partial C(v)} \lVert x - y \rVert = \min_{S \in \calN_*} \, \left\lVert \frac{e(S, x)}{\lVert \eta^S \rVert^2} \eta^S \right\rVert = \min_{S \in \calN_*} \, \frac{\lvert e(S,x) \rvert}{\lVert \eta^S \rVert}. 
\]
Because $x$ belongs to the core, its excess is nonpositive, then 
\[
\min_{y \in \partial C(v)} \lVert x - y \rVert = \max_{S \in \calN_*} \frac{e(S,x)}{\lVert \eta^S \rVert} = \max_{S \in \calN_*} \gamma_S(x),
\]
and we can write 
\[
C_\mu (v) = \arg \min_{x \in \bbX(v)} \max_{S \in \calN_*} \, \gamma_S(x), 
\]
which resembles $LC(v)$. Similarly to the least-core, the failure core is never empty. 

\medskip 

The difference between the least-core and the failure core is the rescaling factor $\lVert \eta^S \rVert^{-1}$. It can be interpreted as a coefficient scaling the excess at the individual level, because $\lVert \eta^S \rVert$ only depends on the cardinality of $S$. If the excess of a $5$-player coalition $S$ is $4$, and the excess of a $2$-player coalition $T$ is $2$, despite the fact that the excess for $S$ is bigger than for $T$, the players in $T$ are more aggrieved than players in $S$. The average excess per player in $T$ is $1$, while the average excess per player in $S$ is $\frac{4}{5} < 1$.

\chapter*{Concluding remarks and perspectives} 
\markboth{Concluding remarks and perspectives}{}

\label{Conclu} 

\addchaptertocentry{Concluding remarks and perspectives}




\setlength{\epigraphwidth}{0.7\textwidth}
\epigraph{\emph{The mathematical idea introduced and developed here (that of a ``balanced set'') promises to be of general interest in the study of finite sets, zero-one matrices, and nonadditive set functions.}}{Lloyd S. Shapley, \emph{On balanced sets and cores} \cite{shapley1967balanced}}

In this final chapter, I briefly recapitulate the main results of the thesis and suggest ideas for further research.

\medskip 

First, in Chapter \ref{ChapterA}, I have defined some new notation to do geometry in the space of preimputations. In particular, I introduced the vectors $\{\eta^S \mid S \in \calN\}$ which are very relevant for the geometrical study of cooperative games, as $\eta^S$ represents the direction towards which a translation profits the most to coalition $S$. If we translate the payment vector $x$ by a side payment $\sigma$, then the change of the payment of coalition $S$ is simply $\langle \eta^S, \sigma \rangle$. This new tool is particularly useful for the construction of projection operators in Chapter 4. Moreover, I have presented the similarities between cores of convex games and different polytopes in various areas of study, summarized in Table \ref{table: convex-polytopes}. 

\medskip 

Eventually, using Kalai and Zemel's characterization of a totally balanced games as a flow game, which is the minimum of a finite set of inessential games, I have made a link between tropical polynomials and cooperative games. This is an interesting new perspective, as the literature of tropical mathematics is rapidly increasing, and can influence cooperative game theory the same way polymatroids and submodular set function theory have inspired the development of convex games theory. 

\medskip 

In Chapter \ref{ChapterB}, mainly based on the paper written with my supervisors \cite{laplace2023minimal}, we have shown that minimal balanced collections are a central notion in cooperative game theory, as well as in other areas of discrete mathematics. As a balanced collection is merely the expression of an organization of the players in $N$ during one unit of time, or more generally one unit of resource among subsets, we believe that many more applications should be possible. 

\medskip 

Just focusing on the domain of cooperative games, the consequences of our results appear to be of primary importance for the computability of many notions like exactness, extendability, etc. Indeed, a blind application of the definition of these notions leads to difficult problems related to polyhedra, limiting their practical applicability. Thanks to my results and the new definition of basic polyhedra, provided minimal balanced collections are generated beforehand (which is possible since they do \emph{not} depend on the considered game or property), these notions can be checked very easily and quickly, as most of the required tests reduce to checking simple linear inequalities. 

\medskip

An interesting perspective would be to study how many minimal balanced collections are generated from each of the four steps of our algorithm based on Peleg's method, and especially to know whether a minimal balanced collection is generated more than once. Also, knowing the set of proper minimal balanced collections, i.e., minimal balanced collections not containing two disjoint coalitions, could greatly improve the speed of the algorithm on proper games, i.e., games with superadditive coalition functions. Another interesting approach would be to study the $1$-skeleton of the polytope of minimal balanced collections, to build efficient optimization algorithms on basic polyhedra. The $1$-skeleton of a polytope is the graph defined on the set of vertices as nodes, and edges of the polytope as edges of the graph. 

\medskip 

In Chapter \ref{ChapterC}, I am blazing a new trail in my investigations of balanced collections. Using Joyal's combinatorial species theory \cite{joyal1981theorie}, I built the species of $k$-uniform hypergraphs of size $p$, which can be of interest in the study of hypergraphic networks, extremal combinatorics, mechanism design, etc. Noticing that a regular hypergraph is virtually a balanced collection, this species together with its generating series can be applied to the study of balanced collections. To continue on these tracks, it would be interesting to study the action of permutations on balanced collections in more depth, as well as the concept of \emph{weighted} species, with bivariate generating series, which may allow us to deduce the generating series of regular hypergraphs, and therefore balanced collections, from the one of uniform hypergraphs. 

\medskip 

Later, I have presented the notion of hyperplane arrangements, and how these objects are relevant in the study of cooperative games, seen as distortions of arrangements. A few results about the facial structure of the core, or the feasible regions are stated, but there is still a lot that could be done in this direction. There can be strong connections with the tropical polynomial view of a totally balanced games. 

\medskip 

Finally, in Chapter \ref{ChapterD}, I have shown that generating minimal balanced collections has also permitted to implement an algorithm that tests core stability, i.e., coincidence between the core and a stable set of a given game. The examples provided have shown that, even if for many cases the answer can be obtained quickly, there are instances where the computation time goes beyond tractability. To tackle this, I looked for \emph{blind spots} in the space of preimputations. Indeed, using the minimal balanced collections, it is possible to quickly identify whether some regions will prevent the core to be stable, and then increase the efficiency of my algorithms. An interesting investigation to perform would be to know whether we can find a similar result, but ensuring core domination instead of its impossibility. 

\medskip 

To conclude this thesis, I have presented a set of new tools based on projection operators. I believe that these operators can be interesting by themselves, especially in the application of cooperative game theory, to know which core element is the closest to a current allocation of a given resource, and therefore reallocate it in the most efficient way. The initial ambition of this study of projectors was to build new procedures to check core stability, or at least subprocedures allowing to exclude some regions from the computations. The idea would be to know the direction of the projection, a linear combination of some vectors $\eta^S$, and then check if, for a given preimputation $x$, there would be a coalition $T \in \phi(x)$ which can support the projection, in the sense that the direction of the projection belongs to the domination cone $\delta_T(x)$. This is also one of the exciting perspectives emerging from this doctoral work, which I intend to continue.


\printbibliography[heading=bibintoc]



\appendix 



\end{document}